**A Global Enhanced Vibrational Kinetic Model for Radio-Frequency Hydrogen Discharges and Application to the Simulation of a High Current Negative Hydrogen Ion Source**

by

Sergey Nikolaevich Averkin

A Dissertation

Submitted to the Faculty

of the

WORCESTER POLYTECHNIC INSTITUTE

in partial fulfillment of the requirements for the

Degree of Doctor of Philosophy

in

Aerospace Engineering

by

________________________________

February 27, 2015

APPROVED:

Dr. Nikolaos A. Gatsonis, Advisor
Professor, Aerospace Engineering Program, Mechanical Engineering Department

Dr. John J. Blandino, Committee Member
Associate Professor, Aerospace Engineering Program, Mechanical Engineering Department

Dr. Lynn Olson, Committee Member
Senior Scientist, Busek Co. Inc., Natick, MA

Dr. Seong-kyun Im, Graduate Committee Representative
Assistant Professor, Aerospace Engineering Program, Mechanical Engineering Department

# Abstract

A Global Enhanced Vibrational Kinetic (GEVKM) model is presented for a new High Current Negative Hydrogen Ion Source (HCNHIS) developed by Busek Co. Inc. and Worcester Polytechnic Institute. The HCNHIS consists of a high-pressure radio-frequency discharge (RFD) chamber in which the main production of high-lying vibrational states of the hydrogen molecules occurs and a bypass system. The RFD chamber is connected via a nozzle to a low-pressure negative hydrogen ion production (NIP) region where negative ions are generated by the dissociative attachment of low energy electrons to rovibrationally excited hydrogen molecules. Two configurations of the HCNHIS have been developed, one with the long NIP region (HCNHIS-1) and a second with a short NIP region (HCNHIS-2). Operation of the HCNHIS covered inlet flow rates from 300 to 3000 sccm and absorbed power of 430-600 W. Experiments using Faraday cups downstream the NIP region have shown the saturation of negative current in the presence of an electron filter. Negative currents attributed to negative hydrogen ions of 2-6 mA were measured for the HCNHIS-1 and 4 μA for the HCNHIS-2.

The GEVKM is developed from moment equations for multi-temperature chemically reacting plasmas derived from the Wang Chang-Uhlenbeck equations for a cylindrical geometry of an inductively coupled discharge chamber. The species included into the model are ground state hydrogen atoms $\mathrm{H}$ and molecules $\mathrm{H}_2$, 14 vibrationally excited hydrogen molecules $\mathrm{H}_2(v)$, $v = 1 - 14$, electronically excited hydrogen atoms $\mathrm{H}(2)$, $\mathrm{H}(3)$, ground state positive ions $\mathrm{H}^+$, $\mathrm{H}_2^+$, $\mathrm{H}_3^+$, ground state negative ions $\mathrm{H}^-$, and electrons $\mathrm{e}$. The power deposition is considered to be primarily due to Joule heating of electrons by RF electric field while the stochastic heating is disregarded. The species temperature in the GEVKM is considered to be uniform in the plasma reactor. The spatial variation of the number densities of the plasma components is assumed to follow the product of two one-dimensional distributions corresponding to infinite long cylinder and two infinite plates. Heuristic expressions derived from exact and numerical solutions of the momentum and continuity equations covering low to high-pressure regimes are used in order to relate the wall, centerfield and average number densities of the plasma components. The volume-averaged steady-state continuity equations coupled with the electron energy equation, the total energy equation and heat transfer to the chamber walls are solved simultaneously in order to ob-tain the volume-averaged number densities of the plasma components, the electron and heavy-particle temperatures as well as the wall temperature. The GEVKM is supplemented by a comprehensive set of surface and volumetric chemical processes governing vibrational and ionization kinetics of hydrogen plasmas. The reaction rates when not available are calculated based on the available cross section-data and fitted to analytical expressions. The input conditions to the GEVKM are the inlet flow rate of the feedstock gas, absorbed power, geometry configuration and material properties of the plasma reactor.

The GEVKM is implemented into a robust computational tool written in Fortran 90. It consists of the non-linear algebraic system of equations solver which utilizes the Newton-Raphson method. The GEVKM includes also a solver framework for the continuity and energy equations with self-consistency checks to guarantee conservation of charge, particles and energy in the system.

The GEVKM is verified and validated in the low-pressure (0.2-100 mTorr) and low absorbed power density (0.053-0.32 W/cm$^3$) regime by comparing simulation results with experimental measurements of the low-pressure negative hydrogen ion source DENISE. The electron temperature and number density predicted by the GEVKM agree very well with the Langmuir probe measurements taken in the DENISE source.

In the intermediate to high-pressure regime (1-100 Torr) and high absorbed power density (8.26-22 W/cm$^3$) the GEVKM is verified and validated by comparisons with the numerical simulations and experimental measurements of a microwave generated plasma reactor. The GEVKM predictions of gas temperature are found in good agreement with the measurements.

The GEVKM is applied to the simulation of the RFD chamber of the HCNHIS in both the HCNHIS-1 and HCNHIS-2 configurations. Analytic boundary conditions are developed for the flow in the bypass tubes based on the Fanno theory with the modifications due to rarefactions effects. The analytical boundary conditions are developed for the nozzle flow based on isentropic flow theory with corrections for high Knudsen numbers effects. These analytical outlet boundary conditions are validated by comparison of the pressures predicted by the GEVKM with the pressure measurements of the HCNHIS-1 RFD chamber undertaken at Busek Co. Inc. The GEVKM is used for simulations of the HCNHIS-2 RFD chamber at inlet flow rate of 1000 sccm and absorbed power of 341 W. The GEVKM predictions of negative hydrogen ions number densities and electron temperatures in the RFD chamber of the HCNHIS-2 are used to estimate the nega-tive hydrogen ion current using the Bohm flux approximation. The estimated negative current compares well with the Faraday Cup measurements and provides additional validation of the model. The GEVKM is used in a parametric investigation of the RFD chamber of the HCNHIS-2 with hydrogen inlet flow rates 5-1000 sccm and absorbed powers 200-1000 W. These simulations examine the effects of inlet flow rate and absorbed power on the production and destruction of vibrationally excited hydrogen molecules, the plasma composition, the production and destruction of negative hydrogen ions, the electron and heavy particles temperature, and the maximum extractable negative hydrogen ion current in the RFD chamber. Simulations show that the inlet flow rate has a major impact on the number densities of plasma species in the RFD chamber. The negative hydrogen ions production and the shape of the vibrational distribution function of hydrogen molecules depend on the inlet flow rate. At high flow rates and discharge pressures the distribution function is near Boltzmann while at low flow rates and discharge pressures the distribution is Bray-like distribution. The form of the distribution function affects the production of negative hydrogen ions through dissociative attachment mechanism. In addition, simulations show that the inlet flow rate affects the electron and heavy particles temperatures in a non-monotonic way. This parametric investigation provides the optimum operational parameters which result in the maximum relative concentration of high-lying vibrationally excited hydrogen molecules responsible for the negative hydrogen ion production.

## Acknowledgments

First of all, I would like to express my sincerest gratitude to my advisor Prof. Nikolaos Gatsonis. His continued guidance, support, encouragement, and inspiration provided throughout the whole period of my research allowed me to gain priceless experience. It was a privilege to work with him; his contribution to this work and my education in general will be always appreciated. In addition, I would like to thank Prof. Gatsonis and his wife Maria Kefallinou for their help and concern to my family and me.

Special thanks are given to my committee members for their time and effort spent on my dissertation.

I also would like to thank Siamak Najafi, Randy Robinson, Barbara Edilberti, Barbara Furman, and Donna M. Hughes.

I would like to extend my gratitude to Dr. Raffaele Potami. Our fruitful discussions and his advices led me through my work at my thesis.

My gratitude is extended to Prof. Alexander Fedorov and Prof. Ivan Egorov for giving the start of my career in Moscow Institute of Physics and Technology.

I would also like to thank my friends from the CFPL laboratory. We spent an awesome time together.

The last but not the least I would like to thank my parents, my son, and my wife for their patience, support, and believing in me.

This work has been partially supported through a Subcontract from Busek Co. Inc. with prime sponsor the Phase I and Phase II DOE SBIR Grant No. DE-SC0004199.

My thesis is dedicated to my son, my wife, and my parents.

# Table of Contents

## List of Figures

## List of Tables

# Nomenclature

| | |
|---|---|
| $\mathbf{A}$ | vector potential |
| $A$ | area |
| $A_{\text{eff}}$ | effective area of positive ion losses |
| $a$ | speed of sound |
| $\mathbf{B}$ | magnetic induction vector |
| $B$ | beam brightness |
| $b$ | mobility |
| $C$ | heat capacity per particle |
| $\mathbf{c}_p$ | particle $p$ peculiar velocity vector |
| $c$ | speed of light |
| $\mathbf{D}$ | electric displacement vector |
| $D$ | diffusion coefficient |
| $D_a$ | ambipolar diffusion coefficient |
| $D^T$ | thermal diffusion coefficient |
| df | duty factor |
| $\mathbf{E}$ | electric field vector |
| $\mathbf{E}_a$ | ambipolar electric field vector |
| $e$ | elementary charge |
| $\mathcal{E}$ | beam energy |
| $\mathbf{F}$ | force vector |
| $f$ | velocity distribution function |
| $\mathbf{G}$ | gravity acceleration vector |
| $g$ | relative velocity |
| $\mathbf{H}$ | the magnetic field vector |
| $H$ | enthalpy per particle |
| $h$ | Planck's constant |
| h | heat transfer coefficient |
| $I$ | current; moment of inertia |
| $i = \sqrt{-1}$ | |
| $\mathbf{J}$ | current density |
| $J$ | collision integral |
| $\mathrm{J}_n$ | Bessel function of the order $n$ |
| $\mathrm{Kn}$ | Knudsen number |
| $k_B$ | Boltzmann constant |
| $k_X$ | reaction rate of $X$ reaction |
| $l$ | rotational quantum number |
| $M$ | Mach number |
| $\mathcal{M}$ | molar mass |
| $m_p$ | mass of a particle $p$ |
| $N$ | number of particles |
| $N_s$ | number of species |
| $\mathcal{N}$ | the number of turns of the coil |
| $\mathbf{n}$ | unit normal vector |
| $n$ | number density |
| $\hat{P}$ | pressure tensor |
| $P$ | power |
| $\mathrm{Pr}$ | Prandtl number |
| $p$ | pressure |
| $Q^{(m)}$ | momentum transfer integral |
| $Q^{(\mathrm{e})}$ | energy transfer integral |
| $Q_p$ | volumetric flow rate of species $p$ |
| $\mathbf{q}$ | heat flux vector |
| $\mathrm{Ra}$ | Rayleigh number |
| $\mathrm{Re}$ | Reynolds number |
| $\mathbf{r}$ | radius-vector |
| $T$ | temperature |
| $T_{p,t}$ | effective temperature of species $p$ and $t$ |
| $T_{\mathrm{w}}$ | wall temperature |
| $T_{\infty}$ | ambient temperature |
| $t$ | time |
| $U$ | internal energy per a particle |
| $\mathbf{u}_p$ | species $p$ average velocity |
| $\mathbf{u}$ | mass-averaged velocity |
| $u_{\mathrm{B}}$ | Bohm velocity |
| $V$ | volume; voltage |
| $\mathbf{v}$ | velocity vector |
| $v$ | velocity; vibrational quantum number |
| $\mathbf{w}$ | diffusion velocity vector |
| $w$ | statistical weight |
| $Z_p$ | charge of species $p$ |
| $r, \upsilon, z$ | cylindrical coordinates |
| $x, y, z$ | Cartesian coordinates |
| $\alpha$ | thermal accommodation coefficient; electronegativity |
| $\beta$ | volume coefficient of expansion |
| $\Gamma$ | particle flux |
| $\gamma$ | isentropic coefficient; temperature ratio |
| $\gamma_{\mathrm{rec}}$ | recombination coefficient |
| $\chi_{01}$ | first zero of the zero order Bessel function |
| $\epsilon$ | emittance; permittivity; emissivity; energy |
| $\eta$ | gas efficiency |
| $\kappa$ | thermal conductivity |
| $\lambda$ | mean free path; Debye length |
| $\mu$ | viscosity; permeability |
| $\mu_{pt}$ | reduced mass of species $p$ and $t$ |
| $\nu$ | collision frequency; stoichiometric coefficient |
| $\Xi$ | nondimensional shape factor |
| $\hat{\pi}$ | viscous stress tensor |
| $\rho$ | mass density |
| $\sigma$ | differential cross section; conductivity |
| $\sigma_v$ | tangential momentum accommodation coefficient |
| $\sigma_{\mathrm{SB}}$ | Stefan–Boltzmann constant |
| $\phi$ | scalar potential |
| $\Omega$ | solid angle |
| $\Omega_p^{(\alpha,\beta)}$ | collision integral of species $p$ |
| $\omega$ | frequency; viscosity index in VHS model |
| $\Re(z)$ | real part of complex number $z$ |

# 1 Introduction

Negative hydrogen ion sources (NHIS), which are the primary interest of this work, have been under development with experimental and modeling efforts for over the three decades (Dudnikov, 2012; Bacal, et al., 2005). They produce ion beams that find applications in a wide range of areas including particle accelerators (Schmidt, 1990; Peters, 2000; Moehs, et al., 2005; Welton, et al., 2010), fusion (Hemsworth, et al., 2009), and medicine (Muramatsu & Kitagawa, 2012). The use of negative ions allowed doubling the energy gained by the ion beam in an electrostatic accelerator. Following the idea proposed more than 60 years ago by Alvarez (1951) negative ions first are accelerated from the ground to a positive potential, then the electrons are stripped from the ions by a foil, thus, producing protons which are again accelerated from the positive to ground potential. In this simple scheme the energy gained by protons is twice the energy gained without using negative ions. In addition, negative hydrogen ions are considered to be an essential part in future thermonuclear reactors (Bacal, 2012; Fantz, et al., 2012). It is known that in order to ignite fusion reactions one requires to heat the gas to very high temperatures. Among possible mechanisms of such heating is the neutral beam injection (NBI), where the plasma in the tokomak is bombarded by a neutral beam of extremely high energy. Although ions are much easier to accelerate to such energies they cannot deeply penetrate into the fusion plasma due to the strong magnetic field confining the plasma. As an alternative, positive or negative ion beams are first accelerated to the required energy and then neutralized producing a neutral beam. The benefit of using negative ions compared to protons or other positive ions is that their neutralization efficiency is much higher at high beam energies (Berkner, et al., 1975). Medical application of NHISs includes the use in cyclotrons in order to produce short-lived isotopes for radionuclide imaging with positron emission tomography and single photon emission computer tomography (Muramatsu & Kitagawa, 2012).

In addition to NHISs the hydrogen plasmas of interest are also found in positive ion sources, RF capacitive and microwave reactors, propulsion devices and have diverse applications in microelectronics (Gottscho, et al., 1990; Chang, et al., 1982), aerospace (Jahn, 1968), material processing (Baklanov, et al., 2001), plasma-enhanced chemical vapor deposition (Hassouni, et al., 1999).

The goal of this dissertation is to develop a comprehensive theoretical and computational model of the chemically reacting plasmadynamics processes in a High Current Negative Hydrogen Ion Source (HCNHIS) shown in Figure 1 under development of a new NHIS developed by Busek Co. Inc. and Worcester Polytechnic Institute (WPI). The primary application of this HCNHIS is in particle accelerators such as the Spallation Neutron Source (SPS) in Oak Ridge National Laboratory.

In this introductory chapter an overview of positive and negative ion sources is presented first, followed by an overview of mechanisms of $H^-$ production and destruction and corresponding NHIS designs used in particle accelerators. Theoretical modeling and numerical approaches used in simulation of the reacting plasmadynamics is presented. The operational concept and characteristics of the new HCNHIS under development is presented and the chapter concludes with the objectives, methods and approaches of this work.

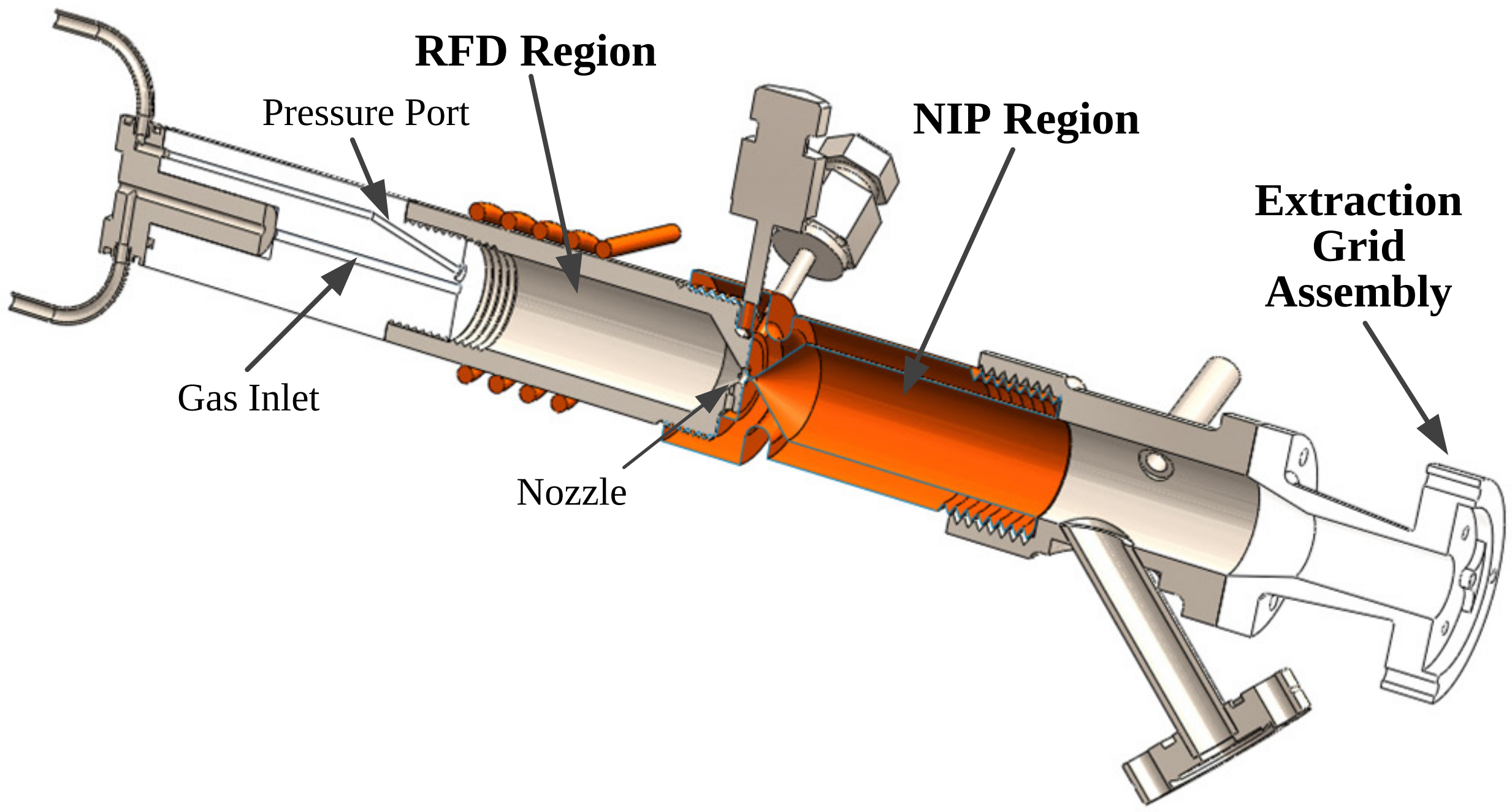

**Figure 1. High Current Negative Hydrogen Ion Source developed by Busek Co. Inc. and Worcester Polytechnic Institute.**

## 1.1 Overview of Positive and Negative Ion Sources Basic Designs and Characteristics

A negative or positive ion source (or specifically plasma ion source) in most cases consists of a plasma generator where charged species are formed and an electrode system called extractor which creates an ion beam with significant directed energy (Brown, 2004; Reiser, 2008). A schematic of a positive ion source with the plasma generated by a radio-frequency discharge is shown in Figure 2(a). The plasma can be produced in the plasma source by radio-frequency or electron cyclotron resonance discharge, hot filaments or any other technique. The voltage to which plasma is biased is called the "extraction voltage" and can reach very high values. The extractor can contain a number of grids. The closest to the plasma is the plasma electrode and usually it is biased to the plasma potential in order to allow acceleration of ions outside the plasma source. There are ion sources that utilize only one electrode (Brown, 2004). In this case, the single electrode is grounded and acceleration of ions takes place inside the electrode sheath within the plasma. The last electrode in a multi-electrode extraction system, such as shown in Figure 2 is the ground electrode. In a simple case shown in Figure 2 the intermediate electrode is the suppression electrode and it is biased to a slightly negative voltage in order to reflect electrons. The extraction system shown in Figure 2(a) is a type of accelerator-decelerator system (Brown, 2004).

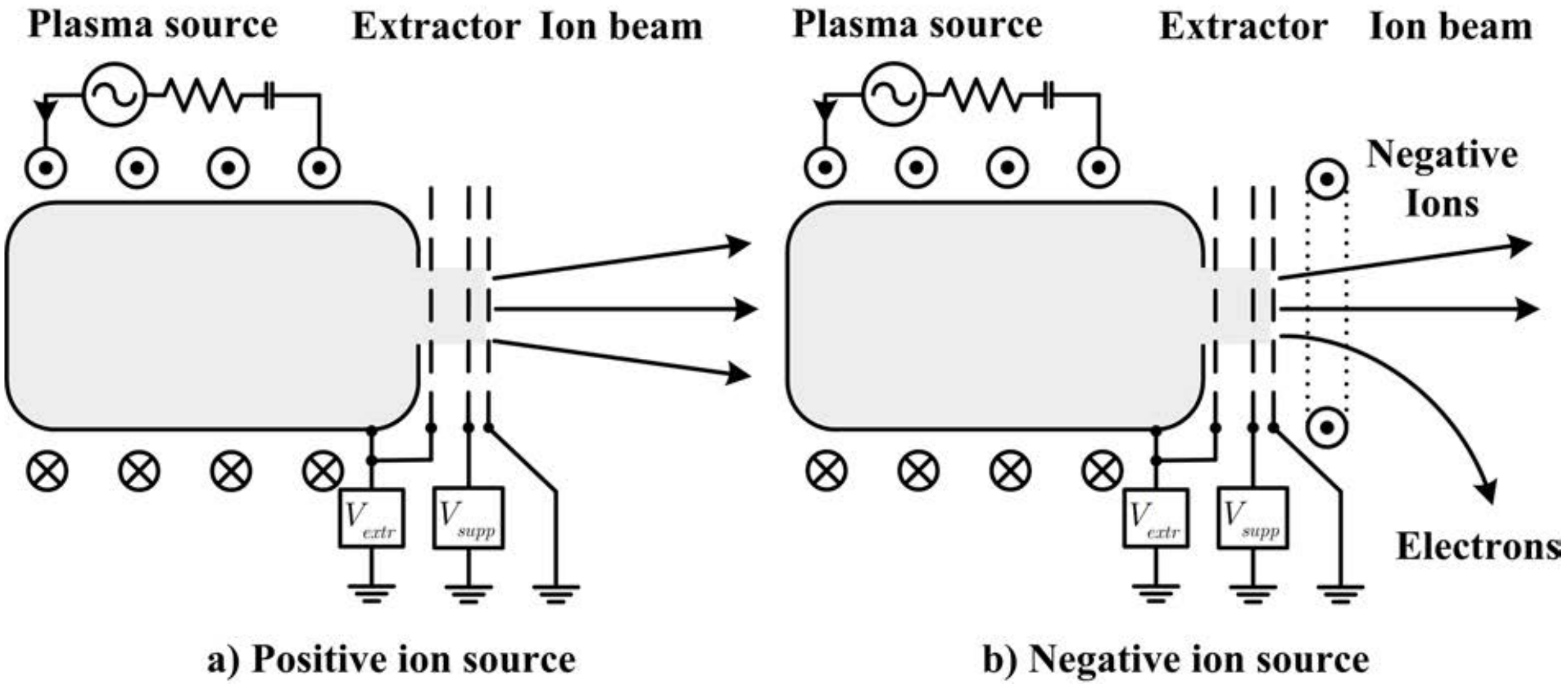

**Figure 2. Schematic of a positive (a) and negative (b) ion source.**

Negative ion sources have some design differences from positive ion sources due to the nature of extracted ions and are shown schematically in Figure 2(b). First, in contrast to positive ion sources the plasma is negatively biased in order to extract negative ions. Second, the plasma consists of three sorts of charged particles: electrons, positive and negative ions. Therefore in addition to electrodes, magnetic filters are often used in order to suppress extracted electrons.

The positive or negative beam properties such as beam current, energy, emittance, and brightness produced by a source depend on the characteristics of the plasma in the chamber and on the extraction system. A brief description of the most important parameters encountered in ion sources below follows the discussion found in the classical text on ion sources by Zhang (1999). The first group of characteristics describes the performance of the extracted beam and includes the beam current, $I$, and the beam energy, $\mathcal{E}$.

- The beam current, $I$, and the beam current density, $\mathbf{J}$, are related by

$$I = \iint_A \mathbf{J} \cdot d\mathbf{A}, \tag{1.1}$$

where $A$ is the area of the ion emission surface. Assuming uniform current density the $\mathrm{H}^-$ current through a circular aperture of radius $R$ from Eq. (1.1) can be written as

$$I_{\mathrm{H}^-} = J_{\mathrm{H}^-} \pi R^2. \tag{1.2}$$

In the volume NHIS of Bacal (2005) extracted through the circular aperture it was experimentally found that the current density could be represented by a thermal flux of $\mathrm{H}^-$ with mass $m_{\mathrm{H}^-}$, temperature $T_{\mathrm{H}^-}$ and number density $n_{\mathrm{H}^-}$ resulting in the following expression for the negative hydrogen ion current

$$I_{\mathrm{H}^-} = n_{\mathrm{H}^-} e \pi R^2 \sqrt{\frac{k_B T_{\mathrm{H}^-}}{2\pi m_{\mathrm{H}^-}}}. \tag{1.3}$$

Depending on the steady or pulsed operation of the ion source the average of peak currents can be used to characterize it. For high energy ion sources typically found in particle accelerators the plasma potential $V_{\mathrm{pl}}$ can be neglected compared to the extraction voltage $V_{\mathrm{extr}}$. In such a case, the beam energy is given by the charge of the extracted ions, $Z$, only

$$\mathcal{E} = eZV_{\mathrm{extr}}. \tag{1.4}$$

- The beam emittance, $\epsilon$, and brightness, $B$, are used to describe the optical properties of ion beams.

Various definitions for the beam emittance are in use (Zhang, 1999; Reiser, 2008; Brown, 2004). In this work we follow Brown (2004). Assuming that the ion beam propagates in the $z$-direction the transverse, laboratory beam emittance in the four-dimensional trace space, $\epsilon_{4,L}$, is determined by

$$\epsilon_{4,L} = \frac{A(x,x',y,y')}{\pi^2} = \frac{1}{\pi^2}\iiiint dxdx'dydy' , \tag{1.5}$$

where $A'$ is the beam four –dimensional area in the transverse direction, $p_x$, $p_y$, $p_z$ are the components of the momentum vector, and $x$, $y$, $x' = p_x/p_z$, $y' = p_y/p_z$ are spatial and angle coordinates transverse to the beam direction. Similar parameters are introduced for two-dimensional trace spaces $(x,x')$ and $(y,y')$ provided that the motion is independent for $x$ and $y$ directions

$$\begin{aligned} \epsilon_{x,L} &= \frac{A(x,x')}{\pi} = \frac{1}{\pi}\iint dxdx', \\ \epsilon_{y,L} &= \frac{A(y,y')}{\pi} = \frac{1}{\pi}\iint dydy'. \end{aligned} \tag{1.6}$$

According to Liouville's theorem in absence of non-conservative forces the six-dimensional particle distribution of negative ions is conserved (Brown, 2004). Thus, the emittance can be used as a characteristic of ion beams.

From the definition (1.5) the emittance depends on the energy of the extracted ion beam. When the energy is increased, the ratios $p_x/p_z$ and $p_y/p_z$ are decreased.

In order to compare sources with different energies the normalized emittance, $\epsilon_n$, is also introduced as

$$\epsilon_n = \beta\gamma\epsilon_L , \tag{1.7}$$

where $\beta = v_{\text{beam}}/c$ and $\gamma = 1/\sqrt{1-\beta^2}$ , $v_{\text{beam}}$ is the beam velocity, $c$ is the speed of light. In cases where relativistic effects are negligible $\gamma \approx 1$ and the definition of normalized emittance reduces to

$$\epsilon_n = \beta\epsilon_L . \tag{1.8}$$

In order to compare emittances based on different distributions of the particles in ion beams, the root mean square (rms) emittance for the two dimensional subspace is introduced (Brown, 2004)

$$\epsilon_{x,rms} = \sqrt{\left\langle x \right\rangle \left\langle x' \right\rangle - \left\langle xx' \right\rangle}\,. \tag{1.9}$$

In the above equation the angular brackets represent the averaging over the particle distribution function in $(x, x')$ space. The root mean square emittance can be used to measure the quality of the ion beams (Reiser, 2008).

Sometimes the $\epsilon_{x,90\%}$ emittance is introduced. In this case the integration in Eq. (1.6) is carried out for 90% of the contour containing the brightest beam. For a Gaussian distribution there is a simple conversion formula between the rms and the 90% emittance (Peters, 2000)

$$\epsilon_{90\%} = 4.6\epsilon_{\mathrm{rms}}\,. \tag{1.10}$$

The normalized rms emittance for a negative hydrogen ion beam extracted from a volume production ion source through an aperture of radius $R$ is given by (Reiser, 2008)

$$\epsilon_{n,\mathrm{rms}} = \frac{R}{2}\sqrt{\frac{k_B T_{\mathrm{H}^-}}{m_{\mathrm{H}^-} c^2}}\,. \tag{1.11}$$

This formula shows the importance of keeping negative hydrogen ions cold in volume ion sources in order to achieve low emittance ion beams. In contrast, for $\mathrm{H}^-$ beams in surface NHIS, extracted from the converter surface with the radius of the emission aperture $R$, converter surface radius $r_c$ and the distance between converter and aperture $d$, the rms emittance depends on the voltage of the converter $V_c$ following Zhang (1999)

$$\epsilon_{n,\mathrm{rms}} = \frac{R r_c}{d}\sqrt{\frac{2eV_c}{m_{\mathrm{H}^-} c^2}}\,. \tag{1.12}$$

- Beam brightness.

The beam emittance is not enough in order to describe the quality of the ion beam. From Eq. (1.11) and Eq. (1.12) it is seen that the emittance can be easily decreased by shrinking the size of the aperture. But from Eq. (1.2) the extracted current is also decreased. Therefore in order to characterize emittance at a given current the brightness, $B$, is introduced. As in the case of emittance there are different definitions of beam brightness (Reiser, 2008).

The brightness of the beam is defined as a mean current density in the transverse phase space for a given energy following Zhang (1999) as

$$B = K\frac{I}{\epsilon_4}. \tag{1.13}$$

In the above equation $K$ is a constant which has different values in different definitions of brightness (Reiser, 2008). In order to compare brightness of ion sources with different energy, the normalized brightness, $B_n$, is introduced as

$$B_{4,n} = K\frac{I}{\epsilon_{4,n}} = \frac{B}{(\beta\gamma)^2}. \tag{1.14}$$

- The gas efficiency, $\eta_g$, of an ion source depends on the plasma source parameters and is defined as the fraction of atoms contained in a desirable ion beam of charge $Z$ to the total consumed number of atoms, $Q_g$. It is given following Zhang (1999) as

$$\eta_g = \frac{I}{eZQ_g}. \tag{1.15}$$

- The ionized power efficiency, $H$, of the ion source is defined as the ratio of the total beam current to the consumed discharge power, $P$, (Zhang, 1999)

$$H = \frac{I}{P}. \tag{1.16}$$

- One of the operational performance parameters of an ion source is its lifetime. It is defined as a time at which the ion source can operate without breakage or maintenance. This characteristic is very important for NHIS dedicated for fusion due to hazardous environments in fusion reactors (Fantz, et al., 2012).
- Duty factor, repetition rate and pulse length.

In order to characterize the performance of pulsed ion sources the duty factor, $\mathrm{df}$, repetition rate, $f$, and pulse length, $t_p$, are introduced. Duty factor specifies the fraction of the operation time of an ion source in percent when the ion source produces ions. Repetition rate and pulse length represent the frequency and the duration of ion pulses. These three parameters are not independent but related to each other through

$$\mathrm{df} = ft_p \times 100\%. \tag{1.17}$$

Additional performance characteristics and parameters used to describe ion sources can be found in Zhang (1999), Brown (2004), and Reiser (2008).

The performance requirements of NHISs vary depending on the intended application. Some of the NHIS requirements related to future particle accelerators are presented in Table 1.

**Table 1. Desired beam parameters of future accelerators adapted from Moehs et al. (2005).**

| Accelerator name | $I_{H^-}$ (mA) | Pulse length (ms) | Repetition rate (Hz) | Duty factor (%) | $\epsilon_{n,rms}\ {}^{x}\!/\!_{y}$ (π mm-mrad) | References |
|---|---|---|---|---|---|---|
| ESS | 70 | 1.2 | 50 | 6 | 0.3 | Gardner et al. (1997) |
| CERN-SPL Phase2 | 30 | 2.8 | 50 | 14 | 0.2 | Gerigk and Vretenar (2002) |
| SNS | 50 | 1 | 60 | 6 | 0.2 | Keller et al. (2002) |
| J-PARC | 50 | 0.5 | 50 | 2.5 | 0.3 | Naito (2002) |
| FNAL Proton Driver | 12/33 | 1/3 | 10/2.5 | 1/0.7 | 0.2 | Foster et al. (2002) |
| LANSCE Upgrade | 40 | 1.0 | 120 | 12 | 0.13/0.13 | Stevens and Fitzgerald (1997) |

## 1.2 Mechanisms of Negative Hydrogen Ion Production and Destruction

There are several mechanisms of negative hydrogen ions formation. Dissociative attachment of electrons to rovibrationally excited hydrogen molecules (DEA) is one of the most efficient volume reactions (Bacal, et al., 2005)

$$\mathrm{e} + \mathrm{H}_2\left(X^1\Sigma_g^+, v\right) \rightarrow \mathrm{H}_2^- \rightarrow \mathrm{H} + \mathrm{H}^- . \tag{1.18}$$

Since its discovery in the late seventies this process was a subject of both theoretical (Wadehra & Bardsley, 1978; Horáček, et al., 2006) and experimental investigations (Allan & Wong , 1978). Recently it was confirmed that this process is one of the dominant volume reac-tion producing negative hydrogen ions (Mosbach, 2005a), (Mosbach, 2005b). The cross sections of electron attachment to vibrational levels $v \geq 5$ are many orders of magnitudes higher than for the first four vibrational states. Therefore, the production of negative ions through this channel should be considered in conjunction with the processes involving excitation and quenching of vibrationally excited hydrogen molecules. The full list of reactions is available in dedicated textbooks and review papers (Janev, et al., 1987; Janev, et al., 2003; Clark & Reiter, 2005; Itikawa, 2007; Yoon, et al., 2008; Fridman, 2008) and they will be outlined below for clarity. The abbreviations of the reaction names used in literature are shown in parentheses (Capitelli & Gorse, 2005; Capitelli, et al., 2011; Bacal, 2012).

1) Electron-vibration energy transfer of low energy electron through $\mathrm{H}_2^-$ (eV process)

$$\mathrm{e} + \mathrm{H}_2\left(X^1\Sigma_g^+, v\right) \to \mathrm{H}_2^- \to \mathrm{e} + \mathrm{H}_2\left(X^1\Sigma_g^+, v'\right). \tag{1.19}$$

2) Electron-vibration energy transfer of high energy electron through the excitation of singlet states and radiative decay (EV)

$$\mathrm{e} + \mathrm{H}_2\left(X^1\Sigma_g^+, v\right) \to \mathrm{H}_2^*\left(\text{singlets}\right) \to \mathrm{e} + \mathrm{H}_2\left(X^1\Sigma_g^+, v'\right) + h\nu\,. \tag{1.20}$$

3) Dissociation of vibrationally excited hydrogen molecule by electron impact (eD)

$$\mathrm{e} + \mathrm{H}_2\left(X^1\Sigma_g^+, v\right) \to \mathrm{e} + 2\mathrm{H}\,. \tag{1.21}$$

4) Ionization of vibrationally excited hydrogen molecule by electron impact (eI)

$$\mathrm{e} + \mathrm{H}_2\left(X^1\Sigma_g^+, v\right) \to 2\mathrm{e} + \mathrm{H}_2^+\,. \tag{1.22}$$

5) Dissociative ionization of vibrationally excited hydrogen molecule by electron impact (eIdiss)

$$\mathrm{e} + \mathrm{H}_2\left(X^1\Sigma_g^+, v\right) \to 2\mathrm{e} + \mathrm{H} + \mathrm{H}^+\,. \tag{1.23}$$

6) Vibration-vibration energy transfer between two vibrationally excited molecules (VV)

$$\mathrm{H}_2\left(X, v\right) + \mathrm{H}_2\left(X, w\right) \to \mathrm{H}_2\left(X, v'\right) + \mathrm{H}_2\left(X, w'\right). \tag{1.24}$$

Usually only single state transitions are considered in numerical models since multi-quantum transitions have much smaller cross sections (Fridman, 2008).

7) Vibration-translation energy transfer by molecule collisions (VTm)

$$\mathrm{H}_2\left(X^1\Sigma_g^+, v\right) + \mathrm{H}_2 \to \mathrm{H}_2\left(X^1\Sigma_g^+, v'\right) + \mathrm{H}_2\,. \tag{1.25}$$

8) Vibration-translation energy transfer by atomic collisions (VTa)

$$\mathrm{H_2}\left(X^1\Sigma_g^+, v\right) + \mathrm{H} \rightarrow \mathrm{H_2}\left(X^1\Sigma_g^+, v'\right) + \mathrm{H}\,. \tag{1.26}$$

9) Dissociation of vibrationally excited hydrogen molecules by molecule impact (Dm)

$$\mathrm{H_2}\left(X^1\Sigma_g^+, v\right) + \mathrm{H_2} \rightarrow 2\mathrm{H} + \mathrm{H_2}\,. \tag{1.27}$$

10) Dissociation of vibrationally excited hydrogen molecules by atom impact (Da)

$$\mathrm{H_2}\left(X^1\Sigma_g^+, v, l\right) + \mathrm{H} \rightarrow 3\mathrm{H}\,. \tag{1.28}$$

11) Ion-electron recombination

$$\mathrm{H_3^+} + \mathrm{e} \rightarrow \mathrm{H_2}(v > 5) + \mathrm{H}(n = 2)\,. \tag{1.29}$$

12) Hydrogen atom recombinative desorption on the wall of the plasma source (RD)

$$\mathrm{H} + \mathrm{H} \xrightarrow{\text{wall}} \mathrm{H_2}\left(X^1\Sigma_g^+, v\right). \tag{1.30}$$

13) Ion recombination on the walls

$$\begin{aligned} &\mathrm{H_2^+} \xrightarrow{\text{wall}} \mathrm{H_2}(v, l), \\ &\mathrm{H_3^+} \xrightarrow{\text{wall}} \mathrm{H_2}(v, l) + \mathrm{H}. \end{aligned} \tag{1.31}$$

14) Vibrational and rotational relaxation in inelastic collisions of vibrationally excited hydrogen molecules with walls

$$\mathrm{H_2}(v, l) \xrightarrow{\text{wall}} \mathrm{H_2}(v', l')\,. \tag{1.32}$$

15) Dissociative chemisorption of rovibrationally excited hydrogen molecules on the walls

$$\mathrm{H_2}(v, l) \xrightarrow{\text{wall}} \left(\mathrm{H_2^*}\right)_{\text{ads}} \rightarrow \mathrm{H} + \mathrm{H}\,. \tag{1.33}$$

16) Hydrogen atom recombination in the volume of the plasma source in the presence of third body

$$M + \mathrm{H} + \mathrm{H} \rightarrow M + \mathrm{H_2}\left(X^1\Sigma_g^+, v\right), \tag{1.34}$$

where $M$ could be electron, atom or molecule.

The above reactions play an important role in the establishment of the vibrational distribution function (VDF) of molecular hydrogen. It was found first theoretically (Capitelli, et al., 1991; Fukumasa, et al., 1992; Berlemont, et al., 1993) and then confirmed experimentally (Mosbach, et al., 2000; Mosbach, et al., 2002) that VDF has a plateau for intermediate vibrational levels $(7 \lesssim v \lesssim 11)$ which primarily produce negative hydrogen ions, thereby deviating from the equilibrium Boltzmann distribution. Thus, in order to take into account in the numerical simulation the $\mathrm{H}^-$ formation due to DEA mechanism it is required to know the population of vibrationally excited states. Therefore vibrational kinetics is an essential part of a theoretical model which includes the volume production of negative hydrogen ions.

Another volume production mechanism of $\mathrm{H}^-$ is the electron dissociative attachment to high-lying Rydberg states of hydrogen molecules

$$\mathrm{e} + \mathrm{H}_2^*\left(HR\right) \rightarrow \mathrm{H}^- + \mathrm{H}(n=2). \tag{1.35}$$

This process was proposed based on the experimental results of $\mathrm{H}^-$ production in laser-exited hydrogen molecules (Pinnaduwage & Christophorou, 1993; Datskos, et al., 1997). This channel was used in kinetic models (Garscadden & Nagpal, 1995; Hassouni, et al., 1998; Hiskes, 1996). The first two simulations showed that the enhancement in $\mathrm{H}^-$ current predictions due to inclusion of this mechanism could be comparable to the current obtained by considering DEA mechanism only or could even outperform it. On the other hand Hiskes concluded that this process doesn't play a significant role compared to dissociative electron attachment to rovibrationally excited hydrogen molecules. Nowadays this mechanism is sometimes considered equally with DEA as a main volume production channel (Dugar-Zhabon, et al., 2013). Nevertheless, since there is no unique opinion on the importance of this process further experimental and theoretical investigations are required (Bacal, 2006).

Another mechanism is the $\mathrm{H}^-$ production by surface effects. In this process, a hydrogen atom or an ion upon striking a surface with a low work function may leave it as a negative ion. The physical mechanism of $\mathrm{H}^-$ formation in this process is an electron shift from the Fermi level of a metal surface to the affinity level of a hydrogen atom by tunneling. Theoretical calculations gave the following formula for the negative hydrogen ion production probability, $P_{\mathrm{H}^-}$, on a surface with the work function, $\phi$, (Rasser, et al., 1982)

$$P_{\mathrm{H}^-} = \frac{2}{\pi} e^{-\frac{\pi\left(\phi - E_a\right)}{2 a v_\perp}}, \tag{1.36}$$

where $E_a$ is the electron affinity for negative ion, $a$ is the exponential decay constant, and $v_\perp$ is the velocity normal to the surface of an emitted ion. The above equation shows strong dependence of the $H^-$ formation on the surface work function. Therefore the surfaces of some of NHIS relying upon this mechanism are covered by alkali or alkaline earth metals in order to lower the work function, such as cesium, which has the lowest work function among all elements (Brown, 2004). The maximum yield of $H^-$ is achieved when the surface is covered by a specific thickness of alkali metal that depends both on the surface and coating materials (Graham, 1980). For $Cs / W$ the coverage of about 0.6 monolayers provides minimum work function of the surface (Brown, 2004). Historically the application of this mechanism along with the use of improved sources geometries in Novosibirsk in early seventies made it possible to achieve very high currents of negative hydrogen ions (Dudnikov, 2012).

Furthermore Dudnikov (2005) proposed that the surface production of negative hydrogen ions is the dominant process in some of the efficient ion sources (high brightness, high $I_{H^-}$ and low ratio of $I_e / I_{H^-}$) in which the formation of $H^-$ was thought to be due to DEA mechanism. The process of surface production in these sources was enhanced due to presence of alkali metal impurities which effectively lowered the work function. The increased performance of some NHISs operating with internal RF antennas coated with enamel containing potassium was explained in such way (Bacal, et al., 2005). However, Bacal (2006) pointed out that the experiments on some of the ion sources relying on volume production did not reveal the presence of these impurities. In addition, the reduced electron number density in the region close to the extraction grids was explained by DEA to vibrationally excited hydrogen molecules produced by RD reaction on the plasma electrode and the walls of the source.

New volume mechanism was proposed by Vogel (2013) where $H^-$ are produced by collisions of electronically excited hydrogen atoms. However, the importance of this mechanism is not clear and additional investigations are required.

It was observed that addition of some admixtures could substantially increase the yield of negative hydrogen ions or decrease the current of coextracted electrons. Cesium is among such substances. Increased negative hydrogen ion current in this case could not be explained by the surface production of $H^-$ ions alone (Bacal, 2012). The main mechanism is still not clear (Bacal, 2006).

There are some other possible channels (Zhang, 1999), which have rather small cross sections:

1. Polar dissociation

$$e + H_2 \rightarrow H^- + H^+ . \tag{1.37}$$

2. Dissociative recombination of hydrogen ions with electrons

$$\begin{aligned} e + H_2^+ &\rightarrow H^- + H^+, \\ e + H_3^+ &\rightarrow H^- + H_2^+. \end{aligned} \tag{1.38}$$

3. Radiative capture

$$e + H \rightarrow H^- + h\nu \, . \tag{1.39}$$

The number density of negative hydrogen ions and, hence, the extracted current depends not only on the production reactions but is also determined by the destruction processes. In other words it is important not only to create negative hydrogen ions but also to keep them in the plasma and finally to extract. Since the electron affinity of hydrogen atom is only 0.7542 eV (Hellborg, 2005) the electron detachment in collisions with other particles has high probability. An outline of some destruction channels of $H^-$ with their abbreviated names in parentheses are given below (Janev, et al., 1987; Janev, et al., 2003; Clark & Reiter, 2005; Bacal, 2012).

1) Mutual neutralization of negative and positive hydrogen ions (MN)

$$\begin{aligned} H^- + H^+ &\rightarrow H(n) + H(1s), \\ H^- + H_2^+ &\rightarrow H_2 + H, \\ H^- + H_2^+ &\rightarrow 3H, \\ H^- + H_3^+ &\rightarrow H_2 + 2H, \\ H^- + H_3^+ &\rightarrow 4H. \end{aligned} \tag{1.40}$$

2) Electron detachment in collisions with positive ions

$$\begin{aligned} H^- + H^+ &\rightarrow H^+ + H + e, \\ H^- + H_2^+ &\rightarrow H_3^+ + e, \\ H^- + H_3^+ &\rightarrow H_2^+ + H_2 + e, \\ H^- + H_3^+ &\rightarrow H_3^+ + H + e. \end{aligned} \tag{1.41}$$

3) Electron detachment in collisions with hydrogen atoms and molecules

$$\begin{aligned} &H^- + H \rightarrow H + H + e, \\ &H^- + H_2 \rightarrow H + H_2 + e, \\ &H^- + H_2 \rightarrow H^+ + H_2 + 2e. \end{aligned} \tag{1.42}$$

4) Associative detachment of the electron in collisions with hydrogen atom (AD)

$$H^- + H \rightarrow H_2(v) + e\,. \tag{1.43}$$

5) Electron detachment in collisions of energetic electrons with negative hydrogen ions (ED)

$$H^- + e \rightarrow H + e + e\,. \tag{1.44}$$

6) Electron detachment in collisions of negative hydrogen ions with vibrationally excited hydrogen molecules (Dem'yanov, et al., 1985)

$$H^- + H_2(v) \rightarrow H + H_2(v-2) + e\,. \tag{1.45}$$

Another process which plays an important role in NHIS operation but does not lead to a destruction of negative ions is the charge exchange reaction,

$$H^-_{\text{fast}} + H_{\text{slow}} \rightarrow H_{\text{fast}} + H^-_{\text{slow}}\,. \tag{1.46}$$

This is an important process because created slow negative ions are useless for accelerator applications due to their low energies. From this point of view the charge exchange reaction has the same effect as previously described destruction processes.

The destruction processes take place inside the ion source and along the path of the ion beam. Therefore it is important not only to provide necessary conditions for $H^-$ production inside the ion source but also to guarantee that negative ions will survive outside of the source. Let’s outline the importance of these reactions and their implications on ion source design.

The significance of a particular destruction reaction in the ion source depends on the conditions there. NHIS are developed in order to maximize production of $H^-$ ions and to minimize their destruction. For example, if the ion source relies on the volume production through DEA channel then it requires both high population of vibrationally excited molecules and low energy electrons. Usually the production of high vibrational levels of hydrogen molecules is attributed to EV mechanism (Zhang, 1999). Meanwhile the demolition of $H^-$ ions by ED reaction is very effective at high electron energies. Therefore, there should be a balance between the production of vibrationally excited molecules and the destruction of negative hydrogen ions due to collisions with these fast electrons. In order to achieve this goal the production region of vibrationally excited molecules is separated in time or in space from the zone where DEA reaction takes place. The space separation of these regions was the first proposed scheme for negative hydrogen ion source relying on DEA reaction (Leung, et al., 1983). In order to separate two regions magnetic field called magnetic filter was used. The time separation could be achieved by using pulsed discharges (Hopkins & Mellon, 1991). In the first phase of the pulse the fast electrons populate high-lying vibrational states of hydrogen molecules. During the afterglow phase the electrons are rapidly cooled and thus are favorable for negative hydrogen ions creation through DEA mechanism and are not effective in the $\mathrm{H}^-$ destruction. Then the pulse is repeated again.

Another important criterion is keeping low pressure in the beam while it is propagating from the source to accelerator or any other facility where it should be utilized. In this case the negative ions are accelerated to very high energies by electric filed and collisions with neutral particles can easily destroy them. In addition, the pressures in the ion source should be as low as possible in order to prevent electron stripping from negative ions. This is also a reason why low electron to negative ion current ratio is required for the $\mathrm{H}^-$ ion beam. Otherwise the ions will be lost very quickly by collisions with electrons. In order to illustrate the relative contributions of the negative hydrogen ion demolition the negative hydrogen ion destruction mean free paths are evaluated. The destruction mean free path of negative hydrogen ions in a particular collision with particles of species $s$ assuming that all the particles have the same drift velocity can be expressed as (Bird, 1994)

$$\lambda_s = \frac{\left\langle v_{\mathrm{H}^-} \right\rangle}{\left\langle \sigma_{T,s}(g_r) g_r \right\rangle n_s}, \tag{1.47}$$

where $\sigma_{T,s}$ is the sum of the total cross sections of all demolition reactions of $\mathrm{H}^-$ and particles of species $s$, $g_r$ is the relative velocity, and $n_s$ is the number density of species $s$, $\left\langle v_{\mathrm{H}^-} \right\rangle$ is the mean thermal speed of the negative hydrogen ions. For Maxwellian distribution function the simplified formula reads as (Appendix A)

$$\lambda_s = \frac{1}{\sqrt{1 + \frac{m_{\mathrm{H}^-}}{m_s}} n_s \int\limits_0^{+\infty} \sigma_{T,s}\left(\sqrt{\frac{2k_B T}{\mu}}\xi\right) \xi^3 e^{-\xi^2} d\xi}. \tag{1.48}$$

Figure 3 shows the destruction mean free paths of $H^-$ for different species, which frequently encountered in hydrogen plasma in the interior of the hydrogen plasma reactors, as a function of their typical concentrations at different temperatures. The values for the cross sections and reaction rates are taken from Hjartarson (2010), Janev et al. (1987), Janev et al. (2003). These mean free paths could be used in order to estimate beam attenuation. The flux density of the $H^-$ beam can be expressed as

$$\Gamma_{H^-}(x) = \Gamma_{H^-}(x=0)e^{-\frac{x}{\lambda_{ds}}}. \tag{1.49}$$

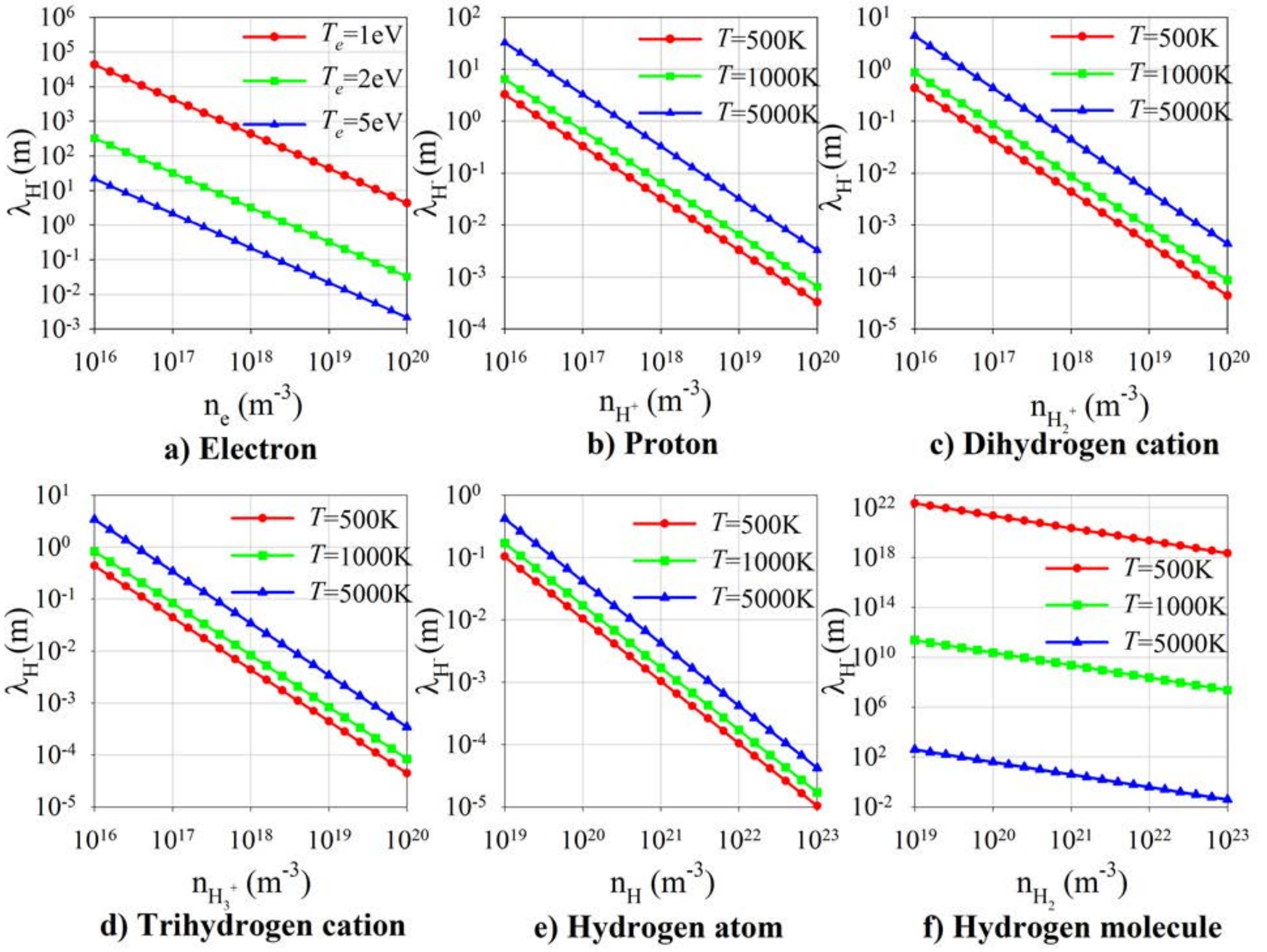

**Figure 3. Comparison of hydrogen anion destruction mean free paths for various plasma species at different temperatures as a function of the plasma species number densities.**

For example, if one requires that $95\%$ of negative hydrogen ion beam would travel from the ion source to an accelerator 1m then from Eq. (1.49) $\lambda_{H^-}$ should be more than $9.75\mathrm{m}$. If on-ly the negative hydrogen ions, neutral particles and electrons escape from the NHIS with temperatures $T_e = 5eV$, $T_{\mathrm{H}^-} = T_{\mathrm{H}_2} = T_{\mathrm{H}} = 1000K$, then according to Figure 3 the number densities should be at most $n_e \leq 3.08 \times 10^{16} m^{-3}, n_{\mathrm{H}} \leq 1.67 \times 10^{17} m^{-3}$ or in terms of pressure if the atomic hydrogen is a result of dissociation and degree of dissociation for molecular hydrogen is assumed to be 1% then $p \leq 0.25Pa$. The above simple estimation shows why most of negative hydrogen ions sources have a strong requirement for operating at very low pressure near the extraction region. For instance, NHIS for the International Thermonuclear Experimental Reactor (ITER) requires the pressure to be below $0.3Pa$ (Fantz, et al., 2012). In addition, it illustrates the destruction efficiency of atomic hydrogen. Meantime the molecular hydrogen doesn't strip electrons from $\mathrm{H}^-$ so effective unless there exist large relative velocities between these species as inside ion beams. It is expected that such a regime will be encountered in the NIP region of the HCNHIS.

## 1.3 Overview of Negative Hydrogen Ion Sources

The review of NHISs shows that designs are driven by the $\mathrm{H}^-$ production mechanism considered as the dominant. The NHISs can be categorized into two large groups based on major mechanism of the $\mathrm{H}^-$ production. The first group constitutes Surface Production Sources (SPS) with $\mathrm{H}^-$ production to be dominated by cesiated surfaces. The second group constitutes Volume Production Sources (VPS) with $\mathrm{H}^-$ be dominated by volumetric processes in particular dissociative electron attachment to rovibrationally excited hydrogen molecules. In most cases, surface and volume mechanisms work together in producing $\mathrm{H}^-$ or in lowering the current of coextracted electrons (Bacal, et al., 2005). A brief overview of NHISs based on volume production is presented following Zhang (1999). For more details about different NHIS developed in the past the interested reader is referred to Prelec and Sluyters (1973), Schmidt (1990), Peters (2000), Welton (2002), Bacal et al. (2005), Moehs et al. (2005), Dudnikov and Johnson (2010), Keller (2010).

1. Duoplasmatron negative ion source.

This source developed in the '60s consists of a chamber with an oxide-coated cathode and a number of electron acceleration grids as shown in Figure 4(a). When the cathode filament emits hot electrons they are accelerated by the grids in the chamber. The gas is introduced by small amounts into the chamber where it becomes ionized. Since the plasma density and electron temperatures are quite high in the central core of the plasma the dissociation mean free path of hydrogen molecules is small. Therefore, the conditions of forming $H^-$ are more favorable on the periphery of the plasma. Extraction in this type of NHIS is carried out from an aperture displaced from the anode symmetry axis as shown in Figure 4(a). The extracted $H^-$ current was below 10 mA and this type of NHIS did not find broad application in particle accelerators.

2. Penning volume negative ion source

This NHIS consists of a cylindrical anode and two cathodes on the ends of the cylinder as shown in Figure 4(b). When an electron is emitted from a cathode it is accelerated by anode and reflected back by the opposite cathode. Their collisions with feedstock gas create excited molecules, atoms and ions. In this case the plasma is divided into a hot core region and the cold surrounding. The ionization and excitation of neutral gas occurs in the central region. The negative ions are in turn produced in a surrounding region where electron temperature is lower and the efficiency of the DEA is higher.

3. Magnetically filtered multicusp volume ion source

This NHIS consists of two regions separated by a magnetic field called magnetic filter as depicted in Figure 4(c). In the first part of the source fast electrons are created by an inductive radio frequency discharge or by filaments and are accelerated to high energies. Then, they ionize, excite and dissociate the neutral gas consisting of molecular hydrogen. High vibrational states are effectively populated by EV collisions which are most effective at the energies exceeding 20 eV (Capitelli, et al., 2002). The electrons in these discharges are characterized by non-Maxwellian distribution function (Bacal, et al., 2005). Vibrationally excited molecules and low energy electrons diffuse through magnetic filter into the second region of the chamber where their dissociative attachment collisions produce negative hydrogen ions.

Additional volumetric NHIS have been proposed (e.g., sheet plasma negative source, reversal ion source, diode source) but they are not widely used in particle accelerators (Zhang, 1999).

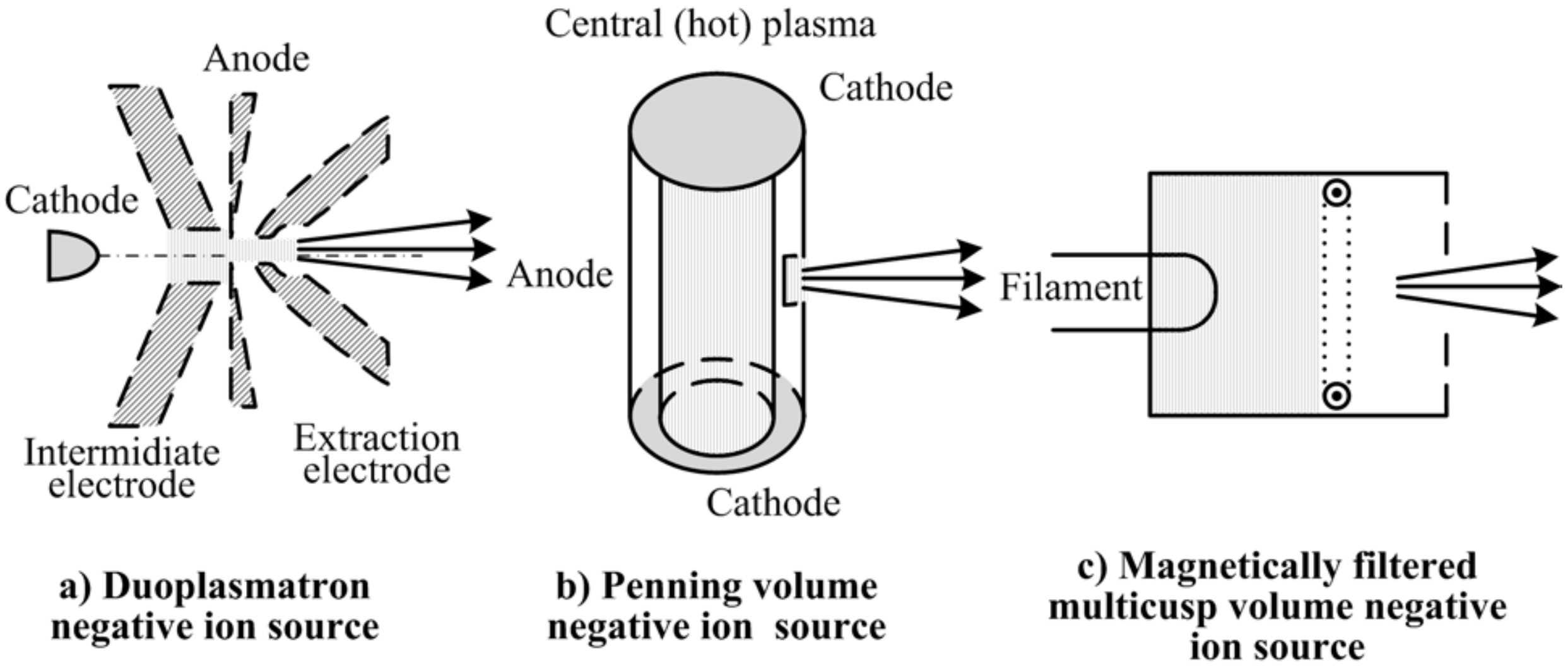

**Figure 4. Volume negative hydrogen ion sources schematics: a) duoplasmatron negative ion source; b) Penning volume negative ion source; c) magnetically filtered multicusp volume negative ion source.**

A brief overview of NHISs based on surface production is presented below following Zhang (1999). Some of the NHIS designs with relevant parameters are outlined in Table 2.

1. Magnetron negative ion source

This NHIS was developed in Novosibirsk by Belchenko et al. (1974). It consists of a cylindrical cathode surrounded by an anode as shown in Figure 5(a). The hydrogen gas is fed from the hole in the anode on the opposite side of the cylinder. Cesium is introduced at one side of the cylinder and usually it is obtained by heating metallic cesium or a mixture of cesium chromate and titanium. A voltage of the order of 150 V is applied between the cathode and anode and allows drawing a current of around 40 A provided by a pulsed DC supply. The magnetic field is applied in the normal direction to the electric field causing electrons to $\mathbf{E}\times\mathbf{B}$ drift in the gap between two electrodes. The positive ions are accelerated to the cathode surface by the electric field in the cathode sheath. The $H^-$ ions are produced on the cathode surface covered by cesium layer when positive ions and neutral particles strike it. The ejected $H^-$ are accelerated by the electric filed in the sheath and extracted from the long slit parallel to cylindrical axes on the anode. Some of the accelerated $H^-$ collide with hydrogen atoms and produce low energy $H^-$ by resonant charge exchange collisions. The main disadvantage of this NHIS is that the extracted energy spectrum of $H^-$ has two distinct peaks. One is due to fast ions accelerated in the sheath and the other is due to the slow ions obtained by charge exchange collisions. In order to increase the low energy peak which is favorable to lower emittance one can increase the pressure and applied voltage.

2. Penning surface-plasma $H^-$ ion sources (Dudnikov source)

This NHIS was developed in the late ‘70s. The geometry is similar to the Penning volume negative source. It consists of a cylindrical anode and two cathodes on both ends of the cylinder as shown in Figure 5(b). The plasma is produced by a pulsed DC potential of about 150 V and 50 A applied between anode and cathodes. The strong magnetic field is applied parallel to the cylinder’s axis causing electrons to spirally move between the two cathodes. The ion particles in the meantime are accelerated toward cathode in the anode sheath and produce negative ions by collisions with cesiated anode surface. The fast negative ions ejected from the cathode exchange charge with slow hydrogen atoms in the bulk of the plasma leading to the low energy negative ions population. These low energy $H^-$ particles diffuse through the hole in the anode. The emittance of the extracted $H^-$ depends on the temperature of $H^-$ allowing this source to achieve a low-emittance negative ion beam.

3. Magnetic cusped surface $H^-$ source

In this type of NHIS the radial and axial magnets form a magnetic bucket confining the plasma as depicted in Figure 5(c). The plasma is created by radio-frequency discharge or by hot filaments. The main feature of this source is that there is a negatively biased concave electrode called a converter. This electrode is covered by a cesium layer which is facing the extractor of the source. Due to the negative voltage it attracts positive ions which then are converted to negative ions. The concave shape of the converter and position of the extraction sleet in front of the converter allows focusing of the negative ion beam outside of the plasma. In order to keep cesium coverage the cesium is injected into the source. The magnetic filter is used in the extraction region to repel extracted electrons.

4. Hollow discharge duoplasmatron $H^-$ source

This NHIS is a duoplasmatron with a hollow rod inserted on the axis as shown in Figure 5(d). The hollow discharge is created inside the rod. Then the plasma diffuses radially from the discharge into the space between the rod and the anode.

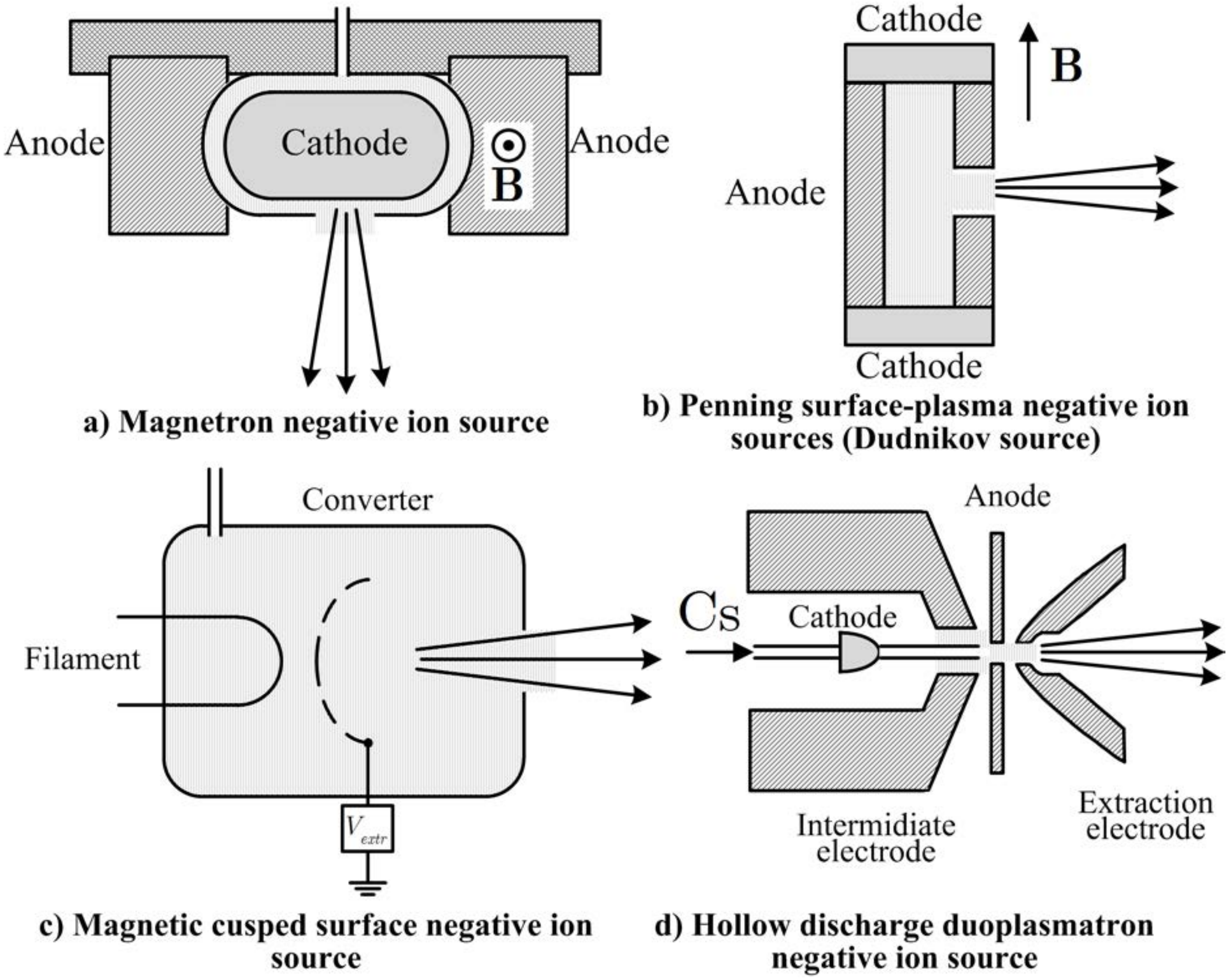

**Figure 5. Surface negative hydrogen ion sources schematics: a) magnetron negative ion source; b) Penning surface-plasma ion source (Dudnikov source); c) magnetic cusped surface ion source; d) hollow discharge duoplasmatron ion source.**

Besides surface and volume negative hydrogen ion sources there are other techniques which were used earlier. Charge-transfer $H^-$ ion source is the source where positive ions are transferred to negative ions by collisions with foils, gas or alkali metal vapors.

The operational and performance characteristics of NHISs developed for particle accelerators are presented in Table 2 which is updated variant of the table from Stockli (2006). The common feature of these sources is the low operational pressure, somewhere below 1 Pa.

**Table 2. Negative hydrogen ion sources with relevant parameters.**

| Facility | Source type | $I_{H^-}$ (mA) | Pulse Length (ms) | Repetition rate (Hz) | $\epsilon_{n,rms}$ (π·mm·rad) | Reference |
|---|---|---|---|---|---|---|
| DESY | Multicusp (external RF) | 40<br>30<br>40 | 0.15 | 8 | 0.25 (90%)<br>0.26 (90%)<br>0.43 (90%) | Moehs et al. (2005)<br>Bacal et al. (2005) |
| Fermilab | magnetron | ~60 | 0.1 | 15 | 0.2/0.3 | Moehs et al. (2005) |
| BNL | magnetron | ~100 | 0.6 | 6.66<br>10 | ~0.4 | Stockli (2006) |
| ISIS | Penning | ~60<br>~35 | 0.5 | 50 | ~1<br>~0.12/0.17 | Faircloth et al. (2006)<br>Thomason et al. (2002) |
| LANSCE | Surface converter | ~18<br>40 | 1 | 120 | ~0.12<br>~0.23 | Sherman et al. (2005) |
| J-PARC | Multicusp LaB6 filament | 20<br>36 | 0.5 | 25 | 0.15/0.18 | Oguri et al. (2006)<br>Oguri et al. (2009) |
| SNS Front-ed | Multicusp (int. RF) | ~20<br>41 | <1 | 1-5 | 0.12/0.14<br>0.25/0.31 | Stockli et al. (2006) |
| SNS Teststand | Multicusp (int. RF) | 33<br>41 | 1.23 | 60<br>10 | 0.18/0.26<br>0.25/0.31 | Stockli et al. (2006) |
| JAERI | Multicusp W-filament | 60<br>72 | 1 | 50 | ~0.21 | Stockli et al. (2004)<br>Oguri et al. (2002) |
| Sumy | Inverse magnetron | ~50<br>~60 | 0.1-1<br>0.4-1.2 | 1-10<br>1-5 | | Kursanov et al. (2005)<br>Baturin et al. (2010) |

## 1.4 High Current Negative Hydrogen Ion Source (HCNHIS)

Busek Co. Inc. and Worcester Polytechnic Institute have proposed a new design for a NHIS using volumetric negative ion production without cesium addition called High Current Negative Hydrogen Ion Source (HCNHIS). This HCNHIS is shown in Figure 6(a) for a baseline configuration and consists of two chambers: the high-pressure radio-frequency discharge chamber (RFD) where the inductive discharge occurs and low-pressure negative hydrogen ion production region (NIP) where the negative ions are mainly produced. The hydrogen gas is fed into the discharge chamber at high pressure (from 30 to 60 Torr) where it is ionized by an inductive RF field. The assumption is that the high pressure in the RFD chamber along with the relatively high neutral temperature allows production of vibrationally excited hydrogen molecule. A similar concept was introduced in Bailey and Garscadden (1980). Also due to the high electron-heavy particle collision rates the electron temperatures are relatively small. Thus, they are more efficient in $H^-$ production and less efficient in their destruction. The hydrogen degree of dissociation in our device is lower compared to conventional source designs. This feature is important because hydrogen atoms are effectively destroying hydrogen ion beams by stripping the electrons and eliminating high energy ions from the beam by charge exchange reactions (Zhang, 1999; Bacal, 2006). The produced hydrogen plasma in the RFD flows into the NIP chamber where additional $H^-$ are primarily produced by DEA mechanism, and from where the extraction takes place. The magnetic field is used to separate electrons from negative ions. Figure 6(a) shows the sketch of the HCNHIS in its baseline configuration and Figure 6(b) shows an alternative setup with a shorter NIP region used in experiments in order to investigate the characteristics of the RFD chamber.

The RF discharge is produced in the RFD chamber by an external coil operating at 27.13 MHz which is a standard frequency for industry and science. The power of the design source is between 0.5-5 kW. The flow rate used in the experiments lies in the range 300-3000 standard cubic centimeters per second (sccm). The resulting pressure in the RFD chamber is from 20-70 Torrs.

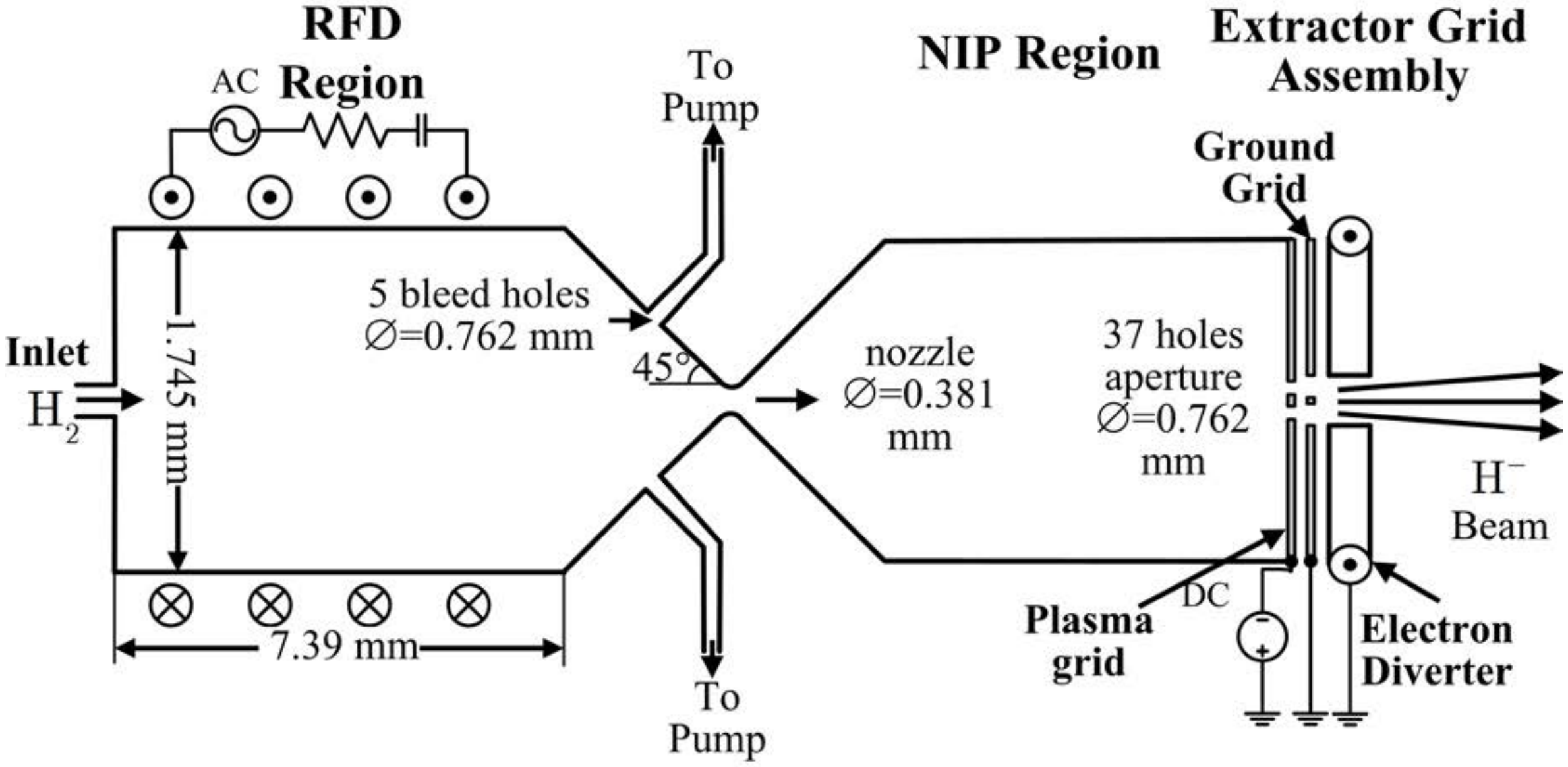

**a) Baseline configuration of the HCNHIS with 5 bypass tubes, extended NIP region, and 37 holes aperture.**

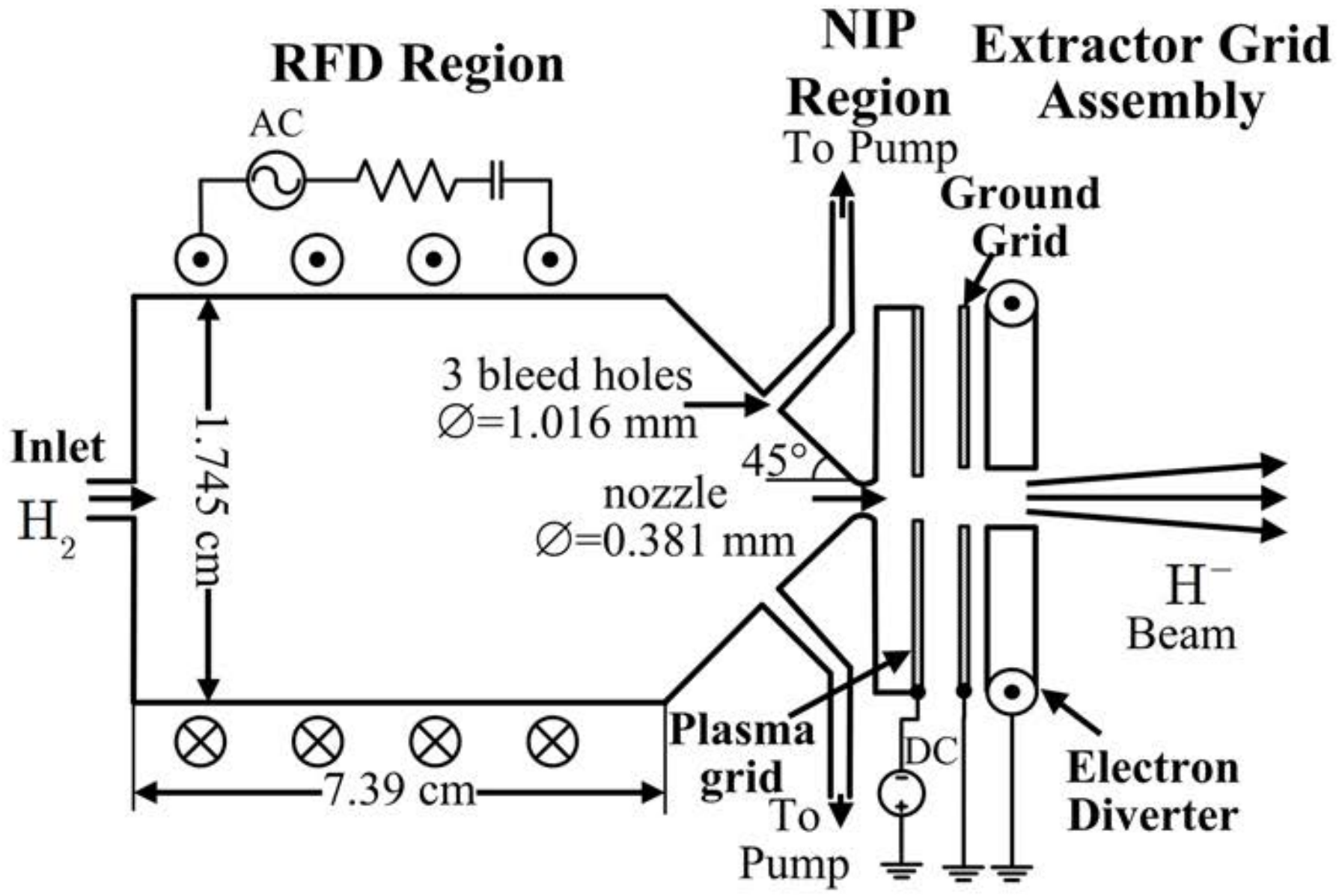

**b) Alternative configuration of the HCNHIS with 3 bypass tubes, reduced NIP region, and single hole aperture.**

**Figure 6. High Current Negative Hydrogen Ion Source schemes: a) baseline configuration with extended NIP region and 5 bypass tubes; b) alternative configuration with reduced NIP region and 3 bypass tubes.**

The baseline design of the HCNHIS allows achieving certain important features specific to high-pressure (compared to existing negative ion sources) plasmas in the RFD chamber:

1. The most important volume reaction that produces negative ions is dissociative electron attachment to vibrationally excited molecules at higher levels. Therefore the operation at high-pressure regime implies that VT and VV reactions may be extremely important in production and destruction of such states.
2. At higher pressures the collision rates between electrons and molecules are considerably higher than in conventional NHIS. Therefore, the expected electron temperature doesn't deviate too much from neutral temperature which under these conditions should be also close to ion temperatures. Thus, electrons are more likely to create negative hydrogen ions other than destroy them. Also negative ion temperature could be much lower than in typical SPS leading to possibly low emittance of the extracted ion beam.
3. High-pressure operation delays the dissociation of hydrogen molecules because recombination reaction play more important role at such conditions which is favorable to negative ion creation in two ways. First, the hydrogen atom density is decreased and since they do not destroy negative ions effectively. Then the efficiency of one of the destruction channels may be lowered. Second, the decreased degree of dissociation leads to a higher availability of vibrationally excited molecules.
4. Bacal et al. (2005) proposed that vibrationally excited hydrogen molecules could be created by atomic hydrogen association at the surfaces. Therefore, this can be another important mechanism of creation of high vibrational states of molecular hydrogen.

However, negative ions are very fragile to collisions because of the low electron affinity. At high pressures collisions with neutral gas may effectively destroy them. In order to understand the importance of different reaction mechanisms and to create the most efficient design a thorough theoretical and experimental investigations are required. A mathematical model which covers a broad range of different regimes: pressures, electron and heavy species temperatures, plasma concentrations, is the topic of this research.

### 1.5 Review of Modeling and Simulation Approaches for Reactive Hydrogen Plasmas

Hydrogen plasmas with chemical processes relevant to those encountered in the HCNHIS have been modeled extensively. The review that follows includes the zero-dimensional kinetic, global, fluid, and kinetic modeling and simulation approaches used for reactive hydrogen plasmas covering the low-pressure to high-pressure regimes from 10 mTorrs to 100 Torrs. The re-view reveals also the advancements needed to address the $H^-$ reactive plasma for the regime of operation and physical scales of the HCNHIS considered in this study.

One of the earliest approaches used for plasmas relevant to NHISs operating at low pressures (10-100 mTorrs) as well as high-pressure plasmas (10-100 Torrs ) useful in diamond deposition, is the zero-order (0D) kinetic model. In this approach the spatial variation of the number densities and temperatures of plasma components is disregarded and all parameters are assumed to be space uniform. The plasma is described by means of master equations representing species continuity equations supplemented by the charge neutrality condition. The rate coefficients are derived by assuming Maxwellian velocity distribution function. Electron and gas temperatures as well as electron number density are input parameters to the model. In general, these equations form the system of first-order ordinary differential equations which could be solved by time marching technique. If only steady-state is of interest then the time dependence terms can be dropped and the resulting system of equations represents nonlinear set of algebraic equations which could be solved by Newton-Raphson method.

Méndez et al. (2006) modeled low and intermediate pressure 0.8-200 Pa (6-1500 mTorr) direct current hydrogen plasma reactor suitable for plasma-enhanced chemical vapor deposition. The model considered atomic hydrogen $H$ and three positive ion species $H^+$, $H_2^+$, $H_3^+$. The concentrations of the species were calculated by solving master equations representing particles continuity equations for $H$, $H_2^+$ and $H_3^+$ supplemented by the charge neutrality condition. The chemistry model included 13 volume and 3 surface chemical reactions. The transport of the species was described by means of the species residence times. The input parameters to the model were the gas and electron temperatures, electron number density and chamber pressure. The actual values used for the electron temperature and number density were based on the double Langmuir probe measurements. For the gas temperature the room temperature of 300K was assumed. The calculated number densities were compared with the experimental predictions and good agreement was established.

In order to include non-equilibrium effects of the internal degrees of freedom the zero-order kinetic models could be further extended. In this case each “species” represents a particular plasma component such as a molecule, atom or an ion in a specific internal state. This approach is also referred to as the state-to-state (Nagnibeda & Kustova, 2009). Once all number densities of internal states are known it is possible to construct the distribution function for the internal degrees of freedom, such as vibrational and rotational. Similar to the simple zero-dimensional kinetic model of ground state species in the state-to-state approach the particle concentrations are described in terms of master equations representing space uniform particles continuity equations.

In Fukumasa and Ohashi (1989), a 0D kinetic model was developed for a tandem NHIS shown in Figure 4(c) and operating at the pressure of 5 mTorr. They considered ground state neutral species, $\mathrm{H}$, $\mathrm{H_2}$, ground state positive and negative ions, $\mathrm{H^+}$, $\mathrm{H_2^+}$, $\mathrm{H_3^+}$, $\mathrm{H^-}$, vibrationally excited molecules in the seventh vibrational level $\mathrm{H_2}(v=7)$, and two electron species representing fast electrons in the first chamber emitted by filament and slow electrons in the second chamber after the magnetic filter. The plasma chemistry model included 27 reactions. The plasma composition was obtained for each chamber of the tandem NHIS separately through a solution of steady-state species conservation equations supplemented by charge neutrality and hydrogen atom conservation conditions. Electron and gas temperatures, fast and slow electrons number densities, pressure in the chamber, the size of the source and location of the magnetic filter which determines the volume of each chamber were considered as input parameters. The model was used to predict $\mathrm{H^-}$ production as a function of the fast and slow electrons number densities and electron temperature in single and two-chamber NHIS configurations. The increased $\mathrm{H^-}$ yield in two-chamber source compared to single chamber was observed. Later (Fukumasa, et al., 1992) this model was extended by taking into consideration full vibrational spectrum of hydrogen molecules $\mathrm{H_2}(v)$, $v = 1 - 14$. Other species were kept the same as in Fukumasa and Ohashi (1989). The chemical model included 31 reactions. Simulation procedures and input parameters to the model were identical to Fukumasa and Ohashi (1989). The model was applied to the tandem NHIS. The vibrational distribution function and the production of $\mathrm{H^-}$ were calculated as a function of the fast and slow electrons number densities and electron temperature.

The main drawback of 0D models described above is that not all of the input parameters are truly independent. For example, electron number density and electron temperature are, in general, a function of the gas pressure and the power absorbed by the plasma (Lieberman & Lichtenberg, 2005). In order to resolve this issue the energy equations for at least electrons and heavy components should be solved. In addition, the rate coefficients are calculated by assuming Maxwellian velocity distribution functions. Although this is usually a good assumption for neutral and heavy ion species, the electron energy distribution function often deviates from the equilibrium. One way of bypassing this difficulty is to use experimental electron energy distribution function. More advanced kinetic models include a 0D electron Boltzmann equation coupled with a set of master equations comprising 0D kinetic model in order to take into account non-equilibrium effects.

Hiskes and his colleagues (Hiskes & Karo, 1984; Hiskes, et al., 1985) investigated the production of $\mathrm{H}^-$ in the tandem high-density hydrogen discharges by 0D kinetic model. The EEDF was assumed Maxwellian with the high energy part modified in order to describe the slowing electrons emitted from hot filaments. The model included full vibrational spectrum of molecular hydrogen. The gas and electron densities as well as system scale length were varied as independent parameters in the parametric study of negative hydrogen ion current densities, while concentrations of positive hydrogen ions ($\mathrm{H}^+$, $\mathrm{H}_2^+$, $\mathrm{H}_3^+$), electron and gas temperatures were fixed during the simulations. This model shares the same drawback as in Fukumasa and Ohashi (1989), Fukumasa et al. (1992): not all of the parameters varied in the simulation are really independent.

In Loureiro and Ferreira (1989) a 0D kinetic model was developed for a low-pressure, moderate current, hydrogen positive column. The species considered in the model were electrons, hydrogen atoms, ground state and 14 vibrationally excited hydrogen molecules $\mathrm{H}_2(v = 0 - 14)$. The steady-state space homogeneous Boltzmann energy equation for the electron energy distribution function in two-term approximation coupled with the steady-state master equations for the vibrationally excited hydrogen molecules representing species continuity equations constituted the governing equations of this 0D kinetic model. The set of reactions considered included eV, eD, VT and VV processes. The neutral gas temperature, reduced electric field, fractional number densities of electrons and hydrogen atoms were input parameters to the model. The model was used to investigate coupling between vibrational distribution function and electron energy distribution function. Non-equilibrium character of VDF was observed. In addition, it was found that the hydrogen atoms play very important role in the quenching of high-lying vibrational levels.

Zero-dimensional kinetic models of interest to this work have also used to predict composition in plasma reactors. In Hassouni et al. (1999) a 0D kinetic model of hydrogen plasma was utilized in order to calculate composition in a diamond deposition hydrogen reactor operating at pressures 1400-11000 Pa (10.5-82.5 Torrs), microwave power 300-2000W corresponding to the averaged microwave power density 4.5-30 W/cm$^3$. Ground state atomic $\mathrm{H}$ and molecular $\mathrm{H_2}$ hydrogen, electronically excited hydrogen atoms $\mathrm{H}(n)$, $n \le 40$ and molecules $\mathrm{H_2}(n)$, $n \le 37$, vibrationally excited hydrogen molecules in the ground electronic state $\mathrm{H_2}(v)$, $1 \le v \le 14$, electrons, positive ions ($\mathrm{H^+}$, $\mathrm{H_2^+}$, $\mathrm{H_3^+}$) and negative ions ($\mathrm{H^-}$) were considered in the model. The set of reactions contained electron-heavy, ground state species reactions, comprehensive collisional radiative processes for electronically excited atomic and molecular hydrogen species, and an extensive set of vibrational kinetics including both VT and VV processes. The governing equations used in the 0D model consisted of the steady-state space homogeneous Boltzmann electron energy equation coupled with the master equations representing the steady-state space homogeneous species continuity equations and the total energy equation supplemented by charge neutrality and the total number density constrains. The wall reactions were taken into account by assuming linear variation of plasma species concentrations in the boundary layer and using experimental measurements of the boundary layers thicknesses. The input parameters to the model were chamber pressure and absorbed microwave power. The calculated gas temperature and hydrogen atom fractions were compared with experimental measurements and good agreement was observed. The model was used in order to get detailed information about non-equilibrium effects in EEDF and VDF. Based on the full description of the plasma simpler model was also proposed. Based on the results of the full 0D model the number of considered species was dramatically reduced to $\mathrm{H_2}$, $\mathrm{H}$, $\mathrm{H}(n=2)$, $\mathrm{H}(n=3)$, $\mathrm{H^+}$, $\mathrm{H_2^+}$, $\mathrm{H_3^+}$, $\mathrm{H^-}$ and electrons. The equilibrium distribution function of vibrationally excited states of molecular hydrogen was assumed. The governing equations were steady-state space homogeneous Boltzmann equation for the electrons, master equations for species, and total energy equation with the additional energy equation for the vibrational temperature. The system was closed by imposing charge neutrality and total number density conditions. The input parameters were identical to the full 0D kinetic model. The electron number densities, electron and gas temperatures, reduced electric fields, and mole frac-tion of hydrogen atoms calculated by both models were compared. The relative difference between predictions of two models was in the range 5-15%. The discrepancy in hydrogen atoms mole fraction was attributed to the exclusion of vibrationally excited states from the simplified model.

One of the main difficulties of 0D kinetic models is the inability to properly take into account non-uniformity in the space distribution of plasma parameters. This problem can be partially solved by using global models. Global models utilize integral forms of plasma species balance equations derived on the volume averaging of conservation equations and are supplemented with charge neutrality and an electron energy equation (Monahan & Turner, 2008; Monahan & Turner, 2009). The boundary fluxes are evaluated based on the heuristic patching of the analytic and numerical solutions of diffusion equations corresponding to the geometry of interest at different pressure conditions. The electron and neutral temperatures are assumed to be uniform in the discharge. For the neutral temperature it is usually assigned constant value which is a reasonable assumption at low pressures. At high pressures additional equation representing neutral energy conservation is used. The reaction rates in the particle balance equations are often calculated by assuming Maxwell-Boltzmann electron energy distribution function (EEDF) and heavy particles distribution functions. However electron non-equilibrium effects sometimes are taken into account by considering specific forms of the electron distribution functions (Gudmundsson, 2001), by using an external Boltzmann solver (Danko, et al., 2013) or by solving Boltzmann equation for electrons (Chen, et al., 1999). Global models have become a standard tool in predicting main plasma parameters such as electron temperature and number density (Lieberman & Lichtenberg, 2005; Chabert & Braithwaite, 2011). Thus, global models are a subset of 0D kinetic models with the ability to estimate surface chemical reactions in a more accurate manner.

Chen et al. (1999) utilized global model for the microwave generated hydrogen plasma operating at pressures 2-60 Torrs, microwave power 50-200W corresponding to the averaged microwave power density 5.5-22 W/cm$^3$. The species considered in the model were molecular and atomic hydrogen, three positive ions ($H^+$, $H_2^+$, $H_3^+$), and electrons. The population of vibrationally excited states was considered to be in the equilibrium with translational and rotational degrees of freedom. The chemistry model consisted of 41 volume and surface reactions. The governing equations were a set of master equations for the number densities of neutrals and ions, a Boltzmann equation for electrons, electron and neutral energy equations supplemented by quasi-neutrality condition. The neutral energy equations had two different forms for different chamber pressures. The gas pressure, absorbed power, and feed flow rate were the input parameters to the model. Calculated gas and electron temperatures, electron number density were compared with experimental measurements and rather good agreement was obtained. The model was used in the parametric study in order to predict the plasma composition as a function of the pressure and absorbed power.

Pagano et al. (2007) simulated magnetically multicusp NHIS by means of global model using state-to-state approach extended not only to vibrationally excited states of molecular hydrogen but also to electrons. The operating conditions were the pressure 1.5 Pa corresponding to 11.25 mTorr, a discharge voltage 100 V, and a discharge current 0.5-10 A. The species considered in the model were ground state atomic and molecular hydrogen, 14 vibrationally excited hydrogen molecules in their fundamental electronic state $\mathrm{H}_2(v)$, electronically excited molecular hydrogen species ($\mathrm{B}^1\Sigma_u^+$, $\mathrm{C}^1\Pi_u$, $\mathrm{D}^1\Pi_u$, $\mathrm{B}'^1\Sigma_u^+$, $\mathrm{D}'^1\Pi_u$, $\mathrm{B}''^1\Sigma_u^+$, and $\mathrm{E\text{-}F}^1\Sigma_u^+$), electronically excited hydrogen atoms legitimate, $n = 2, 3$, positive and negative hydrogen ions ($\mathrm{H}^+$, $\mathrm{H}_2^+$, $\mathrm{H}_3^+$, $\mathrm{H}^-$), and electrons in state-to-state approach. The chemistry model contained 30 volume and surface reactions (not counting state resolved reactions) including comprehensive set of processes describing vibrational and electronic kinetics of hydrogen molecules. The whole range of electron energy was discretized by a set of intervals each representing separate "electron". Then time dependent Boltzmann equation for every such "electron" was considered as a rate equation for different species. The set of Boltzmann equations for each such "electron" was coupled with the volume-averaged time dependent continuity equations supplemented by charge neutrality condition. The input parameters for the model were temperatures for neutral and ionic species, discharge voltage and discharge current. The calculated VDF, electron temperature and number density, and $\mathrm{H}^-$ number density were compared with the experimental measurements of Mosbach (2005b). In this comparison the values for the heavy-particle temperatures, the discharge voltage, and discharge current were chosen based on the experimental results. Rather good qualitative agreement was obtained. In addition, the model was used in order to investigate the cesium effect on hydrogen kinetics.

Zorat et al. (2000) utilized a global model in order to predict plasma composition in the deuterium negative ion source experiment (DENISE). This NHIS operated at pressures 10-100

mTorrs and absorbed power 500-5000 W corresponding to power density 53-530 mW/cm$^3$. The global model included $\mathrm{e}$, $\mathrm{H}$, $\mathrm{H}(2s)$, $\mathrm{H}_2$, $\mathrm{H}^+$, $\mathrm{H}_2^+$, $\mathrm{H}_3^+$ and the first two vibrationally excited $\mathrm{H}_2(v)$ hydrogen molecules species. The chemistry model consisted of 17 volumetric and 6 surface reactions. The heavy species and electron distribution function used to calculate rate equations were Maxwellian. In the global model steady-state particle continuity equations and steady-state electron energy equation supplemented by charge neutrality condition were solved simultaneously in order to obtain a steady-state solution. The system of nonlinear algebraic equations was then solved by Newton-Raphson method. The input parameters for the model were the gas temperature, absorbed power and inlet gas flow rate. The calculated electron temperature and number density were compared with the Langmuir probe measurements of corresponding experimental values and good agreement between the simulation and experiment was obtained. This model was extended subsequently to include the negative ions and the first 9 vibrational levels of hydrogen molecules (Zorat & Vender, 2000). The governing equations, input parameters, and solution technique were similar to the previous model. This extended model was used in the parametric study of the $\mathrm{H}^-$ production in DENISE.

Kimura and Kasugai (2010) developed a global model of electropositive hydrogen plasmas diluted with argon atoms operating at pressures 20-60 mTorr and absorbed power 120 W corresponding to the power density of 79.6 mW/cm$^3$. The model included 8 neutral species: ground, metastable, resonant, 4p state argon atoms, ground state $\mathrm{H}_2$, and three electronically exited hydrogen atoms; it included six charged species: electrons, $\mathrm{Ar}^+$, $\mathrm{ArH}^+$, $\mathrm{H}^+$, $\mathrm{H}_2^+$ and $\mathrm{H}_3^+$. Vibrationally excited hydrogen molecules and negative hydrogen ions were not considered. The chemistry model consisted of 46 surface and volume reactions. The steady-state particle balance and electron energy were solved in order to get plasma composition and electron temperature. The input parameters to the model were gas temperature, absorbed power and inlet gas flow rate. The calculated electron temperatures, electron and atomic hydrogen number densities at pressures of 20, 40 and 60 mTorrs, absorbed power 120 W and argon fraction from 0-50% were compared with Langmuir probe measurements. The agreement was reasonably good.

Hjartarson et al. (2010) developed a global model for a hydrogen discharge diluted with argon operating at pressures 1-100 mTorr and absorbed power 600W corresponding to the power density 0.1 W/cm$^3$. They consider electrons, ground state atoms $\mathrm{Ar}$, $\mathrm{H}$, molecules $\mathrm{H}_2$, positive ions $\mathrm{H}^+$, $\mathrm{H}_2^+$, $\mathrm{H}_3^+$, $\mathrm{Ar}^+$, $\mathrm{ArH}^+$, negative ions $\mathrm{H}^-$, vibrationally excited hydrogen molecules $\mathrm{H}_2(v)$, $v = 1 - 14$, electronically excited argon atoms $\mathrm{Ar}^*$, metastables $\mathrm{Ar}_m$ ($1s^5$ and $1s^3$) and radiatively coupled states $\mathrm{Ar}_r$ ($1s^4$ and $1s^2$). The chemistry model included 69 volume and surface chemical reactions. Similar to Zorat et al. (2000) and Kimura and Kasugai (2010) most of the chemical reactions considered were electron-heavy collisions. VV and VTm processes were disregarded. The general distribution function for electrons (Gudmundsson, 2001) was used in order to calculate volume reaction rates involving electrons. The global model equations including steady-state particle continuity and electron energy equations were solved. The input parameters to the model were gas temperature, absorbed power, inlet gas flow rate, pumping speed of the species out of the chamber, and the parameter in the general distribution function. Langmuir probe measurements of electron temperature and number density as well as plasma potential were compared with the global model predictions. Rather good agreement between experiment and numerical simulation was obtained. In addition, plasma composition was explored by varying pressure, argon dilution, and the atomic hydrogen wall recombination coefficient.

Fluid models have also been developed for hydrogen plasmas. These approaches include continuity, momentum and energy equations under various degrees of simplification supplemented by the Maxwell's equations for the electric and magnetic fields in the plasma (Burgers, 1969; Golant, et al., 1980). The transport coefficients are obtained by assuming specific distribution functions of the plasma components. The main advantage of fluid models is that they can predict the temporal and spatial variation of all plasma properties but range in computational complexity depends on the level of approximation. Fluid models can also be combined with particle-based models to provide hybrid approaches.

Matveyev and Silakov (1995) developed 1D fluid model for prediction of the hydrogen plasma composition, temperatures of heavy particles and electrons as long as vibrational non-equilibrium effects in a pulsed high current low-pressure discharge in a cylindrical tube. The operating parameters of this discharge were discharge current 300 A, initial pressure 1.38 Torr, and the wall temperature 500 K. The model considered detailed chemical reactions involving ground state hydrogen molecules $\mathrm{H}_2$ and atoms $\mathrm{H}$, 14 vibrationally excited hydrogen molecules states $H_2(v)$, electronically excited hydrogen atoms $H(n)$, positive ions ($H^+$, $H_2^+$, $H_3^+$, $H_5^+$), hydrogen anion $H^-$, and electrons. The electron energy distribution function used in the calculations of rate coefficients of chemical reactions involving electron species was assumed to be Maxwellian. The governing equations included unsteady continuity equations for neutrals, ions and electrons, the unsteady equation for mean-mass radial velocity of heavy species, the unsteady heavy particles energy equation, the unsteady radial momentum equation in drift-diffusion approximation for electrons, and the unsteady electron energy equation. In place of solving Maxwell's equations the assumption of ambipolar electric field was used. The input parameters for the model were the initial space distribution of electron and heavy particles number densities and temperatures, wall temperature, and discharge current. The fluid model was used to explore vibrational non-equilibrium effects. It was concluded that correct calculation of gas heating requires detailed consideration of vibrational kinetic.

Hagelaar et al. (2011) developed a 2D multi-fluid and multi-temperature model for the NHIS applicable to ITER experiment. The governing equations included time dependent separate continuity, momentum, and energy equations for neutrals, ions and electrons. For electrons, the drift-diffusion approximation was used for the momentum equation. For neutral species the governing equations were the full Navier-Stokes equations. These equations were coupled with Maxwell's equations for electromagnetic field written for scalar and vector potentials. Fluid boundary conditions were formulated for particle, momentum and energy fluxes for catalytic surfaces including velocity slip and temperature jump effects. The rate coefficient for the volume production and destruction of the particles in continuity equations were derived based on Maxwellian distribution functions with the different temperatures for different plasma species. The transport coefficients in the momentum and energy equations were calculated from the momentum rate coefficients obtained for each pair of particle species. The final system of coupled PDEs was solved by using finite difference method. The plasma was considered to contain ground state molecular hydrogen $H_2$, ground state dihydrogen cation $H_2^+$, and electrons. The chemistry model included only electron-impact ionization of hydrogen molecules and electron-ion recombination at the walls. The model was applied to a single-driver volume connected to an expansion chamber both in Cartesian and in cylindrical coordinates. The RF power absorbed by the electrons was chosen to be 50 kW for a cylindrical case and 50 kW/m for the Cartesian case. It was found that the plasma potential calculated in a self-consistent manner was in good agreement with the classical sheath theory. The calculated plasma densities were strongly non-uniform. In addition, the effects of EF coupling, applied voltage and magnetic filter were studied.

In Boeuf et al. (2011) the above model was applied to NHIS (a single-driver chamber connected to an expansion chamber) developed at IPP-Garching for the ITER project. The plasma composition included electrons, positive and negative ions ($\mathrm{H}^+, \mathrm{H}_2^+, \mathrm{H}_3^+, \mathrm{H}^-$), ground state atoms $\mathrm{H}$ and molecules $\mathrm{H}_2$ as well as electronically excited atoms $\mathrm{H}(n=2)$ and $\mathrm{H}(n=3)$. The chemistry model involved 45 surface and volume reactions. The production of negative ions was estimated based on the assumption that only a fraction of $\mathrm{H}_2$ molecules can take part in a dissociative attachment reaction. Thus, vibrational non-equilibrium was completely ignored. The Maxwell distribution function was used for all species in calculations of reaction rates. Due to low-pressure operation the neutral gas dynamics results obtained by solving full Navier-Stokes were verified by comparison with Direct Simulation Monte Carlo method. The agreement between two approaches was rather good but the velocity distribution function of neutral particles as predicted by DSMC results was highly non-Maxwellian. The model was then applied to NHIS operating at pressures 0.2-0.8 Pa corresponding to 1.5-6 mTorr and absorbed powers 10-80 kW. The calculated plasma density and electron temperature were in a good agreement with the experimental measurements. It was found that calculated molecular hydrogen temperature was about an order of magnitude lower than the temperature of hydrogen atoms which was explained by the electron impact dissociation of hydrogen molecules. In addition, the simulations including magnetic field were performed assuming classical collisional transport. It was observed that the electron temperature decreased in the filter region but then increased outside the filter. Compared to zero-dimensional models where the chemistry model was the key part in the fluid simulations of Hagelaar et al. (2011) and Boeuf et al. (2011) the transport phenomenon was a primary object.

In addition to aforementioned methods the solution of Boltzmann equation by particle methods is used in order to model negative hydrogen ion sources. Such methods include DSMC method developed first for neutral species (Bird, 1994), Particle-in-Cell method (PIC) (Hockney & Eastwood, 1988) and Particle-in-Cell with Monte Carlo collisions (PIC-MCC) methods (Birdsall, 1991). The main idea behind these methods is to replace particle velocity distribution functions by sets of macroparticles representing a phase-space volume. In addition, in the Boltz-mann equation the collisions and motion of the particles are decoupled. This leads to a computational scheme where the computational particles (macroparticles) at every time step are moved in the computational domain according to the Newton's equations of motion which are solved for every particle by leap-frog scheme. After particles displacements the electrostatic Poisson equation is solved in order to obtain electric fields. Then the collisions are performed in a probabilistic manner and the cycle is repeated again (Nanbu, 2000). The space discretization is based on the Debye length in PIC scheme and on the collision mean free path in the DSMC method.

In Kolev et al. (2009) 1D3V (one spatial dimension and three velocity dimensions in the velocity space) PIC-MC model was applied to tandem multicusp NHIS in order to investigate the electron and hydrogen anion transport in the magnetic filter region. The species included into the model were ground state atomic $\mathrm{H}$ and molecular $\mathrm{H_2}$ hydrogen, positive and negative ions ($\mathrm{H^+}$, $\mathrm{H_2^+}$, $\mathrm{H_3^+}$, $\mathrm{H^-}$), and electrons. The chemistry model contained 44 collision processes including electron-heavy elastic and ion-neutral charge exchange collisions. The Coulomb collisions between charged species were disregarded. For calculation of $\mathrm{H^-}$ production it was assumed that the fixed portion of ground state molecules (based on experimental measurements) was in the higher vibrational state ($v > 3$) and these particle took part in DEA reaction. While the operational pressure was 0.3 Pa (2.25 mTorr) the plasma density considered in the simulations was an order of magnitude smaller to allow for PIC-MC computations. The model was used in order to investigate electron and $\mathrm{H^-}$ transport across a localized transverse magnetic field. It was found that the electron transport across the filter followed the classical electron diffusion model. When the bias voltage was applied, the plasma stratified into upstream electropositive and downstream electronegative regions.

In Taccogna et al. (2007) the DSMC model was used in order to simulate an expansion region of NHIS designed by IPP-Garching. Ground state hydrogen atoms and molecules and 14 vibrationally excited molecular hydrogen states constituted the plasma composition. In addition, some electronically excited molecular hydrogen states were included into the model. The actual plasma consisting of positive and negative ions ($\mathrm{H^+}$, $\mathrm{H_2^+}$, $\mathrm{H_3^+}$, $\mathrm{H^-}$), and electrons was considered to be a fixed background. The spatial variations of electron number density as well as constant mole fractions of hydrogen ions were based on the previous global models. The null-collision technique (Nanbu, 2000) was used to describe the elastic and inelastic collisions. The chemistry model included comprehensive set of chemical reactions involving vibrationally excited hydrogen molecules such as VTm and VTa reactions. Due to small pressures in the NHIS VV transitions were neglected. The model was used in order to predict vibrational population of hydrogen molecules at the end of the expansion region.

Taccogna et al. (2008) utilized 1D3V PIC\MC and DSMC in order to simulate an extraction region of NHIS designed by IPP-Garching. The PIC\MC scheme was used for charged particles ($H^+$, $H_2^+$, $H^-$, and electrons). The dynamics of ground state hydrogen atoms and molecules as well as 14 vibrationally excited states of molecular hydrogen were treated by DSMC. The chemistry of the 22 volume reactions (vibrationally resolved reactions are not counted) represented the vibrational and ionization kinetics of hydrogen plasmas. The surface production of negative ions was included into the model. The VDF in the expansion region was calculated by DSMC (Taccogna, et al., 2008) and used as input parameters for the models of the extraction region. The effects of plasma grid bias voltage and magnetic filter were analyzed by the model.

Taccogna et al. (2011) utilized the 2D DSMC model for the expansion and extraction regions of NHIS. The chemistry and chemical processes included into this simulation were similar to Taccogna et al. (2007).

Overall, particle simulation methods have limited applicability to high-pressure plasmas due to mean free paths and Debye length limitations.

## 1.6 Research Goals, Objectives, and Approach

The goal of this dissertation is to develop a comprehensive theoretical and computational model of the chemically reacting plasmadynamics processes in a High Current Negative Hydrogen Ion Source (HCNHIS) under development by Busek Co. Inc. and WPI. The requirements for the model emerge from the unique physical and operational characteristics of the HCNHIS shown in Figure 6.

In order to successfully model the HCNHIS the model should be able to cover following physical scales and processes:

- Operating pressure in the range 10-60 Torr;
- Absorbed power in the range 300-1000 W;

- Species including ground state hydrogen atoms $\mathrm{H}$ and molecules $\mathrm{H}_2$, 14 vibrationally excited hydrogen molecules $\mathrm{H}_2(v)$, $v = 1 - 14$, electronically excited hydrogen atoms $\mathrm{H}(2)$, $\mathrm{H}(3)$, positive ions $\mathrm{H}^+$, $\mathrm{H}_2^+$, $\mathrm{H}_3^+$, anions $\mathrm{H}^-$, and electrons $\mathrm{e}$;
- Volume and surface chemistry;
- Compressible flow phenomena in the bypass and nozzle including rarefaction effects.

To address these requirements and following a comprehensive review of available modeling and simulation approaches, the Objectives and Approach of this dissertation are:

1. Develop a Global Enhanced Vibrational Kinetic Model (GEVKM)
    a. Derive state-to-state moment equations for multi-temperature chemically reacting plasmas based on the Wang Chang-Uhlenbeck Equations;
    b. Review transport properties, transport collision integrals and boundary conditions for the state-to-state moment equations for multi-temperature chemically reacting plasmas;
    c. Include the following species: ground state hydrogen atoms $\mathrm{H}$ and molecules $\mathrm{H}_2$, 14 vibrationally excited molecular hydrogen states $\mathrm{H}_2(v)$, electronically excited hydrogen atoms $\mathrm{H}(n)$, $n = 2,3$, positive hydrogen ions in their ground states $\mathrm{H}^+$, $\mathrm{H}_2^+$, $\mathrm{H}_3^+$, negative hydrogen ions $\mathrm{H}^-$, and electrons;
    d. Include surface and volume chemical reactions governing vibrational and ionization kinetics of hydrogen plasmas;
    e. Develop simplified moment equations coupled with the equation for vector and scalar potential suitable for two-temperature chemically reacting plasmas in cylindrical plasma reactors;
    f. Assuming an isothermal cylindrical plasma develop particle balance equations representing particle continuity equations;
    g. Use electron and total energy equations coupled with the chamber heat transfer for estimation of electron and heavy-particle temperatures as well as wall temperature;
    h. Derive boundary conditions for the nozzle and bypass system of the HCNHIS including compressibility, viscous and rarefaction effects;

2. Implement GEVKM into a robust and computational efficient numerical code.
   a. The code includes a simulation tool that evaluates reaction rates from cross-section data assuming Maxwellian distribution function for colliding partners and fits them into analytical representations;
   b. The code implements Newton-Raphson method for solution of non-linear algebraic equations;
   c. The code implements a solver for continuity and energy equations with the self-consistency checks to guarantee conservation of charge, particles and energy in the system and suitable for simulating arbitrary cylindrical plasma reactors with different plasma composition and chemical reactions of hydrogen plasmas;
   d. The code is configured through text configuration files;
   e. The code is written by using unit testing methodology and a version control system;
3. Perform Validation and Verification
   a. Verify and validate the GEVKM with simulations and measurements in the low-pressure (20-100 mTorr) and low power density (0.053-0.32 W/cm$^3$) negative hydrogen ion source DENISE (Zorat, et al., 2000; Zorat & Vender, 2000);
   b. Verify and validate the GEVKM with simulations and measurements in the intermediate to high pressure (2-60 Torr) and high power density (8.26-22 W/cm3) microwave generated hydrogen plasma (Chen, et al., 1999);
   c. Validate the outlet boundary conditions of the GEVKM by comparison of the simulation results with pressure measurements of the HCNHIS-1 undertaken at Busek Co. Inc.;
   d. Validate the negative hydrogen ion current calculated by the GEVKM with the current measurements in the alternative configuration of the HCNHIS-2 shown in Figure 6(b);
4. Perform a parametric investigation
   a. Investigate the influence of the feedstock gas flow rate and subsequently chamber pressure on:
      i. the production and destruction of vibrationally excited hydrogen molecules and find the optimal parameters;
      ii. the plasma composition in the ion source;

iii. the production and destruction of negative hydrogen ions;
iv. the electron and heavy particles temperature;
v. the maximum extractable negative hydrogen ion current;

b. Investigate the effect of the absorbed power on

vi. the production and destruction of vibrationally excited hydrogen molecules and find the optimal parameters;
vii. the plasma composition in the ion source;
viii. the production and destruction of negative hydrogen ions;
ix. the electron and heavy particles temperature;
x. the maximum extractable negative hydrogen ion current.

The dissertation is organized as follows. In the second chapter the mathematical formulation of moment equations for multi-species multi-temperature chemically reacting partially ionized plasmas is presented. The transport properties, transfer collision integrals, and boundary conditions are reviewed. The third chapter describes GEVKM formulation including electromagnetic model of a cylindrical plasma reactor. In addition, in this chapter the GEVKM is validated and verified in a wide range of pressures and absorbed powers. In the fourth chapter the application of the GEVKM to the HCNHIS is presented. The modifications to the GEVKM in order to take into account more complicated geometry are discussed. The bypass/nozzle model of the RFD chamber outlet including compressible, viscous, and rarefaction effects is then presented. The outlet model is validated by comparing the chamber pressure measured in the HCNHIS to the GEVKM predictions. The plasma composition, vibrational population of hydrogen molecules, negative hydrogen ion production and destruction rates, $H^-$ current are calculated by the GEVKM at different inlet flow rate and absorbed power covering the entire regime of operation of the GEVKM.

# 2 Mathematical Formulation of Multi-fluid, Multi-temperature Chemically Reacting Plasmas in Gas Discharges

The reactive partially ionized hydrogen plasma in the discharge chamber of the HCNHIS contains vibrationally and electronically excited hydrogen molecules, electronically excited hydrogen atoms, a number of different positive hydrogen ions, hydrogen anions, and electrons. Due to the rather high operating pressure (up to 60 Torr), the expected properties in the discharge chamber include high neutral temperature (up to 2000 K), high neutral number density (up to $10^{24}m^{-3}$) and low electron temperature (few eVs). The physical processes in the discharge chamber involve volume and surface chemical reactions, transport phenomena in the bulk, and charged particle interactions with electromagnetic field created by inductive coil. Modeling of these phenomena can resort in kinetic, fluid or hybrid approaches.

In the kinetic modeling approach the plasma particles are considered statistically in terms of velocity distribution functions which under certain conditions (molecular chaos, bimolecular collisions, dilute gas assumption) are governed by the Boltzmann equation and its generalizations for a chemically reacting gas mixtures with internal degrees of freedom which are coupled with Maxwell's equations for self-consistent electromagnetic fields. The kinetic description leads to a system of integro-differential equations that have analytic solutions only in some simplest cases limited to certain geometries and plasma composition. Therefore the kinetic equations usually solved numerically with various methods such as DSMC (Bird, 1994) PIC (Hockney & Eastwood, 1988; Birdsall & Langdon, 2004), PIC-EM, PIC-MC (Birdsall, 1991), and direct solutions of the Boltzmann equation (Aristov, 2001). It should be mentioned that in the electrostatic limit characteristic lengths of the physical phenomena described by these equations include the mean free path for elastic and inelastic collisions and the Debye length. From simplified estimation for a classic Particle-in-Cell approach Debye grid resolution would require up to $10^9$ cells for the discharge chamber of the HCNHIS. These computational requirements make direct solutions of kinetic formulations not attainable.

The fluid formulation represents the processes as conservation laws of the main physical quantities such as mass, momentum and energy. This system of conservation equations requires additional relations in order to be closed. They can be derived directly from the Boltzmann equa-tion with additional assumptions for the distribution function and lead to various levels of approximations (Burgers, 1969; Golant, et al., 1980; Schunk & Nagy, 2000; Zhdanov, 2002). In the case of molecular plasmas due to the presence of internal degrees of freedom of molecules and molecular ions (rotational, vibrational and electronic excitations) the fluid approach requires special treatment. In the simplest case the internal states can be assumed to obey a Boltzmann distribution. In the more general case it is necessary to take into account different excited state as a separate species in an approach that results in separate fluid equations for every considered excited state. This is the basis of the so-called state-to-state approaches (Nagnibeda & Kustova, 2009). In general, fluid approaches lead to systems of unsteady, non-linear PDEs coupled with Maxwell's equations. For the HCNHIS fluid modeling for 21 species including ground state hydrogen molecules and atoms, 14 vibrationally excited hydrogen molecules, 3 positive ions, hydrogen anion, and electrons, under the 5-moment approximation and electrostatic approach would lead to a system of 21 Navier-Stokes type of PDEs with source and sink terms due to inelastic collisions, and a Poisson equation for the electrostatic field. The numerical solution resolving the multiple spatial and time scales (convective, diffusive, elastic and inelastic collisions, electrostatic etc.) is not feasible.

A simplified fluid approach is the so-called "global modeling" which addresses the limitations on computational requirements of fully-resolved fluid approaches. Global models are spatially averaged and provide fundamental plasma parameters and have become a standard tool in predicting main plasma parameters, such as the electron temperature and number density in discharges (Lieberman & Lichtenberg, 2005; Chabert & Braithwaite, 2011). Limitation of global models have been considered by Monahan and Turner (2008) and Monahan and Turner (2009).

In this chapter we begin with a review of the fluid equations for the state-to-state approach as derived from the generalized Boltzmann equations. Then the transport properties, the transport collision integrals, the boundary, and initial conditions applicable for conditions found in the HCNHIS are reviewed.

## 2.1 Wang Chang-Uhlenbeck Equation and Moments of the Distribution Function

The form of fluid equations including multi-fluid multi-temperature equations describing reacting gas mixtures and the derivation of the corresponding transport coefficients from the kinetic equations has been a topic of numerous investigations (Grad, 1949; Chapman & Cowling, 1970; Braginskii, 1965; Burgers, 1969; Ferziger & Kaper, 1972; Schunk, 1977; Golant, et al., 1980; Benilov, 1996; Benilov, 1997; Schunk & Nagy, 2000; Zhdanov, 2002), (Nagnibeda & Kustova, 2009; Meier & Shumlak, 2012; Giovangigli, et al., 2010; Zhang, et al., 2013; Li, et al., 2013) . The starting point of such derivation is the kinetic equation describing the spatial and time evolution of the single-particle velocity distribution function for species $p$ in the excited state $j$. The single-particle velocity distribution function $f_{pj}$ gives the average number of particles in a volume $d\mathbf{r}d\mathbf{v}_{pj} \equiv d^3rd^3v_{pj}$ of the phase space centered at a point $(\mathbf{r},\mathbf{v}_{pj})$ at time $t$ as

$$f_{pj}(t,\mathbf{r},\mathbf{v}_{pj})d\mathbf{r}d\mathbf{v}_{pj} = d^6N_{pj}(t), \tag{2.1}$$

where $N_{pj}$ is the number of particles of species $p$ in the excited state $j$ in a differential phase space volume $d\mathbf{r}d\mathbf{v}_{pj}$.

Partially ionized plasmas including chemical reactions and internal degrees of freedom are described by the generalized Boltzmann equation known as the Wang Chang-Uhlenbeck equation (Kogan, 1969; Zhdanov, 2002; Nagnibeda & Kustova, 2009) for the single-particle velocity distribution function $f_{pj}$

$$\begin{aligned}
&\frac{\partial f_{pj}}{\partial t} + \mathbf{v}_{pj}\cdot\frac{\partial f_{pj}}{\partial \mathbf{r}} + \frac{\mathbf{F}_p}{m_p}\cdot\frac{\partial f_{pj}}{\partial \mathbf{v}_{pj}} = J_{pj}^{\mathrm{el}} + J_{pj}^{\mathrm{int}} + J_{pj}^{\mathrm{chem}},\\
&\mathbf{F}_p = eZ_p\left(\mathbf{E} + \mathbf{v}_{pj}\times\mathbf{B}\right) + m_p\mathbf{G}.
\end{aligned} \tag{2.2}$$

The collision operators for elastic, inelastic, two and three-body chemical reactions in the right hand side have the following form (Nagnibeda & Kustova, 2009)

$$\begin{aligned}
J_{pj}^{\mathrm{el}} &= \sum_{q=1}^{N_s}\sum_{k}\int\left(f_{pj}'f_{qk}' - f_{pj}f_{qk}\right)g_{pq}\sigma\left(g_{pq},\Omega\right)d^2\Omega\, d\mathbf{v}_q,\\
J_{pj}^{\mathrm{int}} &= \sum_{q=1}^{N_s}\sum_{j'kk'}\int\left(f_{pj'}'f_{qk'}'\frac{w_{pj}w_{pk}}{w_{pj'}w_{pk'}} - f_{pj}f_{qk}\right)g_{pq}\sigma_{pqjk}^{j'k'}\left(g_{pq},\Omega\right)d^2\Omega\, d\mathbf{v}_q,\\
J_{pj}^{\mathrm{chem}} &= J_{pj}^{\mathrm{chem},2\rightleftarrows 2} + J_{pj}^{\mathrm{chem},2\rightleftarrows 3},\\
J_{pj}^{\mathrm{chem},2\rightleftarrows 2} &= \sum_{q,p',q'=1}^{N_s}\sum_{j'kk'}\int\left[f_{p'j'}'f_{q'k'}'\frac{w_{pj}w_{pk}}{w_{p'j'}w_{p'k'}}\left(\frac{m_pm_q}{m_{p'}m_{q'}}\right)^3 - f_{pj}f_{qk}\right]g_{pq}\sigma_{pqjk}^{p'q'j'k'}\left(g_{pq},\Omega\right)d^2\Omega\, d\mathbf{v}_q,\\
J_{pj}^{\mathrm{chem},2\rightleftarrows 3} &= \sum_{q,p',q'=1}^{N_s}\sum_{k'}\int\left[f_{pj}'f_{q'}'f_{t'}'h^3w_{pj}\left(\frac{m_p}{m_{q'}m_{t'}}\right)^3 - f_{pj}f_{qk}\right]g_{pq}\sigma_{pqjk}^{\mathrm{diss}}\left(g_{pq},\Omega\right)d^2\Omega\, d\mathbf{v}_q.
\end{aligned} \tag{2.3}$$

In the above expressions $\sigma$ represents differential collision cross section for different elastic and inelastic collision processes and $w_{pj}$ is a statistical weight, $h$ is the Planck constant, $\Omega$ is the solid angle.

The following notation was used in Eq. (2.3) for the distribution functions

$$\begin{aligned} f_{pj} &= f_{pj}(\mathbf{r}, \mathbf{v}_{pj}, t), \\ f'_{pj} &= f_{pj}(\mathbf{r}, \mathbf{v}'_{pj}, t). \end{aligned} \tag{2.4}$$

The first collision operator in Eq. (2.3) describes the elastic collisions between different species in different internal states as in the conventional Boltzmann equation. The second operator corresponds to inelastic collisions leading to excitation and quenching of internal states (in general rotational, vibrational or electronic) which could be written as the following chemical reaction

$$X_p\left(\mathbf{v}_{pj}, j\right) + X_q\left(\mathbf{v}_{qk}, k\right) \leftrightarrows X_p\left(\mathbf{v}'_{pj'}, j'\right) + X_q\left(\mathbf{v}'_{qk'}, k'\right). \tag{2.5}$$

The last two operators $J_{pj}^{\mathrm{chem},2\rightleftarrows 2}$ and $J_{pj}^{\mathrm{chem},2\rightleftarrows 3}$ represent exchange and dissociation reactions which could be written as

$$\begin{aligned} X_p\left(\mathbf{v}_{pj}, j\right) + X_q\left(\mathbf{v}_{qk}, k\right) &\leftrightarrows X_{p'}\left(\mathbf{v}'_{p'j'}, j'\right) + X_{q'}\left(\mathbf{v}'_{q'k'}, k'\right), \\ X_p\left(\mathbf{v}_{pj}, j\right) + X_q\left(\mathbf{v}_{qk}, k\right) &\leftrightarrows X_p\left(\mathbf{v}'_{pj}, j\right) + X_{q'}\left(\mathbf{v}'_{q'}\right) + X_{q'}\left(\mathbf{v}'_{t'}\right). \end{aligned} \tag{2.6}$$

In the above dissociation reaction it is assumed that the dissociating particle is a diatomic molecule. Moreover the cross sections are supposed to be independent from the internal state of the colliding partner $X_p\left(\mathbf{v}_{pj}, j\right)$ and the state of this partner remains the same during collisions. In addition to dissociation this reaction also covers the case of electron impact ionization of atomic species.

The electromagnetic field in the Wang Chang-Uhlenbeck equation (2.2) is self-consistently determined from Maxwell's equations

$$\begin{aligned} \nabla \cdot \mathbf{D} &= e\sum_{p=1}^{N_s} Z_p n_p, \\ \nabla \cdot \mathbf{B} &= 0, \\ \nabla \times \mathbf{E} &= -\frac{\partial \mathbf{B}}{\partial t}, \\ \nabla \times \mathbf{H} &= \mathbf{J} + \frac{\partial \mathbf{D}}{\partial t}, \end{aligned} \tag{2.7}$$

where $\mathbf{E}$ is the electric field vector, $\mathbf{D}$ is the electric displacement vector, $\mathbf{B}$ is the magnetic induction vector, $\mathbf{H}$ is the magnetic field vector, the total current density $\mathbf{J}$ is represented by the sum of the plasma conduction current density

$$\mathbf{J}_{\text{cond}} = \sum_{p=1}^{N_s} eZ_p n_p \mathbf{u}_p , \tag{2.8}$$

the polarization current density $\mathbf{J}_{\text{pol}}$, the magnetization current density $\mathbf{J}_{\text{magn}}$, and the external current density (for example, current in the coil)

$$\mathbf{J} = \mathbf{J}_{\text{cond}} + \mathbf{J}_{\text{pol}} + \mathbf{J}_{\text{magn}} + \mathbf{J}_{\text{ext}}. \tag{2.9}$$

In the vacuum the constitutive relations between vectors $\mathbf{D}$ and $\mathbf{E}$, $\mathbf{H}$ and $\mathbf{B}$ are

$$\begin{aligned} \mathbf{D} &= \epsilon_0 \mathbf{E}, \\ \mathbf{B} &= \mu_0 \mathbf{H}, \end{aligned} \tag{2.10}$$

where $\epsilon_0$ and $\mu_0$ are permittivity and permeability of free space respectively, for a medium Eq. (2.10) are

$$\begin{aligned} \mathbf{D} &= \epsilon \mathbf{E}, \\ \mathbf{B} &= \mu \mathbf{H}. \end{aligned} \tag{2.11}$$

Knowing the distribution functions the macroscopic properties can be calculated by weighted averaging over all velocity space. The number density of species $p$ particles in $j$ excited state can be calculated as

$$n_{pj}(\mathbf{r},t) = \int f_{pj}(\mathbf{r},\mathbf{v}_{pj},t) d\mathbf{v}_{pj} . \tag{2.12}$$

The number density of all particles of species $p$ can be obtained by summing the number densities of all excited states

$$n_p = \sum_j n_{pj} . \tag{2.13}$$

The density of the species $p$ is related to its number density by

$$\rho_p = m_p n_p. \tag{2.14}$$

The density of the mixture is then calculated as

$$\rho = \sum_{p=1}^{N_s} m_p n_p = mn, \tag{2.15}$$

where $m$ is the average particle mass in the mixture and $n = \sum_{p=1}^{N_s} n_p$ is the total number density of the mixture.

In general the average of a property $\mathbf{B}_{pj}(\mathbf{r}, \mathbf{v}_p, t)$ is defined by the following integral

$$\left\langle \mathbf{B}_{pj}(\mathbf{r}, t) \right\rangle = \frac{1}{n_{pj}(\mathbf{r}, t)} \int \mathbf{B}_{pj}(\mathbf{r}, \mathbf{v}_{pj}, t) f_{pj}(\mathbf{r}, \mathbf{v}_{pj}, t) d^3 v_{pj}. \tag{2.16}$$

The average velocity of species $p$ particles in an excited state $j$ is given by

$$\mathbf{u}_{pj}(\mathbf{r}, t) = \frac{1}{n_{pj}(\mathbf{r}, t)} \int \mathbf{v}_{pj} f_{pj}(\mathbf{r}, \mathbf{v}_{pj}, t) d^3 v_{pj} = \left\langle \mathbf{v}_{pj} \right\rangle. \tag{2.17}$$

The averaged velocity of species $p$ in all excited states is given by

$$\mathbf{u}_p = \frac{1}{n_p} \sum_j n_{pj} \mathbf{u}_{pj}. \tag{2.18}$$

The mass-averaged velocity of a mixture as a whole is

$$\mathbf{u} = \frac{1}{mn} \sum_{p=1}^{N_s} m_p n_p \mathbf{u}_p. \tag{2.19}$$

Species $p$ diffusion velocity is defined as

$$\mathbf{w}_{pj} = \mathbf{u}_{pj} - \mathbf{u}. \tag{2.20}$$

Particles peculiar (sometimes called thermal or random) velocity can be introduced with respect to mass-averaged velocity

$$\mathbf{c}_{pj} = \mathbf{v}_{pj} - \mathbf{u} \tag{2.21}$$

or with respect to species averaged velocity

$$\mathbf{c}^*_{pj} = \mathbf{v}_{pj} - \mathbf{u}_{pj} = \mathbf{c}_{pj} + \mathbf{w}_{pj}. \tag{2.22}$$

There are important identities for averaged values of $\mathbf{c}_{pj}$ and $\mathbf{c}^*_{pj}$

$$\begin{aligned} \left\langle \mathbf{c}_{pj} \right\rangle &= \mathbf{w}_{pj}, \\ \left\langle \mathbf{c}^*_{pj} \right\rangle &= \mathbf{0}, \end{aligned} \tag{2.23}$$

and for diffusion velocities $\mathbf{w}_{pj}$

$$\sum_{pj} m_p n_{pj} \mathbf{w}_{pj} = \mathbf{0}. \tag{2.24}$$

Physically significant moments of the distribution function can be derived by using as a reference velocity either species or mass-averaged velocity. The quantities derived by using species averaged velocity will be denoted with star in order to distinguish them from macroscopic properties obtained by means of mass-averaged velocity. The species translational energy per particle (temperature), pressure tensor, scalar pressure, viscous stress tensor, and translational heat flux vector in the mass-averaged system are

$$\begin{aligned}
&U_{\mathrm{tr},p}(\mathbf{r},t) = \frac{3}{2} k_B T_{\mathrm{tr},p}(\mathbf{r},t) = \sum_j \frac{3}{2} k_B T_{\mathrm{tr},pj} = \sum_j \left\langle \frac{1}{2} m_p c_{pj}^2 \right\rangle, \\
&\hat{\mathrm{P}}_p(\mathbf{r},t) = \sum_j \hat{\mathrm{P}}_{pj} = \sum_j n_{pj} \left\langle m_p \mathbf{c}_{pj} \mathbf{c}_{pj} \right\rangle, \\
&p_p(\mathbf{r},t) = \sum_j p_{pj} = \frac{1}{3} \sum_j n_{pj} \left\langle m_p c_{pj}^2 \right\rangle, \\
&\hat{\pi}_p(\mathbf{r},t) = \sum_j \left( p_{pj} \hat{\mathrm{I}} - \hat{\mathrm{P}}_{pj} \right) = \sum_j n_{pj} \left\langle \frac{1}{3} m_p c_{pj}^2 \hat{\mathrm{I}} - m_p \mathbf{c}_{pj} \mathbf{c}_{pj} \right\rangle, \\
&\mathbf{q}_{\mathrm{tr},p}(\mathbf{r},t) = \sum_j \mathbf{q}_{\mathrm{tr},pj} = \sum_j n_{pj} \left\langle \frac{1}{2} m_p c_{pj}^2 \mathbf{c}_{pj} \right\rangle.
\end{aligned} \tag{2.25}$$

The same quantities calculated with respect to the species averaged velocities are

$$\begin{aligned}
&U^*_{\mathrm{tr},p}(\mathbf{r},t) = \frac{3}{2} k_B T^*_{\mathrm{tr},p}(\mathbf{r},t) = \sum_j \frac{3}{2} k_B T^*_{\mathrm{tr},pj} = \sum_j \left\langle \frac{1}{2} m_p c_{pj}^{*\,2} \right\rangle, \\
&\hat{\mathrm{P}}^*_p(\mathbf{r},t) = \sum_j \hat{\mathrm{P}}^*_{pj} = \sum_j n_{pj} \left\langle m_p \mathbf{c}^*_{pj} \mathbf{c}^*_{pj} \right\rangle, \\
&p^*_p(\mathbf{r},t) = \sum_j p^*_{pj} = \frac{1}{3} \sum_j n_{pj} \left\langle m_p c_{pj}^{*\,2} \right\rangle, \\
&\hat{\pi}^*_p(\mathbf{r},t) = \sum_j \hat{\pi}^*_{pj} = \sum_j n_{pj} \left\langle \frac{1}{3} m_p c_{pj}^{*\,2} \hat{\mathrm{I}} - m_p \mathbf{c}^*_{pj} \mathbf{c}^*_{pj} \right\rangle, \\
&\mathbf{q}^*_{\mathrm{tr},p}(\mathbf{r},t) = \sum_j \mathbf{q}^*_{pj} = \sum_j n_{pj} \left\langle \frac{1}{2} m_p c_{pj}^{*\,2} \mathbf{c}^*_{pj} \right\rangle.
\end{aligned} \tag{2.26}$$

The relationship between these two sets of parameters is

$$
\begin{aligned}
T_{\mathrm{tr},p} &= T^*_{\mathrm{tr},p} + \frac{1}{3k_B} m_p w_p^2,\\
\hat{\mathrm{P}}_p &= \hat{\mathrm{P}}^*_p + n_p m_p \mathbf{w}_p \mathbf{w}_p,\\
p_p &= p^*_p + \frac{1}{3} n_p m_p w_p^2,\\
\hat{\pi}_p &= \hat{\pi}^*_p + \frac{1}{3} n_p m_p w_p^2 \hat{\mathrm{I}} - n_p m_p \mathbf{w}_p \mathbf{w}_p,\\
\mathbf{q}_{\mathrm{tr},p} &= \mathbf{q}^*_{\mathrm{tr},p} + \frac{3}{2} p^*_p \mathbf{w}_p + \mathbf{w}_p \cdot \hat{\mathrm{P}}^*_p + \frac{1}{2} n_p m_p \mathbf{w}_p w_p^2.
\end{aligned}
\tag{2.27}
$$

In addition, the molecules could have excited states such as rotational and vibrational. Also atoms and molecules may be electronically excited. In chemical reactions it is the total particle energy that is conserved during collision compared to kinetic energy conservation in elastic collisions. In order to take this into account the internal energy and the internal energy heat fluxes should be introduced. The average internal energy is

$$
U_{\mathrm{int},p}(\mathbf{r},t) = U^*_{\mathrm{int},p}(\mathbf{r},t) = \frac{1}{n_p} \sum_j \epsilon_{pj} n_{pj}. \tag{2.28}
$$

The internal energy heat flux could be introduced with respect to mass-averaged velocity as

$$
\mathbf{q}_{\mathrm{int},p}(\mathbf{r},t) = \sum_j \epsilon_{pj} n_{pj} \left\langle \mathbf{c}_{pj} \right\rangle = \sum_j \epsilon_{pj} n_{pj} \mathbf{w}_{pj} \tag{2.29}
$$

or species averaged velocity as

$$
\mathbf{q}^*_{\mathrm{int},p}(\mathbf{r},t) = \sum_j \epsilon_{pj} n_{pj} \left\langle \mathbf{c}^*_{pj} \right\rangle = \mathbf{0}. \tag{2.30}
$$

Finally, all the properties in Eq. (2.26) can be introduced for the mixture as a whole with respect to mass-averaged velocity as

$$
\begin{aligned}
U_{\mathrm{tr}} &= \frac{1}{n}\sum_{p=1}^{N_s} n_p U_{\mathrm{tr},p}, \\
\hat{\mathrm{P}} &= \sum_{p=1}^{N_s} \hat{\mathrm{P}}_p, \\
p &= \sum_{p=1}^{N_s} p_p, \\
\hat{\pi} &= \sum_{p=1}^{N_s} \hat{\pi}_p, \\
\mathbf{q}_{\mathrm{tr}} &= \sum_{p=1}^{N_s} \mathbf{q}_{\mathrm{tr},p}, \\
U_{\mathrm{int}} &= \frac{1}{n}\sum_{p=1}^{N_s} n_p U_{\mathrm{int},p}, \\
\mathbf{q}_{\mathrm{int}} &= \sum_{p=1}^{N_s} \mathbf{q}_{\mathrm{int},p},
\end{aligned}
\tag{2.31}
$$

and similarly with respect to species averaged velocities

$$
\begin{aligned}
U^*_{\mathrm{tr}} &= \frac{1}{n}\sum_{p=1}^{N_s} n_p U^*_{\mathrm{tr},p}, \\
\hat{\mathrm{P}}^* &= \sum_{p=1}^{N_s} \hat{\mathrm{P}}^*_p, \\
p^* &= \sum_{p=1}^{N_s} p^*_p, \\
\hat{\pi}^* &= \sum_{p=1}^{N_s} \hat{\pi}^*_p, \\
\mathbf{q}^*_{\mathrm{tr}} &= \sum_{p=1}^{N_s} \mathbf{q}^*_{\mathrm{tr},p}, \\
U^*_{\mathrm{int}} &= \frac{1}{n}\sum_{p=1}^{N_s} n_p U^*_{\mathrm{int},p}, \\
\mathbf{q}^*_{\mathrm{int}} &= \sum_{p=1}^{N_s} \mathbf{q}^*_{\mathrm{int},p}.
\end{aligned}
\tag{2.32}
$$

The total energy and heat fluxes of the particles are given with respect to mass-averaged velocity

$$
\begin{aligned}
U &= U_{\mathrm{tr}} + U_{\mathrm{int}}, \\
\mathbf{q} &= \mathbf{q}_{\mathrm{tr}} + \mathbf{q}_{\mathrm{int}},
\end{aligned}
\tag{2.33}
$$

and with respect to species averaged velocities

$$U^* = U^*_{\text{tr}} + U^*_{\text{int}},$$
$$\mathbf{q}^* = \mathbf{q}^*_{\text{tr}} + \mathbf{q}^*_{\text{int}}. \tag{2.34}$$

The species $p$ particle internal energy $\epsilon_{pj}$ in general includes contribution from rotational, vibrational, electronically excited states and the energy of formation. By making the rigid rotor approximation for rotational degrees of freedom and using the anharmonic model based on Morse potential for vibrational degrees of freedom, the internal energy can be written as

$$\epsilon_{pj} = \frac{h^2}{8\pi^2 I_p} l(l+1) + hc\left[\omega_{p,\text{e}}\left(v + \frac{1}{2}\right) - \omega_{p,\text{e}} x_{p,\text{e}}\left(v + \frac{1}{2}\right)^2\right] + \epsilon^{\text{elec}}_{pm} + \epsilon_{p0}, \tag{2.35}$$

where $I_p$ is the moment of inertia of a molecule of species $p$ with respect to its rotation axis, $l$ is the rotational quantum number, $c$ is the speed of light in vacuum, $\omega_{p,\text{e}}$ and $x_{p,\text{e}}$ are the spectroscopic constants in Morse potential model characterizing the frequency of vibrations and their anharmonicity (Nagnibeda & Kustova, 2009), $v$ is the vibrational quantum number, $\epsilon^{\text{elec}}_{pm}$ is the energy of electronic excitation, $\epsilon_{p0}$ is the formation energy.

Following equations (2.25) and (2.26) the equations of state for particles of species $p$ in excited state $j$ can be expressed with respect to mass-averaged and species averaged velocities as

$$p_{pj} = n_{pj} k_B T_{\text{tr},pj},$$
$$p^*_{pj} = n_{pj} k_B T^*_{\text{tr},pj}. \tag{2.36}$$

The system (2.2), (2.7) involves integro-differential equations coupled with the Maxwell's equations. Analytical solutions are available for simple cases such as collisionless, 1-d flows. The direct simulation of the generalized Boltzmann equation (2.2) can be based on particle simulation methods such as Direct Simulation Monte Carlo method (DSMC) developed first for flows of neutral rarefied gases (Bird, 1994) and Particle-in-Cell method (Birdsall, 1991; Hockney & Eastwood, 1988) with various types of approximations for the collisions. Depending on the level of approximation of Maxwell's equations, these particle-solvers are supplemented with either an electrostatic or electromagnetic solver. As we mentioned in the introduction the direct solution of the Boltzmann-Maxwell equations for the HCNHIS is computationally very expensive.

## 2.2 Multi and Single Fluid Equations for State-to-state Approach

Another modeling approach is based on deriving moments (e.g. mass, momentum, total energy etc.) of the Boltzmann equation and obtaining fluid equations for every species $p$ in a state $j$. In order to derive these equations we multiply Eq. (2.2) by a function $\psi_{pj}\left(\mathbf{v}_{pj}\right)$ and integrate over the velocity space to get

$$\begin{aligned}&\int \psi_{pj}\frac{\partial f_{pj}}{\partial t}d^3v_{pj}+\int \psi_{pj}\mathbf{v}_{pj}\cdot\frac{\partial f_{pj}}{\partial \mathbf{r}}d^3v_{pj}\\&+\int \psi_{pj}\frac{\mathbf{F}_p}{m_p}\cdot\frac{\partial f_{pj}}{\partial \mathbf{v}_{pj}}d^3v_{pj}=\int \psi_{pj}\left(J_{pj}^{\mathrm{el}}+J_{pj}^{\mathrm{int}}+J_{pj}^{\mathrm{chem}}\right)d^3v_{pj}.\end{aligned} \tag{2.37}$$

The terms in the left hand side of the above expression can be simplified as follows

$$\begin{aligned}&\int \psi_{pj}\frac{\partial f_{pj}(\mathbf{r},\mathbf{v}_{pj},t)}{\partial t}d^3v_p=\frac{\partial}{\partial t}\left(n_{pj}\left\langle\psi_{pj}\right\rangle\right),\\&\int \psi_{pj}\mathbf{v}_{pj}\cdot\frac{\partial f_{pj}(\mathbf{r},\mathbf{v}_{pj},t)}{\partial \mathbf{r}}d^3v_{pj}=\frac{\partial}{\partial \mathbf{r}}\left(n_{pj}\left\langle\psi_{pj}\mathbf{v}_{pj}\right\rangle\right)\equiv\nabla\cdot\left(n_{pj}\left\langle\psi_{pj}\mathbf{v}_{pj}\right\rangle\right),\\&\int \psi_{pj}\frac{\mathbf{F}_p}{m_p}\cdot\frac{\partial f_{pj}}{\partial \mathbf{v}_{pj}}d^3v_{pj}=-\frac{eZ_p}{m_p}n_{pj}\left(\left\langle\frac{\partial\psi_{pj}}{\partial \mathbf{v}_{pj}}\right\rangle\cdot\mathbf{E}+\left\langle\frac{\partial\psi_{pj}}{\partial \mathbf{v}_{pj}}\cdot\left[\mathbf{v}_{pj}\times\mathbf{B}\right]\right\rangle\right)-n_{pj}\left\langle\frac{\partial\psi_{pj}}{\partial \mathbf{v}_{pj}}\right\rangle\cdot\mathbf{G}.\end{aligned} \tag{2.38}$$

The right hand side of Eq. (2.37) is called the transfer collision integral and represents the change of quantity $\psi_{pj}\left(\mathbf{v}_{pj}\right)$ due to collisions. Its evaluation presents one of the main difficulties while dealing with chemically reacting mixtures described by the Wang Chang-Uhlenbeck equation. For simplicity let us denote the transfer collision integral as follows

$$\int \psi_{pj}(\mathbf{v}_{pj})\left(J_{pj}^{\mathrm{el}}+J_{pj}^{\mathrm{int}}+J_{pj}^{\mathrm{chem}}\right)d^3v_{pj}=\frac{\delta\left(n_{pj}\psi_{pj}\right)}{\delta t}. \tag{2.39}$$

Finally, the moment equation (2.37) can be written as

$$\begin{aligned}&\frac{\partial}{\partial t}\left(n_{pj}\left\langle\psi_{pj}\right\rangle\right)+\nabla\cdot\left(n_{pj}\left\langle\psi_{pj}\mathbf{v}_{pj}\right\rangle\right)-\frac{eZ_p}{m_p}n_{pj}\left[\left\langle\frac{\partial\psi_{pj}}{\partial \mathbf{v}_{pj}}\right\rangle\cdot\mathbf{E}+\left\langle\frac{\partial\psi_{pj}}{\partial \mathbf{v}_{pj}}\cdot\left(\mathbf{v}_{pj}\times\mathbf{B}\right)\right\rangle\right]\\&-n_{pj}\left\langle\frac{\partial\psi_{pj}}{\partial \mathbf{v}_{pj}}\right\rangle\cdot\mathbf{G}=\frac{\delta\left(n_{pj}\psi_{pj}\right)}{\delta t}.\end{aligned} \tag{2.40}$$

Physically significant moments of the Wang Chang-Uhlenbeck equation (2.2) are obtained when the collision invariants (the microscopic properties conserved during a collision) such as $m_p$, $m_p\mathbf{v}_{pj}$, $\frac{1}{2}m_p v_{pj}^2 + \epsilon_{pj}$ are used as a function $\psi_{pj}$. Then Eq. (2.40) can be applied to each of the collision invariants resulting in

$$
\begin{aligned}
&\frac{\partial}{\partial t}\left(m_p n_{pj}\right) + \nabla\cdot\left(m_p n_{pj}\mathbf{u}_{pj}\right) = \frac{\delta}{\delta t}\left(n_{pj} m_p\right),\\
&\frac{\partial}{\partial t}\left(n_{pj}\left\langle m_p\mathbf{v}_{pj}\right\rangle\right) + \nabla\cdot\left(n_{pj}\left\langle m_p\mathbf{v}_{pj}\mathbf{v}_{pj}\right\rangle\right) - eZ_p n_{pj}\left(\mathbf{E}+\mathbf{u}_{pj}\times\mathbf{B}\right)\\
&-m_p n_{pj}\mathbf{G} = \frac{\delta}{\delta t}\left(n_{pj} m_p \mathbf{v}_{pj}\right),\\
&\frac{\partial}{\partial t}\left(n_{pj}\left\langle \frac{m_p v_{pj}^2}{2} + \epsilon_{pj}\right\rangle\right) + \nabla\cdot\left(n_{pj}\left\langle\left(\frac{m_p v_{pj}^2}{2} + \epsilon_{pj}\right)\mathbf{v}_{pj}\right\rangle\right) - eZ_p n_{pj}\mathbf{u}_{pj}\cdot\mathbf{E}\\
&-m_p n_{pj}\mathbf{u}_{pj}\cdot\mathbf{G} = \frac{\delta}{\delta t}\left(n_{pj}\frac{m_p v_{pj}^2}{2} + n_{pj}\epsilon_{pj}\right).
\end{aligned}
\tag{2.41}
$$

After using the following identities

$$
\begin{aligned}
&n_{pj}\left\langle m_p\mathbf{v}_{pj}\mathbf{v}_{pj}\right\rangle = m_p n_{pj}\mathbf{u}_{pj}\mathbf{u}_{pj} + \hat{\mathrm{P}}_{pj}^*,\\
&n_{pj}\left\langle \frac{m_p v_{pj}^2}{2} + \epsilon_{pj}\right\rangle = n_{pj}\frac{m_p u_{pj}^2}{2} + \frac{3}{2}n_{pj}k_B T_{\mathrm{tr},pj}^* + n_{pj}\epsilon_{pj},\\
&n_{pj}\left\langle\left(\frac{m_p v_{pj}^2}{2} + \epsilon_{pj}\right)\mathbf{v}_{pj}\right\rangle = n_{pj}\frac{m_p u_{pj}^2}{2}\mathbf{u}_{pj} + \frac{3}{2}n_{pj}k_B T_{\mathrm{tr},pj}^*\mathbf{u}_{pj} + n_{pj}\epsilon_{pj}\mathbf{u}_{pj} + \mathbf{u}_{pj}\cdot\hat{\mathrm{P}}_{pj}^* + \mathbf{q}_{\mathrm{tr},pj}^*
\end{aligned}
\tag{2.42}
$$

the system (2.41) can be rewritten as

$$
\begin{aligned}
&\frac{\partial n_{pj}}{\partial t} + \nabla\cdot\left(n_{pj}\mathbf{u}_{pj}\right) = \frac{\delta n_{pj}}{\delta t},\\
&\frac{\partial}{\partial t}\left(n_{pj} m_p\mathbf{u}_{pj}\right) + \nabla\cdot\left(n_{pj} m_p\mathbf{u}_{pj}\mathbf{u}_{pj}\right) + \nabla\cdot\hat{\mathrm{P}}_{pj}^* - eZ_p n_{pj}\left(\mathbf{E}+\mathbf{u}_{pj}\times\mathbf{B}\right)\\
&-m_p n_{pj}\mathbf{G} = \frac{\delta}{\delta t}\left(n_{pj} m_p\mathbf{v}_{pj}\right),\\
&\frac{\partial}{\partial t}\left[n_{pj}\left(\frac{m_p u_{pj}^2}{2} + \frac{3}{2}k_B T_{\mathrm{tr},pj}^* + \epsilon_{pj}\right)\right] + \nabla\cdot\left[n_{pj}\mathbf{u}_{pj}\left(\frac{m_p u_{pj}^2}{2} + \frac{3}{2}k_B T_{\mathrm{tr},pj}^* + \epsilon_{pj}\right) + \mathbf{u}_{pj}\cdot\hat{\mathrm{P}}_{pj}^* + \mathbf{q}_{\mathrm{tr},pj}^*\right]\\
&-eZ_p n_{pj}\mathbf{u}_{pj}\cdot\mathbf{E} - m_p n_{pj}\mathbf{u}_{pj}\cdot\mathbf{G} = \frac{\delta}{\delta t}\left(n_{pj}\frac{m_p v_{pj}^2}{2} + n_{pj}\epsilon_{pj}\right).
\end{aligned}
\tag{2.43}
$$

By taking the dot product of the momentum equation of the system (2.43) and $\mathbf{u}_{pj}$, multiplying the continuity equation of (2.43) by $\epsilon_{pj} - \frac{1}{2} m_p u_{pj}^2$ and subtracting the sum of resulting equations from the energy equation of the system (2.43) it is possible to eliminate the kinetic energy of the bulk motion and the internal energy of the species $p$ in a state $j$ fluid. The resulting energy equation becomes

$$\frac{\partial}{\partial t}\left(\frac{3}{2} n_{pj} k_B T^*_{\mathrm{tr},pj}\right) + \nabla \cdot \left(\frac{3}{2} n_{pj} k_B T^*_{\mathrm{tr},pj} \mathbf{u}_{pj} + \mathbf{q}^*_{\mathrm{tr},pj}\right) + \hat{\mathrm{P}}^*_{pj} : \nabla \mathbf{u}_{pj} = \frac{\delta}{\delta t}\left(n_{pj} \frac{m_p c^{*\,2}_{pj}}{2}\right), \tag{2.44}$$

where $\frac{\delta}{\delta t}\left(n_{pj} \frac{m_p c^{*\,2}_{pj}}{2}\right) = \int \frac{m_p c^{*\,2}_{pj}}{2} \left(J^{\mathrm{el}}_{pj} + J^{\mathrm{int}}_{pj} + J^{\mathrm{chem}}_{pj}\right) d^3 v_{pj}$ .

In a similar way, by multiplying the continuity equation of (2.43) by $m_p \mathbf{u}_{pj}$ and subtracting the result from the momentum equation of (2.43) it can be written as

$$\begin{aligned} &m_p n_{pj} \frac{\partial \mathbf{u}_{pj}}{\partial t} + m_p n_{pj} \left(\mathbf{u}_{pj} \cdot \nabla\right) \mathbf{u}_{pj} + \nabla \cdot \hat{\mathrm{P}}^*_{pj} - e Z_p n_{pj} \left(\mathbf{E} + \mathbf{u}_{pj} \times \mathbf{B}\right) \\ &- m_p n_{pj} \mathbf{G} = \frac{\delta}{\delta t}\left(n_{pj} m_p \mathbf{c}^*_{pj}\right). \end{aligned} \tag{2.45}$$

Thus, the fluid equations can be written in the following form

$$\begin{aligned} &\frac{\partial n_{pj}}{\partial t} + \nabla \cdot \left(n_{pj} \mathbf{u}_{pj}\right) = \frac{\delta n_{pj}}{\delta t}, \\ &m_p n_{pj} \frac{\partial \mathbf{u}_{pj}}{\partial t} + m_p n_{pj} \left(\mathbf{u}_{pj} \cdot \nabla\right) \mathbf{u}_{pj} + \nabla \cdot \hat{\mathrm{P}}^*_{pj} - e Z_p n_{pj} \left(\mathbf{E} + \mathbf{u}_{pj} \times \mathbf{B}\right) \\ &- m_p n_{pj} \mathbf{G} = \frac{\delta}{\delta t}\left(n_{pj} m_p \mathbf{c}^*_{pj}\right), \\ &\frac{\partial}{\partial t}\left(\frac{3}{2} n_{pj} k_B T^*_{\mathrm{tr},pj}\right) + \nabla \cdot \left(\frac{3}{2} n_{pj} k_B T^*_{\mathrm{tr},pj} \mathbf{u}_{pj} + \mathbf{q}^*_{\mathrm{tr},pj}\right) + \hat{\mathrm{P}}^*_{pj} : \nabla \mathbf{u}_{pj} = \frac{\delta}{\delta t}\left(n_{pj} \frac{m_p c^{*\,2}_{pj}}{2}\right). \end{aligned} \tag{2.46}$$

By splitting the pressure tensor into scalar pressure and viscous stress tensor the above system can be recast as

$$\frac{\partial n_{pj}}{\partial t}+\nabla\cdot\left(n_{pj}\mathbf{u}_{pj}\right)=\frac{\delta n_{pj}}{\delta t},$$
$$m_p n_{pj}\frac{\partial \mathbf{u}_{pj}}{\partial t}+m_p n_{pj}\left(\mathbf{u}_{pj}\cdot\nabla\right)\mathbf{u}_{pj}+\nabla p^*_{pj}-\nabla\cdot\hat{\pi}^*_{pj}-eZ_p n_{pj}\left(\mathbf{E}+\mathbf{u}_{pj}\times\mathbf{B}\right)$$
$$-m_p n_{pj}\mathbf{G}=\frac{\delta}{\delta t}\left(n_{pj}m_p\mathbf{c}^*_{pj}\right), \tag{2.47}$$
$$\frac{\partial}{\partial t}\left(\frac{3}{2}n_{pj}k_B T^*_{\mathrm{tr},pj}\right)+\nabla\cdot\left(\frac{5}{2}n_{pj}k_B T^*_{\mathrm{tr},pj}\mathbf{u}_{pj}+\mathbf{q}^*_{\mathrm{tr},pj}\right)+\mathbf{u}_{pj}\cdot\nabla p^*_{pj}-\hat{\pi}^*_{pj}:\nabla\mathbf{u}_{pj}=\frac{\delta}{\delta t}\left(n_{pj}\frac{m_p c^{*\,2}_{pj}}{2}\right).$$

It is also necessary to obtain an equation for the mixture as a whole. Summing up the fluid equations (2.43) over all species and internal states, and using the following identities representing collision invariance of $m_p$, $m_p\mathbf{v}_{pj}$, $\frac{1}{2}m_p v^2_{pj}+\epsilon_{pj}$

$$\sum_{pj}\frac{\delta}{\delta t}\left(n_{pj}m_p\right)=0,$$
$$\sum_{pj}\frac{\delta}{\delta t}\left(n_{pj}m_p\mathbf{v}_{pj}\right)=0, \tag{2.48}$$
$$\sum_{pj}\frac{\delta}{\delta t}\left(n_{pj}\frac{m_p v^2_{pj}}{2}+n_{pj}\epsilon_{pj}\right)=0,$$

the fluid equations for the plasma as a whole can be written as

$$\frac{\partial}{\partial t}\left(mn\right)+\nabla\cdot\left(mn\mathbf{u}\right)=0,$$
$$\frac{\partial}{\partial t}\left(mn\mathbf{u}\right)+\nabla\cdot\left(nm\mathbf{u}\mathbf{u}\right)+\nabla\cdot\left(\hat{\mathrm{P}}^*+\sum_{pj}n_{pj}m_p\mathbf{w}_{pj}\mathbf{w}_{pj}\right)$$
$$-\sum_{pj}eZ_p n_{pj}\left(\mathbf{E}+\mathbf{u}_{pj}\times\mathbf{B}\right)-mn\mathbf{G}=\mathbf{0},$$
$$\frac{\partial}{\partial t}\left[n\left(\frac{mu^2}{2}+U^*_{\mathrm{tr}}+U^*_{\mathrm{int}}\right)+\sum_{pj}n_{pj}\frac{m_p w^2_{pj}}{2}\right]$$
$$+\nabla\cdot\left[\left(n\frac{mu^2}{2}+nU^*_{\mathrm{tr}}+nU^*_{\mathrm{int}}+\sum_{pj}n_{pj}\frac{m_p w^2_{pj}}{2}\right)\mathbf{u}\right]$$
$$+\nabla\cdot\left[\mathbf{u}\cdot\left(\hat{\mathrm{P}}^*+\sum_{pj}n_{pj}m_p\mathbf{w}_{pj}\mathbf{w}_{pj}\right)+\sum_{pj}n_{pj}\frac{m_p}{2}w^2_{pj}\mathbf{w}_{pj}\right]$$
$$+\nabla\cdot\left(\sum_{pj}n_{pj}U^*_{\mathrm{tr},p}\mathbf{w}_{pj}+\mathbf{q}_{\mathrm{int}}+\sum_{pj}\mathbf{w}_{pj}\cdot\hat{\mathrm{P}}^*_{pj}+\mathbf{q}^*_{\mathrm{tr}}\right)-\sum_{pj}eZ_p n_{pj}\mathbf{u}_{pj}\cdot\mathbf{E}-mn\mathbf{u}\cdot\mathbf{G}=0. \tag{2.49}$$

The above system can be also written by using parameters introduced with respect to the mass-averaged velocity as

$$
\begin{aligned}
&\frac{\partial}{\partial t}\left(mn\right)+\nabla\cdot\left(mn\mathbf{u}\right)=0,\\
&\frac{\partial}{\partial t}\left(mn\mathbf{u}\right)+\nabla\cdot\left(nm\mathbf{u}\mathbf{u}\right)+\nabla\cdot\hat{\mathrm{P}}-\sum_{pj}eZ_p n_{pj}\left(\mathbf{E}+\mathbf{u}_{pj}\times\mathbf{B}\right)-mn\mathbf{G}=\mathbf{0},\\
&\frac{\partial}{\partial t}\left[n\frac{mu^2}{2}+n\left(U_{\mathrm{tr}}+U_{\mathrm{int}}\right)\right]+\nabla\cdot\left[n\frac{mu^2}{2}\mathbf{u}+n\left(U_{\mathrm{tr}}+U_{\mathrm{int}}\right)\mathbf{u}+\mathbf{u}\cdot\hat{\mathrm{P}}+\mathbf{q}_{\mathrm{int}}+\mathbf{q}_{\mathrm{tr}}\right]\\
&-\sum_{pj}eZ_p n_{pj}\mathbf{u}_{pj}\cdot\mathbf{E}-mn\mathbf{u}\cdot\mathbf{G}=0.
\end{aligned}
\tag{2.50}
$$

Applying procedures similar to those used to derive Eqs. (2.46) the above system can be further reduced to

$$
\begin{aligned}
&\frac{\partial}{\partial t}\left(mn\right)+\nabla\cdot\left(mn\mathbf{u}\right)=0,\\
&mn\frac{\partial\mathbf{u}}{\partial t}+nm\left(\mathbf{u}\cdot\nabla\right)\mathbf{u}+\nabla\cdot\hat{\mathrm{P}}-\sum_{pj}eZ_p n_{pj}\left(\mathbf{E}+\mathbf{u}_{pj}\times\mathbf{B}\right)-mn\mathbf{G}=\mathbf{0},\\
&\frac{\partial}{\partial t}\left[n\left(U_{\mathrm{tr}}+U_{\mathrm{int}}\right)\right]+\nabla\cdot\left[n\left(U_{\mathrm{tr}}+U_{\mathrm{int}}\right)\mathbf{u}+\mathbf{q}_{\mathrm{int}}+\mathbf{q}_{\mathrm{tr}}\right]\\
&+\hat{\mathrm{P}}:\nabla\mathbf{u}-\sum_{pj}eZ_p n_{pj}\mathbf{w}_{pj}\cdot\mathbf{E}=0.
\end{aligned}
\tag{2.51}
$$

## 2.3 Transport Properties of Partially Ionized Plasmas in a State-to-state Approach for Internal Degrees of Freedom

The system of 5 scalar fluid equations (2.43) (or the equivalent system (2.47) is not closed because it contains 13 unknown variables, namely: $n_{pj}$, 3 components of $\mathbf{u}_{pj}$ and $\mathbf{q}^*_{\mathrm{tr},pj}$, and 6 components of $\hat{\mathrm{P}}^*_{pj}$. Note that $T^*_{pj}$ is excluded from the list of unknown variables since it can be written as a function of $n_{pj}$ and $p^*_{pj}$ from the equation of state (2.36). The other difficulty associated with the moment approach is the evaluation of the transfer collision integrals in Eq. (2.39) that depend on the particular type of distribution function.

In order to close the systems of multi-fluid, multi-temperature equations such as (2.43) or (2.47) it is necessary to assume specific forms of the distribution function and use them to calculate transport properties and collision integrals. A number of different approaches have been developed including the Chapman-Enskog method (Chapman & Cowling, 1970), its modifications for two-temperature plasmas (Ferziger & Kaper, 1972), its extension for vibrationally and chemically non-equilibrium flows with the same translational temperature for all species (Nagnibeda & Kustova, 2009; Capitelli, et al., 2013), and Grad's method (Grad, 1949; Burgers, 1969; Schunk, 1977). We summarize these approaches here for the sake of completeness.

In the Chapman-Enskog method the distribution function is expanded in a power series containing a small parameter (Nagnibeda & Kustova, 2009; Capitelli, et al., 2013)

$$f_{pj} = f_{0pj}\left(1 + \sum_{k=1}^{\infty} \phi_{pj}^{(k)} \epsilon^k\right), \tag{2.52}$$

where $f_{0pj} = n_{pj}\left(\frac{m_p}{2k_B T}\right)^{3/2} e^{-\frac{m_p c_p^2}{2k_B T}}$ is the equilibrium distribution function, $\epsilon$ is a formal small parameter and $T$ is the temperature assumed to be the same for all species. Then the expansion (2.52) is substituted into the Boltzmann equation (2.2) where the inverse of a small parameter $\epsilon$ appears in the right hand side in front of one of the collision integrals. Essentially the Maxwellian distribution function $f_{0pj}$ in Eq. (2.52) is a solution of an integral equation for the elastic collision kernel (2.3) corresponding to zero-order in $\epsilon$. The solution of the integral equations for higher orders of $\epsilon$ gives the higher-order corrections $\phi_{pj}^{(k)}$ for the distribution function $f_{pj}$ which combined with Eqs. (2.25) provide transport coefficients. In Chapman-Enskog's original work on rarefied gases the small parameter $\epsilon$ was set to be the Knudsen number. Extensions for two-temperature plasmas assigned $\epsilon$ to the electron to heavy-particle mass ratio (Ferziger & Kaper, 1972) and for vibrational-chemical non-equilibrium high enthalpy flows assigned epsilon to the ratio of elastic mean collision time to inelastic one (Nagnibeda & Kustova, 2009). The Chapman-Enskog approach is valid only for relatively small deviations from the equilibrium state where the resulting distribution function is written in terms of the first moments (number density, velocity and temperature).

In the Grad's method the distribution function is expanded in the infinite series of Hermite polynomials around a zero-order approximation function. The typical choice of this approximation function is a drifting Maxwellian distribution with mass-averaged or species-averaged drift velocities depending on a particular configuration and in general different temperatures (Zhdanov, 2002). The former results in the single-fluid approximation while the latter in the mul-ti-fluid approximation. In order to close the infinite set of equations, the expansion is broken at some order. Then the expansion is substituted into Boltzmann equation (2.2) and integrated over velocity space in order to obtain the system of equations for the coefficients of the expansion. The common examples are 5-, 8-, 10-, 13- and 20-moment approximations named by the number of moments used (Schunk, 1977).

In this work, it is assumed that the drift velocities of the plasma species in the discharge chamber of the HCNHIS are much smaller than the species thermal velocities

$$w_p \ll \sqrt{\frac{8k_B T_p}{\pi m_p}}. \tag{2.53}$$

Therefore, all terms containing second-order drift velocity in Eq. (2.27) are neglected leading to the following expressions

$$\begin{aligned} T_{\mathrm{tr},p} &= T^*_{\mathrm{tr},p}, \\ \hat{\mathrm{P}}_p &= \hat{\mathrm{P}}^*_p, \\ p_p &= p^*_p, \\ \hat{\pi}_p &= \hat{\pi}^*_p, \\ \mathbf{q}_{\mathrm{tr},p} &= \mathbf{q}^*_{\mathrm{tr},p} + \frac{3}{2} p^*_p \mathbf{w}_p + \mathbf{w}_p \cdot \hat{\mathrm{P}}^*_p. \end{aligned} \tag{2.54}$$

It is also assumed that the distribution function of each molecular species in the bulk of the plasma can be represented by the drifting Maxwellian

$$f^0_{pj}\left(\mathbf{r}, \mathbf{v}_{pj}, t\right) = n_{pj} \left(\frac{m_p}{2\pi k_B T^*_{\mathrm{tr},pj}}\right)^{3/2} e^{-\frac{m_p\left(\mathbf{v}_{pj}-\mathbf{u}_{pj}\right)^2}{2k_B T^*_{\mathrm{tr},pj}}} \frac{1}{Q^{\mathrm{rot}}_p} g_l e^{-\frac{\epsilon_l}{k_B T^*_{\mathrm{tr},pj}}}, \tag{2.55}$$

where $Q^{\mathrm{rot}}_{pj} = \sum_{\mathrm{rot}} g_l e^{-\frac{\epsilon_l}{k_B T^*_{\mathrm{tr},pj}}}$ is the partition function for rotational degree of freedom, the multi-index $j$ describing internal degrees of freedom in this case does not include rotational contribution. For atoms the distribution function does not contain part related to rotational degree of freedom

$$f^0_{pj}\left(\mathbf{r}, \mathbf{v}_{pj}, t\right) = n_{pj} \left(\frac{m_p}{2\pi k_B T^*_{\mathrm{tr},pj}}\right)^{3/2} e^{-\frac{m_p\left(\mathbf{v}_{pj}-\mathbf{u}_{pj}\right)^2}{2k_B T^*_{\mathrm{tr},pj}}}. \tag{2.56}$$

This approximation results to zero viscous stress tensor and heat flux vector defined with respect to the species averaged velocities

$$\begin{aligned} \hat{\pi}^{*}_{pj} &= 0, \\ \mathbf{q}^{*}_{pj} &= \mathbf{0}. \end{aligned} \tag{2.57}$$

The viscous stress tensor and heat flux vector using the mass-averaged velocities are

$$\begin{aligned} \hat{\pi}_{p} &= 0, \\ \mathbf{q}_{\mathrm{tr},p} &= \frac{5}{2} p^{*}_{p} \mathbf{w}_{p}. \end{aligned} \tag{2.58}$$

The assumption of the negligible viscosity in the bulk of the plasma is often used while describing gas discharges (Golant, et al., 1980).

Assuming that the translational temperature exceeds the characteristic rotational temperature of the molecular species $p$ in the internal state $j$ it is possible to calculate the average internal energy of this molecular species as (Vincenti & Kruger, 1967)

$$U_{\mathrm{int},pj} = n_{pj} \left( \epsilon'_{pj} + k_B T_{pj,\mathrm{tr}} \right), \tag{2.59}$$

where

$$\epsilon'_{pj} = hc \left[ \omega_{p,\mathrm{e}} \left( v + \frac{1}{2} \right) - \omega_{p,\mathrm{e}} x_{p,\mathrm{e}} \left( v + \frac{1}{2} \right)^2 \right] + \epsilon^{\mathrm{elec}}_{pm} + \epsilon_{p0} \tag{2.60}$$

represents internal energy contributions except rotational energy.

With the assumption (2.53) the distribution function (2.55) can be approximated by (Zhdanov, 2002)

$$f_{pj} \left( \mathbf{r}, \mathbf{v}_{pj}, t \right) = n_{pj} \left( \frac{m_p}{2\pi k_B T_{pj}} \right)^{3/2} e^{-\frac{m_p \left( \mathbf{v}_{pj} - \mathbf{u} \right)^2}{2 k_B T_{pj}}} \left[ 1 - \frac{m_p}{k_B T_{pj}} \mathbf{w}_{pj} \cdot \left( \mathbf{v}_{pj} - \mathbf{u} \right) \right]. \tag{2.61}$$

The system of multi-fluid equations that can be derived using Eq. (2.56) is called quasi-hydrodynamic approximation.

In order to take into account viscous and heat transfer effects it is necessary to incorporate additional terms in the expression for the distribution function. Benilov (1997) suggested to use in the Eq. (2.43) or (2.47) the transport coefficients derived for a simple gas given, for example, by Bird et al. (2002) or Ferziger and Kaper (1972). The heat flux is written in terms of the temperature gradient

$$\mathbf{q}^{*}_{\mathrm{tr},pj} = \kappa_{\mathrm{tr},pj} \nabla T^{*}_{pj}, \tag{2.62}$$

where heat conductivity corresponding to translational motion is expressed as

$$\kappa_{\mathrm{tr},pj} = \frac{25}{16} \frac{k_B T_{pj}}{\Omega^{(2,2)}_{pj}} c_v . \tag{2.63}$$

The term $\Omega^{(i,j)}_{pj}$ in the above expression is given by

$$\Omega^{(\alpha,\beta)}_{pj} = \sqrt{\frac{k_B T_{pj,tk}}{2\pi \mu_{pt}}} \int_0^\infty x^{2\alpha+3} e^{-x^2} \Sigma^{(\beta)}_{pj,tk} \left( x \sqrt{\frac{2k_B T_{pj,tk}}{\mu_{pt}}} \right) dx \,, \tag{2.64}$$

where $\mu_{pt} = \dfrac{m_p m_t}{m_p + m_t}$ is the reduced mass, $T_{pj,tk} = \dfrac{m_t T_{pj} + m_p T_{tk}}{m_p + m_t}$ is the effective temperature of two species, and the transport cross section is written as

$$\Sigma^{\beta}_{pj,tk} \left( g_{pj,st} \right) = \int_0^\pi \sigma_{pj,tk} \left( g_{pj,st}, \chi \right) \left( 1 - \cos^\beta \chi \right) \sin \chi d\chi \,. \tag{2.65}$$

The viscous stress tensor is given by

$$\hat{\pi}^{*}_{pj} = -\mu \left[ \nabla \mathbf{u}_{pj} + \left( \nabla \mathbf{u}_{pj} \right)^T \right] - \left( \frac{2}{3} \mu - \kappa \right) \left( \nabla \cdot \mathbf{u}_{pj} \right) \hat{\mathrm{I}}, \tag{2.66}$$

where viscosity is written by

$$\mu = \frac{5}{8} \frac{k_B T_{pj}}{\Omega^{(2,2)}_{pj}} . \tag{2.67}$$

## 2.4 Transfer Collision Integrals for Chemically Reacting Flows

The calculation of the transport integrals even in the simplest case of Maxwellian distribution functions with arbitrary drift velocities and species temperatures for a simple two body chemical reaction is quite cumbersome (Benilov, 1997). Inclusion of higher order approximations gives rise to much more complicated expressions. As it was noted in Nagnibeda and Kustova (2009) the first-order corrections to the rate coefficients in the continuity equations are typically small unless very large gradients present in the flow field. Therefore these corrections are neglected in the continuity equations in the present model for the HCNHIS.

The case of elastic collisions has been addressed by many researchers (Braginskii, 1965; Burgers, 1969; Schunk, 1977; Golant, et al., 1980; Benilov, 1997). For Maxwellian distribution functions with different drift velocities and species temperatures as in the case of Eq. (2.56) the appropriate terms were derived by Benilov (1997)

$$
\begin{aligned}
&\frac{\delta}{\delta t}\left(n_{pj}\right)_{\text{elast}} = 0,\\
&\frac{\delta}{\delta t}\left(n_{pj} m_{p} \mathbf{v}_{pj}\right)_{\text{elast}} = -\sum_{t=1}^{N_s}\sum_{k}\frac{n_{pj} n_{tk} k_B T_{pj,tk}}{n D_{pj,tk}}\left(\mathbf{u}_{pj}-\mathbf{u}_{tk}\right),\\
&\frac{\delta}{\delta t}\left(n_{pj}\frac{m_p v_{pj}^2}{2}+n_{pj}\epsilon_{pj}\right)_{\text{elast}} = \frac{\delta}{\delta t}\left(n_{pj}\frac{m_p v_{pj}^2}{2}\right)_{\text{elast}}\\
&= -\sum_{t=1}^{N_s}\sum_{k}\frac{n_{pj} n_{tk} k_B}{n D_{pj,tk}\left(m_p + m_t\right)}\left[3\frac{Q_{pj,tk}^{(e)}}{Q_{pj,tk}^{(m)}} k_B T_{pj,tk}\left(T_{pj}-T_{tk}\right)+\left(m_p T_{tk}\mathbf{u}_{pj}+m_t T_{pj}\mathbf{u}_{tk}\right)\cdot\left(\mathbf{u}_{pj}-\mathbf{u}_{tk}\right)\right],
\end{aligned}
\tag{2.68}
$$

where the effective diffusion coefficient $D_{pj,tk}$ is given by

$$
D_{pj,tk} = \frac{3\pi}{16}\sqrt{\frac{2k_B T_{pj,tk}}{\pi\mu_{pt}}}\frac{1}{nQ_{pj,tk}^{(m)}}. \tag{2.69}
$$

The effective temperature and reduced mass of two species $p$ and $t$ in internal states $j$ and $k$ are defined respectively

$$
\begin{aligned}
T_{pj,tk} &= \frac{m_t T_{pj} + m_p T_{tk}}{m_p + m_t},\\
\mu_{pt} &= \frac{m_p m_t}{m_p + m_t}.
\end{aligned}
\tag{2.70}
$$

The velocity averaged elastic momentum and energy transfer collision cross section and species relative Mach number are

$$
\begin{aligned}
Q_{pj,kt}^{(m)} &= e^{-M_{pj,tk}^2}\int_0^\infty x^5 e^{-x^2}\sigma_{pj,tk}^{(m)}\left(x\sqrt{\frac{2k_B T_{pj,tk}}{\mu_{pt}}}\right)F^{(m)}\left(2M_{pj,tk}x\right)dx,\\
Q_{pj,kt}^{(e)} &= e^{-M_{pj,tk}^2}\int_0^\infty x^5 e^{-x^2}\sigma_{pj,tk}^{(m)}\left(x\sqrt{\frac{2k_B T_{pj,tk}}{\mu_{pt}}}\right)F^{(n)}\left(2M_{pj,tk}x\right)dx,\\
M_{pj,tk} &= \sqrt{\frac{\mu_{pt}}{2k_B T_{pj,tk}}}\left|\mathbf{u}_{pj}-\mathbf{u}_{tk}\right|,
\end{aligned}
\tag{2.71}
$$

where $\sigma^{(m)}_{pj,tk} = \Sigma^{(1)}_{pj,tk}$ is the momentum transport collision cross section given by Eq. (2.65).

In the expressions (2.71) the functions $F^{(m)}(y)$ and $F^{(n)}(y)$ are defined as

$$\begin{aligned} F^{(m)}(y) &= 3\frac{y\cosh(y)-\sinh(y)}{y^3}, \\ F^{(n)}(y) &= \frac{\sinh(y)}{y}. \end{aligned} \tag{2.72}$$

For binary exchange chemical reactions

$$X_p\left(\mathbf{v}_{pj}, j\right) + X_t\left(\mathbf{v}_{tk}, k\right) \to X_{p'}\left(\mathbf{v}'_{p'j'}, j'\right) + X_{t'}\left(\mathbf{v}'_{t'k'}, k'\right), \tag{2.73}$$

the transfer collision integrals become more complicated. For example, the losses of the particles of species $p$ in an excited state $j$ due to binary reaction (2.73) with the particles of species $t$ in an excited state $k$ (see also Appendix A for the details of the rate coefficients for continuity equations derivation, numerical calculation and fitting) have been derived by Benilov (1996)

$$\begin{aligned} \frac{\delta}{\delta t}\left(n_{pj}\right)_{\text{chem,destr}} &= -k_{pj,tk} n_{pj} n_{tk}, \\ \frac{\delta}{\delta t}\left(n_{pj} m_p \mathbf{v}_{pj}\right)_{\text{chem,destr}} &= -k_{pj,tk} n_{pj} n_{tk} m_p \left[\mathbf{u}_{pj} - \frac{\mu_{pt} T_{pj}}{m_p T_{pj,tk}}\left(\frac{2}{3}\frac{Q^{(m)}_{pj,tk}}{Q^{(n)}_{pj,tk}} - 1\right)\left(\mathbf{u}_{tk} - \mathbf{u}_{pj}\right)\right], \\ \frac{\delta}{\delta t}\left(n_{pj}\frac{m_p v^2_{pj}}{2} + n_{pj}\epsilon_{pj}\right)_{\text{chem,destr}} &= \frac{\delta}{\delta t}\left(n_{pj}\frac{m_p v^2_{pj}}{2}\right)_{\text{chem}} + \epsilon_{pj}\frac{\delta}{\delta t}\left(n_{pj}\right)_{\text{chem,destr}}, \end{aligned} \tag{2.74}$$

$$\begin{aligned} \frac{\delta}{\delta t}\left(n_{pj}\frac{m_p v^2_{pj}}{2}\right)_{\text{chem,destr}} &= -k_{pj,tk} n_{pj} n_{tk}\left[\frac{3k_B T_{pj}}{2} + \frac{m_p u^2_{pj}}{2}\right. \\ &+ \frac{\mu_{pt} T_{pj}}{m_p T_{pj,tk}}\left(\frac{2}{3}\frac{Q^{(e)}_{pj,tk}}{Q^{(n)}_{pj,tk}} + \frac{2}{3}M^2_{pj,tk}\left(1 - \frac{4}{3}\frac{Q^{(m)}_{pj,tk}}{Q^{(n)}_{pj,tk}}\right) - 1\right)\frac{3k_B T_{pj}}{2} \\ &\left. - \frac{\mu_{pt} T_{pj}}{m_p T_{pj,tk}}\left(\frac{2}{3}\frac{Q^{(m)}_{pj,tk}}{Q^{(n)}_{pj,tk}} - 1\right) m_p \mathbf{u}_{pj}\left(\mathbf{u}_{pj} - \mathbf{u}_{tk}\right)\right], \end{aligned}$$

where $k_{pj,tk} = 4\sqrt{\frac{2k_B T_{pj.tk}}{\pi\mu_{pt}}} Q^{(n)}_{pj,tk}$, $Q^{(n)}_{pj,tk} = e^{-M^2_{pj,tk}}\int_0^\infty x^3 e^{-x^2}\sigma^{(n)}_{pj,tk}\left(x\sqrt{\frac{2k_B T_{pj,tk}}{\mu_{pt}}}\right) F^{(n)}\left(2M_{pj,tk}x\right)dx$,

and $\sigma^{(n)}_{pj,tk} = \Sigma^{0}_{pj,tk}$. For the derivation of the expressions for $k_{pj,tk}$ see also Appendix A.

For the production of particles of species $p'$ in an excited state $j'$ in a reaction (2.73) the transfer collision integrals are (Benilov, 1996)

$$\begin{aligned}
&\frac{\delta}{\delta t}\left(n_{p'j'}\right)_{\text{chem,prod}} = k_{pj,tk} n_{pj} n_{tk},\\
&\frac{\delta}{\delta t}\left(n_{p'j'} m_{p'} \mathbf{v}_{p'j'}\right)_{\text{chem,prod}} = k_{pj,tk} n_{pj} n_{tk} \frac{m_{p'}}{m_p + m_t}\left[m_p \mathbf{u}_{pj} + m_t \mathbf{u}_{tk}\right]\\
&-k_{pj,tk} n_{pj} n_{tk} \frac{m_{p'}}{m_p + m_t}\left[\mu_{pt} \frac{T_{pj} - T_{tk}}{T_{pj,tk}}\left(\frac{2}{3}\frac{Q^{(m)}_{pj,tk}}{Q^{(n)}_{pj,tk}} - 1\right) + \frac{2}{3}\frac{\tilde{Q}^{(m)}_{pj,tk}}{Q^{(n)}_{pj,tk}} m_{t'}\sqrt{\frac{\mu_{pt}}{\mu_{p't'}}}\right]\left(\mathbf{u}_{tk} - \mathbf{u}_{pj}\right),\\
&\frac{\delta}{\delta t}\left(n_{p'j'}\frac{m_{p'} v^2_{p'j'}}{2} + n_{p'j'}\epsilon_{p'j'}\right)_{\text{chem,prod}} = \frac{\delta}{\delta t}\left(n_{p'j'}\frac{m_{p'} v^2_{p'j'}}{2}\right)_{\text{chem,prod}} + \epsilon_{p'j'}\frac{\delta}{\delta t}\left(n_{p'j'}\right)_{\text{chem,prod}},\\
&\frac{\delta}{\delta t}\left(n_{p'j'}\frac{m_{p'} v^2_{p'j'}}{2}\right)_{\text{chem,prod}} = k_{pj,tk} n_{pj} n_{tk}\frac{m_{p'}}{m_p + m_t}\left[\frac{3k_B\left(T_{pj} + T_{tk}\right)}{2} + \frac{m_p u^2_{pj} + m_t u^2_{tk}}{2}\right]\\
&+k_{pj,tk} n_{pj} n_{tk}\frac{m_{p'}}{m_p + m_t}\left(\frac{2}{3}\frac{Q^{(e)}_{pj,tk}}{Q^{(n)}_{pj,tk}} + \frac{2}{3}M^2_{pj,tk}\left(1 - \frac{4}{3}\frac{Q^{(m)}_{pj,tk}}{Q^{(n)}_{pj,tk}}\right) - 1\right)\frac{3k_B\mu_{pt}}{2T_{pj,tk}}\left(\frac{m_p T^2_{pj} + m_t T^2_{tk}}{m_p + m_t}\right)\\
&-k_{pj,tk} n_{pj} n_{tk}\frac{m_{p'}}{m_p + m_t}\left(\frac{2}{3}\frac{Q^{(m)}_{pj,tk}}{Q^{(n)}_{pj,tk}} - 1\right)\mu_{pt}\left(\mathbf{u}_{tk} - \mathbf{u}_{pj}\right)\frac{T_{pj}\mathbf{u}_{pj} - T_{tk}\mathbf{u}_{tk}}{T_{pj,tk}}\\
&+k_{pj,tk} n_{pj} n_{tk}\frac{m_{t'}}{m_p + m_t}E_{\text{th}} - k_{pj,tk} n_{pj} n_{tk}\frac{m_{p'} - m_{t'}}{m_p + m_t}\frac{Q^{(e)}_{pj,tk}}{Q^{(n)}_{pj,tk}}k_B T_{pj,tk}
\end{aligned} \tag{2.75}$$

$$\begin{aligned}
&-k_{pj,tk} n_{pj} n_{tk}\frac{m_{p'} m_{t'}}{m_p + m_t}\frac{2}{3}\frac{\tilde{Q}^{(m)}_{pj,tk}}{Q^{(n)}_{pj,tk}}\sqrt{\frac{\mu_{pt}}{\mu_{p't'}}}m_{t'}\left(\mathbf{u}_{tk} - \mathbf{u}_{pj}\right)\cdot\frac{m_p T_{tk}\mathbf{u}_{pj} + m_t T_{pj}\mathbf{u}_{tk}}{(m_p + m_t)T_{pj,tk}}\\
&+k_{pj,tk} n_{pj} n_{tk}\frac{m_{p'} m_{t'}}{(m_p + m_t)^2}\frac{2}{3}\frac{\tilde{Q}^{(m)}_{pj,tk}}{Q^{(n)}_{pj,tk}}\sqrt{\frac{\mu_{pt}}{\mu_{p't'}}}3k_B\left(T_{pj} - T_{tk}\right),
\end{aligned}$$

where $E_{\text{th}} = \epsilon_{pj} + \epsilon_{tk} - \epsilon_{p'j'} - \epsilon_{t'k'}$ is the threshold energy of the reaction. The velocity-averaged collision cross sections for $E_{\text{th}} \geq 0$ are

$$\begin{aligned}
\tilde{Q}^{(m)}_{pj,tk} &= \int_0^\infty x^4\sqrt{x^2 + \frac{E_{\text{th}}}{k_B T_{pj,tk}}}e^{-x^2}\sigma^{(m)}_{pj,tk}\left(x\sqrt{\frac{2k_B T_{pj,tk}}{\mu_{pt}}}\right)F^{(m)}\left(2M_{pj,tk}x\right)dx,\\
\tilde{Q}^{(e)}_{pj,tk} &= \int_0^\infty x^4\sqrt{x^2 + \frac{E_{\text{th}}}{k_B T_{pj,tk}}}e^{-x^2}\sigma^{(m)}_{pj,tk}\left(x\sqrt{\frac{2k_B T_{pj,tk}}{\mu_{pt}}}\right)F^{(n)}\left(2M_{pj,tk}x\right)dx,
\end{aligned} \tag{2.76}$$

and for $E_{\text{th}} < 0$ are

$$\tilde{Q}^{(m)}_{pj,tk} = \int\limits_{\sqrt{-\frac{E_{\text{th}}}{k_B T_{pj,tk}}}}^{\infty} x^4 \sqrt{x^2 + \frac{E_{\text{th}}}{k_B T_{pj,tk}}} e^{-x^2} \sigma^{(m)}_{pj,tk} \left( x \sqrt{\frac{2k_B T_{pj,tk}}{\mu_{pt}}} \right) F^{(m)} \left( 2M_{pj,tk} x \right) dx,$$

$$\tilde{Q}^{(e)}_{pj,tk} = \int\limits_{\sqrt{-\frac{E_{\text{th}}}{k_B T_{pj,tk}}}}^{\infty} x^4 \sqrt{x^2 + \frac{E_{\text{th}}}{k_B T_{pj,tk}}} e^{-x^2} \sigma^{(m)}_{pj,tk} \left( x \sqrt{\frac{2k_B T_{pj,tk}}{\mu_{pt}}} \right) F^{(n)} \left( 2M_{pj,tk} x \right) dx, \tag{2.77}$$

It should be noted that Eqs. (2.74) are not only applicable to a binary exchange reaction (2.73) but also to any chemical reaction with two reactants. The equations (2.75) are only valid for products of the chemical reaction (2.73). The velocity-averaged collision cross sections $Q^{(n)}_{pj.tk}$, $Q^{(m)}_{pj.tk}$, $Q^{(e)}_{pj.tk}$, $\tilde{Q}^{(m)}_{pj.tk}$, and $\tilde{Q}^{(e)}_{pj.tk}$ in Eqs. (2.68), (2.74), and (2.75) depend on the effective temperature of the two species $T_{pj,tk}$ and on their relative Mach number $M_{pj,tk}$. In the limit of $M_{pj,tk} = 0$ they reduce to $\Omega^{(i,j)}_{pj}$ given by Eq. (2.64) with appropriate coefficients $i$ and $j$.

Similar expressions can be derived to other chemical reactions given by Eqs. (2.6). The first of these reactions is a particular case of more general binary exchange reaction considered above.

The complexity of the collision transfer integrals derived so far can be alleviated in a number of very important practical cases. Consider plasma containing neutral and ion particles at the same temperature $T_h$ and electrons at the temperature $T_e \neq T_h$ for which the condition represented by Eq. (2.53) is valid. Also, assume that $T_e \gg T_h$ which is a good approximation in the plasmas in NHISs. Then all particle collisions including chemically reacting ones can be separated into several groups: elastic collisions of particles of the same species, heavy particles-electrons collisions including elastic and chemically reacting collisions, elastic and chemically reacting collisions between heavy particles. The first group of reactions does not have contributions in the fluid equations. However, these collisions are very important in bringing particle distribution functions to Maxwellian distributions given by Eqs. (2.55) and (2.56). For the second group, the corresponding terms for neutral or ion particles denoted by the subscript $h$ undergoing elastic collisions with electrons given by Eq. (2.68) can be simplified to

$$
\begin{aligned}
&\frac{\delta}{\delta t}\left(n_h\right)_{\text{elast}} = 0,\\
&\frac{\delta}{\delta t}\left(n_h m_h \mathbf{v}_h\right)_{\text{elast}} = \nu_{he}^{\text{elast}} n_h m_e \left(\mathbf{u}_e - \mathbf{u}_h\right),\\
&\frac{\delta}{\delta t}\left(n_h \frac{m_h v_h^2}{2} + n_h \epsilon_h\right)_{\text{elast}} = \frac{\delta}{\delta t}\left(n_h \frac{m_h v_h^2}{2}\right)_{\text{elast}} = 3\frac{m_e}{m_h}\nu_{he}^{\text{elast}} n_h k_B \left(T_e - T_h\right),
\end{aligned}
\tag{2.78}
$$

where $\nu_{he}^{\text{elast}} = \frac{16}{3}\sqrt{\frac{k_B T_e}{2\pi m_e}} Q_{h,e}^{(e)} n_e$ is the elastic electron-heavy particle collision frequency and the following approximations were used

$$
\begin{aligned}
&T_{h,e} = \frac{m_e T_h + m_h T_e}{m_h + m_e} \simeq T_e,\\
&\mu_{eh} = \frac{m_h m_e}{m_h + m_e} \simeq m_e,\\
&Q_{h,e}^{(e)} = Q_{h,e}^{(m)},\\
&Q_{h,e}^{(e)} = \int_0^\infty x^5 e^{-x^2} \sigma_{h,e}^{(m)}\left(x\sqrt{\frac{2k_B T_e}{m_e}}\right) dx.
\end{aligned}
\tag{2.79}
$$

For the electrons these terms are

$$
\begin{aligned}
&\frac{\delta}{\delta t}\left(n_e\right)_{\text{elast}} = 0,\\
&\frac{\delta}{\delta t}\left(n_e m_e \mathbf{v}_e\right)_{\text{elast}} = \nu_{eh}^{\text{elast}} n_e m_e \left(\mathbf{u}_h - \mathbf{u}_e\right),\\
&\frac{\delta}{\delta t}\left(n_e \frac{m_e v_e^2}{2}\right)_{\text{elast}} = 3\frac{m_e}{m_h}\nu_{eh}^{\text{elast}} n_e k_B \left(T_h - T_e\right),
\end{aligned}
\tag{2.80}
$$

where $\nu_{eh}^{\text{elast}} = \frac{16}{3}\sqrt{\frac{k_B T_e}{2\pi m_e}} Q_{h,e}^{(e)} n_h$.

Using the same assumptions as for deriving Eq. (2.78) the loss terms for the heavy particles $h$ undergoing the binary chemically reacting collisions with electrons can be simplified to

$$\begin{aligned}
\frac{\delta}{\delta t}\left(n_h\right)_{\text{chem,destr}} &= -k_{h,e} n_h n_e, \\
\frac{\delta}{\delta t}\left(n_h m_h \mathbf{v}_h\right)_{\text{chem,destr}} &= -k_{h,e} n_h n_e m_h \mathbf{u}_h, \\
\frac{\delta}{\delta t}\left(n_h \frac{m_h v_h^2}{2} + n_h \epsilon_h\right)_{\text{chem,destr}} &= \frac{\delta}{\delta t}\left(n_h \frac{m_h v_h^2}{2}\right)_{\text{chem}} + \epsilon_h \frac{\delta}{\delta t}\left(n_h\right)_{\text{chem,destr}}, \\
\frac{\delta}{\delta t}\left(n_h \frac{m_h v_h^2}{2}\right)_{\text{chem,destr}} &= -k_{h,e} n_h n_e \frac{3k_B T_h}{2}.
\end{aligned} \tag{2.81}$$

For the electrons the corresponding loss terms are

$$\begin{aligned}
\frac{\delta}{\delta t}\left(n_e\right)_{\text{chem,destr}} &= -k_{e,h} n_e n_h, \\
\frac{\delta}{\delta t}\left(n_e m_e \mathbf{v}_e\right)_{\text{chem,destr}} &= -k_{e,h} n_e n_h m_e \left[\mathbf{u}_e - \frac{T_e}{T_h}\left(\frac{2}{3}\frac{Q_{e,h}^{(m)}}{Q_{e,h}^{(n)}} - 1\right)\left(\mathbf{u}_h - \mathbf{u}_e\right)\right], \\
\frac{\delta}{\delta t}\left(n_e \frac{m_e v_e^2}{2}\right)_{\text{chem,destr}} &= -k_{e,h} n_e n_h \frac{3k_B T_e}{2}\left[1 + \frac{T_e}{T_h}\left(\frac{2}{3}\frac{Q_{e,h}^{(e)}}{Q_{e,h}^{(n)}} - 1\right)\right],
\end{aligned} \tag{2.82}$$

where $Q_{e,h}^{(m)}$ and $Q_{e,h}^{(e)}$ are given by Eqs. (2.79) and $Q_{e,h}^{(n)} = \int_0^\infty x^3 e^{-x^2} \sigma_{e,h}^{(m)}\left(x\sqrt{\frac{2k_B T_e}{m_e}}\right) dx$.

If the particles of species $h$ and $p$ are produced in the collisions of the heavy particles $t$ and electrons then the simplified production terms are given from Eq. (2.75) as

$$\begin{aligned}
\frac{\delta}{\delta t}\left(n_h\right)_{\text{chem,prod}} &= k_{t,e} n_t n_e, \\
\frac{\delta}{\delta t}\left(n_h m_h \mathbf{v}_h\right)_{\text{chem,prod}} &= k_{t,e} n_t n_e m_h \mathbf{u}_t \\
-k_{t,e} n_t n_e \frac{m_h}{m_t}&\left[m_e \frac{T_t - T_e}{T_e}\left(\frac{2}{3}\frac{Q_{t,e}^{(m)}}{Q_{t,e}^{(n)}} - 1\right) + \frac{2}{3}\frac{\tilde{Q}_{t,e}^{(m)}}{Q_{t,e}^{(n)}} m_t \sqrt{\frac{m_e}{\mu_{hp}}}\right]\left(\mathbf{u}_e - \mathbf{u}_t\right), \\
\frac{\delta}{\delta t}\left(n_h \frac{m_h v_h^2}{2} + n_h \epsilon_h\right)_{\text{chem,prod}} &= \frac{\delta}{\delta t}\left(n_h \frac{m_h v_h^2}{2}\right)_{\text{chem,prod}} + \epsilon_h \frac{\delta}{\delta t}\left(n_h\right)_{\text{chem,prod}}, \\
\frac{\delta}{\delta t}\left(n_h \frac{m_h v_h^2}{2}\right)_{\text{chem,prod}} &= k_{t,e} n_t n_e \left[\frac{m_p}{m_t}\frac{3k_B\left(T_t + T_e\right)}{2} + \frac{m_h}{m_t}\left(\frac{2}{3}\frac{Q_{t,e}^{(e)}}{Q_{t,e}^{(n)}} - 1\right)\frac{3k_B m_e T_t^2}{2T_e}\right] \\
+k_{t,e} n_t n_e &\left[\frac{m_p}{m_t} E_{\text{th}} - \frac{m_h - m_p}{m_t}\frac{Q_{t,e}^{(e)}}{Q_{t,e}^{(n)}} k_B T_e + \frac{m_h m_p}{m_t^2}\frac{2}{3}\frac{\tilde{Q}_{t,e}^{(m)}}{Q_{t,e}^{(n)}}\sqrt{\frac{m_e}{\mu_{hp}}} 3k_B\left(T_t - T_e\right)\right].
\end{aligned} \tag{2.83}$$

Consider finally the destruction terms of the heavy particle $h$ undergoing a chemically reacting collision with another heavy particle $p$ then from Eq. (2.74) and using aforementioned simplified assumptions this equation can be rewritten as

$$\begin{aligned}
&\frac{\delta}{\delta t}\left(n_h\right)_{\text{chem,destr}} = -k_{hp} n_h n_p, \\
&\frac{\delta}{\delta t}\left(n_h m_h \mathbf{v}_h\right)_{\text{chem,destr}} = -k_{hp} n_h n_p m_h \left[\mathbf{u}_h - \frac{\mu_{hp}}{m_h}\left(\frac{2}{3}\frac{Q_{h,p}^{(m)}}{Q_{h,p}^{(n)}} - 1\right)\left(\mathbf{u}_p - \mathbf{u}_h\right)\right], \\
&\frac{\delta}{\delta t}\left(n_h \frac{m_h v_h^2}{2} + n_h \epsilon_h\right)_{\text{chem,destr}} = \frac{\delta}{\delta t}\left(n_h \frac{m_h v_h^2}{2}\right)_{\text{chem}} + \epsilon_h \frac{\delta}{\delta t}\left(n_h\right)_{\text{chem,destr}}, \\
&\frac{\delta}{\delta t}\left(n_h \frac{m_h v_h^2}{2}\right)_{\text{chem,destr}} = -k_{hp} n_h n_p \frac{3 k_B T_h}{2}\left[1 + \frac{\mu_{hp}}{m_h}\left(\frac{2}{3}\frac{Q_{h,p}^{(e)}}{Q_{h,p}^{(n)}} - 1\right)\right].
\end{aligned} \tag{2.84}$$

For elastic collisions of two heavy particles $h$ and $p$ the collision terms corresponding to elastic collisions using assumptions stated above can be written as

$$\begin{aligned}
&\frac{\delta}{\delta t}\left(n_h\right)_{\text{elast}} = 0, \\
&\frac{\delta}{\delta t}\left(n_h m_h \mathbf{v}_h\right)_{\text{elast}} = \nu_{hp}^{\text{elast}} n_h \mu_{hp}\left(\mathbf{u}_p - \mathbf{u}_h\right), \\
&\frac{\delta}{\delta t}\left(n_{pj} \frac{m_p v_{pj}^2}{2}\right)_{\text{elast}} = \nu_{hp}^{\text{elast}} n_h \frac{m_h m_p}{(m_h + m_p)^2}\left(m_h \mathbf{u}_h + m_p \mathbf{u}_p\right)\cdot\left(\mathbf{u}_p - \mathbf{u}_h\right),
\end{aligned} \tag{2.85}$$

where $\nu_{hp}^{\text{elast}} = \frac{16}{3} Q_{h,p}^{(m)} \sqrt{\frac{k_B T_h}{2\pi\mu_{hp}}} n_p$.

And the simplified production terms of heavy particles of species $h$ (the second product is also a heavy particle but of species $p$) in a binary exchange reaction of particles of species $t$ and $q$ are

$$
\begin{aligned}
&\frac{\delta}{\delta t}\left(n_h\right)_{\text{chem,prod}} = k_{t,q} n_t n_q, \\
&\frac{\delta}{\delta t}\left(n_h m_h \mathbf{v}_{pj}\right)_{\text{chem,prod}} = k_{t,q} n_t n_q \frac{m_h}{m_t + m_{qt}}\left[m_t \mathbf{u}_t + m_q \mathbf{u}_q\right] \\
&-k_{t,q} n_t n_q \frac{m_h m_p}{m_t + m_q} \frac{2}{3} \frac{\tilde{Q}_{t,q}^{(m)}}{Q_{t,q}^{(n)}} \sqrt{\frac{\mu_{tq}}{\mu_{hp}}}\left(\mathbf{u}_{tk} - \mathbf{u}_{pj}\right), \\
&\frac{\delta}{\delta t}\left(n_h \frac{m_h v_h^2}{2} + n_h \epsilon_h\right)_{\text{chem,prod}} = \frac{\delta}{\delta t}\left(n_h \frac{m_h v_h^2}{2}\right)_{\text{chem,prod}} + \epsilon_h \frac{\delta}{\delta t}\left(n_h\right)_{\text{chem,prod}}, \\
&\frac{\delta}{\delta t}\left(n_h \frac{m_h v_h^2}{2}\right)_{\text{chem,prod}} = k_{t,q} n_t n_q \left[\frac{m_h}{m_t + m_{qt}} 3 k_B T_h + \frac{m_h}{m_t + m_{qt}}\left(\frac{2}{3} \frac{Q_{pj,tk}^{(e)}}{Q_{pj,tk}^{(n)}} - 1\right) \frac{3 k_B \mu_{pt}}{2} T_h\right] \\
&+k_{t,q} n_t n_q \left[\frac{m_p}{m_t + m_q} E_{\text{th}} - \frac{m_h - m_p}{m_t + m_q} \frac{Q_{t,q}^{(e)}}{Q_{t,q}^{(n)}} k_B T_h\right].
\end{aligned}
\tag{2.86}
$$

In general different chemical reactions may occur in the plasma, therefore, the most general form of species continuity equation is

$$
\frac{\partial n_{pj}}{\partial t} + \nabla \cdot \left(n_{pj} \mathbf{u}_{pj}\right) = \sum_{r=1}^{N_r} \left(\nu_{r,pj} - \nu'_{r,pj}\right) k_r \prod_{p=1}^{N_s} \prod_j n_{pj}^{\nu_{r,pj}}, \tag{2.87}
$$

where the production and destruction terms on the right hand side are written for the general chemical reaction

$$
\sum_{p=1}^{N_s} \sum_j \nu'_{r,pj} X_{pj} \to \sum_{p=1}^{N_s} \sum_j \nu_{r,pj} X_{pj}, \tag{2.88}
$$

where $N_r$ is the number of chemical reactions included into the model, $\nu_{r,pj}$, $\nu'_{r,pj}$ are the production and destruction stoichiometric coefficients respectively of species $p$ in excited state $j$ participating in the reaction $r$, $k_r$ is the reaction rate coefficient, $N_s$ is the number of species taken into account.

If instead of using a Maxwellian distribution function, the corrections in the form of Chapman-Enskog expansion are used, the transfer collision integrals get additional terms. For example, in addition to the friction force, the thermo force associated with the gradient of the species temperature appears in the momentum equations even for elastic collisions (Golant, et al., 1980).

Frequently, the transfer collision integrals are calculated based on the simplified collision integral approximations such as the Fokker-Planck (Schunk, 1977) or the Bhatnagar-Gross-Krook model (BGK) (Burgers, 1969; Lieberman & Lichtenberg, 2005; Chabert & Braithwaite, 2011). In addition, phenomenological model for the transfer collision integrals are also sometimes encountered in the literature (Li, et al., 2013).

As for the ternary reactions or other reactions involving more than two particles as products or reactants their contribution to the momentum and energy equations should be calculated based on the corresponding collision integral. In practice, however, these contributions are disregarded. In part this assumption can be justified due to the fact that elastic collision cross sections are typically much large compared to ternary reactions ones. Since the destruction and production terms are proportional to the product of number densities the ternary reactions start playing significant role only at high pressures. Also from microscopic point of view in a volume atomic recombination reaction such as Eq. (1.34) the role of the third body is to take off the excess of the energy and momentum in order to satisfy conservation laws.

Finally, for electrons the corresponding collision transfer integrals including elastic and inelastic collisions with heavy particles become

$$\begin{aligned}
&\frac{\delta}{\delta t}\left(n_e\right) = \sum_{r=1}^{N_r}\left(\nu_{r,e} - \nu'_{r,e}\right) k_r \prod_{p=1}^{N_s} n_p^{\nu_{r,p}}, \\
&\frac{\delta}{\delta t}\left(n_e m_e \mathbf{u}_e\right) = \sum_{h=1}^{N_h} \overline{\nu}_{e,h} n_e m_e \left(\mathbf{u}_h - \mathbf{u}_e\right), \\
&\frac{\delta}{\delta t}\left(n_e \frac{m_e v_e^2}{2}\right) = \sum_{h=1}^{N_h} 3 \frac{m_e}{m_h} \nu_{eh}^{\text{elast}} n_e k_B \left(T_h - T_e\right) + \sum_{r=1}^{N_r} E_{r,\text{th}} k_r \prod_{p=1}^{N_s} n_p^{\nu_{r,p}}.
\end{aligned} \tag{2.89}$$

where $N_h$ is the number of heavy particles (atoms, molecules and ions) in the plasma, $\overline{\nu}_{e,h}$ is the average collision frequency of electrons and heavy particles $h$ including elastic and inelastic contributions. In the above expression for the electron energy collision transfer integral the term proportional to the difference of the temperatures for inelastic collisions was neglected in comparison with the last term.

For the heavy particles the collision transfer integrals can be represented as

$$
\begin{aligned}
&\frac{\delta}{\delta t}\left(n_h\right) = \sum_{r=1}^{N_r}\left(\nu_{r,e} - \nu'_{r,e}\right) k_r \prod_{p=1}^{N_s} n_p^{\nu_{r,p}}, \\
&\frac{\delta}{\delta t}\left(n_h m_h \mathbf{v}_h\right) = \sum_{p=1}^{N_s} \overline{\nu}_{h,p} n_h \mu_{hp}\left(\mathbf{u}_p - \mathbf{u}_h\right), \\
&\frac{\delta}{\delta t}\left(n_{pj}\frac{m_p v_{pj}^2}{2}\right) = 3\frac{m_e}{m_h}\nu_{eh}^{\text{elast}} n_e k_B\left(T_e - T_h\right) + \sum_{r=1}^{N_r} E_{r,\text{th}} k_r \prod_{\substack{p=1\\ p\neq e}}^{N_s} n_p^{\nu_{r,p}}.
\end{aligned}
\tag{2.90}
$$

## 2.5 Electromagnetic Equations

The fluid equations derived in the Section 2.2 are supplemented by Maxwell's equations (2.7) for the electromagnetic fields. By introducing scalar $\phi$ and vector $\mathbf{A}$ potentials by

$$
\begin{aligned}
&\mathbf{E} = -\frac{\partial \mathbf{A}}{\partial t} - \nabla\phi, \\
&\mathbf{B} = \nabla\times\mathbf{A},
\end{aligned}
\tag{2.91}
$$

the Maxwell's equations (2.7) can be written as

$$
\begin{aligned}
&-\nabla\cdot\left(\epsilon\frac{\partial \mathbf{A}}{\partial t}\right) - \nabla\cdot\left(\epsilon\nabla\phi\right) = e\sum_{p=1}^{N_s} Z_p n_p, \\
&\nabla\times\frac{1}{\mu}\nabla\times\mathbf{A} = \mathbf{J} - \epsilon\frac{\partial^2\mathbf{A}}{\partial t^2} - \epsilon\frac{\partial}{\partial t}\left(\nabla\phi\right).
\end{aligned}
\tag{2.92}
$$

After some manipulations involving vector calculus identities Eqs. (2.92) can be rewritten as

$$
\begin{aligned}
&-\left(\frac{\partial \mathbf{A}}{\partial t} + \nabla\phi\right)\cdot\nabla\epsilon - \epsilon\nabla\cdot\left(\frac{\partial \mathbf{A}}{\partial t} + \nabla\phi\right) = e\sum_{p=1}^{N_s} Z_p n_p, \\
&\nabla\left(\nabla\cdot\mathbf{A}\right) - \nabla^2\mathbf{A} + \mu\left(\nabla\frac{1}{\mu}\right)\times\mathbf{A} = \mu\mathbf{J} - \epsilon\mu\frac{\partial^2\mathbf{A}}{\partial t^2} - \epsilon\mu\frac{\partial}{\partial t}\left(\nabla\phi\right),
\end{aligned}
\tag{2.93}
$$

By using Lorenz gauge condition (Jackson, 1999)

$$
\nabla\cdot\mathbf{A} + \epsilon\mu\frac{\partial\phi}{\partial t} = 0 \tag{2.94}
$$

and assuming space uniform permeability $\mu$ and permittivity $\epsilon$ of the medium, the Maxwell's equations become equations for scalar and vector potentials

$$
\begin{aligned}
&\epsilon\mu\frac{\partial^2\phi}{\partial t^2}-\nabla^2\phi=\frac{e}{\epsilon}\sum_{p=1}^{N_s}Z_p n_p,\\
&\epsilon\mu\frac{\partial^2\mathbf{A}}{\partial t^2}-\nabla^2\mathbf{A}=\mu\mathbf{J}.
\end{aligned}
\tag{2.95}
$$

If instead of the Lorenz gauge condition the Coulomb gauge condition is used

$$
\nabla\cdot\mathbf{A}=0\,, \tag{2.96}
$$

then the Maxwell's equations (2.93) with the assumption of space uniform the permeability $\mu$ and the permittivity $\epsilon$ of the medium become

$$
\begin{aligned}
&\nabla^2\phi=-\frac{e}{\epsilon}\sum_{p=1}^{N_s}Z_p n_p,\\
&\epsilon\mu\frac{\partial^2\mathbf{A}}{\partial t^2}-\nabla^2\mathbf{A}=\mu\mathbf{J}-\epsilon\mu\frac{\partial}{\partial t}\left(\nabla\phi\right).
\end{aligned}
\tag{2.97}
$$

An important case corresponds to a situation when the displacement current can be neglected. Equations (2.97) can be written then as

$$
\begin{aligned}
&\nabla^2\phi=-\frac{e}{\epsilon}\sum_{p=1}^{N_s}Z_p n_p,\\
&\epsilon\mu\frac{\partial^2\mathbf{A}}{\partial t^2}-\nabla^2\mathbf{A}=\mu\mathbf{J}.
\end{aligned}
\tag{2.98}
$$

In addition to above equations another useful relation can be derived. Taking the time derivative of the Gauss's law (the first equation of Eqs. (2.7)) and applying the divergence operator to the Ampere's law (the fourth equation of Eqs. (2.7)) the continuity equation for charges and current can be obtained

$$
e\sum_{p=1}^{N_s}Z_p\frac{\partial n_p}{\partial t}+\nabla\cdot\mathbf{J}=0. \tag{2.99}
$$

## 2.6 Initial and Boundary Conditions

The system of multi-fluid (2.43) and Maxwell's (2.97) equations need to be supplemented with proper initial and boundary conditions.

Wall chemical reactions are taken into account by imposing a catalytic wall boundary condition (Scott, 1992; Hagelaar, et al., 2011). This boundary condition expresses the balance between particles incident from the plasma to the wall, potential absorption of particles by the wall due to chemical reactions and the emission of particles created on the wall. The flux of species $p$ in the internal state $j$ at the wall is written as

$$\Gamma_{pj}\Big|_{\mathrm{wall}} = \Gamma_{pj,\mathrm{w}} - \left(1 - \gamma_{pj,pj,\mathrm{w}}\right)\Gamma_{pj,\mathrm{w}} - \sum_t \sum_k \gamma_{tk,pj,\mathrm{w}} \Gamma_{tk,\mathrm{w}}, \qquad (2.100)$$

where $\Gamma_{pj}\Big|_{\mathrm{wall}}$ is the average flux of the particles at the wall, $\Gamma_{pj,\mathrm{w}}$ represents the flux of the particles of species $p$ in the internal state $j$ incident to the wall, $\gamma_{tk,pj,\mathrm{w}}$ is the sticking coefficient of the wall reaction of species $t$ in an internal state $k$ producing the particles of species $p$ in the internal state $j$. If $p = t$ and $j = k$ in the above expression then this situation corresponds to a flux of reflected particles (not fully absorbed catalytic wall). From the kinetic point of view Eq. (2.100) represents the modification of the distribution function due to the flux of particles from the wall.

Positive ion species $X^+$ are usually considered to be fully absorbed and neutralized by a wall. Assuming that ions do not undergo collisions in the sheath, which is equivalent to the assumption that the sheath thickness is much smaller than the ion mean free path, their flux to the wall is

$$\Gamma_{X^+}\Big|_{\mathrm{wall}} = \Gamma_{X^+,s}, \qquad (2.101)$$

where $\Gamma_{X^+,s}$ is the flux of positive ions at the edge of the sheath. Moreover for ions to reach the wall their velocity at the sheath edge should exceed the Bohm speed (Lieberman & Lichtenberg, 2005) resulting in the Bohm flux given by

$$\Gamma_{X^+,s} = n_{X^+,s} u_{\mathrm{B},X^+}, \qquad (2.102)$$

where $u_{\mathrm{B},X^+} = \sqrt{\dfrac{k_B T_e}{m_{X^+}}}$ is the Bohm speed of positive ions in the absence of negative ions and $n_{X^+,s}$ is the number density at the edge of the sheath. The modifications of the Bohm velocity in electronegative plasma are discussed in the Chapter 3.

At high pressures Eq. (2.102) is usually replaced by the Schottky boundary condition (Chabert & Braithwaite, 2011)

$$n_{X^+}\Big|_{\mathrm{wall}} = 0. \qquad (2.103)$$

The flux of electrons to a wall at the floating potential can be found by direct integration of the half of Maxwellian distribution (Chabert & Braithwaite, 2011)

$$\Gamma_e\Big|_{\mathrm{wall}} = \Gamma_{e,\mathrm{s}} = \frac{n_{e,s} v_{e,\mathrm{th}}}{4} e^{-\frac{eV_{\mathrm{sh}}}{k_B T_e}}, \tag{2.104}$$

where $V_{\mathrm{sh}}$ is the voltage drop across the sheath.

Neutrals are typically not fully absorbed by the walls so it can be assumed that their distribution function is a perturbed Maxwellian (Hagelaar, et al., 2011). If there is only one surface reaction that leads to neutral species $X$ destruction and there is no other surface production processes, Eq. (2.100) can be simplified to (Chantry, 1987)

$$\Gamma_X\Big|_{\mathrm{wall}} = \frac{2\gamma_{X,\mathrm{rec}}}{2-\gamma_{X,\mathrm{rec}}} \frac{1}{4} n_{X,\mathrm{w}} v_{X,\mathrm{th}}, \tag{2.105}$$

where $n_{X,\mathrm{w}}$ is the number density of the species $p$ in a state $j$ at the wall and

$$v_{X,\mathrm{th}} = \sqrt{\frac{8k_B T_X}{\pi m_X}} \tag{2.106}$$

is the thermal velocity, and $\gamma_{X,\mathrm{rec}}$ is the recombination coefficient of neutral species. The incident flux in this case is given by

$$\Gamma_{X,\mathrm{inc}} = \frac{2}{2-\gamma_{X,\mathrm{rec}}} \frac{1}{4} n_{X,\mathrm{w}} v_{X,\mathrm{th}}, \tag{2.107}$$

while the reflected flux is

$$\Gamma_{X,\mathrm{ref}} = \frac{2(1-\gamma_{X,\mathrm{rec}})}{2-\gamma_{X,\mathrm{rec}}} \frac{1}{4} n_{X,\mathrm{w}} v_{X,\mathrm{th}}. \tag{2.108}$$

If, however, there are other chemical reactions such as neutralization of ions leading to the flux of neutrals from the walls then there is interdependence between incident particle flux $\Gamma_{pj,\mathrm{w}}$ and particle fluxes from the walls $\Gamma_{tk,\mathrm{w}}$ (Hagelaar, et al., 2011). In order to resolve this issue it is assumed that these fluxes act independently and, thus, the total flux of neutral particles $\Gamma_X\Big|_{\mathrm{wall}}$ can be found from Eq. (2.100), (2.101), and (2.105) as

$$\Gamma_X\Big|_{\text{wall}} = \frac{2\gamma_{X,\text{rec}}}{2-\gamma_{X,\text{rec}}}\frac{1}{4}n_{X,\text{w}}v_{X,\text{th}} - \sum_p n_{p,s}u_{\text{B},p}\,. \tag{2.109}$$

In continuum flow regimes (Knudsen number $\text{Kn}<0.01$) the no-slip boundary condition is imposed for the velocity component of the mass-averaged velocity $\mathbf{u}$ parallel to the stationary wall

$$u_{\parallel}\Big|_{\text{wall}} = 0. \tag{2.110}$$

For the slip regime ($0.01<\text{Kn} \le 0.1$) the slip boundary conditions should be used instead. The general slip condition at the stationary wall (Karniadakis, et al., 2005) can be written assuming that the jumps in the velocities of different species are independent of each other as

$$u_{pj,\parallel}\Big|_{\text{wall}} = \frac{2-\sigma_{pj,v}}{\sigma_{pj,v}}\frac{\lambda_{pj}}{1-b\lambda_{pj}}\frac{\partial u_{pj,\parallel}}{\partial n}\Bigg|_{\text{wall}} + \frac{3}{4}\frac{\mu_{pj}}{\rho_{pj}T_{pj,\text{w}}}\frac{\partial T_{pj}}{\partial s}\Bigg|_{\text{wall}}, \tag{2.111}$$

where $\sigma_{pj,v}$ is the tangential momentum accommodation coefficient, $n$ is the outward normal direction to the surface, $s$ is the tangential direction along the surface, $\lambda_{pj}$ is the mean free path of species $p$ in an excited state $j$, $b$ is a phenomenological parameter, $\mu_{pj}$ is the viscosity, $\rho_{pj}$ is the density. A similar expression may be written for the species velocity component parallel to the wall by utilizing the species momentum conservation equation (Hagelaar, et al., 2011).

For the normal component of the mass-averaged velocity in the absence of the deposition or sputtering from the surface of the stationary wall, the impermeability boundary condition is usually used

$$u_{\perp}\Big|_{\text{wall}} = 0. \tag{2.112}$$

The boundary conditions for the temperatures are typically Dirichlet and Neumann boundary conditions respectively

$$\begin{aligned} T_{pj}\Big|_{\text{wall}} &= T_{pj,\text{w}}, \\ -\kappa_{pj}\frac{\partial T_{pj}}{\partial \mathbf{n}}\Bigg|_{\text{wall}} &= \mathbf{q}_{pj,\text{w}}. \end{aligned} \tag{2.113}$$

The above expressions are only valid for the continuum regime in the absence of chemical reactions. In cases where rarefaction effects play an important role the fixed wall temperature condition should be substituted by a temperature jump boundary condition (Karniadakis, et al., 2005; Hagelaar, et al., 2011) which is written assuming that the temperature jumps for different species are independent of each other and there is no creation or destruction of particles on the wall as

$$q_{pj}\Big|_{\text{wall}} = -\kappa_{pj}\frac{\partial T_{pj}}{\partial n}\Bigg|_{\text{wall}} = \frac{2\alpha_{pj}}{2-\alpha_{pj}}\left(C_{p,pj} - \frac{1}{2}k_B\right)\frac{1}{4}n_{pj}v_{\text{th},pj}\left(T_{pj}\Big|_{\text{wall}} - T_{\text{w}}\right), \tag{2.114}$$

where $\alpha_{pj}$ is the thermal accommodation coefficient of species $p$ in an excited state $j$, $C_{p,pj}$ is the heat capacity per particle at constant pressure. The thermal accommodation coefficient of translational and rotational degrees of freedom of hydrogen molecules was experimentally measured by Leroy et al. (1997) for different surfaces and its typical values are of the order of 0.1.

In the presence of chemical reactions the expression for the heat flux $q_{pj}$ of the particles of species $p$ in an internal state $j$ to the wall can be written similarly to Eq. (2.100) (Hagelaar, et al., 2011)

$$q_{pj}\Big|_{\text{wall}} = \Gamma_{pj,\text{w}}\bar{U}_{pj} - q_{pj,\text{w}}, \tag{2.115}$$

where $\bar{U}_{pj}$ is the mean energy of incident particles, $q_{pj,\text{w}}$ is the heat flux of the particles of species $p$ in an internal state $j$ from the surfaces. This flux is composed of the flux of reflected particles and the flux due to chemical reactions. By making additional approximations related to the shape of the distribution functions it is possible then to estimate the mean energy term and wall heat fluxes.

For ions and electrons the estimation of $q_{pj}\Big|_{\text{wall}}$ is rather straightforward. Neglecting secondary electron emission from the walls and assuming that ions fully recombine implies that Eq. (2.115) for these species does not contain terms $q_{pj,\text{w}}$ corresponding to the heat flux from the surfaces. Then the only unknown variable is the mean incident energy $\bar{U}_{pj}$. For electrons assuming that they obey Maxwellian distribution with the temperature $T_e$ the mean incident energy is (Chabert & Braithwaite, 2011)

$$\bar{U}_e = 2k_B T_e. \tag{2.116}$$

Ions on the other hand reach the plasma edge with the Bohm velocity (in the absence of collisions in the sheath) and then are further accelerated by the sheath. In addition to kinetic en-ergy ions also bring their ionization energy (Pekker & Hussary, 2014) resulting in the following expression for $\bar{U}_{X^+}$

$$\bar{U}_{X^+} = eV_{\mathrm{sh}} + \frac{m_{X^+} u_{\mathrm{B},X^+}^2}{2} + U_{\mathrm{iz},X^+}, \tag{2.117}$$

where $U_{\mathrm{iz},X^+}$ is the ionization energy of $X^+$. In the above equation the contribution of the thermal motion of ions was neglected since the ion temperature is typically much smaller than the electron one.

In a case of trace neutral species $X$ such as vibrationally or electronically excited hydrogen molecules or atoms the situation is more complicated. If there are no chemical reactions (for example, for ground state hydrogen molecules) then Eq. (2.114) is valid. In a more general case it is necessary to estimate the mean incident energy $\bar{U}_X$ and heat fluxes $q_{pj,\mathrm{w}}$ from the wall due to surface chemical reactions and particle reflection from the surfaces. It was shown in Hagelaar et al. (2011) that there is an implicit coupling between energy $\bar{U}_X$ and $q_{pj,\mathrm{w}}$ in Eq. (2.115) similar to the situation with the neutral particle fluxes discussed earlier in this section. In order to eliminate the coupling it is assumed in this work that neutral molecules and atoms coming from the walls due to chemical reactions are in full thermal accommodation with the walls and the flux of neutral particles reflected from the walls is given by expression similar to Eq. (2.114). Thus, using Eq. (2.107) for incident flux and Eq. (2.108) for reflected one the coupling expressed in Eq. (2.115) is eliminated, and the heat flux of neutral species $X$ to the wall can be written as

$$\begin{aligned} q_X\Big|_{\mathrm{wall}} = & \frac{2-2\left(1-\gamma_{X,\mathrm{rec}}\right)\left(1-\alpha_X\right)}{1+\left(1-\gamma_{X,\mathrm{rec}}\right)\left(1-\alpha_X\right)} \frac{1}{4} n_X v_{X,\mathrm{th}} \left[\left(C_{p,X} - \frac{1}{2}k_B\right)T_X + U_{\mathrm{rec},X}\right] \\ & - \frac{2\left(1-\gamma_{X,\mathrm{rec}}\right)}{1+\left(1-\gamma_{X,\mathrm{rec}}\right)\left(1-\alpha_X\right)} \frac{2\alpha_X}{2-\gamma_{X,\mathrm{rec}}} \frac{1}{4} n_{X,\mathrm{w}} v_{X,\mathrm{th}} \left[\left(C_{p,X} - \frac{1}{2}k_B\right)T_{\mathrm{w}} + U_{\mathrm{rec},X}\right] \\ & - \sum_p n_{p,s} u_{\mathrm{B},p} \left[\left(C_{p,X} - \frac{1}{2}k_B\right)T_{\mathrm{w}} + U_{\mathrm{rec},X}\right], \end{aligned} \tag{2.118}$$

where $U_{\mathrm{rec},X}$ represents the formation energy which is the dissociation energy in a case of atomic recombination or the vibrational energy in a case of wall quenching of vibrationally excited molecules.

The total energy flux is obtained from Eqs. (2.115) written for all species by summing them up and taking into account the radiation from the plasma to the wall resulting in (Kersten, et al., 2001)

$$q\Big|_{\mathrm{wall}} = \sum_{p=1}^{N_s} q_p\Big|_{\mathrm{wall}} + q_{\mathrm{rad,w}}\,, \tag{2.119}$$

where $q_{\mathrm{rad,w}}$ is the heat flux associated with the radiation from the plasma to the wall. The simplified expression for $q_{\mathrm{rad,w}}$ is given in the grey-body approximation (Kersten, et al., 2001) as

$$q_{\mathrm{rad,w}} = \sigma_{\mathrm{SB}}\left(\epsilon_{\mathrm{rad}} T_{\mathrm{rad}}^4 - \epsilon_{\mathrm{w}} T_{\mathrm{w}}^4\right), \tag{2.120}$$

where $\sigma_{\mathrm{SB}} = \dfrac{\pi^2 k_B^4}{60\hbar^3 c^2}$ is the Stefan–Boltzmann constant, $\epsilon_{\mathrm{rad}}$ and $T_{\mathrm{rad}}$ are emittance and temperature of radiation source respectively, $\epsilon_{\mathrm{w}}$ is the wall emissivity. However, direct application of Eq. (2.120) is limited due to the lack of emissivity coefficient $\epsilon_{\mathrm{rad}}$ for non-equilibrium plasmas.

For subsonic/supersonic inlet and outlet neutral flows in the continuum regime the boundary conditions could be imposed based on characteristic theory (Thompson, 1990; Poinsot & Lele, 1992).

The boundary conditions for the electric and magnetic fields are given in terms of Dirichlet and Neumann boundary conditions respectively for scalar

$$\begin{aligned} \phi\Big|_{\mathrm{wall}} &= \phi_w, \\ \frac{\partial \phi}{\partial \mathbf{n}}\bigg|_{\mathrm{wall}} &= \mathbf{E}_w \end{aligned} \tag{2.121}$$

and vector potentials

$$\begin{aligned} \mathbf{A}\Big|_{\mathrm{wall}} &= \mathbf{A}_w, \\ \frac{\partial \mathbf{A}}{\partial n}\bigg|_{\mathrm{wall}} &= \frac{\partial \mathbf{A}_w}{\partial n}. \end{aligned} \tag{2.122}$$

# 3 Global Enhanced Vibrational Kinetic Model Formulation

The fluid equations (2.43), (2.47) derived in Chapter 2 coupled with Maxwell's equations (2.7) and appropriate boundary conditions can be used in order to calculate the fluid variables of the HCNHIS plasma. However, as it was pointed out in Chapter 1 modeling of the negative ion production requires knowledge of the vibrational population of $H_2$. Even under the assumption of rotational equilibrium it is necessary to develop fluid equations (continuity, momentum, energy) for each vibrational degree of freedom of $H_2$. In addition, similar equations should be developed for hydrogen ions, electronically excited atomic and molecular states, and electrons. The overall system of partial differential equations for this multi-species plasma is computationally non tractable and needs to be reduced following two approaches. Under the first approach, we consider a reduced set of species and chemical reactions and consequently solve the coupled fluid and Maxwell's equations following Hagelaar et al. (2011) and Boeuf et al. (2011). The main advantage of this approach is that it provides the spatial distribution of the plasma species. Under the second approach, we consider the full set of species and chemical reactions but make simplifying approximations regarding the space variation of the plasma variables resulting in global models (Lee, et al., 1997). While the spatial variation of plasma parameters is an important aspect of the HCNHIS modeling the knowledge of the non-equilibrium vibrational distribution of hydrogen molecules allows more accurate prediction of the $H^-$ formation. The latter is most important at these early stages of the HCNHIS development.

In this chapter we present a multi-fluid model for the HCNHIS based on fluid equations derived in Section 2.2 and proceed with the derivation of the Global Enhanced Vibrational Kinetic Model (GEVKM). The derivation includes the model for the bulk of the plasma which allows estimation of the central to averaged number density ratio and the sheath model. Bulk and sheath models are used to estimate sheath edge to center number density ratios of positive ions. These two ratios play an important role in the estimation of the discharge parameters since they take into account non-uniformity effects of the particle number densities compared to homogeneous zero-dimensional kinetic models. For neutral trace species (hydrogen atoms, vibrationally and electronically excited hydrogen molecules) the transport is assumed to be diffusive. The hydrogen plasma chemistry in the HCNHIS is reviewed and presented. The Global Enhanced Vi-brational Kinetic Model (GEVKM) is verified and validated by comparisons with the global and kinetic models developed by other researchers and by the comparison with experiments covering wide range of parameters such as the absorbed power density and the chamber pressure. The cases cover operating pressures from 0.02-100 Torr which includes the regime of operation of the HCNHIS.

The material of this chapter can be also found in Averkin et al. (2015a).

## 3.1 Multi-fluid Model of the HCNHIS Discharge Region

In this section multi-fluid model of the HCNHIS is formulated as a starting point for the derivation of the Global Enhanced Vibrational Kinetic Model.

### 3.1.1 General Assumptions

The GEVKM is developed for a cylindrical geometry of an inductively coupled discharge or a plasma reactor shown in Figure 7. The chamber has the radius $R$ and the length $L$. The species included in the model are ground state hydrogen atoms $\mathrm{H}$ and molecules $\mathrm{H}_2$, 14 vibrationally excited hydrogen molecules $\mathrm{H}_2(v)$, $v = 1 - 14$, electronically excited hydrogen atoms $\mathrm{H}(2)$, $\mathrm{H}(3)$, ground state positive ions $\mathrm{H}^+$, $\mathrm{H}_2^+$, $\mathrm{H}_3^+$, ground state anions $\mathrm{H}^-$, and electrons $\mathrm{e}$. Chemical reactions are mostly binary collisions with the inclusion of hydrogen recombination ternary reactions which could be important at high enough operating pressures. Electrons and translational-rotational degrees of freedom of heavy particles (ions, atoms and molecules) are assumed to obey equilibrium distribution functions of the form (2.55) for molecules and molecular ions and (2.56) for atoms, atomic ions, and electrons. The heavy particles are assumed to have the same temperature $T_h$. The electron temperature is $T_e$. The rotational degrees of freedom of $\mathrm{H}_2^+$ and $\mathrm{H}_3^+$ are considered to be in equilibrium with the translational degree of freedom while vibrational and electronic excitations are neglected. The power deposition is assumed to be primarily due to Joule heating of electrons by RF electric field while the stochastic heating is disregarded.

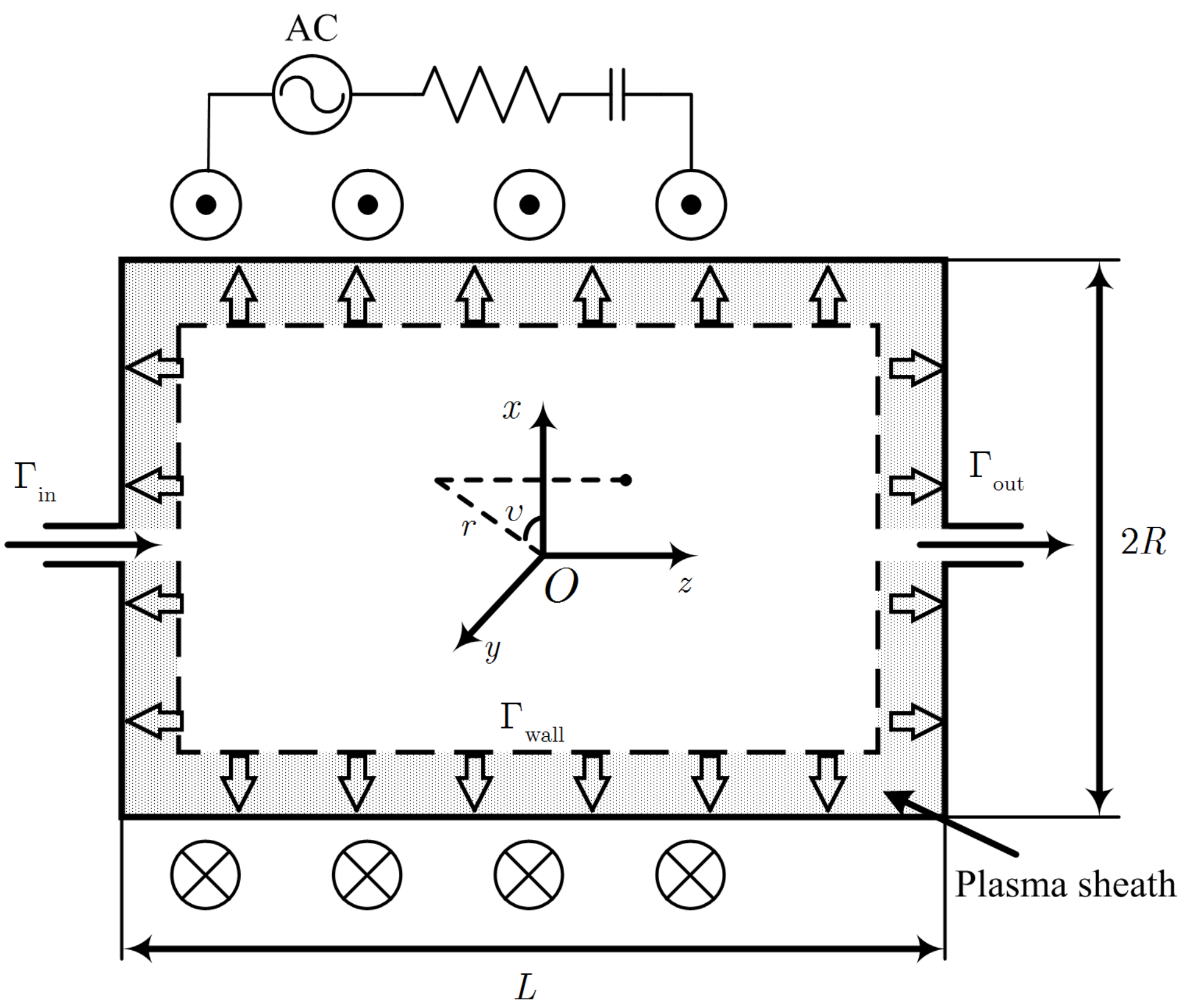

**Figure 7. Geometry of a cylindrical inductively coupled plasma reactor.**

### 3.1.2 Multi-Fluid Model for the HCNHIS Discharge Region

With the assumptions made in the previous section the unknown variables in the multi-fluid equations are reduced to the number densities and three velocity components of 23 species, and two temperatures, $T_e$ and $T_h$. Following Eq. (2.87) the species continuity equation for heavy species $p$ and electrons $e$ is

$$\frac{\partial n_p}{\partial t} + \nabla \cdot \left(n_p \mathbf{u}_p\right) = \sum_{r=1}^{N_r} \left(\nu_{r,p} - \nu'_{r,p}\right) k_r \prod_{p=1}^{N_s} n_p^{\nu_{r,p}}. \tag{3.1}$$

The electron momentum conservation equation (2.47) with transfer collision integrals derived in Section 2.4 is

$$m_e n_e \frac{\partial \mathbf{u}_e}{\partial t} + m_e n_e \left(\mathbf{u}_e \cdot \nabla\right) \mathbf{u}_e + \nabla p_e + e n_e \left(\mathbf{E} + \mathbf{u}_e \times \mathbf{B}\right) = \sum_{h=1}^{N_h} \overline{\nu}_{e,h} n_e m_e \left(\mathbf{u}_h - \mathbf{u}_e\right), \tag{3.2}$$

The heavy particles momentum equation (2.47) with transfer collision integrals derived in Section 2.4 is

$$m_p n_p \frac{\partial \mathbf{u}_p}{\partial t} + m_p n_p \left(\mathbf{u}_p \cdot \nabla\right)\mathbf{u}_p + \nabla p_p - eZ_p n_p \left(\mathbf{E} + \mathbf{u}_p \times \mathbf{B}\right) = \sum_{q=1}^{N_s} \overline{\nu}_{p,q} n_p \mu_{pq} \left(\mathbf{u}_q - \mathbf{u}_p\right). \quad (3.3)$$

The electron energy equation (2.43) is

$$\begin{aligned} &\frac{\partial}{\partial t}\left(\frac{3}{2}k_B T_e n_e\right) + \nabla \cdot \left(\frac{5}{2} n_e \mathbf{u}_e k_B T_e\right) - e n_e \mathbf{u}_e \cdot \mathbf{E} \\ &= \sum_{h=1}^{N_h} 3\frac{m_e}{m_h} \nu_{eh}^{\text{elast}} n_e k_B \left(T_h - T_e\right) + \sum_{r=1}^{N_r} E_{r,\text{th}} k_r \prod_{p=1}^{N_s} n_p^{\nu_{r,p}}. \end{aligned} \quad (3.4)$$

The energy equation of the mixture as a whole (2.50) is

$$\begin{aligned} &\frac{\partial}{\partial t}\left(n\frac{mu^2}{2} + n\left(U_{\text{tr}} + U_{\text{int}}\right)\right) + \nabla \cdot \left[n\frac{mu^2}{2}\mathbf{u} + n\left(U_{\text{tr}} + U_{\text{int}}\right)\mathbf{u} + p\mathbf{u} + \mathbf{q}_{\text{int}} + \mathbf{q}_{\text{tr}}\right] \\ &-\sum_{p=1}^{N_s} eZ_p n_p \mathbf{u}_p \cdot \mathbf{E} = 0. \end{aligned} \quad (3.5)$$

The equations of state for each component of the plasma are given by Eqs. (2.36)

$$\begin{aligned} p_p &= n_p k_B T_h \ (p \neq e), \\ p_e &= n_e k_B T_e. \end{aligned} \quad (3.6)$$

### 3.1.3 Electromagnetic Model for the HCNHIS Discharge Region

The fluid equations (3.1)-(3.6) are still not closed since they contain unknown variables $\mathbf{E}$ and $\mathbf{B}$ which should be obtained in a self-consistent manner by solving Maxwell's equations coupled with the fluid equations. In the cylindrical plasma reactor shown in Figure 7 the current in the coil which flows in azimuthal direction $\upsilon$ produces time varying magnetic field in the $z$-direction. In addition there is an electrostatic component of electric field. Thus the total electric field can be represented as

$$\mathbf{E} = \mathbf{E}_{\text{a}} + \mathbf{E}_{\text{rf}}, \quad (3.7)$$

where from Eq. (2.91) the electrostatic field contribution is

$$\mathbf{E}_{\text{a}} = -\nabla \phi \quad (3.8)$$

and the radio-frequency component is

$$\mathbf{E}_{\text{rf}} = -\frac{\partial \mathbf{A}}{\partial t}. \quad (3.9)$$

If capacitive coupling is neglected which is a good approximation for high electron number densities expected in the HCNHIS (Chabert & Braithwaite, 2011) then there is only one component of the external current and the vector potential is in the $v$ direction. Then Eq. (2.98) becomes

$$\epsilon\mu\frac{\partial^2 A_v}{\partial t^2} - \nabla^2 A_v + \frac{1}{r^2}A_v = \mu J_v^{\text{tot}}, \tag{3.10}$$

where $J_v^{\text{tot}} = J_v^{\text{cond}} + J_v^{\text{coil}}$ is the total current.

Using phasor representation the vector potential, the electric field and the current can be written as

$$\begin{aligned} A_v &= \Re\left(\tilde{A}_v(\mathbf{r})e^{i\omega t}\right), \\ E_{\text{rf},v} &= \Re\left(\tilde{E}_{\text{rf},v}(\mathbf{r})e^{i\omega t}\right), \\ J_v &= \Re\left(\tilde{J}_v(\mathbf{r})e^{i\omega t}\right), \end{aligned} \tag{3.11}$$

where $i = \sqrt{-1}$, $\omega$ is the applied frequency, the tilde denotes complex amplitudes of the corresponding variables, and $\Re(z)$ is the real part of a complex number $z$.

With Eq. (3.11) the time dependence in Eq. (3.10) can be eliminated and the equation for the complex amplitudes becomes

$$\nabla^2\tilde{A}_v + \left(\omega^2\epsilon\mu - \frac{1}{r^2}\right)\tilde{A}_v = -\mu\tilde{J}_v^{\text{tot}}. \tag{3.12}$$

From Eqs. (3.9) and (3.11) $\tilde{E}_{\text{rf},v} = -i\omega\tilde{A}_v$ and Eq. (3.12) can be written in terms of $\tilde{E}_{\text{rf},v}$ as

$$\nabla^2\tilde{E}_{\text{rf},v} + \left(\omega^2\epsilon\mu - \frac{1}{r^2}\right)\tilde{E}_{\text{rf},v} = i\omega\mu\tilde{J}_v^{\text{tot}}. \tag{3.13}$$

The equation for the electrostatic potential is given by Eq. (2.98)

$$\nabla^2\phi = -\frac{e}{\epsilon}\sum_{p=1}^{N_s} Z_p n_p. \tag{3.14}$$

Equations (3.12), (3.13) and (3.14) contain yet unknown parameters $\epsilon$ and $\mu$. For the plasma in the HCNHIS the permeability $\mu$ is that of vacuum

$$\mu = \mu_0. \tag{3.15}$$

In order to obtain the plasma permittivity $\epsilon$ it is necessary to consider the electron momentum equation (3.2). For the azimuthal velocity component $u_{e,\upsilon}$ due to the symmetry of the plasma reactor and the assumption that the ions do not respond to radio-frequency electric field, the heavy species velocity components in the collision term and the pressure gradient as well as the electron inertia term can be neglected in Eq. (3.2) leading to the following equation

$$-m_e n_e \frac{\partial u_{e,\upsilon}}{\partial t} + e n_e E_{\mathrm{rf},\upsilon} = n_e m_e u_{e,\upsilon} \sum_{h=1}^{N_h} \overline{\nu}_{e,h}. \tag{3.16}$$

Using phasor representation as in Eq. (3.11) the above equation can be transformed into Ohm's law with the complex conductivity as

$$\tilde{J}_{e,\upsilon} = \sigma_{\mathrm{cond}} \tilde{E}_{\mathrm{rf},\upsilon}, \tag{3.17}$$

where $\sigma_{\mathrm{cond}} = \left( i\omega + \sum_{h=1}^{N_h} \overline{\nu}_{e,h} \right)^{-1} \epsilon_0 \omega_{pe}^2$ is plasma conductivity in the cold plasma approximation (Lymberopoulos & Economou, 1995) and $\omega_{pe} = \sqrt{\frac{e^2 n_e}{\epsilon_0 m_e}}$ is the plasma frequency.

The plasma can also be considered as a dielectric medium with the complex permittivity $\epsilon_{\mathrm{pl}}$ defined as (Lieberman & Lichtenberg, 2005; Chabert & Braithwaite, 2011)

$$\epsilon_{\mathrm{pl}} = \epsilon_0 \left[ 1 + \omega_{pe}^2 \left( i\omega \sum_{h=1}^{N_h} \overline{\nu}_{e,h} - \omega^2 \right)^{-1} \right]. \tag{3.18}$$

By using Eqs. (3.7) and (3.17) and noting that $\mathbf{E}_{\mathrm{rf}} \perp \mathbf{E}_a$ Eq. (3.4) can be written as

$$\begin{aligned} &\frac{\partial}{\partial t}\left( \frac{3}{2} k_B T_e n_e \right) + \nabla \cdot \left( \frac{5}{2} n_e \mathbf{u}_e k_B T_e \right) - e n_e \mathbf{u}_e \cdot \mathbf{E}_{\mathrm{a}} \\ &= \sum_{h=1}^{N_h} 3 \frac{m_e}{m_h} \nu_{eh}^{\mathrm{elast}} n_e k_B \left( T_h - T_e \right) + \sum_{r=1}^{N_r} E_{r,\mathrm{th}} k_r \prod_{p=1}^{N_s} n_p^{\nu_{r,p}} + \frac{P_{\mathrm{abs}}}{V}, \end{aligned} \tag{3.19}$$

where the term $e n_e \mathbf{u}_e \cdot \mathbf{E}_a$ represents the work of the electrostatic field. The term $P_{\mathrm{abs}}$ is the averaged absorbed power over period of oscillation by the electrons from the radio-frequency electric field and is given by

$$\frac{P_{\text{abs}}(\mathbf{r})}{V} = \frac{\omega}{2\pi}\int_0^{\frac{2\pi}{\omega}} en_e u_{e,v}(\mathbf{r},t) E_{\text{rf},v}(\mathbf{r},t)dt = \frac{1}{2}\Re\left(\tilde{J}_v(\mathbf{r})^* \tilde{E}_{\text{rf},v}(\mathbf{r})\right), \tag{3.20}$$

where the asterisk implies a complex conjugate.

The multi-fluid equations (3.1)-(3.6) from Section 3.1.2 coupled with Eqs. (3.12) and (3.14) and boundary conditions from Section 2.6 form a system of 23 Navier-Stokes like PDEs coupled with the PDEs (3.12) and (3.14) for the electromagnetic field. In the subsequent section, a reduction is accomplished to achieve a computationally tractable Global Enhanced Vibrational Kinetic Model (GEVKM) for the HCNHIS that preserves the physical processes.

## 3.2 Global Enhanced Vibrational Kinetic Model for the HCNHIS

The species temperature in the GEVKM is considered to be uniform in the plasma reactor which is true for the bulk of the plasma except the regions near the walls where most of RF power is deposited. We begin with the derivation of the transport models in the bulk and the derivation of wall flux models, considering the continuity and momentum equations.

### 3.2.1 Continuity and Momentum Equations for the Bulk

The spatial variation of the number densities of the plasma components is assumed to follow a distribution expressed as

$$n_p(\mathbf{r}) = n_{p,0}\Xi_{p,R}(r)\Xi_{p,L}(z), \tag{3.21}$$

where $n_{p,0}$ is the number density in the center of the discharge, $\Xi_{p,R}(r)$ and $\Xi_{p,L}(z)$ are the non-dimensional shape factors for species $p$ number density in the $r$- and $z$-directions respectively. The origin of the reference frame coincides with the center of the discharge as shown in Figure 7. The shape factors in Eq. (3.21) are estimated based on analytic and heuristic solutions of continuity and momentum equations for different regimes of operation from low-pressure (collisionless) to high-pressure (collisional) regimes (Lieberman & Lichtenberg, 2005).

The governing equations in GEVKM include the steady-state continuity equations for heavy species based on Eq. (3.1)

$$\nabla\cdot\left(n_p\mathbf{u}_p\right) = \sum_{r=1}^{N_r}\left(\nu_{r,p} - \nu'_{r,p}\right)k_r\prod_{p=1}^{N_s} n_p^{\nu_{r,p}}\ . \tag{3.22}$$

Instead of solving the electrons continuity equation the quasi-neutrality condition for charged species in the bulk is used

$$\sum_{p=1}^{N_h} Z_p n_p - n_e = 0, \tag{3.23}$$

which for hydrogen plasma takes the form

$$n_e + n_{\mathrm{H}^-} = n_{\mathrm{H}^+} + n_{\mathrm{H}_2^+} + n_{\mathrm{H}_3^+}. \tag{3.24}$$

In essence, the quasi-neutrality condition is a manifestation that the electron Debye length in the discharge is much smaller compared to the dimensions of the discharge chamber as follows from the estimation (Lieberman & Lichtenberg, 2005)

$$\frac{1}{n_e}\left|\sum_{p=1}^{N_s} Z_p n_p\right| \lesssim \frac{\lambda_{De,0}^2}{L^2}, \tag{3.25}$$

where $\lambda_{De,0} = \sqrt{\frac{\epsilon_0 k_B T_e}{n_{0,e} e^2}}\sqrt{\frac{1}{1+\gamma\alpha_0}}$ is the electron Debye length modified due to the presence of negative ions $\mathrm{H}^-$ (Chabert & Braithwaite, 2011), $\gamma = T_e / T_h$, $\alpha_0 = n_{\mathrm{H}^-,0} / n_{e,0}$ is the electronegativity in the center of the reactor, and $L$ is the characteristic length (radius or length of the discharge chamber in a case of the HCNHIS). The condition $\lambda_{De,0}^2 / L^2 \ll 1$ should be checked *a posteriori* once the solution of the problem is obtained.

The steady-state heavy species momentum equation (3.3) is written as

$$m_p n_p \left(\mathbf{u}_p \cdot \nabla\right)\mathbf{u}_p + \nabla p_p - eZ_p n_p \mathbf{E}_a = \sum_{q=1}^{N_s} \overline{\nu}_{p,q} n_p \mu_{pq} \left(\mathbf{u}_q - \mathbf{u}_p\right), \tag{3.26}$$

where it is assumed that ions do not respond to radio-frequency electric field and the effect of the magnetic field is negligible compared to other terms. Also due to the symmetry of the plasma reactor all vectors in Eq. (3.26) do not contain azimuthal components.

Similarly to Eq. (3.26) the steady-state electrons momentum equation (3.2) is written as

$$m_e n_e \left(\mathbf{u}_e \cdot \nabla\right)\mathbf{u}_e + \nabla p_e + en_e \mathbf{E}_a = \sum_{h=1}^{N_h} \overline{\nu}_{e,h} n_e m_e \left(\mathbf{u}_h - \mathbf{u}_e\right). \tag{3.27}$$

The momentum equations (3.26) and (3.27), species continuity equations (3.22) and quasi-neutrality condition (3.23) will be used in the next subsections in the derivation of the bulk plasma transport models for different pressure regimes.

Averaging the continuity equations (3.22) over the discharge volume provides

$$Q_{p,\mathrm{out}} - Q_{p,\mathrm{in}} + \frac{A}{V}\Gamma_{p,\mathrm{wall}} = \sum_{r=1}^{N_r} k_r \Theta_r \prod_{p=1}^{N_s} \left(\nu_{r,p} - \nu'_{r,p}\right) \overline{n}_p^{\nu_{r,p}}, \tag{3.28}$$

where $Q_{p,\mathrm{out}}$ is the particle flow rate leaving the discharge by pumping or through the nozzle or orifice depending on particular configuration of the plasma reactor, $Q_{p,\mathrm{in}}$ is the particle flow rate into the discharge chamber, $\Gamma_{p,\mathrm{wall}}$ is the particles flux at the walls which is, in general, not zero due to wall chemical reactions, $V$ is the entire volume of the plasma reactor, $A$ is the internal area of the plasma reactor.

The volume-averaged number density $\overline{n}_p$ is given as

$$\overline{n}_p = \frac{1}{V}\int_V n_p(\mathbf{r})d\mathbf{r} = n_{p,0}\theta_p, \tag{3.29}$$

where

$$\theta_p = \Lambda_{R,p}\Lambda_{L,p}, \tag{3.30}$$

and

$$\begin{aligned} \Lambda_{R,p} &= \frac{2}{R^2}\int_0^R r\Xi_{R,p}(r)dr, \\ \Lambda_{L,p} &= \frac{1}{L}\int_{-L/2}^{L/2} \Xi_{L,p}(z)dz. \end{aligned} \tag{3.31}$$

The nondimensional coefficients $\Theta_r \sim \mathrm{O}(1)$ are similar to those obtained in Monahan and Turner (2008), Monahan and Turner (2009) and given by

$$\Theta_r = \frac{2}{R^2 L}\frac{\int_0^R \left[\prod_{j=1}^{N_s} \Xi_{R,j}^{\nu_{r,j}}\left(r\right)\right] r dr \int_{-L/2}^{L/2} \left[\prod_{j=1}^{N_s} \Xi_{L,j}^{\nu_{r,j}}\left(z\right)\right] dz}{\prod_{j=1}^{N_s} \theta_j^{\nu_{r,j}}}. \tag{3.32}$$

The inlet flow rate $Q_{p,\mathrm{in}}$ is usually due to the feedstock gas pumping into the discharge chamber. In such case this term is only present in the ground state molecular hydrogen continuity equation. Usually the inlet flow rates $Q_{p,\mathrm{in}}$ in Eq. (3.28) are given in the units of standard cubic centimeters per minute (sccm) or standard liters per minute (slm). The conversion between sccm or slm and m$^{-3}$s$^{-1}$ is done by using the following coefficients that are given as (Goebel & Katz, 2008)

$$Q_{\mathrm{in}}\left[\mathrm{m}^{-3}\mathrm{s}^{-1}\right] = 4.477962\times 10^{17}\,\frac{Q_{\mathrm{in}}\left[\mathrm{sccm}\right]}{V} = 4.477962\times 10^{20}\,\frac{Q_{\mathrm{in}}\left[\mathrm{slm}\right]}{V}. \tag{3.33}$$

The outlet flow rate $Q_{p,\mathrm{out}}$ depends on the particular configuration of the discharge chamber. Examples of orifice and pumps are given in Zorat et al. (2000) and Hjartarson et al. (2010) respectively.

The expression for the wall fluxes $\Gamma_{p,\mathrm{wall}}$ can be written in general as

$$\Gamma_{p,\mathrm{wall}} = \frac{1}{A}\int_0^R\int_0^{2\pi}\left(\Gamma_p\Big|_{z=-L/2} + \Gamma_p\Big|_{z=L/2}\right) r d\upsilon dr + \frac{1}{A}\int_{-L/2}^{L/2}\int_0^{2\pi}\Gamma_p\Big|_{r=R} R d\upsilon dz\,. \tag{3.34}$$

Using the symmetry of the plasma profile and the boundary conditions from Section 2.6 Eq. (3.34) can be rewritten as

$$\Gamma_{p,\mathrm{wall}} = \frac{2\pi R^2}{A} u_{p,R,\mathrm{w}} n_{p,0} \Xi_{L,p}\left(\frac{L}{2}\right)\Lambda_{R,p} + \frac{2\pi RL}{A} u_{p,L,\mathrm{w}} n_{p,0} \Xi_{R,p}(R)\Lambda_{L,p}\,, \tag{3.35}$$

where $u_{R,p,\mathrm{wall}}$ and $u_{L,p,\mathrm{wall}}$ are characteristic wall velocities of species $p$ at the side and cylinder parts of the walls respectively. The characteristic velocity for neutral species recombining on the wall can be found from Eq. (2.105)

$$u_{p,R,\mathrm{wall}} = u_{p,L,\mathrm{wall}} = \frac{\gamma_{p,\mathrm{rec}}}{2\left(2-\gamma_{p,\mathrm{rec}}\right)} v_{p,\mathrm{th}}\,. \tag{3.36}$$

For the ions the characteristic velocity is the individual Bohm velocity as can be found from Eq. (2.102).

Equation (3.35) for ions can be rewritten in the following expression (Monahan & Turner, 2009)

$$\Gamma_{X^+,\mathrm{wall}} = \left(\frac{2\pi R^2}{A} h_{L,X^+}\Lambda_{R,\mathrm{X}^+} + \frac{2\pi RL}{A} h_{R,\mathrm{X}^+}\Lambda_{L,X^+}\right) u_{\mathrm{B},X^+} n_{0,X^+}\,, \tag{3.37}$$

where the nondimensional factors $h_{R,X^+}$ and $h_{L,X^+}$ commonly used in global models are given as

$$h_{L,X^+} = \Xi_{L,X^+}\left(\frac{L}{2}\right),$$
$$h_{R,X^+} = \Xi_{R,X^+}(R). \tag{3.38}$$

With the help of Eqs. (3.30) and (3.38), Eq. (3.37) can be rewritten as

$$\Gamma_{X^+,\text{wall}} = \left(\frac{2\pi R^2}{A}\frac{h_{L,X^+}}{\Lambda_{L,X^+}} + \frac{2\pi RL}{A}\frac{h_{R,X^+}}{\Lambda_{R,X^+}}\right) u_{\text{B},X^+}\overline{n}_{X^+}. \tag{3.39}$$

It should be noted that the expressions for the wall fluxes of charged particles usually found in the literature (Lieberman & Lichtenberg, 2005; Chabert & Braithwaite, 2011) are different from Eq. (3.37) as was pointed out by Monahan and Turner (2009) due to the lack of coefficients $\Lambda_{R,X^+}$ and $\Lambda_{R,X^+}$. The determination of the nondimensional shape-factors $\Xi_{R,X^+}(r)$ and $\Xi_{L,X^+}(z)$ depend on the pressure regime in the plasma reactor and, in general, can be obtained numerically. However, with some additional simplifications there exist closed form solutions that are widely used in global models (Chabert & Braithwaite, 2011). The next subsections consider the derivation of the bulk transport models applicable for the estimation of the nondimensional shape-factors.

#### *3.2.1.1 High-Pressure Regime*

Scaling analysis of Eq. (3.26) shows that at high pressures the inertia term can be neglected if

$$u \ll \sqrt{\frac{k_B T_h}{m_p}},$$
$$\overline{\nu} \gg \frac{u}{L}, \tag{3.40}$$

where $u$ is the characteristic speed of the ion flow in the bulk, $L$ is the characteristic dimension of the plasma reactor, and $\overline{\nu}$ is the characteristic collision frequency.

The momentum equation becomes

$$\nabla p_p - eZ_p n_p \mathbf{E}_a = \sum_{q=1}^{N_s} \overline{\nu}_{p,q} n_p \mu_{pq}\left(\mathbf{u}_q - \mathbf{u}_p\right). \tag{3.41}$$

If in addition only collisions with the background gas (for the HCNHIS it is molecular hydrogen) are considered and the velocity of the background gas is assumed to be much smaller than the velocity of the ions or neutral trace species, Eq. (3.41) can be further simplified to the so-called drift-diffusion approximation (Golant, et al., 1980)

$$\Gamma_p = n_p \mathbf{u}_p = b_p n_p \mathbf{E}_a - D_p \nabla n_p - D_p^T n_p \nabla \ln T_h, \tag{3.42}$$

where index $q$ represents the background gas, $b_p = \frac{eZ_p}{\overline{\nu}_{p,q} \mu_{pq}}$ is the mobility of charged species $p$,

$D_p = \frac{k_B T_h}{\overline{\nu}_{p,q} \mu_{pq}}$ is the diffusion coefficient of the species $p$ in a background gas $q$, and

$D_p^T = \frac{k_B T_h}{\overline{\nu}_{p,q} \mu_{pq}}$ is the thermal diffusion coefficient of the species $p$ in a background gas $q$.

If the temperature of heavy particles $T_h$ is uniform, a typical assumption in global models, then the last term in Eq. (3.42) drops out resulting in

$$\Gamma_p = n_p \mathbf{u}_p = b_p n_p \mathbf{E}_a - D_p \nabla n_p. \tag{3.43}$$

For neutral particles Eq. (3.43) does not contain the electric force term.

For electrons the steady-state momentum equation (3.2) for the non-azimuthal velocity components is written as

$$\nabla p_e + e n_e \mathbf{E}_a = \sum_{h=1}^{N_h} \overline{\nu}_{e,h} n_e m_e \left( \mathbf{u}_h - \mathbf{u}_e \right), \tag{3.44}$$

where the electron inertia is neglected (Sentis, 2014). If in addition electrons are isothermal and collide mostly with the stationary background as expressed in Eq. (3.42) then Eq. (3.44) becomes

$$\Gamma_e = n_e \mathbf{u}_e = -b_e n_e \mathbf{E}_a - D_e \nabla n_e, \tag{3.45}$$

where $b_e = \frac{e}{\overline{\nu}_{e,q} m_e}$ is the electron mobility, the index $q$ represents background gas, and

$D_e = \frac{k_B T_e}{\overline{\nu}_{e,q} m_e}$ is the electron diffusion coefficient.

In order to determine the electrostatic field $\mathbf{E}_a$ Eq. (3.14) should be solved. If however the congruent assumption (Lieberman & Lichtenberg, 2005) of the current in the bulk of the plasma is used i.e. the total conduction current is zero in the bulk of the plasma or in other words there is no built up of the space charge (this assumption does not hold for the sheath on the walls) then

$$\sum_{p=1}^{N_h} Z_p \Gamma_p - \Gamma_e = 0. \tag{3.46}$$

Substituting Eqs. (3.43) and (3.45) into Eq. (3.46) we obtain

$$\mathbf{E}_a = \frac{\sum_{p=1}^{N_h} Z_p D_p \nabla n_p - D_e \nabla n_e}{\sum_{p=1}^{N_h} Z_p b_p n_p + b_e n_e}. \tag{3.47}$$

Substituting Eq. (3.47) back into Eqs. (3.43) and (3.45) we obtain for the ion species

$$\Gamma_p = \frac{b_p n_p \sum_{q=1, q\neq p}^{N_h} Z_q D_q \nabla n_q}{\sum_{q=1}^{N_h} Z_q b_q n_q + b_e n_e} - \frac{b_p n_p D_e}{\sum_{q=1}^{N_h} Z_q b_q n_q + b_e n_e} \nabla n_e + D_p \left( \frac{Z_p b_p n_p}{\sum_{q=1}^{N_h} Z_q b_q n_q + b_e n_e} - 1 \right) \nabla n_p \tag{3.48}$$

and for electrons

$$\Gamma_e = -\frac{b_e n_e \sum_{p=1}^{N_h} Z_p D_p \nabla n_p}{\sum_{p=1}^{N_h} Z_p b_p n_p + b_e n_e} + D_e \left( \frac{b_e n_e}{\sum_{p=1}^{N_h} Z_p b_p n_p + b_e n_e} - 1 \right) \nabla n_e. \tag{3.49}$$

From the above two equations the ambipolar tensor (Rosenau & Turkel, 1985) can be introduced

$$\left(\hat{D}_a\right)_{ij} = \begin{cases} \left( \sum_{k=1}^{N_h} Z_k b_k n_k + b_e n_e \right)^{-1} b_i n_i Z_j D_j, i \neq j \\ D_i \left[ \left( \sum_{k=1}^{N_h} Z_k b_k n_k + b_e n_e \right)^{-1} Z_i b_i n_i - 1 \right], i = j, \end{cases} \tag{3.50}$$

which relates particle fluxes to the gradients of species number densities as

$$\Gamma_i = \left(\hat{D}_a\right)_{ij} \nabla n_j, \tag{3.51}$$

where the summation of the repeated indexes is implied due to Einstein summation convention. In general, the ambipolar tensor is not diagonal but under certain conditions it can be diagonalized.

For the HCNHIS, three positive ions and one negative ion are considered. It is assumed that electrons and negative ions are in Boltzmann equilibrium (Lieberman & Lichtenberg, 2005) which implies that the collision terms can be neglected compared to pressure and electric terms in the electron (3.44) and negative ion momentum (3.41) equations resulting in

$$\begin{aligned} n_e &= n_{e,0} e^{\frac{e\phi}{k_B T_e}}, \\ n_- &= n_{-,0} e^{-\frac{eZ_-\phi}{k_B T_h}}, \end{aligned} \tag{3.52}$$

where the subscript "$-$"refers to the negative ion species, the potential is chosen to be zero in the center of the discharge resulting in the number densities $n_{e,0}$ and $n_{-,0}$ for electrons and negative ions respectively as expected from Eq. (3.21).

In addition, the space distributions of positive ions are assumed to have the same shape

$$\frac{\nabla n_p}{n_p} = \frac{\nabla n_+}{n_+}. \tag{3.53}$$

Then using Eqs. (3.23), (3.52), and (3.53) the ambipolar diffusion coefficients for positive ions of species $p$ can be written using Eq. (3.50) as

$$D_{a,p} = D_p \frac{1 + Z_p\gamma + Z_-(Z_- - Z_p)\alpha\gamma}{1 + Z_-^2\alpha\gamma}, \tag{3.54}$$

where $\gamma = T_e / T_h$, $Z_-$ is the charge of the negative ion, $\alpha = n_e / n_-$ is the electronegativity of the plasma. Similar expression for only one positive and one negative single charged ions is given in Lieberman and Lichtenberg (2005). The positive ions particle fluxes become

$$\Gamma_p = -D_{a,p} \nabla n_p. \tag{3.55}$$

Substituting Eq. (3.55) into Eq. (3.22) one gets steady-state diffusion equation for positive ions or for neutrals (if $D_{a,p}$ is replaced by $D_p$)

$$\nabla \cdot \left( D_{a,p} \nabla n_p \right) = -\sum_{r=1}^{N_r} \left( \nu_{r,p} - \nu'_{r,p} \right) k_r \prod_{j=1}^{N_s} n_j^{\nu_{r,j}}. \tag{3.56}$$

Since it was assumed that $T_e$ and $T_h$ are uniform in the discharge the above equation gives positive ions (or neutral particles) space distribution at given $T_e$ and $T_h$. In the global models Eq. (3.56) is not explicitly solved but its solution in limiting simple cases is used in order to es-timate the nondimensional shape factors in Eq. (3.21). For this purpose consider a case of a single positive ion for which the production occurs in the volume of the discharge with the constant effective collision frequency $\overline{\nu}_{\mathrm{iz}}$ (here the charge neutrality is assumed) and the losses are to the wall then Eq. (3.56) is written as

$$\nabla \cdot \left( D_{a,p} \nabla n_p \right) = -\overline{\nu}_{\mathrm{iz}} n_p \,. \tag{3.57}$$

The solution of Eq. (3.57) under assumption of constant ambipolar diffusion coefficient with Schottky boundary condition (2.103) is (Lieberman & Lichtenberg, 2005)

$$n_p = n_{0,p} \mathrm{J}_0 \left( \frac{\chi_{01} r}{R} \right) \cos\left( \frac{\pi z}{L} \right), \tag{3.58}$$

where $\chi_{01} \simeq 2.405$ is the first zero of the zero order Bessel function $\mathrm{J}_0\left(\chi\right)$.

The importance of the Eq. (3.58) is that it can be used to estimate nondimensional shape-factors $\Xi_{R,p}$ and $\Xi_{L,p}$, introduced by Eq. (3.21)

$$\begin{aligned} \Xi_{R,p}(r) &= \mathrm{J}_0\left( \frac{\chi_{01} r}{R} \right), \\ \Xi_{L,p}(z) &= \cos\left( \frac{\pi z}{L} \right). \end{aligned} \tag{3.59}$$

For the neutral trace species (atoms, electronically and vibrationally excited molecules) Eq. (3.40) is valid for low speed flows. Therefore, in plasma reactors where the gas flow is subsonic this condition is valid no matter which pressure regime is considered. If the dissociation or excitation collision frequency $\overline{\nu}_{\mathrm{ex}}$ is constant throughout the reactor then Eq. (3.57) is valid for neutrals as well resulting in

$$\nabla \cdot \left( D_n \nabla n_n \right) = -\overline{\nu}_{\mathrm{ex}} n_n \,, \tag{3.60}$$

where subscript $n$ refers to neutral trace species and the ambipolar diffusion coefficient $D_{a,p}$ is replaced by the conventional diffusion coefficient $D_n$.

### *3.2.1.2 Low-Pressure Regime*

At lower pressures the particle continuity equations are still represented in the form expressed by Eq. (3.22). In the low-pressure regime, scaling analysis of Eq. (3.22) shows that if

$$u \gg \sqrt{\frac{k_B T_h}{m_p}},$$
$$\overline{\nu} \ll \frac{u}{L}. \tag{3.61}$$

then pressure and collision terms in Eq. (3.26) can be neglected resulting in

$$m_p n_p \left(\mathbf{u}_p \cdot \nabla\right)\mathbf{u}_p = eZ_p n_p \mathbf{E}_a. \tag{3.62}$$

If, in addition, electrons and negative ions are in Boltzmann equilibrium then Eq. (3.52) applies

$$n_e = n_{e,0} e^{\frac{e\phi}{k_B T_e}},$$
$$n_- = n_{-,0} e^{-\frac{eZ_-\phi}{k_B T_h}}. \tag{3.63}$$

In fact, Eq. (3.62) represents a conservation of energy of newly created positive ions accelerated by the electric field. Integrating Eq. (3.62) along the trajectory of ions gives

$$m_p v_p^2(x', x) = 2eZ_p \left(\phi(x') - \phi(x)\right), \tag{3.64}$$

where $\phi(x')$ is the potential at the point $x'$ where the ion was born and $\phi(x)$ is the potential at some point $x$, and $v_p(x', x)$ is the velocity of an individual ion born at $x'$ accelerated to the point $x$.

Similar to what was done in Section 3.2.1.1 it is assumed that there is only one positive ion and no negative ions present in the plasma and that the production of ions occurs in the volume of the discharge with constant effective collision frequency $\overline{\nu}_{\text{iz}}$ as in the Eq. (3.57). For a one-dimensional case the ion flux created in the volume can be written as (Tonks & Langmuir, 1929)

$$x^\beta dn_p = \frac{x'^\beta \overline{\nu}_{\text{iz}} n(x') dx'}{v_p(x', x)}, \tag{3.65}$$

where $\beta$ takes the values 0, 1, 2 for parallel plane, cylindrical and spherical geometries respectively, $x$ and $x'$ represent the corresponding coordinates for the plane, cylindrical and spherical geometries.

Combining Eqs. (3.63), (3.64), and (3.65) yields

$$x^{\beta} e^{\frac{e\phi(x)}{k_B T_e}} = \sqrt{\frac{m_p}{2eZ_p}} \int_0^x e^{\frac{e\phi(x')}{k_B T_e}} \frac{x'^{\beta} \overline{\nu}_{\text{iz}} dx'}{\sqrt{\phi(x') - \phi(x)}}. \tag{3.66}$$

Equation (3.66) has an analytic solution for a parallel plate case in terms of Dawson functions (Harrison & Thompson, 1959) but the general case requires numerical solution. Once the potential is found, the number density and velocity can be recovered from Eqs. (3.63) and (3.64) respectively. In this case the shape-factors $\Xi$ in Eq. (3.21) could be evaluated numerically.

#### *3.2.1.3 Intermediate Pressure Regime*

The two cases considered so far cover the collisionally dominated and collisionless plasmas. The regime of operation between these two is referred to as the intermediate pressure regime (Lieberman & Lichtenberg, 2005; Chabert & Braithwaite, 2011). Scaling analysis shows that if

$$\begin{aligned} &\frac{u}{L} \ll \overline{\nu}, \\ &\frac{k_B T_h}{m_p} \ll \overline{\nu} L u, \end{aligned} \tag{3.67}$$

then the particle momentum equation can be simplified to

$$\Gamma_p = n_p \mathbf{u}_p = \frac{eZ_p}{\overline{\nu}_{p,q} \mu_{pq}} n_p \mathbf{E}_a = b_p n_p \mathbf{E}_a, \tag{3.68}$$

As in the low-pressure limit, the collision frequency $\overline{\nu}_{p,q}$ in Eq. (3.68) was derived in Section 2.4 under the assumption (2.53) which is different from the assumptions given by Eq. (3.67). As a reminder the collision frequency in Eq. (3.68) is considered to be between ion $p$ and the background gas $q$ species and from Eq. (2.85) is written as

$$\overline{\nu}_{pq}^{\text{elast}} = \frac{16}{3} Q_{p,q}^{(m)} \sqrt{\frac{k_B T_h}{2\pi \mu_{pq}}} n_q, \tag{3.69}$$

where $Q_{p,q}^{(m)} = \int_0^{\infty} x^5 e^{-x^2} \sigma_{p,q}^{(m)} \left( x \sqrt{\frac{2k_B T_h}{\mu_{pq}}} \right) dx$.

If the stricter condition compared to Eq. (3.67) is used

$$\sqrt{\frac{k_B T_h}{m_p}} \ll u\,, \tag{3.70}$$

then the collision frequency in Eq. (3.68) can be written (Benilov, 1997) as

$$\overline{\nu}_{pq}^{\mathrm{elast}} = \overline{Q}_{p,q}^{(m)}(u_p) u_p n_q\,, \tag{3.71}$$

where $\overline{Q}_{p,q}^{(m)}(u_p)$ is the velocity-averaged cross section. In this case the ions mobility is given by

$$b_p = \frac{eZ_p}{\overline{Q}_{p,q}^{(m)}(u_p) u_p n_q \mu_{pq}}\,, \tag{3.72}$$

which corresponds to the expression for the variable mobility regime found in the literature (Lieberman & Lichtenberg, 2005; Chabert & Braithwaite, 2011).

If the plasma is composed of one positive species and no negative species and the ions are produced by ionization in the bulk, then continuity and momentum equations for positive ions become

$$\begin{aligned} \nabla \cdot \left(n_p \mathbf{u}_p\right) &= \overline{\nu}_{\mathrm{iz}} n_p, \\ n_p \mathbf{u}_p &= \frac{eZ_p}{\overline{Q}_{p,q}^{(m)}(u_p) u_p n_q \mu_{pq}} n_p \mathbf{E}_a. \end{aligned} \tag{3.73}$$

If, in addition, the electrons are in Boltzmann equilibrium and the background gas has a uniform density in the reactor then the system of the continuity and momentum equations from Eqs. (3.73) becomes

$$\begin{aligned} \nabla \cdot \left(n_p \mathbf{u}_p\right) &= \overline{\nu}_{\mathrm{iz}} n_p, \\ n_p \mathbf{u}_p &= -\frac{eZ_p}{\overline{Q}_{p,q}^{(m)}(u_p) n_q \mu_{pq}} \frac{k_B T_e}{e} \frac{\nabla n_p}{n_p}. \end{aligned} \tag{3.74}$$

The above system can be solved numerically or analytically for a one-dimensional geometry (Lieberman & Lichtenberg, 2005) in order to obtain the shape-factors $\Xi$.

#### *3.2.1.4 Heuristic Particle Fluxes to the Walls for the High-Intermediate-Low Pressure Regime*

In the Sections 3.2.1.1-3.2.1.3 a number of models were obtained for different pressure regimes in the plasma reactor. These models are not explicitly solved in this work. The only values used in the simulations are the factors given by Eq. (3.38). In global models these factors are joined smoothly by using heuristic expression derived based on the simplified fluid models and in some cases verified by PIC simulations (Monahan & Turner, 2009).

More generally the positive ions $X^+$ wall loses can be estimated by patching low and high-pressure solutions using the heuristic expression of the form (3.39)

$$\Gamma_{X^+,\mathrm{wall}} = \frac{A_{\mathrm{eff},X^+}}{A} u_{\mathrm{B},X^+} \overline{n}_{X^+}, \tag{3.75}$$

In the above equation the Bohm velocity of $X^+$ ions modified due to presence of the negative hydrogen ions is (Lieberman & Lichtenberg, 2005)

$$u_{\mathrm{B},X^+} = \sqrt{\frac{k_B T_e \left(1+\alpha_s\right)}{m_{X^+}\left(1+\alpha_s \gamma\right)}}, \tag{3.76}$$

where $\alpha_s = n_{\mathrm{H}^-,s} / n_{e,s}$ is the electronegativity at the sheath edge.

The electronegativity at the sheath edge $\alpha_s$ is related to the electronegativity in the bulk $\alpha_0$ by (Lieberman & Lichtenberg, 2005)

$$\frac{\alpha_s}{\alpha_0} = \exp\left[\frac{\left(1+\alpha_s\right)\left(1-\gamma\right)}{2\left(1+\alpha_s\gamma\right)}\right]. \tag{3.77}$$

In a low-pressure hydrogen plasma it is usually $\alpha_0 \ll 1$ and $\gamma \gg 1$ and from Eq. (3.77) $\alpha_s \simeq 0$. Therefore at low electronegativity the solution of Eq. (3.77) can be omitted or replaced by a heuristic solution like it was done in Thorsteinsson and Gudmundsson (2010). At higher pressures, however, the temperature ratio could be relatively low $\gamma \simeq 1$, thus, making $\alpha_s \approx \alpha_b$. It should be emphasized that Eq. (3.77) was derived based on the Boltzmann equilibrium for both electrons and negative ions.

In Eq. (3.75) $A_{\mathrm{eff},X^+}$ is the effective area of ion losses given by

$$A_{\mathrm{eff},X^+} = 2\pi R^2 \frac{h_{L,X^+}}{\Lambda_{L,X^+}} + 2\pi R L \frac{h_{R,X^+}}{\Lambda_{R,X^+}}, \tag{3.78}$$

where $h_{L,X^+}$ and $h_{R,X^+}$ are the nondimensional factors introduced in Eqs. (3.38) and $\Lambda_{R,X^+}$ and $\Lambda_{L,X^+}$ are given by Eq. (3.31). These factors are given by the following expressions which were found by heuristically patching low and high-pressure solutions of the simple plasma fluid models discussed earlier (Hjartarson, et al., 2010)

$$\begin{aligned} h_R &= \frac{1}{1+\alpha_0}\left[0.80\left(4+\frac{\eta R}{\lambda_{X^+}}+\left(\frac{0.80 R u_{B,X^+}}{\chi_{01}\mathrm{J}_1\left(\chi_{01}\right)D_{a,X^+}}\right)^2\right)+h_c^2\right]^{-1/2}, \\ h_L &= \frac{1}{1+\alpha_0}\left[0.86\left(3+\frac{\eta L}{2\lambda_{X^+}}+\left(\frac{0.86 L u_{B,X^+}}{\pi D_{a,X^+}}\right)^2\right)+h_c^2\right]^{-1/2}, \end{aligned} \tag{3.79}$$

where $h_c = \left(1+\alpha_0\right)\left[\gamma^{1/2}+\gamma^{1/2}\left(\frac{15}{56}\frac{\eta^2 v_{X^+,\mathrm{th}}}{k_{\mathrm{rec}}\lambda_{X^+}}\right)^{1/2}\frac{n_{X^+}}{n_{\mathrm{H}^-}^{3/2}}\right]^{-1}$ is a scaling factor for a one-region flat-topped electronegative profile, $\mathrm{J}_1\left(\chi\right)$ is the first order Bessel function, $\eta = 2T_{X^+}\big/(T_{X^+}+T_{\mathrm{H}^-}) = 1$ for a special case of unique heavy particles temperature considered in this work, $k_{\mathrm{rec}}$ is the rate coefficient for mutual neutralization of the positive $X^+$ and negative $\mathrm{H}^-$ ions in the bulk of the plasma, $\lambda_{X^+}$ is the positive ion mean free path, ambipolar diffusion coefficient $D_{a,X^+}$ is given by Eq. (3.54).

Another challenge is the estimation of the coefficients $\Lambda_{R,p}$ and $\Lambda_{L,p}$. Monahan and Turner (2009) gave the following expressions for these terms for the low-pressure regime based on the low-pressure solution of Tonks and Langmuir (1929)

$$\begin{aligned} \Lambda_{R,X^+} &\approx 0.70, \\ \Lambda_{L,X^+} &\approx 0.85. \end{aligned} \tag{3.80}$$

On the other hand, from Eq. (3.58) it can be shown that in the high-pressure limit these coefficients become

$$\Lambda_{R,X^+} = \frac{2}{R^2}\int_0^R r\mathrm{J}_0\left(\frac{\chi_{01} r}{R}\right)dr = \frac{2}{\chi_{01}}\mathrm{J}_1\left(\chi_{01}\right),$$
$$\Lambda_{L,X^+} = \frac{1}{L}\int_{-L/2}^{L/2}\cos\left(\frac{\pi z}{L}\right)dz = \frac{2}{\pi}. \tag{3.81}$$

Following the idea of Danko et al. (2013) Eqs. (3.80) and (3.81) can be heuristically patched together in order to have a solution covering from low to high-pressure regime written as

$$\Lambda_{R,X^+} = 0.70\frac{b_R}{1+b_R} + \frac{2}{\chi_{01}}\mathrm{J}_1\left(\chi_{01}\right)\frac{1}{1+b_R},$$
$$\Lambda_{L,X^+} = 0.85\frac{b_L}{1+b_L} + \frac{2}{\pi}\frac{1}{1+b_L}, \tag{3.82}$$

where $b_R = 2\frac{\lambda_{X^+}}{R}\frac{T_e}{T_h}$ and $b_L = 2\frac{\lambda_{X^+}}{L}\frac{T_e}{T_h}$.

The neutral particles $X$ wall losses to the wall are considered to be diffusive but not fully absorbing; therefore, by solving Eq. (3.60) with the boundary conditions given by Eq. (2.105) they can be represented as (Chantry, 1987)

$$\Gamma_{X,\mathrm{wall}} = \left[\frac{\Lambda^2}{D_X} + \frac{2V\left(2-\gamma_{X,\mathrm{rec}}\right)}{Av_{\mathrm{th},X}\gamma_{X,\mathrm{rec}}}\right]^{-1}\frac{V}{A}\overline{n}_X, \tag{3.83}$$

where $\Lambda = \left[\left(\pi/L\right)^2 + \left(\chi_{01}/R\right)^2\right]^{-1/2}$, $\gamma_{X,\mathrm{rec}}$ is the recombination coefficient in a case of atomic hydrogen or the quenching coefficient in a case of vibrationally excited hydrogen molecules, and $D_X$ is the diffusion coefficient of the species $X$. It should be noted that in discharges with low degree of dissociation the diffusion of neutral trace species occurs mostly in presence of the background gas (molecular hydrogen in the case of the HCNHIS) as can be seen from Eq. (2.69). Therefore, the diffusion coefficient $D_X$ can be treated as the binary diffusion coefficient between neutral species $X$ and the background gas (the ground state hydrogen molecules $\mathrm{H}_2$ in the case of the HCNHIS).

### 3.2.2 Electron Energy Equation

Assuming steady-state condition the electron energy equation (3.19) can be written as

$$\nabla \cdot \left( \frac{5}{2} n_e \mathbf{u}_e k_B T_e \right) - e n_e \mathbf{u}_e \cdot \mathbf{E}_\mathrm{a} = \sum_{h=1}^{N_h} 3 \frac{m_e}{m_h} \nu_{eh}^{\mathrm{elast}} n_e k_B \left( T_h - T_e \right) + \sum_{r=1}^{N_r} E_{r,\mathrm{th}} k_r \prod_{p=1}^{N_s} n_p^{\nu_{r,p}} + \frac{P_\mathrm{abs}}{V}. \tag{3.84}$$

The electron energy equation (3.84) averaged over the discharge volume can be written as

$$\frac{5}{2} k_B T_e Q_{e,\mathrm{out}} - \frac{5}{2} k_B T_{e,\mathrm{in}} Q_{e,\mathrm{in}} + \frac{1}{V} \int_A \frac{5}{2} \Gamma_e k_B T_e dA + \frac{1}{V} \int_V e \Gamma_e \cdot \nabla \phi dV = k_B \left( T_h - T_e \right) \sum_{h=1}^{N_h} 3 \frac{m_e}{m_h} k_{eh}^{\mathrm{elast}} \Theta_h \bar{n}_h \bar{n}_e + \sum_{r=1}^{N_r} E_{r,\mathrm{th}} k_r \Theta_r \prod_{p=1}^{N_s} \bar{n}_p^{\nu_{r,p}} + \frac{P_\mathrm{abs}}{V}, \tag{3.85}$$

where $k_{eh}^{\mathrm{elast}}$ is the elastic rate coefficient of electron collisions with heavy particles of species $h$ and $k_r$ is the inelastic rate of the chemical reaction $r$ with electron participation respectively, $Q_{e,\mathrm{out}}$ and $Q_{e,\mathrm{in}}$ represent the flow rates of electrons out and into the plasma reactor respectively, $T_{e,\mathrm{in}}$ is the temperature of electrons injected into the reactor (for example, from filaments).

In order to integrate first integral in the left hand side of the Eq. (3.85) it is necessary to use boundary conditions of a type given by Eq. (2.115). These boundary conditions can be found by taking into account the plasma sheath located near the surfaces of the plasma reactor and expressing the fluxes to the walls. In the further analysis the capacitive coupling will be neglected which is good approximation for high electron number densities. Also we assume that all ions are single charged and that the potential at the edge of the sheath with respect to the wall of the reactor is $V_\mathrm{sh}$ then the flux of electrons to the sheath can be found by integrating Maxwellian distribution (Chabert & Braithwaite, 2011)

$$\Gamma_{e,s} = \frac{n_{e,s} v_{e,\mathrm{th}}}{4} e^{-\frac{eV_\mathrm{sh}}{k_B T_e}}. \tag{3.86}$$

On the other hand the flux of positive ions of species $p$ is given by Eq. (2.102)

$$\Gamma_{p,s} = n_{p,s} u_{\mathrm{B},p} = n_{p,s} \sqrt{\frac{k_B T_e \left(1 + \alpha_s\right)}{m_p \left(1 + \alpha_s \gamma\right)}}, \tag{3.87}$$

and the negative ions flux is given similarly to electrons as

$$\Gamma_{-,s} = \frac{n_{-,s} v_{-,\mathrm{th}}}{4} e^{-\frac{eV_{\mathrm{sh}}}{k_B T_e}} . \tag{3.88}$$

If the walls of the reactor are insulated (floating wall) then the net current flowing to the sheath should be zero resulting in the following implicit algebraic relation for the sheath potential

$$\sum_{p=1}^{N_h^+} \Gamma_{p,s} - \Gamma_{-,s} - \Gamma_{e,s} = 0 . \tag{3.89}$$

If the negative ions are absent from the sheath edge then $\alpha_s = 0$ and the sheath voltage is given by

$$V_{\mathrm{sh}} = \frac{k_B T_e}{e} \ln\left( \frac{4\overline{u}_{\mathrm{B}}}{v_{e,\mathrm{th}}} \right), \tag{3.90}$$

where $\overline{u}_{\mathrm{B}} = \left( \sum_{p=1}^{N_h^+} n_{p,s} \right)^{-1} \sum_{p=1}^{N_h^+} n_{p,s} u_{\mathrm{B},p}$ is the average Bohm velocity.

If there are negative ions at the edge of the sheath the following approximate expression can be used (Thorsteinsson & Gudmundsson, 2010)

$$V_{\mathrm{sh}} = \ln\left[ 4 \frac{\overline{u}_{\mathrm{B}}}{v_{e,\mathrm{th}}} \frac{1+\alpha_s}{1+\alpha_s \left( v_{\mathrm{H}^-,\mathrm{th}} / v_{e,\mathrm{th}} \right)} \right] \frac{k_B T_e}{e} . \tag{3.91}$$

Equations (3.90) and (3.91) contain average Bohm velocities which require the knowledge of the ions number densities at the edge of the sheath. These number densities can be recovered from Eqs. (3.75) and (3.87). If the assumption expressed in Eq. (3.53) is used, then the Bohm average velocity can be written as

$$\overline{u}_{\mathrm{B}} = \left( \sum_{p=1}^{N_h^+} n_{p,0} \right)^{-1} \sum_{p=1}^{N_h^+} n_{p,0} u_{\mathrm{B},p} . \tag{3.92}$$

If the reactor walls are biased then the sheath potential can be approximated by the bias potential $V_{\mathrm{bias}}$ as

$$V_{\mathrm{sh}} = V_{\mathrm{bias}} . \tag{3.93}$$

Using Eqs. (3.89) and (3.91), assuming that the negative ions current does not contribute to the current to the wall the electron energy equation (3.85), and imposing boundary conditions from Section 2.6 the volume-averaged electron energy equation can be cast as

$$\begin{aligned}&\frac{5}{2}k_B T_e Q_{e,\mathrm{out}}-\frac{5}{2}k_B T_{e,\mathrm{in}} Q_{e,\mathrm{in}}+\frac{A}{V}\sum_{p=1}^{N_h^+}\Gamma_{p,\mathrm{wall}}\left(2k_B T_e+eV_p+eV_{\mathrm{sh}}\right)\\&=k_B\left(T_h-T_e\right)\sum_{h=1}^{N_h}3\frac{m_e}{m_h}k_{eh}^{\mathrm{elast}}\Theta_h\bar{n}_h\bar{n}_e+\sum_{r=1}^{N_r}E_{r,\mathrm{th}}k_r\Theta_r\prod_{p=1}^{N_s}\bar{n}_p^{\nu_{r,p}}+\frac{P_{\mathrm{abs}}}{V}.\end{aligned} \tag{3.94}$$

where

$$V_p=\frac{1}{2}\frac{1+\alpha_s}{1+\alpha_s\gamma_-}\frac{k_B T_e}{e} \tag{3.95}$$

is the plasma potential modified due to the presence of negative ions (Chabert & Braithwaite, 2011).

### 3.2.3 Total Energy Equation

The remaining unknown variable is the temperature of heavy particles, $T_h$. In the low-pressure plasma the gas heating is usually small and the gas temperature is postulated rather than calculated during the simulation. At higher temperatures the effect of gas heating is more pronounced and cannot be disregarded. In order to close the system of equations of GEVKM another energy equation is required. In principle, if all heavy species have different temperatures it is necessary to include energy equations for all of them. In two-temperature fluid simulations the energy equation for heavy particles is often used (e.g. Baeva et al. (2012)). Another approach is to utilize a total energy equation of the plasma as a whole (Hassouni, et al., 1999). The advantage of this approach is that there is no terms associated with the particle collisions due to the energy conservation in every collision. Using the definitions of average parameters from Section 2.1 and assumptions about distribution functions from Eqs. (2.53), (2.55), and (2.56) the total energy equation in a steady-state approximation can be written from Eq. (3.5) as

$$\nabla\cdot\left[\sum_{p=1}^{N_s}\left(\frac{m_p u_p^2}{2}+U_{\mathrm{int},p}+\frac{5}{2}k_B T_{\mathrm{tr},p}\right)n_p\mathbf{u}_p+\mathbf{q}_{\mathrm{cond}}\right]-\sum_{p=1}^{N_s}eZ_p n_p\mathbf{u}_p\cdot\mathbf{E}=0, \tag{3.96}$$

where $U_{\text{int},p}$ is given by Eq. (2.59). It should be noted that this equation does not take into account the losses associated with the radiation and, thus, is valid only for optically thick plasmas. Also even though according to the assumption of the distribution functions given by Eq. (2.55) and (2.56) the conductive heat flux term $\mathbf{q}_{\text{cond}}$ should be zero this term is retained in the Eq. (3.96). The reason is that the conductive heat flux may contribute up to 25% of the total power lost to the walls (Leroy, et al., 1997; Rousseau, et al., 2004).

Using Eq. (3.7) and assuming that only electrons respond to RF field Eq. (3.96) is further transformed to

$$\nabla \cdot \left[ \sum_{p=1}^{N_s} \left( \frac{m_p u_p^2}{2} + H_p \right) n_p \mathbf{u}_p + \mathbf{q}_{\text{cond}} \right] - \sum_{p=1}^{N_s} e Z_p n_p \mathbf{u}_p \cdot \mathbf{E}_a = \frac{P_{\text{abs}}}{V}, \tag{3.97}$$

where

$$H_p = U_{\text{int},p} + \frac{5}{2} k_B T_p \tag{3.98}$$

is the species enthalpy per particle. Using the definition of the formation enthalpy $H_{p,0}$ at a reference temperature $T_{p,0}$ the species enthalpy can be rewritten as

$$H_p = H_{p,0} + C_{p,p} k_B \left( T_p - T_{p,0} \right), \tag{3.99}$$

where $C_{p,p}$ is the heat capacity per particle at constant pressure. The species formation enthalpy per molecule including ion species can be found in Lias et al. (1988) or by using Eq. (2.59).

The total energy equation (3.97) averaged over the discharge volume can be written

$$\begin{aligned} &\sum_{p=1}^{N_s} \left( \frac{m_p u_{p,\text{out}}^2}{2} + H_{p,\text{out}} \right) Q_{p,\text{out}} - \sum_{p=1}^{N_s} \left( \frac{m_p u_{p,\text{in}}^2}{2} + H_{p,\text{in}} \right) Q_{p,\text{in}} \\ &+ \frac{A}{V} \sum_{p=1}^{N_s} \left( \frac{m_p u_{p,\text{wall}}^2}{2} + H_{p,\text{wall}} \right) \Gamma_{p,\text{wall}} + \frac{A}{V} q_{\text{cond}} \Big|_{\text{wall}} - \frac{1}{V} \int_V \sum_{p=1}^{N_s} e Z_p n_p \mathbf{u}_p \cdot \mathbf{E}_a dV = \frac{P_{\text{abs}}}{V}. \end{aligned} \tag{3.100}$$

Under the assumption that the total current is zero in the plasma reactor the term representing work done by the ambipolar filed should be set to zero in Eq. (3.100) resulting in

$$\sum_{p=1}^{N_s} \left( \frac{m_p u_{p,\text{out}}^2}{2} + H_{p,\text{out}} \right) Q_{p,\text{out}} - \sum_{p=1}^{N_s} \left( \frac{m_p u_{p,\text{in}}^2}{2} + H_{p,\text{in}} \right) Q_{p,\text{in}} + \frac{A}{V} q_{\text{wall}} = \frac{P_{\text{abs}}}{V}. \tag{3.101}$$

In the above equation the term

$$\frac{A}{V}q_{\mathrm{wall}}=\frac{A}{V}\sum_{p=1}^{N_s}\left(\frac{m_p u_{p,\mathrm{wall}}^2}{2}+H_{p,\mathrm{wall}}\right)\Gamma_{p,\mathrm{wall}}+\frac{A}{V}q_{\mathrm{cond}}\Big|_{\mathrm{wall}} \tag{3.102}$$

represents the heat flow to the walls.

If the variation of the wall temperature or the heat flux at the wall is known from experiments then by using Eq. (2.119) in the total energy equation (3.101) one can obtain the heavy-particle temperature. Often such data are unknown and it is required to consider the heat transfer problem as well. In Chen et al. (1999) the heavy-particle temperature was assumed to be the same as the wall temperature. Then Eq. (3.101) reduces to

$$\begin{aligned}&\frac{P_{\mathrm{abs}}}{V}-\sum_{p=1}^{N_s}\left(\frac{m_p u_{p,\mathrm{out}}^2}{2}+H_{p,\mathrm{out}}\right)Q_{p,\mathrm{out}}+\sum_{p=1}^{N_s}\left(\frac{m_p u_{p,\mathrm{in}}^2}{2}+H_{p,\mathrm{in}}\right)Q_{p,\mathrm{in}}\\&=\frac{2\pi R}{V}\left(\mathrm{h}_{\mathrm{side}}R+\mathrm{h}_{\mathrm{cyl}}L\right)(T_h-T_\infty)+\epsilon_{\mathrm{emis}}\sigma_{\mathrm{SB}}\frac{A}{V}\left(T_h^4-T_\infty^4\right),\end{aligned} \tag{3.103}$$

where $T_\infty$ is the ambient temperature,

$$\mathrm{h}_{\mathrm{side}}=\frac{\kappa}{2R}\left(0.68+0.67\mathrm{Ra}^{1/4}\left[1+\left(\frac{0.492}{\mathrm{Pr}}\right)^{9/16}\right]^{4/9}\right) \tag{3.104}$$

is the convective heat-transfer coefficient from the side walls of the plasma reactor to the vacuum chamber assuming that the plasma reactor is placed horizontally (Holman, 2010),

$$\mathrm{h}_{\mathrm{cyl}}=\frac{\kappa}{2R}\left[0.6+0.387\left(\mathrm{Ra}\left\{1+\left(\frac{0.559}{\mathrm{Pr}}\right)^{9/16}\right\}^{-16/9}\right)^{1/6}\right]^2 \tag{3.105}$$

is the convective heat transfer coefficient from the outer cylindrical wall of the plasma reactor to the vacuum chamber, $\mathrm{Pr}=c_p\mu/\kappa$ is the Prandtl number, $\mathrm{Ra}=\rho^2 g\beta\left(T_h-T_\infty\right)8R^3c_p/(\mu\kappa)$ is the Rayleigh number, $\beta=1/T$ is the volume coefficient of expansion assuming ideal gas law, $\epsilon_{\mathrm{emis}}$ is the emissivity of the plasma reactor.

The advantage of Chen's approach is that it eliminates all complexities associated with the plasma-wall heat transfer. In a more general case, the conductive heat transfer from the plasma to the wall as well as enthalpic losses should be taken into account. Meyyappan and Govindan (1995) considered the heat transfer as the series of the resistance network which in turn is an application of the simplified versions of the boundary conditions from Section 2.6. Using Eq. (2.119) the total energy equation (3.101) becomes

$$\begin{aligned} &\frac{P_{\text{abs}}}{V} - \sum_{p=1}^{N_s}\left(\frac{m_p u_{p,\text{out}}^2}{2} + H_{p,\text{out}}\right)Q_{p,\text{out}} + \sum_{p=1}^{N_s}\left(\frac{m_p u_{p,\text{in}}^2}{2} + H_{p,\text{in}}\right)Q_{p,\text{in}} \\ &= \frac{A}{V}\sum_{p=1}^{N_s} q_p\Big|_{\text{wall}} + \frac{A}{V} q_{\text{cond}}\Big|_{\text{wall}} + \frac{P_{\text{rad}}}{V}. \end{aligned} \tag{3.106}$$

In the above equation all terms are discussed in detail in Section 2.6. Also compared to Eq. (3.101) Eq. (3.106) contains radiation term and as such valid irrespective of the optical thickness of the plasma.

In order to find the conductive heat flux term $q_{\text{cond}}\Big|_{\text{wall}}$ we follow the approach of Rousseau et al. (2004) assuming that the conductive heat flux is due to molecular hydrogen conduction only which is good approximation at low degrees of dissociation. In a case of a non-uniform gas temperature the energy equation for molecular hydrogen can be written from Eq. (2.47) as

$$\nabla \cdot \left(\kappa_{\text{H}_2} \nabla T_h\right) = -\frac{P_{\text{abs,H}_2}}{V}, \tag{3.107}$$

where $\kappa_{\text{H}_2}$ is the heat conductivity of molecular hydrogen and $P_{\text{abs,H}_2}$ represents the power deposited into molecular hydrogen by chemical reactions. Assuming that the heat conductivity $\kappa_{\text{H}_2}$ and power deposited in the hydrogen gas $P_{\text{abs,H}_2}$ are uniform in the reactor Eq. (3.107) becomes

$$\nabla^2 T_h = -\frac{P_{\text{abs,H}_2}}{V\kappa_{\text{H}_2}}. \tag{3.108}$$

In principle Eq. (3.108) with the boundary conditions given by Eq. (2.113) can be solved to get the temperature distribution in the plasma reactor. It should be stressed that even at pressure 1 Torr the rarefaction effect does play an important role (Leroy, et al., 1997). This is due to small values of the thermal accommodation coefficient of hydrogen molecules which are of the order of 0.1 for typical materials. It can be seen from the Baule formula (Goodman, 1980) which gives the approximation to the thermal accommodation coefficient as

$$\alpha_X = \frac{2.4\mu}{(1+\mu)^2}, \tag{3.109}$$

where $\mu = m_X / m_\mathrm{w}$ is the mass ratio of the gas molecules or atoms $X$ with the mass $m_X$ impinging the surface molecules with the mass $m_\mathrm{w}$.

Therefore temperature jump boundary conditions given by Eq. (2.114) as

$$q_\mathrm{cond}\Big|_\mathrm{wall} = -\kappa_{\mathrm{H}_2} \frac{\partial T_h}{\partial n}\Big|_\mathrm{wall} = \frac{2\alpha_{\mathrm{H}_2}}{2-\alpha_{\mathrm{H}_2}}\left(C_{p,\mathrm{H}_2} - \frac{1}{2}k_B\right)\frac{1}{4} n_{\mathrm{H}_2} v_{\mathrm{th},\mathrm{H}_2}\left(T_h\Big|_\mathrm{wall} - T_\mathrm{w}\right) \tag{2.114}$$

are used.

Assume the following decomposition of the conductive heat flux for a cylindrical geometry given in Figure 7 as

$$q_\mathrm{cond}\Big|_\mathrm{wall} = q_{\mathrm{cond},R}\Big|_\mathrm{wall} + q_{\mathrm{cond},L}\Big|_\mathrm{wall}, \tag{3.110}$$

where $q_{\mathrm{cond},R}\Big|_\mathrm{wall}$ the conductive heat flux to the wall in the infinite cylinder of radius $R$ and $q_{\mathrm{cond},L}\Big|_\mathrm{wall}$ is the conductive heat flux to the walls between two parallel plates separated by the distance $L$.

The expressions of the conductive heat flux to the wall in the infinite cylinder and between two parallel plates as a function of the volume-averaged temperature $\overline{T}_h$ and the wall temperature $T_\mathrm{w}$ can be found from the solutions of the one dimensional energy equations given from Eq. (3.108) with the boundary conditions (2.114) as

$$\begin{aligned} q_{\mathrm{cond},L}\Big|_\mathrm{wall} &= 2\mathrm{h}_L T_\mathrm{w}\left[\frac{\mathrm{h}_L}{\mathrm{h}_\mathrm{Kn}} + \frac{\overline{T}_h}{T_\mathrm{w}} - \sqrt{\left(\frac{\mathrm{h}_L}{\mathrm{h}_\mathrm{Kn}}\right)^2 + 2\frac{\mathrm{h}_L}{\mathrm{h}_\mathrm{Kn}}\frac{\overline{T}_h}{T_\mathrm{w}} + 1}\right], \\ q_{\mathrm{cond},R}\Big|_\mathrm{wall} &= \mathrm{h}_R T_\mathrm{w}\left[\frac{\mathrm{h}_R}{\mathrm{h}_\mathrm{Kn}} + \frac{\overline{T}_h}{T_\mathrm{w}} - \sqrt{\left(\frac{\mathrm{h}_R}{\mathrm{h}_\mathrm{Kn}}\right)^2 + 2\frac{\mathrm{h}_R}{\mathrm{h}_\mathrm{Kn}}\frac{\overline{T}_h}{T_\mathrm{w}} + 1}\right], \end{aligned} \tag{3.111}$$

where

$$
\begin{aligned}
\mathrm{h}_L &= \frac{6\kappa_{\mathrm{H}_2}}{L}, \\
\mathrm{h}_{\mathrm{Kn}} &= \frac{2\alpha_{\mathrm{H}_2}}{2-\alpha_{\mathrm{H}_2}}\left(C_{p,\mathrm{H}_2} - \frac{1}{2}k_B\right)\frac{1}{4}n_{\mathrm{H}_2}\sqrt{\frac{8k_B T_{\mathrm{w}}}{\pi m_{\mathrm{H}_2}}}, \\
\mathrm{h}_R &= \frac{4\kappa_{\mathrm{H}_2}}{R}.
\end{aligned} \tag{3.112}
$$

It should be noted that in the derivation of Eqs. (3.111) the assumption

$$
\frac{1}{T_{\mathrm{w}}}\left(T_h\big|_{\mathrm{wall}} - T_{\mathrm{w}}\right) \ll 1 \tag{3.113}
$$

was used. In the GEVKM the volume averaged temperature $\bar{T}_h$ in (3.111) is replaced by the temperature $T_h$.

It is also necessary to include an additional expression in order to determine the unknown wall temperature $T_{\mathrm{w}}$. The simplest way is to rewrite Eq. (3.103) as

$$
\begin{aligned}
&\frac{P_{\mathrm{abs}}}{V} - \sum_{p=1}^{N_s}\left(\frac{m_p u_{p,\mathrm{out}}^2}{2} + H_{p,\mathrm{out}}\right)Q_{p,\mathrm{out}} + \sum_{p=1}^{N_s}\left(\frac{m_p u_{p,\mathrm{in}}^2}{2} + H_{p,\mathrm{in}}\right)Q_{p,\mathrm{in}} \\
&= \frac{2\pi R}{V}\left(\mathrm{h}_{\mathrm{side}}R + \mathrm{h}_{\mathrm{cyl}}L\right)(T_{\mathrm{w}} - T_\infty) + \epsilon_{\mathrm{emis}}\sigma_{\mathrm{SB}}\frac{A}{V}\left(T_{\mathrm{w}}^4 - T_\infty^4\right).
\end{aligned} \tag{3.114}
$$

### 3.2.4 Simplified Electromagnetic Model for the HCNHIS Discharge Region

In order to find the time-averaged power deposition $P_{\mathrm{abs}}$ it is necessary to solve Eq. (3.12). In cylindrical coordinates this equation can be written as

$$
\frac{1}{r}\frac{\partial}{\partial r}\left(r\frac{\partial \tilde{A}_v}{\partial r}\right) + \frac{\partial^2 \tilde{A}_v}{\partial z^2} + \left(\omega^2\epsilon_{\mathrm{pl}}\mu_0 - \frac{1}{r^2} - i\omega\mu\sigma_{\mathrm{cond}}\right)\tilde{A}_v = -\mu\tilde{J}_v^{\mathrm{coil}}. \tag{3.115}
$$

Assuming long cylinder filled with the uniform plasma density the above equation reduces to

$$
\frac{1}{r}\frac{\partial}{\partial r}\left(r\frac{\partial \tilde{A}_v}{\partial r}\right) + \left(\omega^2\epsilon_{\mathrm{pl}}\mu_0 - \frac{1}{r^2} - i\omega\mu\sigma_{\mathrm{cond}}\right)\tilde{A}_v = -\mu\tilde{J}_v^{\mathrm{coil}} \tag{3.116}
$$

The solution of Eq. (3.116) is given in terms of Bessel function (Chabert & Braithwaite, 2011) allowing to get the power deposited in the plasma as

$$P_{\text{abs}} = \frac{\pi \mathcal{N}^2}{L\omega\epsilon_0} \Re\left[\frac{i\frac{\omega}{c} r_0 \mathrm{J}_1\left(\frac{\omega}{c}\sqrt{\epsilon_p} R\right)}{\sqrt{\epsilon_p}\mathrm{J}_0\left(\frac{\omega}{c}\sqrt{\epsilon_p} R\right)}\right] I_{\text{coil}}^2, \tag{3.117}$$

where $\mathcal{N}$ is the number of turns of the coil, $I_{\text{coil}}$ is the current in the coil.

The solution given by Eq. (3.117) neglects effects of capacitive coupling and as such is only applicable to high electron densities in the plasma reactor (Chabert & Braithwaite, 2011).

### 3.2.5 Chemical Reactions in the HCNHIS

The chemical reactions included into the GEVKM are listed in Tables 3-9. Electron collisions with ground state species are shown in Table 3. These reactions include various types of ionization, dissociation, and neutralization processes. The heavy-particle collisions in ground states are summarized in Table 4. These are molecular hydrogen dissociation due to collisions with molecules and atoms, charge exchange, and ion conversion reactions. Due to the interest in the application of the GEVKM in simulation of high-pressure negative hydrogen ion source an extensive set of $\mathrm{H}^-$ destruction processes due to collisions with heavy particles is considered separately in Table 5. Compared to conventional $\mathrm{H}^-$ sources operating at low pressures hydrogen anion destruction due to an electron detachment in collisions with heavy particles may be dominant in our device compared to $\mathrm{H}^-$ destruction by energetic electrons. Table 6 and Table 7 present volumetric chemical reactions involving production and destruction of vibrationally excited hydrogen molecules which was discussed in details in Section 1.2 of Chapter 1. These reactions include vibrational excitation of hydrogen molecules by electron impact (eV and EV). Generally, it is assumed that the population of high-lying vibrational states is due to collisions with energetic electrons (EV reaction). In addition, single quantum transitions between vibrationally excited hydrogen molecules are considered (VV and VTm processes). In a case of atom-hydrogen molecules collisions the multi-quantum transitions are included. In addition, Table 7 includes dissociative electron attachment to rovibrationally excited hydrogen molecules, which is one of the most important channels in $\mathrm{H}^-$ production. Table 8 shows reactions involving electronically excited hydrogen atoms. According to Hassouni et al. (1999) they can be one of the major mechanisms of $\mathrm{H}_3^+$ production, which is often a dominant species in the discharge. Table 9 summarizes wall chemical reactions. These are atomic hydrogen recombination, quenching of vibrationally excited hydrogen molecules and recombination of positive ions. It was also proposed that surface recombination of atomic hydrogen may effectively produce high-lying vibrational states of molecular hydrogen (Hall, et al., 1988). However, at high pressures the wall reactions are usually not the dominating processes. Therefore, this mechanism is disregarded in this study.

All reaction rates are taken from the cited references if available or calculated assuming Maxwellian distribution functions. The details of the calculation procedures are given in Appendix A. The numerical values for the reaction rates used in this work are listed in Appendix B.

Wall chemical reaction rates from Table 9 depend on recombination and quenching coefficients which are wall-material specific and their actual values are external parameters to the simulation. The quenching coefficients $\gamma_{v\to v'}$ of vibrationally excited states in this work are assumed independent of the wall material. It is achieved by assuming that vibrationally excited particles are always deexcited in collisions with the walls resulting in the following conditions for $\gamma_{v\to v'}$

$$\sum_{v'=0}^{v-1} \gamma_{v\to v'} = 1. \qquad (3.118)$$

The actual values of $\gamma_{v\to v'}$ in this work are based on the vibrational distribution of hydrogen molecules reflected from the walls found by Hiskes and Karo (1989) by using the molecular trajectories calculations. Fitting their nondimensional vibrational distribution functions the following analytic expressions for $\gamma_{v\to v'}$ are obtained in this work

$$\gamma_{v\to v'} = \frac{w_0 - \alpha \times (v' - v) \times e^{-\beta (v'-v)^2}}{1.0 + 12.0 \times e^{-1.5v}}, \qquad (3.119)$$

where

$$\begin{aligned} w_0 &= 0.065 \times (v - 3.2) \times e^{0.1\times(-(v-4.2)^2 - v)} + 0.045 \times (1.0 - e^{3.0-v}), \\ \alpha &= 3.5 \times (v - 0.1)^{-2.2} + 0.01, \\ \beta &= 0.025 \times (3.2 - 0.79 \times \tan^{-1}(v - 7.2) \times 1.9). \end{aligned} \qquad (3.120)$$

The error of the fitting formulas given by Eq. (3.119) does not exceed 15%.

The wall quenching coefficient of electronically excited hydrogen atoms $\gamma_{\mathrm{H}(n)}$ is set to 1 in all simulations due to lack of the data.

In addition, the electronic excitations of the hydrogen molecules by the electron impact are taken into account in the electron energy equation. The electronic states with corresponding threshold energies are given in Table 10. The cross-sections data for these processes are taken from Yoon et al. (2008).

**Table 3. Electron collisions with ground state species.**

| Reactions | References |
|---|---|
| 1. $e + H_2 \rightarrow e + H_2^+ + e$ | Hjartarson et al. (2010) |
| 2. $e + H_2 \rightarrow e + H + H$ | Capitelli et al. (2002) |
| 3. $e + H_2 \rightarrow H + H^-$ | Hjartarson et al. (2010) |
| 4. $e + H_2 \rightarrow e + H^+ + H + e$ | Janev et al. (1987) |
| 5. $e + H \rightarrow e + H^+ + e$ | Hjartarson et al. (2010) |
| 6. $e + H^- \rightarrow e + H + e$ | Hjartarson et al. (2010) |
| 7. $e + H_2^+ \rightarrow H + H$ | Hjartarson et al. (2010) |
| 8. $e + H_2^+ \rightarrow e + H + H^+$ | Hjartarson et al. (2010) |
| 9. $e + H_3^+ \rightarrow 3H$ | Hjartarson et al. (2010) |
| 10. $e + H_3^+ \rightarrow H + H_2$ | Hjartarson et al. (2010) |
| 11. $e + H_3^+ \rightarrow H_2^+ + H^-$ | Hjartarson et al. (2010) |
| 12. $e + H_3^+ \rightarrow e + H + H + H^+$ | Hjartarson et al. (2010) |

**Table 4. Collisions of ground state heavy species.**

| Reactions | References |
|---|---|
| 13. $H_2 + H_2 \rightarrow H_2 + 2H$ | Martin et al. (1998) |
| 14. $H + H_2 \rightarrow H + 2H$ | Capitelli et al. (2002) |
| 15. $H_2 + H^+ \rightarrow H_3^+ + h\nu$ | Gerlich and Horning (1992) |
| 16. $H_2 + H^+ \rightarrow H_2^+ + H$ | Phelps (1990) |
| 17. $H_2 + H_2^+ \rightarrow H + H_3^+$ | Hjartarson et al. (2010) |
| 18. $H + H_2^+ \rightarrow H_2 + H^+$ | Hjartarson et al. (2010) |

**Table 5. Destruction collisions of negative hydrogen ions with ground state neutrals and positive ions.**

| Reactions | References |
|---|---|
| 19. $H^- + H \rightarrow e + H_2$ | Janev et al. (1987) |
| 20. $H^- + H \rightarrow e + H + H$ | Janev et al. (1987) |
| 21. $H^- + H_2 \rightarrow H + H_2 + e$ | Janev et al. (2003) |
| 22. $H^- + H^+ \rightarrow H + H$ | Janev et al. (2003) |
| 23. $H^- + H^+ \rightarrow H_2^+ + e$ | Janev et al. (2003) |
| 24. $H^- + H_2^+ \rightarrow H_2 + H$ | Fukumasa (1989) |
| 25. $H^- + H_2^+ \rightarrow 3H$ | Hjartarson et al. (2010) |
| 26. $H^- + H_2^+ \rightarrow H_3^+ + e$ | Janev et al. (2003) |
| 27. $H^- + H_3^+ \rightarrow H_2 + 2H$ | Fukumasa (1989) |
| 28. $H^- + H_3^+ \rightarrow H_2 + H_2$ | Matveyev and Silakov (1995) |
| 29. $H^- + H_3^+ \rightarrow 4H$ | Hjartarson et al. (2010) |

**Table 6. Three body collisions of heavy species.**

| Reactions | References |
|---|---|
| 30. $H_2 + H + H \rightarrow H_2 + H_2(v = 14)$ | Matveyev and Silakov (1995) |
| 31. $H + H + H \rightarrow H + H_2(v = 14)$ | Matveyev and Silakov (1995) |
| 32. $H^+ + 2H_2 \rightarrow H_3^+ + H_2$ | Matveyev and Silakov (1995) |

**Table 7. Collisions involving vibrationally excited hydrogen molecules.**

| Reactions | References |
|---|---|
| 33. $\mathrm{e} + \mathrm{H}_2\left(X^1\Sigma_g^+, v\right) \rightarrow$ $\mathrm{H}_2^- \rightarrow \mathrm{e} + \mathrm{H}_2\left(X^1\Sigma_g^+, v'\right)$ | Capitelli et al. (2002) |
| 34. $\mathrm{e} + \mathrm{H}_2\left(X^1\Sigma_g^+, v\right) \rightarrow$ $\mathrm{e} + \mathrm{H}_2\left(B^1\Sigma_u^+, C^1\Pi_u\right) \rightarrow$ $\mathrm{e} + \mathrm{H}_2\left(X^1\Sigma_g^+, v'\right) + h\nu$ | Capitelli et al. (2002) |
| 35. $\mathrm{e} + \mathrm{H}_2(v) \rightarrow \mathrm{H}_2^- \rightarrow$ $\mathrm{e} + \mathrm{H}_2\left(X^1\Sigma_g^+\right) \rightarrow \mathrm{e} + \mathrm{H} + \mathrm{H}$ | Capitelli et al. (2002) |
| 36. $\mathrm{e} + \mathrm{H}_2(v) \rightarrow \mathrm{H}_2^- \rightarrow$ $\mathrm{e} + \mathrm{H}_2\left(b^3\Sigma_g^+\right) \rightarrow \mathrm{e} + \mathrm{H} + \mathrm{H}$ | Capitelli et al. (2002) |
| 37. $\mathrm{e} + \mathrm{H}_2(v) \rightarrow$ $\mathrm{e} + \mathrm{H}_2\left(b^3\Sigma_u^+, a^3\Sigma_g^+, c^3\Pi_u, \mathrm{singlets}\right) \rightarrow$ $\mathrm{e} + \mathrm{H} + \mathrm{H}$ | Capitelli et al. (2002) |
| 38. $\mathrm{e} + \mathrm{H}_2(v) \rightarrow \mathrm{H}_2^- \rightarrow \mathrm{H} + \mathrm{H}^-$ | Celiberto et al. (2001) |
| 39. $\mathrm{H}_2(v) + \mathrm{H}_2(w) \rightarrow \mathrm{H}_2(v+1) + \mathrm{H}_2(w-1)$ | Matveyev and Silakov (1995) |
| 40. $\mathrm{H}_2(v) + \mathrm{H}_2(w) \rightarrow \mathrm{H}_2(v \pm 1) + \mathrm{H}_2(w)$ | Matveyev and Silakov (1995) |
| 41. $\mathrm{H}_2 + \mathrm{H}_2(v) \rightarrow \mathrm{H}_2 + 2\mathrm{H}$ | Matveyev and Silakov (1995) |
| 42. $\mathrm{H} + \mathrm{H}_2(v) \rightarrow \mathrm{H} + \mathrm{H}_2(v')$ | Capitelli et al. (2002) |
| 43. $\mathrm{H} + \mathrm{H}_2(v) \rightarrow \mathrm{H} + 2\mathrm{H}$ | Gorse and Capitelli (1987) |
| 44. $\mathrm{H}^+ + \mathrm{H}_2(v) \rightarrow \mathrm{H} + \mathrm{H}_2^+$ | Janev et al. (2003) |
| 45. $\mathrm{H}^- + \mathrm{H}_2(v) \rightarrow \mathrm{H}_2(v-2) + \mathrm{H} + \mathrm{e}, (2 \leq v \leq 6)$ | Dem'yanov et al. (1985) |

**Table 8. Collisions involving electronically excited hydrogen atoms.**

| Reactions | References |
|---|---|
| 46. $e + H \rightarrow e + H(n = 2,3)$ | Janev et al. (1987) |
| 47. $e + H_2 \rightarrow e + H + H(n = 2,3)$ | Janev et al. (1987) |
| 48. $e + H(n = 2,3) \rightarrow e + H^+ + e$ | Janev et al. (1987) |
| 49. $H(n = 3) \rightarrow H(n = 2) + h\nu$ | Johnson (1972) |
| 50. $H(n = 2,3) + H_2 \rightarrow H_3^+ + e$ | Glass-Maujean (1989) |
| 51. $H(n = 2,3) + H_2 \rightarrow 3H$ | Glass-Maujean (1989) |

**Table 9. Wall chemical reactions.**

| Reactions | Reaction rates |
|---|---|
| 52. $H + H + \text{wall} \rightarrow H_2$ | $\left[\frac{\Lambda^2}{D_H} + \frac{2V\left(2-\gamma_H\right)}{A v_{H,th}\gamma_H}\right]^{-1}$ |
| 53. $H_2(v) + \text{wall} \rightarrow H_2(v)$ | $\sum_{v'=0}^{v-1}\left[\frac{\Lambda^2}{D_{H_2(v)}} + \frac{2V\left(2-\gamma_{v\rightarrow v'}\right)}{A v_{H_2(v),th}\gamma_{v\rightarrow v'}}\right]^{-1}$ |
| 54. $H^+ + \text{wall} \rightarrow H$ | $\frac{A_{eff,H^+}}{V} u_{B,H^+}$ |
| 55. $H_2^+ + \text{wall} \rightarrow H_2$ | $\frac{A_{eff,H_2^+}}{V} u_{B,H_2^+}$ |
| 56. $H_3^+ + \text{wall} \rightarrow H + H_2$ | $\frac{A_{eff,H_3^+}}{V} u_{B,H_3^+}$ |
| 57. $H(n) + \text{wall} \rightarrow H$ | $\left[\frac{\Lambda^2}{D_{H(n)}} + \frac{2V\left(2-\gamma_{H(n)}\right)}{A v_{H(n),th}\gamma_{H(n)}}\right]^{-1}$ |

**Table 10. Electronic states of hydrogen molecules used in the electron and total energy equations.**

| Electronically excited states | Threshold energy (eV) |
|---|---|
| $H_2(b^3\Sigma_u^+)$ | 10.45 |
| $H_2(a^3\Sigma_g^+)$ | 11.79 |
| $H_2(c^3\Pi_u)$ | 11.76 |
| $H_2(B^1\Sigma_u^+)$ | 11.18 |
| $H_2(C^1\Pi_u)$ | 12.29 |
| $H_2(E,F^1\Sigma_g^+)$ | 12.3 |
| $H_2(e^3\Sigma_u^+)$ | 13.23 |

### 3.2.6 Summary of the GEVKM and Procedures for Numerical Solution

The GEVKM model consists of $N_s - 1$ particle continuity equations for each heavy species $p$ given by Eq. (3.28)

$$Q_{p,\text{out}} - Q_{p,\text{in}} + \frac{A}{V}\Gamma_{p,\text{wall}} = \sum_{r=1}^{N_r} k_r \Theta_r \prod_{p=1}^{N_s}\left(\nu_{r,p} - \nu'_{r,p}\right)\overline{n}_p^{\nu_{r,p}}\ , \tag{3.121}$$

from which the electron number density is excluded by using quasi-neutrality condition in the bulk (3.23)

$$\sum_{p=1}^{N_h} Z_p \overline{n}_p - \overline{n}_e = 0, \tag{3.122}$$

an electron energy equation given by Eq. (3.94)

$$\begin{aligned}&\frac{5}{2}k_B T_e Q_{e,\text{out}} - \frac{5}{2}k_B T_{e,\text{in}} Q_{e,\text{in}} + \frac{A}{V}\sum_{p=1}^{N_h^+}\Gamma_{p,\text{wall}}\left(2k_B T_e + eV_p + eV_{\text{sh}}\right)\\ &= k_B\left(T_h - T_e\right)\sum_{h=1}^{N_h} 3\frac{m_e}{m_h}k_{eh}^{\text{elast}}\Theta_h \overline{n}_h \overline{n}_e + \sum_{r=1}^{N_r} E_{r,\text{th}} k_r \Theta_r \prod_{p=1}^{N_s}\overline{n}_p^{\nu_{r,p}} + \frac{P_{\text{elec}}}{V} + \frac{P_{\text{abs}}}{V},\end{aligned} \tag{3.123}$$

a total energy equation given by Eq. (3.106)

$$\frac{P_{\text{abs}}}{V} - \sum_{p=1}^{N_s}\left(\frac{m_p u_{p,\text{out}}^2}{2} + H_{p,\text{out}}\right)Q_{p,\text{out}} + \sum_{p=1}^{N_s}\left(\frac{m_p u_{p,\text{in}}^2}{2} + H_{p,\text{in}}\right)Q_{p,\text{in}}$$
$$= \frac{A}{V}\sum_{p=1}^{N_s} q_p\Big|_{\text{wall}} + \frac{A}{V} q_{\text{cond}}\Big|_{\text{wall}} + \frac{P_{\text{elec}}}{V} + \frac{P_{\text{rad}}}{V}, \tag{3.124}$$

and a heat transfer equation given by Eq. (3.114)

$$\frac{P_{\text{abs}}}{V} - \sum_{p=1}^{N_s}\left(\frac{m_p u_{p,\text{out}}^2}{2} + H_{p,\text{out}}\right)Q_{p,\text{out}} + \sum_{p=1}^{N_s}\left(\frac{m_p u_{p,\text{in}}^2}{2} + H_{p,\text{in}}\right)Q_{p,\text{in}}$$
$$= \frac{2\pi R}{V}\left(\text{h}_{\text{side}} R + \text{h}_{\text{cyl}} L\right)\left(T_{\text{w}} - T_{\infty}\right) + \epsilon_{\text{emis}}\sigma_{\text{SB}}\frac{A}{V}\left(T_{\text{w}}^4 - T_{\infty}^4\right). \tag{3.125}$$

The wall flux terms in the above equations are presented in Section 3.2.1.4. It should be noted that electron energy equation (3.123) and total energy equation (3.124) include losses associated with the excitation of electronic states of hydrogen molecules given in Table 10 and are expressed as

$$P_{\text{elec}} = V\sum_{r=1}^{N_{r,\text{elec}}} E_{r,\text{th,elec}} k_r \overline{n}_e \overline{n}_{\text{H}_2}\ . \tag{3.126}$$

In the above equation it is assumed for simplicity that the power losses associated with the excitation of electronic states of hydrogen molecules are lost to the walls. It is done in such way because these electronic states are not included in species continuity equations of the GEVKM.

The radiation term in Eq. (3.124) is given as

$$P_{\text{rad}} = V\left(k_{15} n_{\text{H}} n_{\text{H}^+}\Delta H_{15} + k_{49} n_{\text{H}(3)}\Delta H_{49} + \sum_{v=0}^{14}\sum_{v'=0}^{14} k_{34}(v,v') n_e n_{\text{H}_2(v)}\Delta H_{34,v\to v'}\right), \tag{3.127}$$

where $k_{15}$, $k_{49}$, and $k_{34}(v,v')$ are rate coefficients of reactions 15, 49, and 34 given in Table 3-Table 9 respectively, $\Delta H_{15}$, $\Delta H_{49}$, and $\Delta H_{34,v\to v'}$ are reaction enthalpy of the respective reactions. In essence, in Eq. (3.127) and (3.124) it is assumed that the reaction enthalpy is taken away by the photons to the walls.

In addition to the above equations it is necessary to link the properties at the center of the reactor to the volume-averaged values through Eqs. (3.29) and (3.32). A thorough analysis of different approximations was given in Monahan and Turner (2009). The following additional assumptions are used in this work. The background gas number density (molecular hydrogen in the case of the HCNHIS) is assumed to be uniform in the plasma reactor leading to $\Theta_r = 1$ in the reactions where background gas is one of the reactants. For the mutual neutralization of positive and negative ions at low electronegativity the coefficients $\Theta_r \approx 3/2$ (Monahan & Turner, 2009). For other reactions it is assumed that $\Theta_r \approx 1$.

Diffusion coefficients in Eqs. (3.79) and (3.83) are calculated based on the collision integrals $\Omega_X^{(\alpha,\beta)}$ reported in Capitelli et al. (2013). The mean free path of heavy particles in Eq. (3.79) and (3.82) is found from the diffusion coefficient $D_X$ as

$$\lambda_X = D_X \sqrt{\frac{8m_X}{\pi k_B T_h}}. \tag{3.128}$$

This choice of the mean free path is motivated by the fact that the expressions for sheath edge to center number density ratios are derived from the momentum equation where diffusion coefficient is present in the collision term (see Section 2.4).

Thus, GEVKM becomes a system of $N_s + 3$ nonlinear algebraic equations with $N_s + 3$ unknowns. This system is iteratively solved by using Newton-Raphson method (Teukolsky, et al., 1996). The GEVKM is implemented in the simulation code written in Fortran 90 language. This implementation consists of the solver of an arbitrary system of non-linear algebraic equations by Newton-Raphson method, the continuity and energy equations framework with the self-consistency checks to guarantee conservation of charge, particles and energy in the system and suitable for simulating arbitrary cylindrical plasma reactors with different plasma composition and chemical reactions of hydrogen plasmas.

The computational tool is configured by text configuration files for the greater flexibility. The reliability of the computational code is achieved by enhanced unit testing and usage of the version control system *git* (Chacon & Straub, 2014). The calculation and analytical fitting of the rate coefficients for the chemical reactions given in Table 3-Table 9 is implemented in an independent computational tool. The details of the rate calculation and fitting procedures are discussed in Appendix A. The analytical expression for the rate coefficients are given in Appendix B.

## 3.3 Verification and Validation of the Global Enhanced Vibrational Kinetic Model

In the low-pressure regime (0.02-0.1 Torr) charged particles are created in the volume of the discharge and lost to the walls. The energy balance is primarily between electrons heated by the RF field and the electron-ion losses to the walls of the discharge. Neutral gas temperature is typically close to the standard temperature under these conditions. At high pressures (10-100 Torrs) the charged particles are also created in the volume of the discharge, however, while moving to the walls they experience many collisions making volume recombination reactions more dominant than the surface chemical reactions. The volume processes under these conditions play a major role. The energy gained by electrons is transferred partially to heat the neutral gas, excite its internal degrees of freedom and to ionize molecules and atoms. In these cases $T_h$ is usually much closer to $T_e$ and can substantially deviate from the standard temperature. Therefore the comparisons with the experiments and simulations covering these regimes verify and validate the GEVKM and its implementation for the entire regime of pressure operation, from low to high pressures.

### 3.3.1 Low-Pressure and Low Absorbed Power Density Regime

For the validation and verification of the GEVKM at low pressures the results of the present model are compared with the simulation predictions and experimental measurements in $H^-$ source Deuterium Negative Ion Source Experiment (DENISE) (Zorat, et al., 2000; Zorat & Vender, 2000). DENISE is a cylindrical tandem multicusp negative hydrogen ion source where the plasma is produced by internal coil surrounded by dielectric. The volume of the source is 9409.07 cm$^3$. Typical operating parameters covered in the work of Zorat and his colleagues (Zorat, et al., 2000; Zorat & Vender, 2000) are the pressures from 2-100 mTorr and absorbed power 500-3000 W. The absorbed power corresponds to an averaged power density 0.053-0.32 W/cm$^3$. In Zorat et al. (2000) and Zorat and Vender (2000) a global model solver (GMS) was applied to DENISE in order to investigate basic plasma parameters of the hydrogen plasma. The GMS simulation results were compared to Langmuir probe measurements of electron temperature and number density. For validation-verification of the GEVKM with the results of Zorat et al. (2000) and Zorat and Vender (2000) all chemical reactions in Tables 3-9 are used. In Zorat et al. (2000) and Zorat and Vender (2000) the effective area $A_{\text{eff}}$ in Eq. (3.75) was increased by the area of the internal antenna while the source volume was decreased by the volume of the antenna and the same modifications are used in the GEVKM.

The wall material parameters used in the GEVKM simulations are listed in Table 11. The thermal accommodation coefficient of molecular hydrogen is taken from Leroy et al. (1997). For atomic hydrogen the experimental value is not available. One could estimate it from Eq.(3.109). In this work the thermal accommodation coefficient is calculated by using the expression obtained in Song and Yovanovich (1987) by correlating experimental data of different materials and wall temperatures which is written as

$$\alpha_X = \exp\left(-0.57\frac{T_\mathrm{w}-273}{273}\right)\frac{\mathcal{M}_X^*}{6.8+\mathcal{M}_\mathrm{w}} + \left[1-\exp\left(-0.57\frac{T_\mathrm{w}-273_0}{273}\right)\right]\frac{2\mu}{(1+\mu)^2}, \quad (3.129)$$

where $T_\mathrm{w}$ is the wall temperature in Kelvins, $\mathcal{M}_\mathrm{w}$ is the wall molar mass expressed in g/mole, $\mathcal{M}_X^*$ is given by

$$\mathcal{M}_X^* = \begin{cases} \mathcal{M}_X, & X \text{ is monoatomic,} \\ 1.4\mathcal{M}_X, & X \text{ is diatomic,} \end{cases} \quad (3.130)$$

and $\mathcal{M}_X$ is the molar mass of the gas $X$ expressed in g/mole, the rest of the variables have the same meaning as in the Baule formula given by Eq. (3.109). The estimated molar mass of 55.4 g/mole is used for the stainless steel in calculating the thermal accommodation coefficient of $\mathrm{H}$.

**Table 11. Parameters used in the GEVKM simulations of the negative hydrogen ion source DENISE.**

| Parameters | $\gamma_{\mathrm{H,rec}}$ | $\alpha_{\mathrm{H_2}}$ | $\alpha_{\mathrm{H}}$ | $\epsilon_{\mathrm{SS}}$ |
|---|---|---|---|---|
| Numerical values | 0.1 | 0.13 | Eq. (3.129) | 0.5 |

For the quenching of vibrationally excited hydrogen molecules the coefficients given by Eq. (3.119) are used. For the wall emissivity a typical value for the stainless steel is chosen (Holman, 2010).

The outflow of neutral particles is considered to be through the orifice of area $A_\mathrm{or}$ as in Zorat et al. (2000) and is given by

$$Q_{\mathrm{out},p} = \frac{1}{4} n_p v_{p,\mathrm{th}} \frac{A_{\mathrm{or}}}{V}. \tag{3.131}$$

The charged species are assumed to be lost to the walls and as such their outflow through the orifice is neglected.

In the GEVKM simulations $T_e$, $T_h$ and $T_{\mathrm{w}}$ are obtained from the electron energy (3.94), the total energy (3.106), and the wall heat transfer (3.114) equations in contrast to Zorat et al. (2000) where only $T_e$ was calculated and a constant value of 0.05 eV for the heavy particles temperature was assigned.

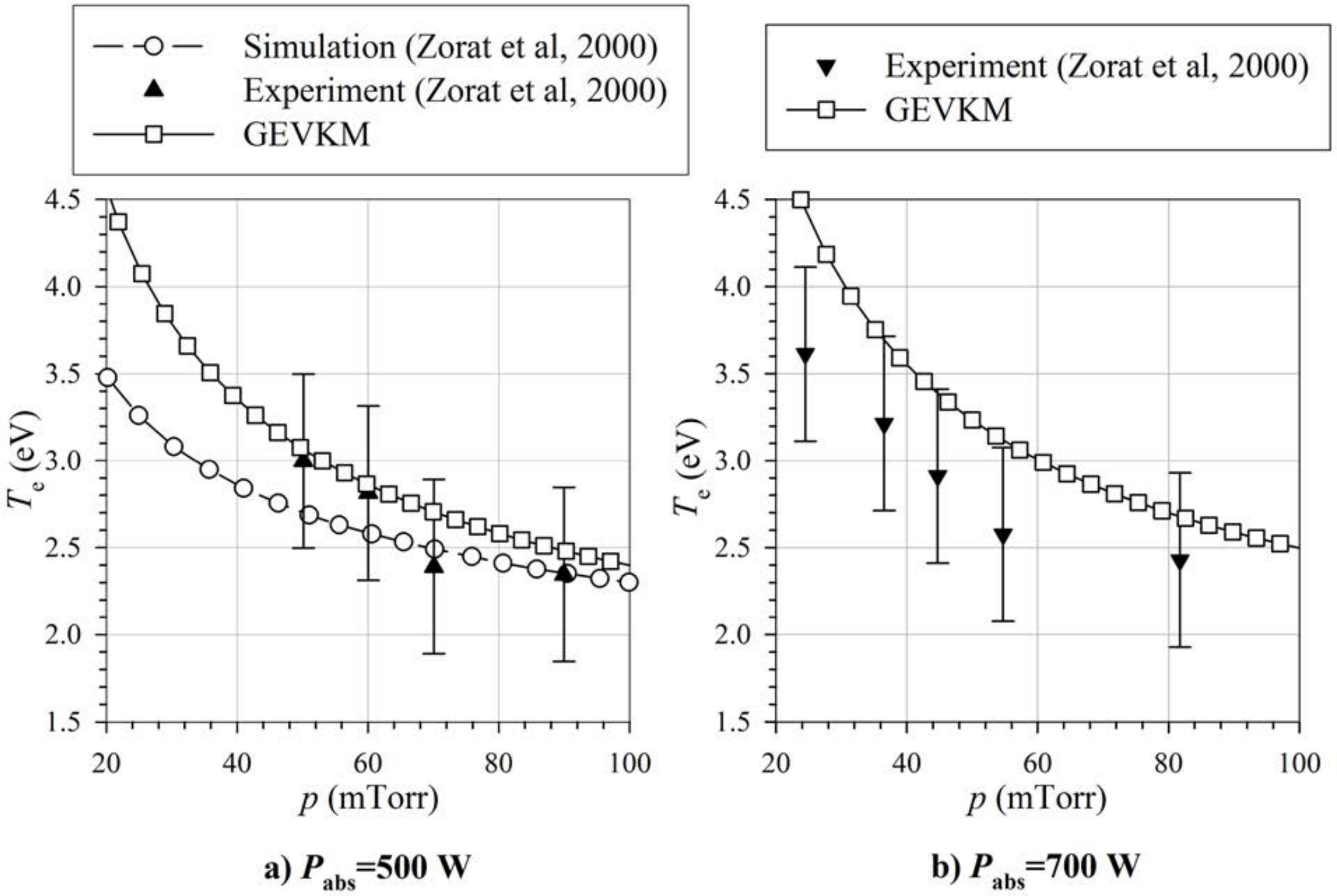

**Figure 8. Electron temperature in the negative hydrogen ion source DENISE as a function of the chamber pressure at the absorbed power 500 W (a) and 700 W (b).**

The results of the GEVKM simulations are compared with the Langmuir probe measurements and GMS simulations of Zorat et al. (2000) in Figure 8 and Figure 9. The electron temperature is plotted in Figure 8 as a function of the chamber pressure at different absorbed powers.

For the entire regime of chamber pressures considered the GEVKM predictions are within the experimental uncertainty and are close to the GMS predictions. Also the electron temperature calculated by the GEVKM better corresponds to the experimental results compared to GMS solver. The electron number density as a function of chamber pressure at different absorbed powers is plotted on Figure 9. The electron number density obtained by the GEVKM is close to the measurements in the DENISE. Although there is a difference between the GEVKM predictions and experimental measurements at high pressures the overall agreement is very good. More importantly, the GEVKM prediction follows the dependence in the electron number density on the discharge pressure, in contrast to the GMS results.

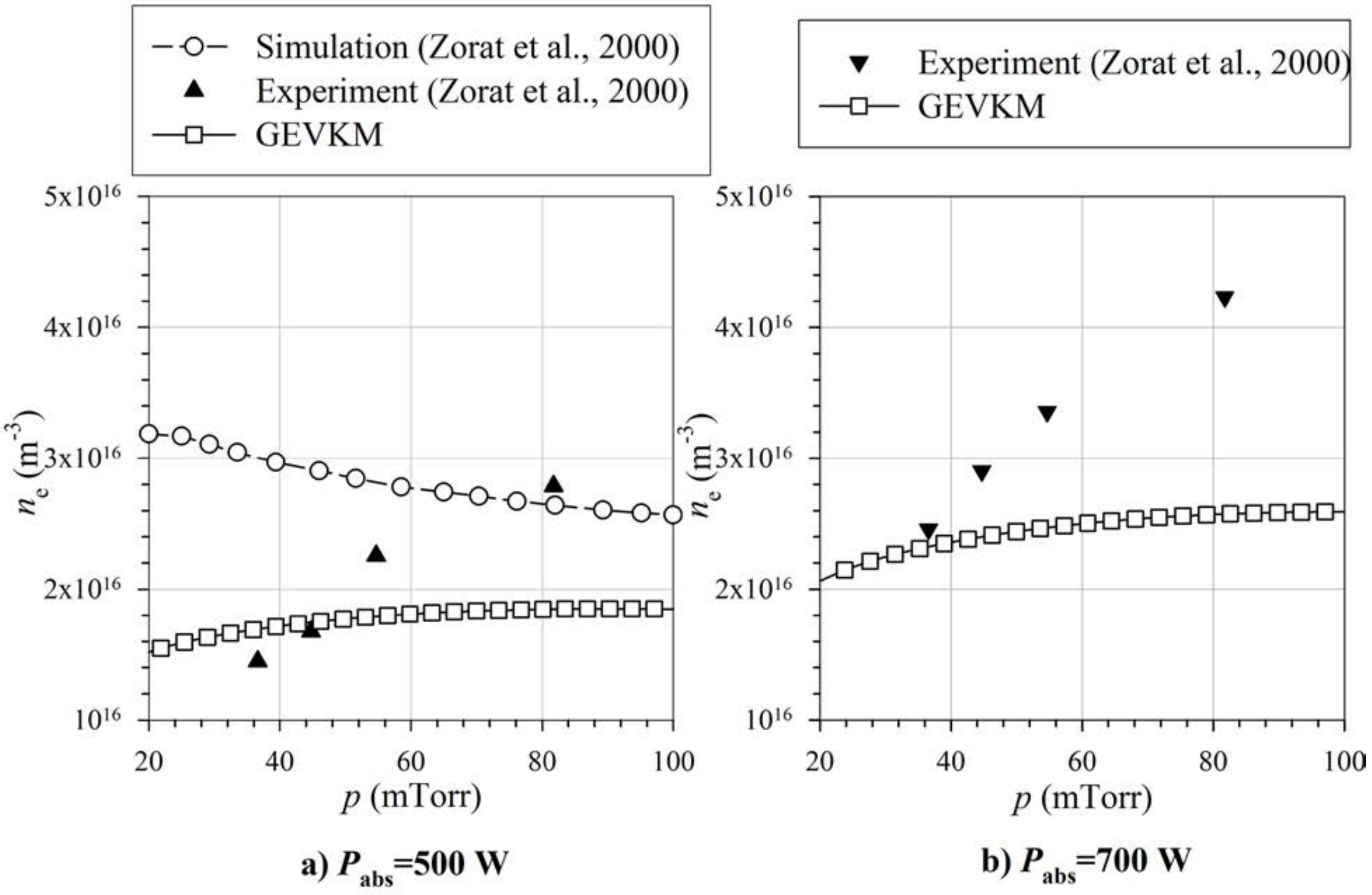

**Figure 9. Electron number density in the negative hydrogen ion source DENISE as a function of the chamber pressure at the absorbed power 500 W (a) and 700 W (b).**

Heavy particles and wall temperatures predictions by the GEVKM are plotted in Figure 10(a) as a function of the discharge pressure. For reference, the gas temperature assumed in Zorat et al. (2000) is also shown. The wall temperature is shown to be insensitive to the pressure variation for the range considered, while the gas temperature decreases as the discharge pressure increases. This behavior of the heavy particles temperature is associated with the small values of the thermal accommodation coefficient of molecular hydrogen. At low discharge pressures the wall collisions are dominant and low thermal accommodation coefficient results in inefficient energy exchange between gas molecules and the walls. As the pressure increases the particle-wall collision rate also increases and the temperature drops. It should be noted that at these pressures (20-100 mTorrs) the heat conduction term in Eq. (3.111) is much smaller than the term corresponding to the temperature jump.

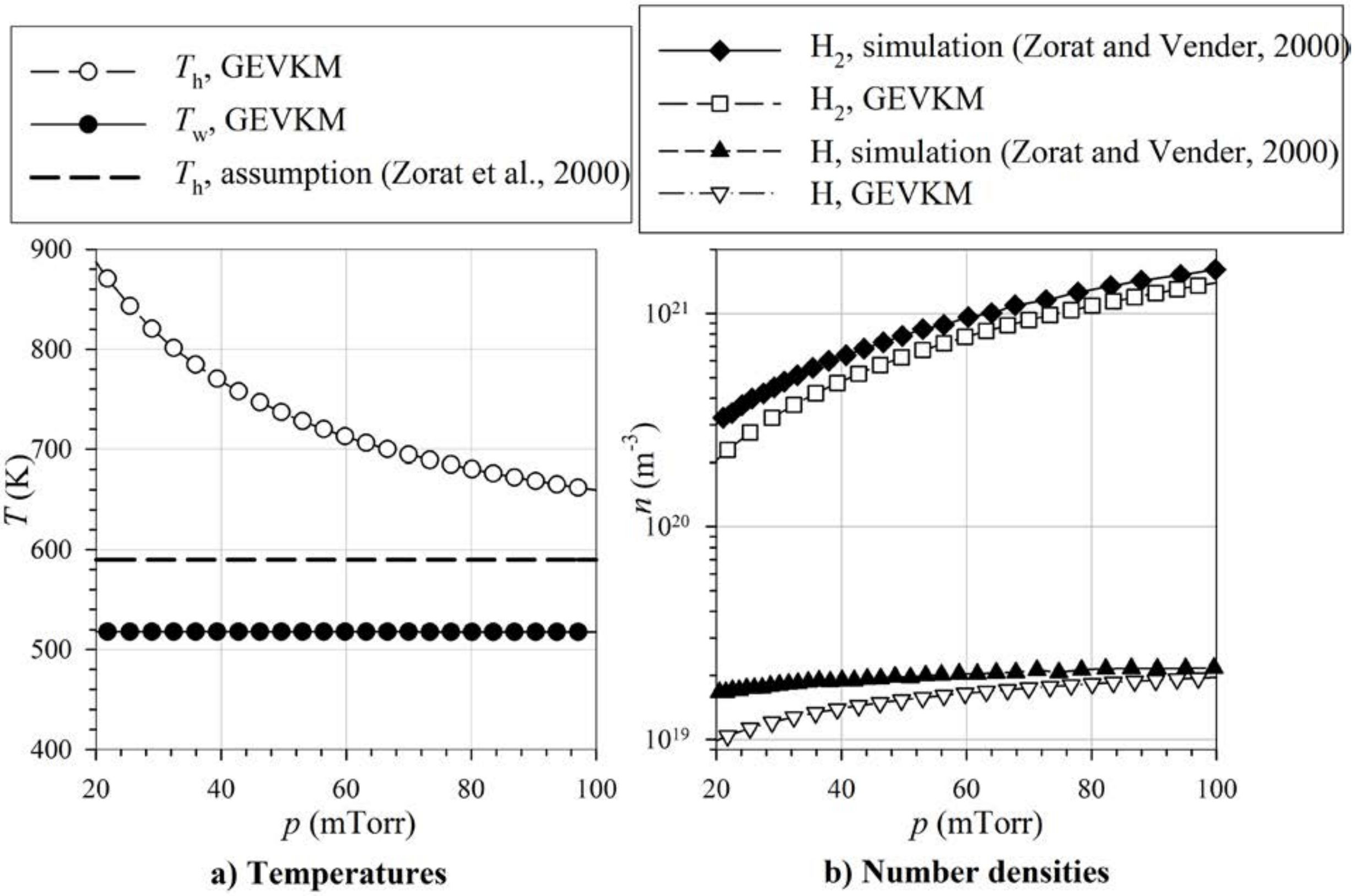

**Figure 10. Heavy particles and wall temperatures in the DENISE calculated by the GEVKM as a function of the chamber pressure at the absorbed power 500 W and the gas temperature assumed by Zorat et al. (2000) (a); number densities of ground states neutral particles calculated by the GEVKM and Zorat and Vender (2000) (b).**

The number densities of molecular and atomic hydrogen as a function of discharge pressure are plotted in Figure 10(b). The difference between molecular hydrogen number densities predicted by the GEVKM and by the global model of Zorat et al. (2000) is due to the higher heavy particles temperature calculated by the GEVKM. The same argument applies to atomic hydrogen number density.

Positive ions number densities are presented in Figure 11. The difference between the GEVKM and Zorat et al. (2000) predictions are in part due to the temperature difference as explained above but more importantly is associated with the differences in the wall fluxes estimations. In Zorat et al. (2000) the wall flux is estimated from the heuristic solution for the intermediate pressure regime (Lieberman & Lichtenberg, 2005) while more complicated expressions (3.39), (3.79) covering from low to high-pressure regime are used in this work. In addition, the GEVKM simulations show that the $H_3^+$ production through the ionization of electronic states of hydrogen atoms (reaction 50 in Table 8) which was not considered in Zorat et al. (2000) contributes around 40% of the overall $H_3^+$ production.

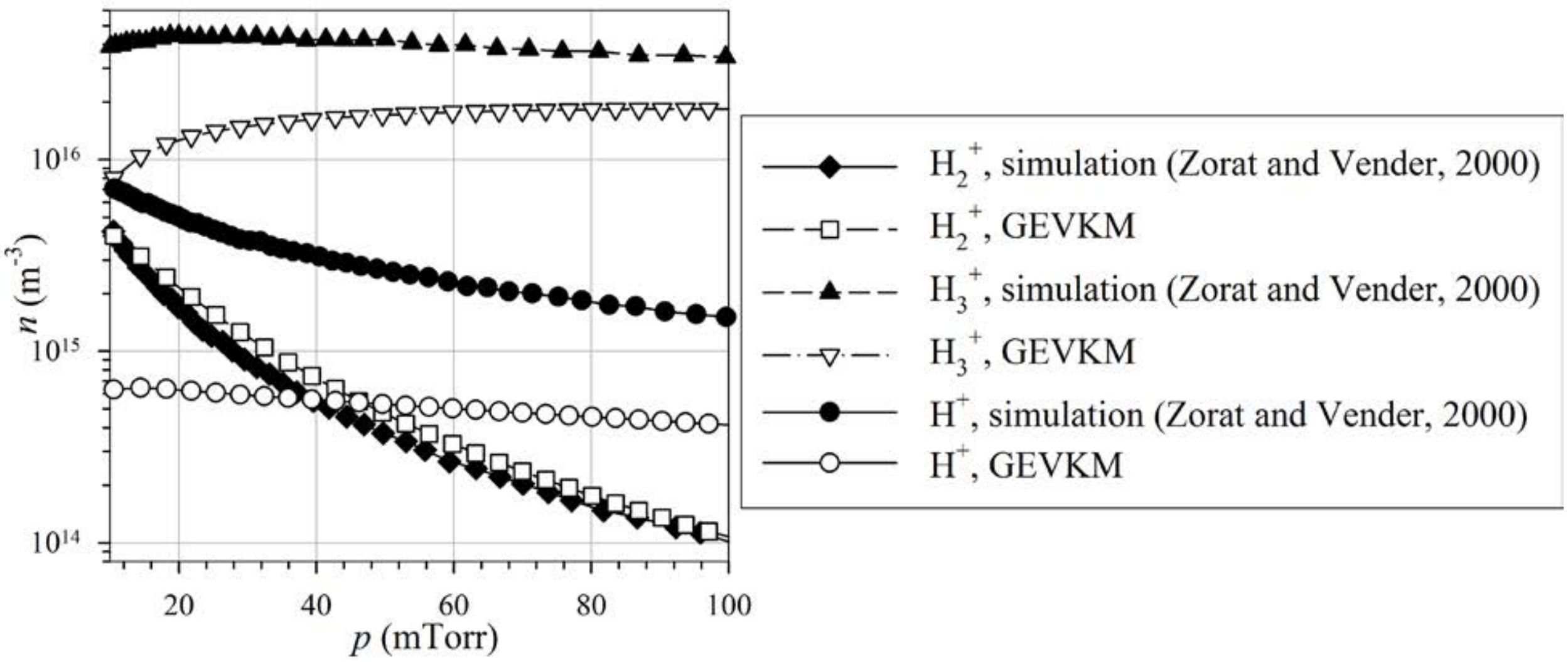

**Figure 11. Positive ions number densities in the negative hydrogen ion source DENISE as a function of the chamber pressure at fixed absorbed power 500 W.**

### 3.3.2 Intermediate and High-Pressure and High Absorbed Power Density Regime

Additional verification and validation of the GEVKM model is accomplished by simulations of a microwave-generated plasma reactor operating at intermediate to high pressures (Chen, et al., 1999). The volume of the reactor is 9.08 cm$^3$, the pressure ranges from 1-100 Torr and the absorbed power is from 75-200 W resulting in the power density in the range of 8.26-22

W/cm$^3$. The chemical reactions used in the GEVKM are those outlined in Tables 3-9. The wall material parameters used in the GEVKM simulations are listed in Table 12.

**Table 12. Parameters used in the GEVKM simulations of the microwave plasma reactor.**

| Parameters | $\gamma_{\mathrm{H,rec}}$ | $\alpha_{\mathrm{H}_2}$ | $\alpha_{\mathrm{H}}$ | $\epsilon_{\mathrm{quartz}}$ |
|---|---|---|---|---|
| Numerical values | 0.1 | Eq. (3.129)<br>1 | Eq. (3.129) | 0.85 |

Figure 12 shows the comparison of the heavy particles temperature calculated by the global model which includes Boltzmann equation solver (Chen, et al., 1999), experimental measurements (Chen, et al., 1999), and GEVKM predictions at fixed absorbed power 200W as a function of the chamber pressure. In addition to the heavy species temperature obtained using the thermal accommodation coefficient from Eq. (3.129), the plot shows GEVKM results at complete thermal accommodation with $\alpha_{\mathrm{H}_2} = 1$. The heavy-particle temperature predicted by the GEVKM has a non-monotonic behavior. It decreases as the discharge pressure increases at pressures 1-8 Torr, and then increases. The GEVKM predicts qualitatively the dependence of the heavy-particle temperature on the discharge pressure in contrast to the global model of Chen et al. (1999). Quantitatively there is a difference of 300 K between measured and predicted by the GEVKM heavy-particle temperature at discharge pressure of 2 Torr. This difference is much smaller when complete thermal accommodation assumed. The discrepancy between the GEVKM predictions and the experiments can be attributed to the substantial heating of hydrogen atoms in the electron impact dissociation collisions and low efficiency of energy transfer from molecules to walls due to the low thermal accommodation coefficient. Atom heating can be seen from the difference between the threshold energy of 10 eV for ground state hydrogen molecules electronic excitation through which dissociation proceeds and the dissociation energy of hydrogen molecule of 4.52 eV. Even at pressures around 1 Torr the molecular and atomic hydrogen could have temperatures that differ by 200-400 K (Chabert, et al., 1998). However, the heavy-particle temperature predicted by the GEVKM corresponds to an average heavy-particle temperature. Therefore the GEVKM predictions of the wall and heavy-particle temperatures provide an envelope of the actual gas temperature. At higher pressures the heat conduction term dominates over temperature jump and the temperature increases with pressure increase. The wall temperature predicted by the GEVKM in Figure 12 is very close to the measurements of the gas temperature in Chen et al. (1999) at low pressures while the heavy-particle temperature is close to the gas temperature of Chen et al. (1999) at high pressures. Overall, the agreement between the GEVKM predictions and measurements of Chen et al. (1999) is very good.

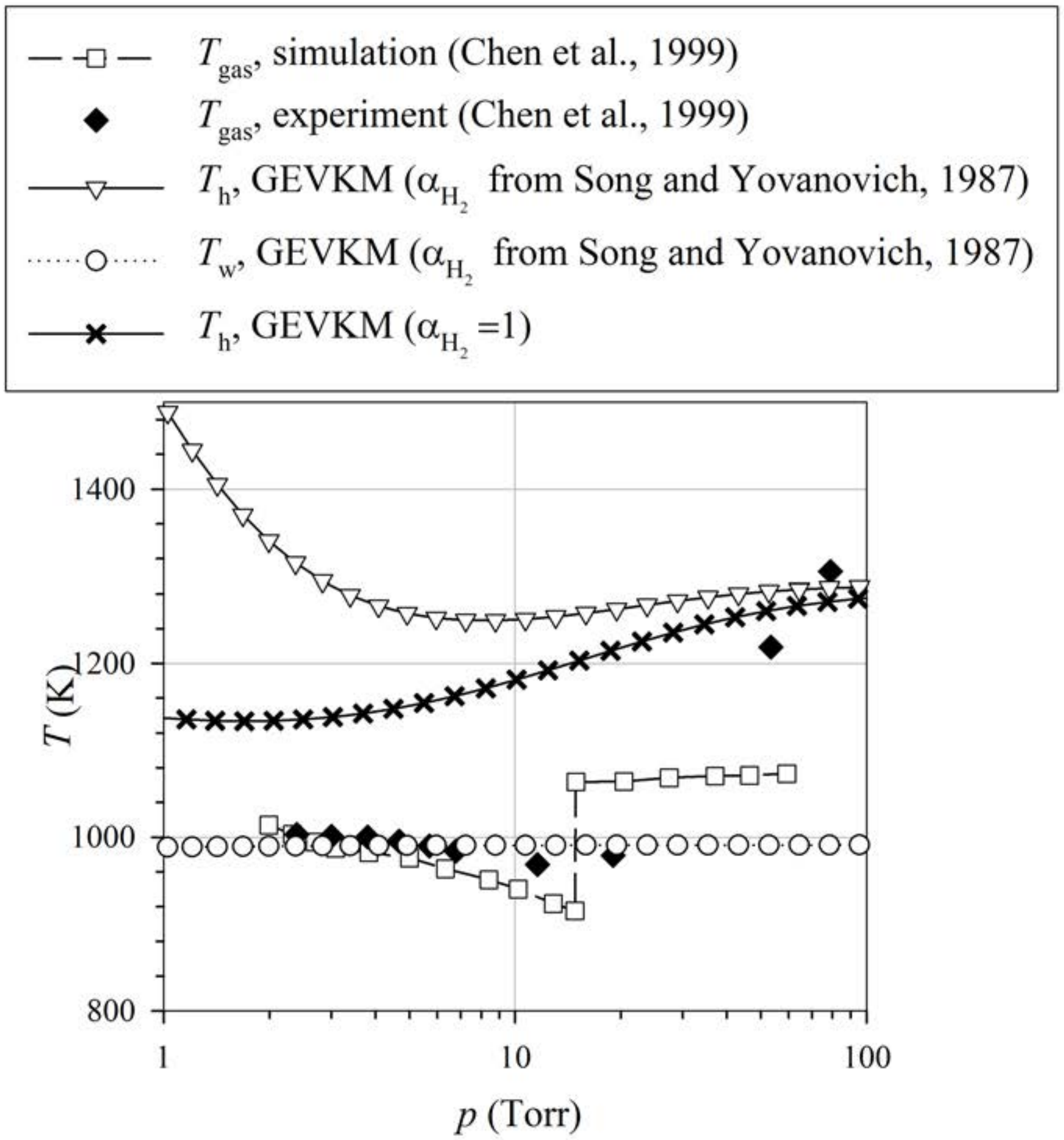

**Figure 12. Heavy-particle and wall temperatures in the microwave plasma reactor as a function of the chamber pressure at fixed absorbed power of 200 W and inlet flow rate 100 sccm.**

Figure 13 presents the heavy-particle temperature predicted by the global model of Chen et al. (1999), the experimental measurements by Chen et al. (1999) and GEVKM predictions at fixed discharge pressure of 10 Torr as a function of the absorbed power. The experimentally measured gas temperature is in close agreement with the wall temperature predicted by the GEVKM while the predicted heavy-particle temperature is larger by 40-200 K.

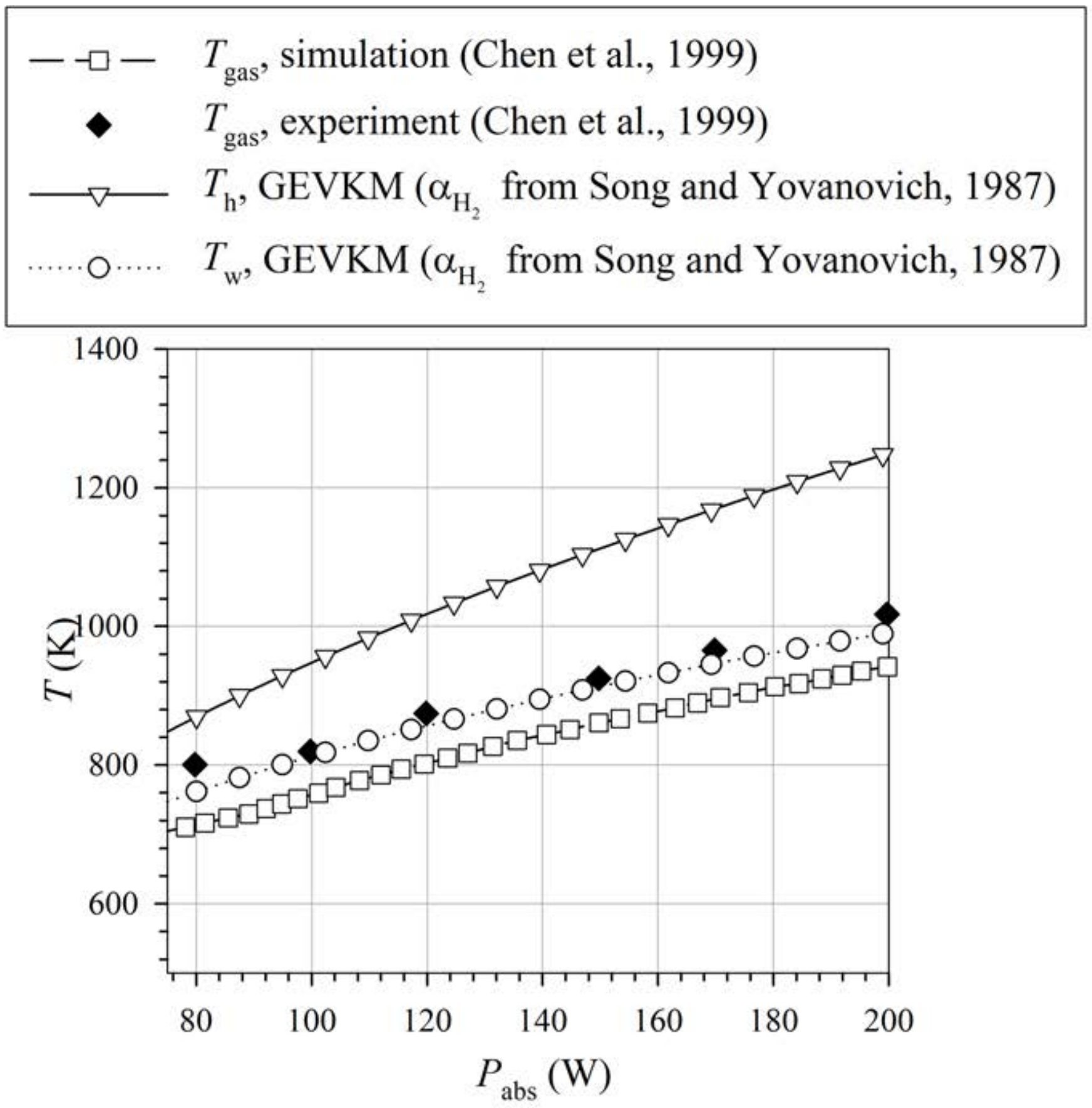

**Figure 13. Heavy-particle and wall temperatures in the microwave plasma reactor as a function of the absorbed power at fixed inlet hydrogen flow rate of 100 sccm and chamber pressure 10 Torr.**

Electron number density and temperature as a function of the discharge pressure are plotted in Figure 14(a) and Figure 14(b) respectively. The GEVKM predictions are close to the global model results of Chen et al. (1999). The slight difference in values is probably due to the difference in the heavy-particle temperatures. Also the electron temperature predicted by the GEVKM is about 0.5 eV smaller than the overpredictied values of Chen et al. (1999), as noted in Chen et al. (1999).

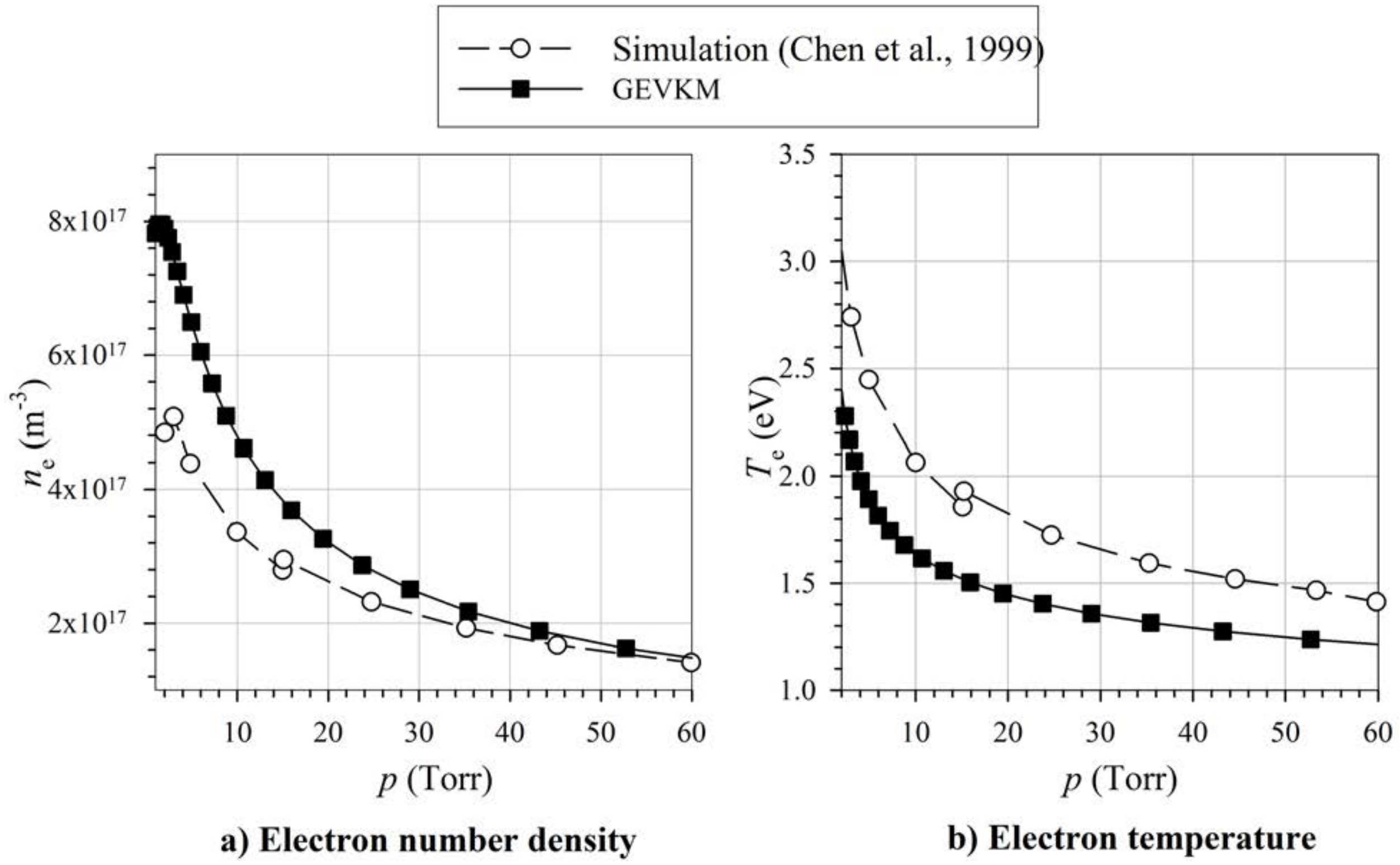

**Figure 14. Electron number density (a) and electron temperature (b) in the microwave plasma reactor as a function of the chamber pressure at fixed absorbed power of 200 W and inlet flow rate 100 sccm.**

Figure 15 shows the positive-ion densities as a function of the discharge pressure. There is a good agreement between the GEVKM predictions and the global model results of Chen et al. (1999). The difference in $H^+$ is due to the models used for the positive ion surface neutralization. Chen et al. (1999) used high-pressure diffusion flux model for all pressures while the GEVKM the expressions (3.39), (3.79) are used covering the low to high-pressure regime.

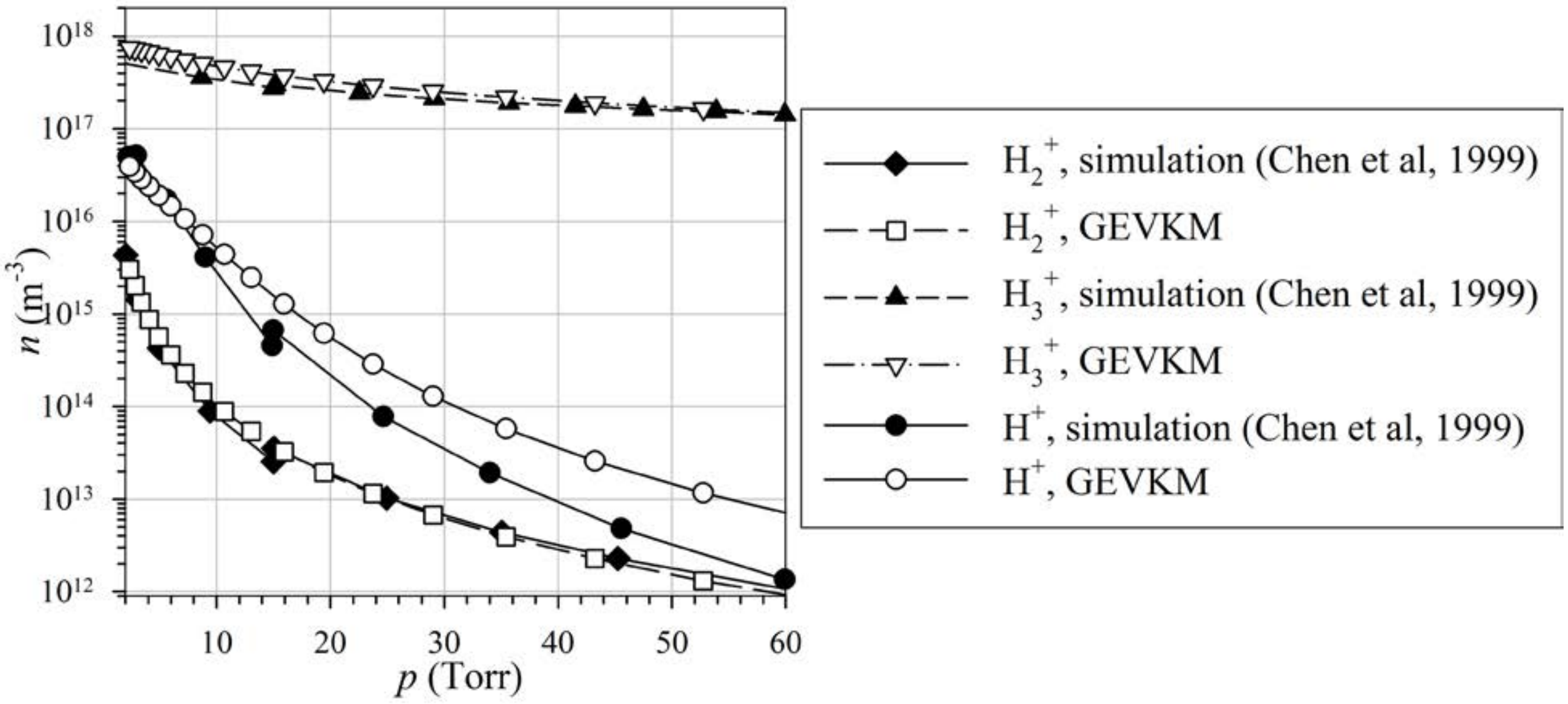

**Figure 15. Positive ions number densities in the microwave plasma reactor as a function of the chamber pressure at fixed absorbed power of 200 W and inlet flow rate 100 sccm.**

# 4 Application of the Global Enhanced Vibrational Kinetic Model to the High Current Negative Hydrogen Ion Source

The implementation of the GEVKM to the HCNHIS and its validation is discussed in this chapter. Compared to the idealized cylindrical geometry considered in Chapter 3, the HCNHIS designs used during development include complex bypass regions, the Negative Ion Production (NIP) region as well as the extraction system. In order to alleviate these complexities the GEVKM is supplemented with models that provide the outlet particle fluxes for the baseline and alternative HCNHIS design configurations considered. Validation of the inlet\outlet model is performed by comparisons of the numerical simulation results with measurements of the pressure inside the RFD chamber under various inlet and absorbed power conditions. Validation of the GEVKM is performed by comparisons of negative ion current predictions with the negative current measurements obtained in the downstream region of the extractor. Finally, the GEVKM is used in a parametric study of performance of the HCNHIS alternative design in order to investigate the dependence of the production and destruction of $H^-$ on main parameters such as feedstock gas flow rate, absorbed power, and bypass geometries.

The material of this chapter can be also found in Averkin et al. (2015b).

## 4.1 Review of the Experimental Investigation and Negative Current Measurements for Various HCNHIS Configurations

A series of experimental investigations were performed during the design iterations of the HCNHIS. These experiments included the effects of inlet flow rate, input power, bypass system, and NIP region (Olson, et al., 2012; Gatsonis , et al., 2012; Blandino, et al., 2012; Averkin, et al., 2012; Averkin, et al., 2014; Averkin & Gatsonis, 2014; Averkin, et al., 2015b; Averkin, et al., 2015a; Taillefer, et al., 2015).

The baseline HCNHIS configuration (referred in the text as the HCNHIS-1) used in the experiments is composed of the RFD chamber connected to the NIP region by a nozzle as shown in Figure 16. In order to decrease the flow rate from the high-pressure chamber, a system of five bypass tubes connected to a vacuum chamber is used. The electrons and vibrationally excited molecules exiting from the nozzle enter the NIP region where most of the production of negative ions takes place. The negative current is then extracted by extraction grid assembly, which consists of a pair of electrodes: a negatively biased plasma grid and a ground grid as shown in Figure 16. The plasma grid has 37 holes of the same diameter 0.381 mm and the ground grid has 37 holes of the same diameter 0.305 mm. In addition, the electron diverter is placed downstream of the grids to deflect the trajectory of electrons as shown in Figure 16.

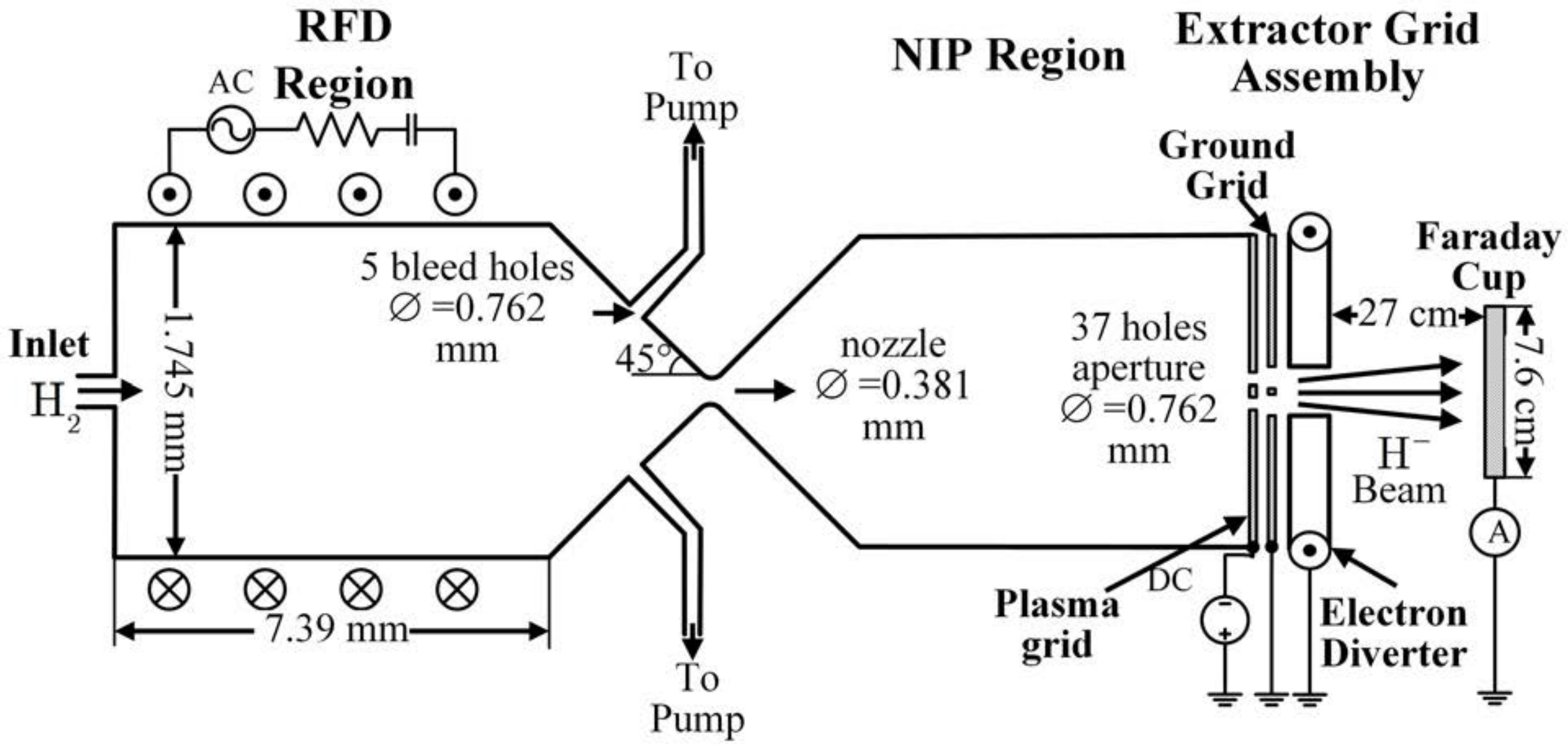

**Figure 16. Experimental setup of $H^-$ current measurements in the HCNHIS-1 with 5 by-pass tubes, extended NIP region, and 37 holes aperture.**

Experiments were performed for the inlet flow rates from 300 to 3000 sccm and absorbed power of 430-600 W as summarized in Table 13. During operation of the HCNHIS-1 the pressure in the RFD chamber was measured by the pressure port located near the inlet as shown in Figure 1. The pressure in the RFD chamber as a function of the absorbed power at constant inlet flow rate 3000 sccm is shown in Figure 17.

**Table 13. Operating parameters of the HCNHIS-1 and HCNHIS-2.**

| Parameters | HCNHIS-1 | HCNHIS-2 |
|---|---|---|
| $Q_{in}$(sccm) | 300-3000 | 1000 |
| $P_{abs}(W)$ | 430-600 | 341 |

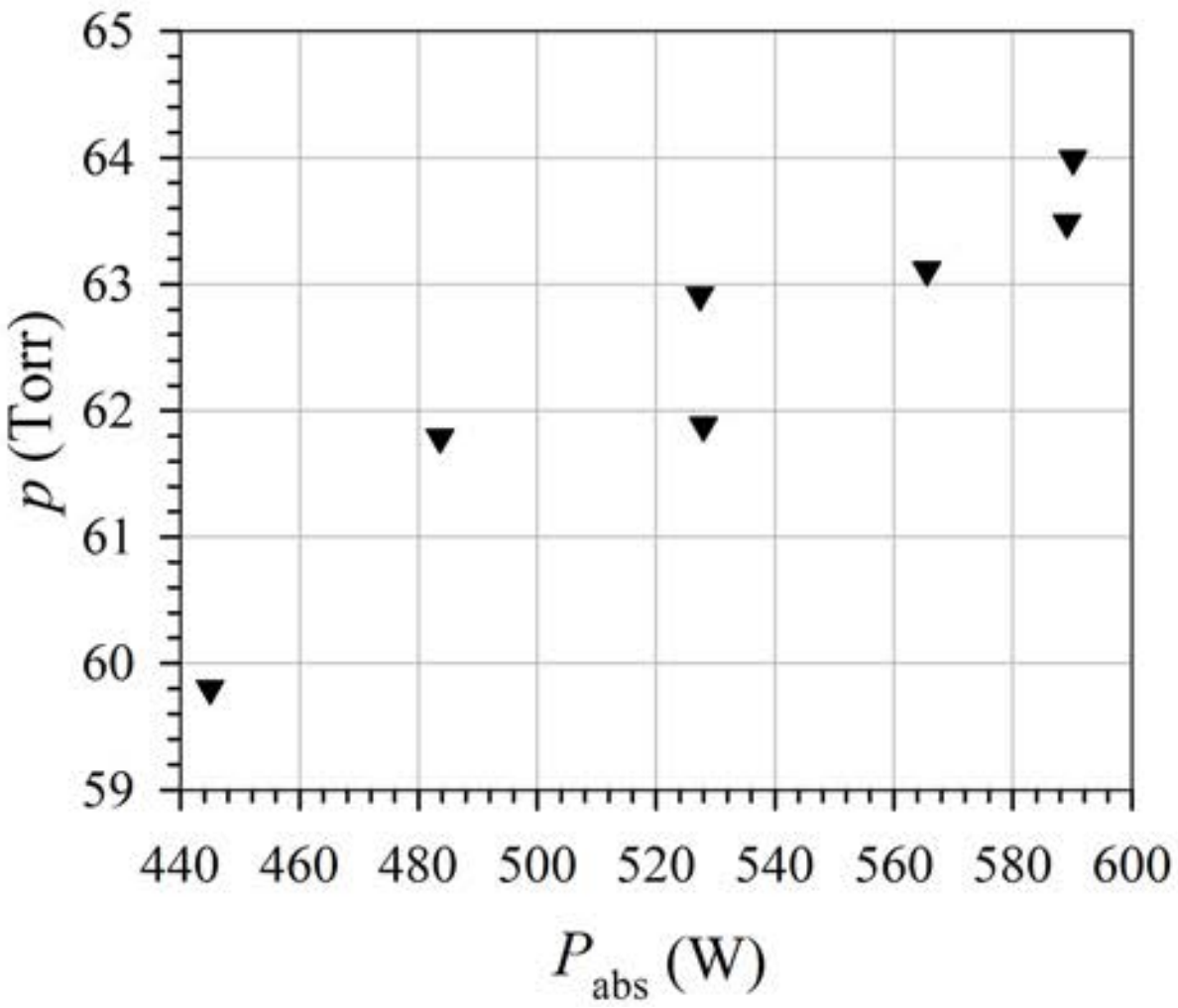

**Figure 17. Pressure in the HCNHIS-1 as a function of the absorbed power at fixed inlet flow rate 3000 sccm.**

In order to investigate the negative ion production ability of the HCNHIS-1 an experimental setup was used to measure the negative current at the downstream region of the NIP region as shown in Figure 16. A 7.6-cm diameter Faraday cup was placed 27 cm downstream from the extraction region of the HCNHIS.

The measured negative current obtained by a number of experiments is presented in Figure 18. Figure 18(a) shows the dependence of the negative current measured by the Faraday cup on the applied magnetic field of the electron diverter. As the magnetic field increases the negative current saturates and the saturated value can be attributed to the negative hydrogen ion current. Figure 18(b) shows the negative current as a function of the absorbed power. In this case the inlet flow rate of hydrogen is fixed at 3000 sccm, the magnetic diverter was also turned off. Therefore, this negative current is the sum of electron and negative ion currents. As it is seen from the plot the increase in the absorbed power results in nearly linear increase in the current.

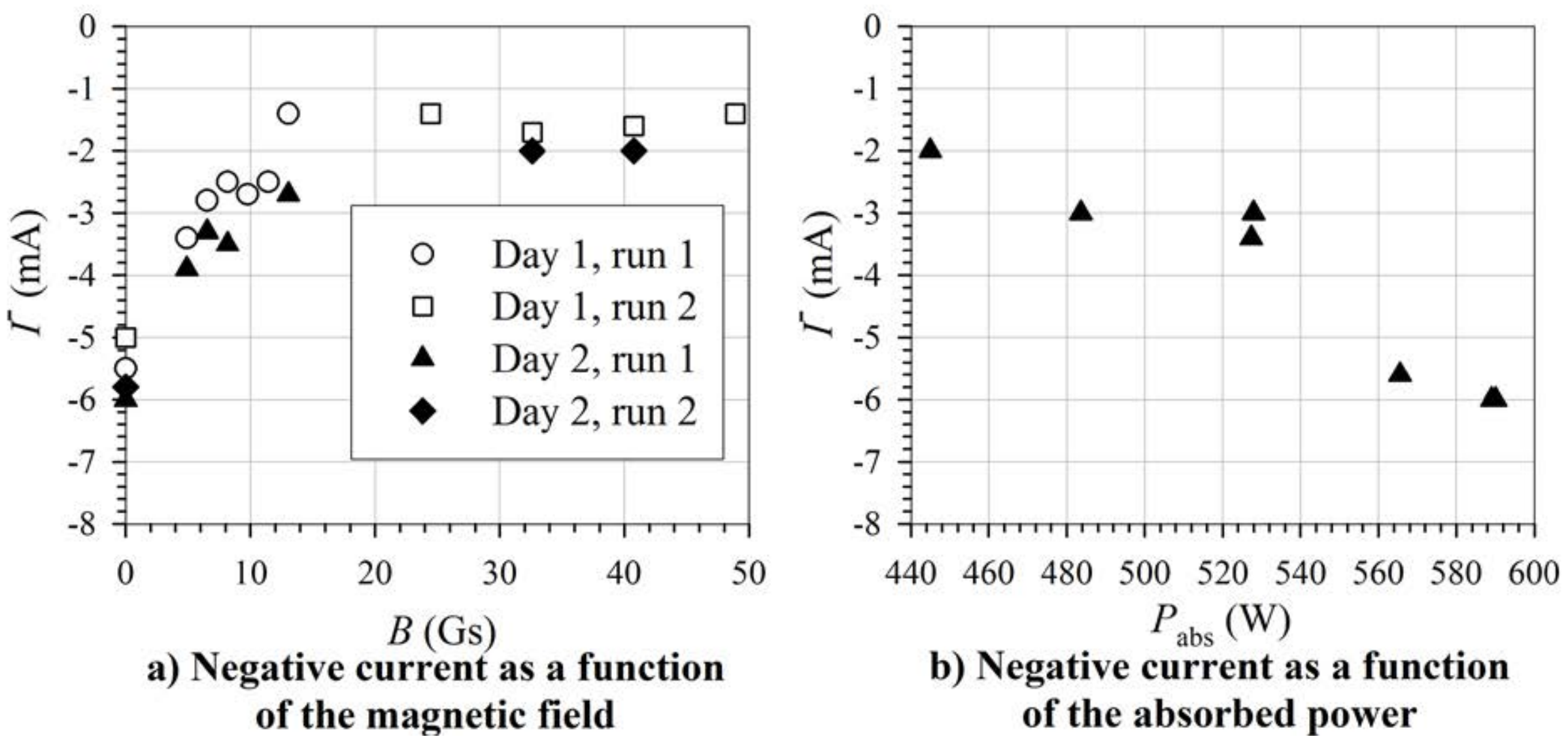

**Figure 18. Negative current in the HCNHIS-1 with multi-hole aperture extraction system at fixed absorbed power 590 W as a function of the magnetic field (a) and as a function of the absorbed power (b) at fixed inlet flow rate 3000 sccm.**

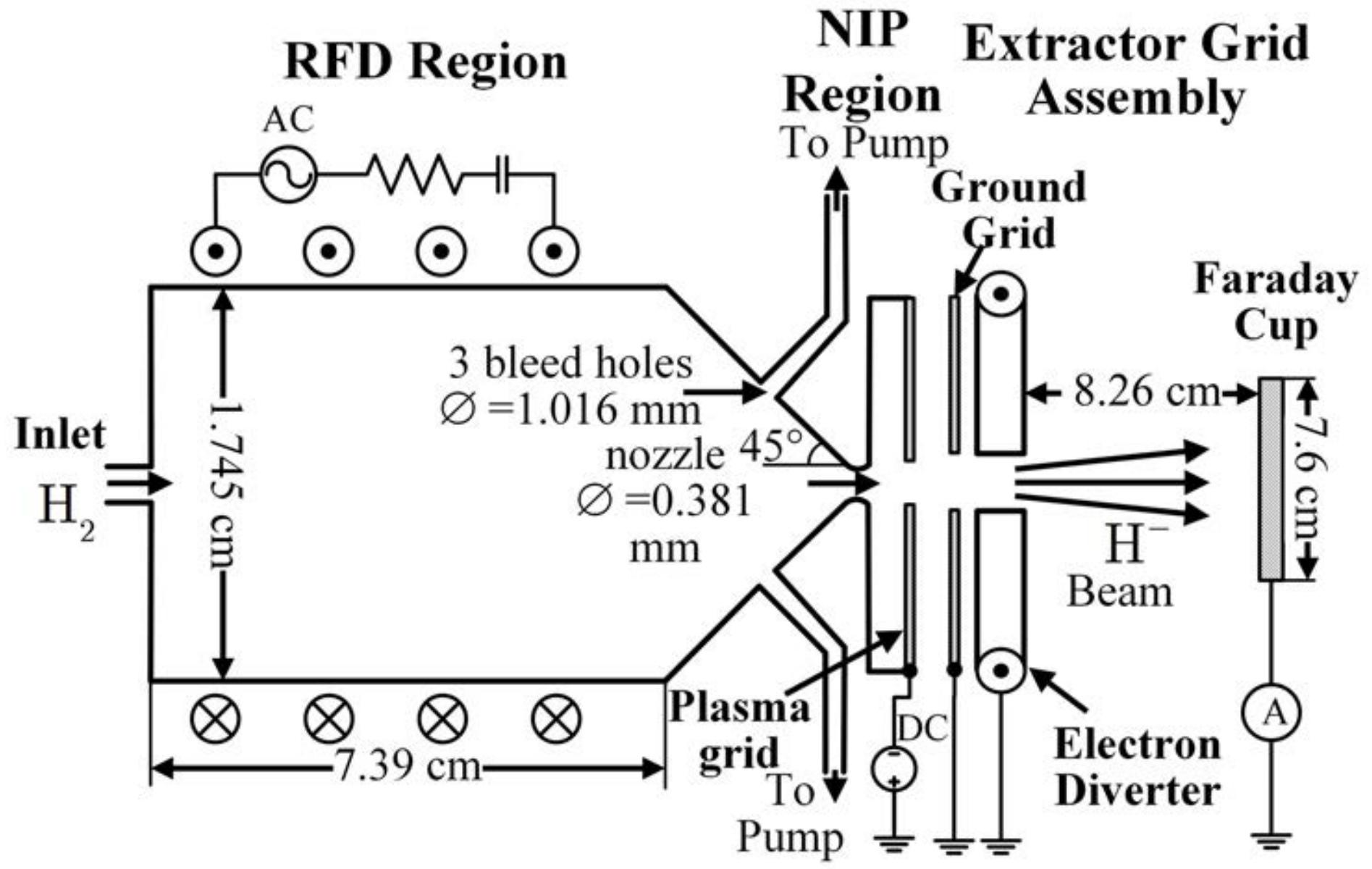

**Figure 19. Experimental setup of $H^-$ current measurements in the HCNHIS-2 with 3 by-pass tubes, reduced NIP region, and single-hole aperture.**

The alternative design configuration also used in the experiments is referred to as HCNHIS-2 and is depicted schematically in Figure 19. This configuration includes 3 bypass tubes with increased diameters compared to the HCNHIS-1 resulting in a slight change in the total bypass area. More importantly, the NIP region of the HCNHIS-2 is reduced to a tiny region just before screen and ground grids. The operating parameters used in the experiment are listed in Table 13. The negative current measurements were performed in a similar manner compared to the HCNHIS-1. The 7.6 cm Faraday cup was placed 8.26 cm downstream the electron diverter with the extraction taking place from a single-hole aperture. The diameters of the hole in the plasma and ground grids are 0.914 mm and 1.5 mm respectively. The negative current as a function of the magnetic field in the electron diverter is shown in Figure 20. Similar to the HCNHIS-1 the negative current decreases as the magnetic field increases and then saturates to a value attributed to the negative hydrogen ion current.

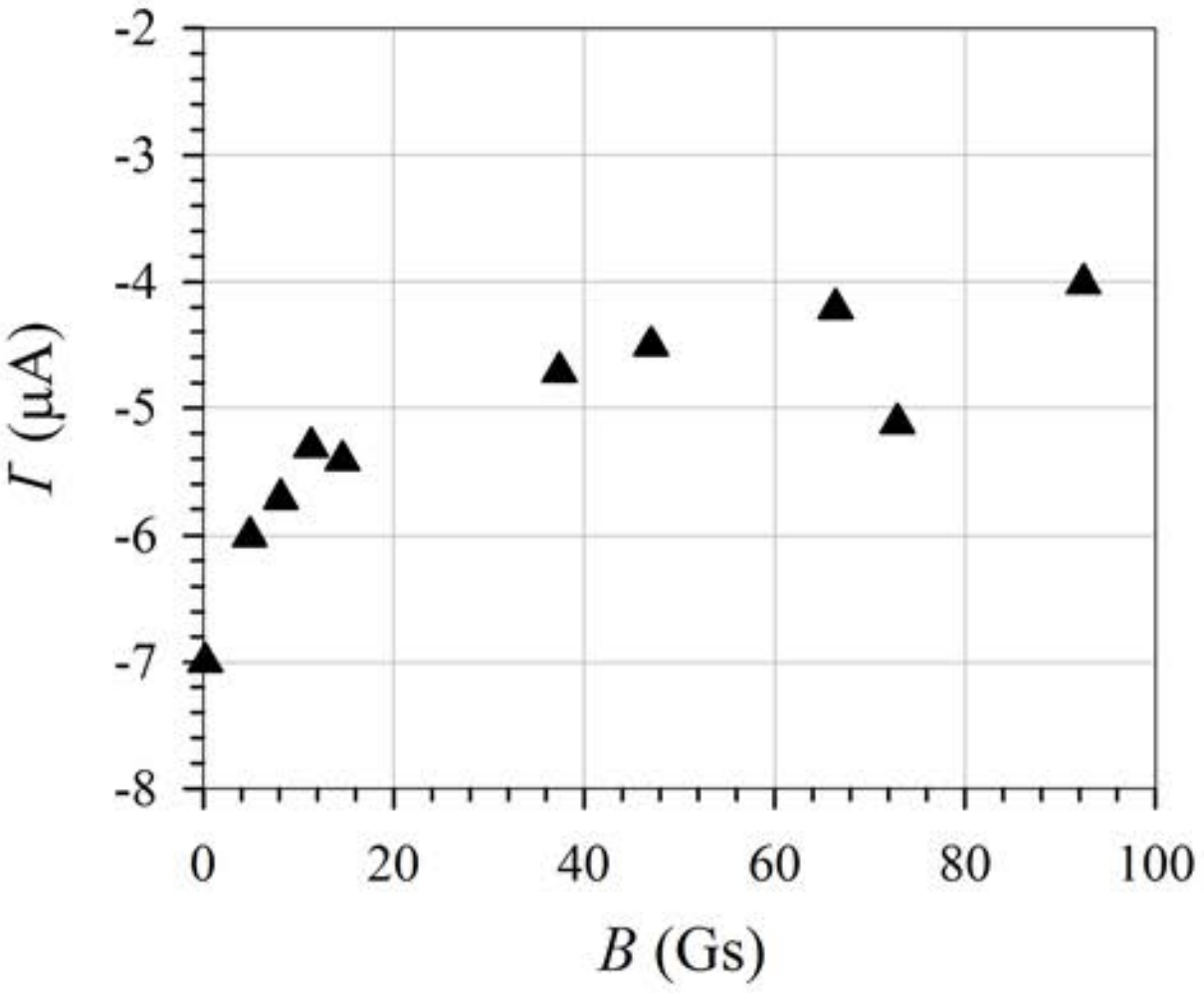

**Figure 20. Negative current in the HCNHIS-2 with single-hole aperture as a function of the magnetic field in the diverter at fixed inlet flow rate of 1000 sccm and absorbed power 341 W.**

### 4.2 Modeling and Simulation of the HCNHIS

This section summarizes the modeling equations, chemical reactions, inlet/outlet conditions, and other modifications needed for the implementation of the GEVKM to the simulation of the HCNHIS-1 and the HCNHIS-2 designs shown in Figure 16 and Figure 19 respectively.

#### 4.2.1 GEVKM of the RFD Chamber

The RFD chamber is simulated by the GEVKM and the results are used as input parameters to a NIP region model in order to obtain predictions of the negative hydrogen ion current.

The HCHNIS-1 and HCHNIS-2 configurations used during the design iterations and the negative ion experiments involve a complex bypass system with tubes. In order to implement the GEVKM to the HCNHIS-1 and HCNHIS-2, inlet and outlet boundary conditions of the RFD nozzle and bypass system are needed. Such boundary conditions provide the particle fluxes in the continuity (3.28), electron energy (3.94) and total energy (3.101) equations.

The HCNHIS RFD chamber is represented in the GEVKM by the effective cylindrical plasma reactor, which accounts for the volume of the cylindrical and conical parts of the HCNHIS. The effective cylinder configuration is shown in Figure 7. The bypass system is excluded from the GEVKM simulation but the fluxes associated with the bypass are included. The radius of the effective cylinder in the GEVKM is set to the radius of the cylindrical part of the HCNHIS. The length of the effective cylinder is calculated from the volume of the RFD chamber of the HCNHIS and the radius of the effective cylinder.

The species included in the GEVKM model of the RFD chamber are ground state hydrogen atoms $\mathrm{H}$ and molecules $\mathrm{H}_2$, 14 vibrationally excited hydrogen molecules $\mathrm{H}_2(v)$, $v = 1 - 14$, electronically excited hydrogen atoms $\mathrm{H}(2)$, $\mathrm{H}(3)$, ground state positive ions $\mathrm{H}^+$, $\mathrm{H}_2^+$, $\mathrm{H}_3^+$, ground state anions $\mathrm{H}^-$, and electrons $\mathrm{e}$. The input parameters to the GEVKM are the geometry configuration, the wall material parameters such as emissivity, momentum and thermal accommodation coefficients, the molecular hydrogen inlet flow rate, and the absorbed power. The results provided by the GEVKM include the steady-state spatially averaged species composition, the electron temperature, the heavy species temperature, the wall temperature, particle fluxes to the NIP region. The full chemical reaction model is incorporated as shown in Table 3-Table 9. The spatially averaged number densities of neutral and ion species are obtained from the solution of spatially averaged steady species continuity equations

$$Q_{p,\mathrm{out}} - Q_{p,\mathrm{in}} + \frac{A}{V}\Gamma_{p,\mathrm{wall}} = \sum_{r=1}^{N_r} k_r \Theta_r \prod_{p=1}^{N_s} \left(\nu_{r,p} - \nu'_{r,p}\right) \overline{n}_p^{\nu_{r,p}} \ . \tag{3.28}$$

The term $Q_{p,\mathrm{in}}$ in the above equation designates the inlet flow rate in Figure 7 and $Q_{p,\mathrm{out}}$ designates the outlet flow rate through the nozzle and bypass system, shown in Figure 7. These terms are discussed in details in the next section. The quasi-neutrality condition is used for electron number density estimation

$$n_e + n_{H^-} = n_{H^+} + n_{H_2^+} + n_{H_3^+}\,. \tag{3.24}$$

The electron temperature is obtained as a solution of the electron energy equation

$$\begin{aligned} &\frac{5}{2}k_B T_e Q_{e,\text{out}} - \frac{5}{2}k_B T_{e,\text{in}} Q_{e,\text{in}} + \frac{A}{V}\sum_{p=1}^{N_h^+}\Gamma_{p,\text{wall}}\left(2k_B T_e + eV_p + eV_{\text{sh}}\right) \\ &= k_B\left(T_h - T_e\right)\sum_{h=1}^{N_h} 3\frac{m_e}{m_h} k_{eh}^{\text{elast}}\Theta_h \overline{n}_h \overline{n}_e + \sum_{r=1}^{N_r} E_{r,\text{th}} k_r \Theta_r \prod_{p=1}^{N_s}\overline{n}_p^{\nu_{r,p}} + \frac{P_{\text{elec}}}{V} + \frac{P_{\text{abs}}}{V}. \end{aligned} \tag{3.123}$$

The solution of the total energy equation provides the heavy species temperature

$$\begin{aligned} &\frac{P_{\text{abs}}}{V} - \sum_{p=1}^{N_s}\left(\frac{m_p u_{p,\text{out}}^2}{2} + H_{p,\text{out}}\right)Q_{p,\text{out}} + \sum_{p=1}^{N_s}\left(\frac{m_p u_{p,\text{in}}^2}{2} + H_{p,\text{in}}\right)Q_{p,\text{in}} \\ &= \frac{A}{V}\sum_{p=1}^{N_s} q_p\Big|_{\text{wall}} + \frac{A}{V} q_{\text{cond}}\Big|_{\text{wall}} + \frac{P_{\text{elec}}}{V} + \frac{P_{\text{rad}}}{V}, \end{aligned} \tag{3.124}$$

The wall temperature is obtained from the solution of the heat transfer equation

$$\begin{aligned} &\frac{P_{\text{abs}}}{V} - \sum_{p=1}^{N_s}\left(\frac{m_p u_{p,\text{out}}^2}{2} + H_{p,\text{out}}\right)Q_{p,\text{out}} + \sum_{p=1}^{N_s}\left(\frac{m_p u_{p,\text{in}}^2}{2} + H_{p,\text{in}}\right)Q_{p,\text{in}} \\ &= \frac{2\pi R}{V}\left(\text{h}_{\text{side}} R + \text{h}_{\text{cyl}} L\right)(T_{\text{w}} - T_\infty) + \epsilon_{\text{emis}}\sigma_{\text{SB}}\frac{A}{V}\left(T_{\text{w}}^4 - T_\infty^4\right). \end{aligned} \tag{3.114}$$

Equations (3.28), (3.24), (3.123), (3.124), and (3.114) represent a system of 26 non-linear algebraic equations and are solved simultaneously by Newton-Raphson method (Teukolsky, et al., 1996).

### 4.2.2 Inlet and Outlet Fluxes in the RFD Chamber

The inlet feedstock gas flux is usually given in terms of a flow rate with the units of sccm or slm. The conversion between sccm or slm to $m^{-3}s^{-1}$ is done by using Eq. (3.33).

The operating parameters in the RFD chamber of the HCNHIS-1 shown in Table 13 result in the chamber pressure in the range 6.7-64.3 Torr with corresponding Knudsen numbers in the throat and bypass tube entry in the range 0.0175-0.2 and 0.009-0.1 respectively. For the RFD chamber of the HCNHIS-2 the operating parameters listed in Table 13 result in the chamber pressure 20.3 Torr, the Knudsen numbers in the throat and bypass tube entry are 0.065 and 0.02 respectively. The flow therefore is expected to cover from the continuum to slip regime for the bypass and slip to transition regime for the nozzle (Karniadakis, et al., 2005) in both setups and as such the outlet fluxes have to account for these flow regimes. In order to evaluate the outlet fluxes, the flow in the RFD chamber, nozzle as well as the bypass system is modeled as a cold neutral gas ignoring plasma effects.

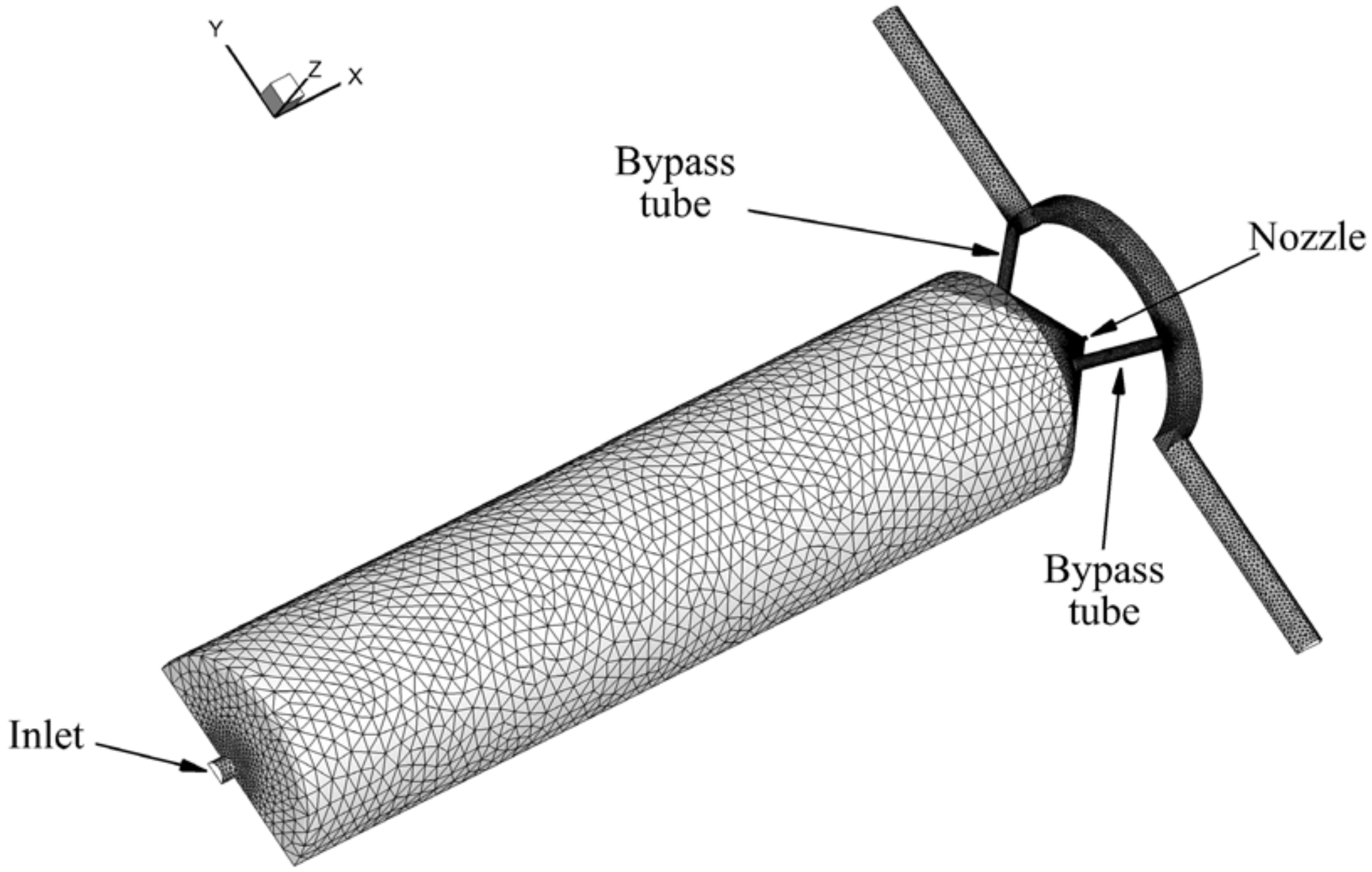

**Figure 21. Unstructured three-dimensional mesh used in the compressible inviscid cold pure hydrogen flow simulation of the RFD chamber, nozzle and bypass system of the HCNHIS-2.**

The continuum flow regime in the RFD chamber of the HCNHIS-2 setup is investigated with an inlet flow rate 1000 sccm as shown in Table 13 due to the lack of the experimental measurements of the chamber pressure in this device. The full 3d geometry of the RFD chamber and bypass system shown in Figure 19 is considered and the hydrogen flow is modeled with the 3d compressible inviscid equations using ANSYS Fluent (ANSYS, Inc., 2012). Due to the symmetry of the RFD chamber and the bypass system only half of the device is simulated with a mesh shown in Figure 21. The boundary conditions include symmetry at the symmetry plane of the chamber, specified mass flow rate at the inlet, and pressure boundary conditions at the outlet of bypasses and the nozzle. The solid walls are modeled as impermeable to the flow. The simulation is performed at inlet flow rate 1000 sccm and gas temperature 300 K.

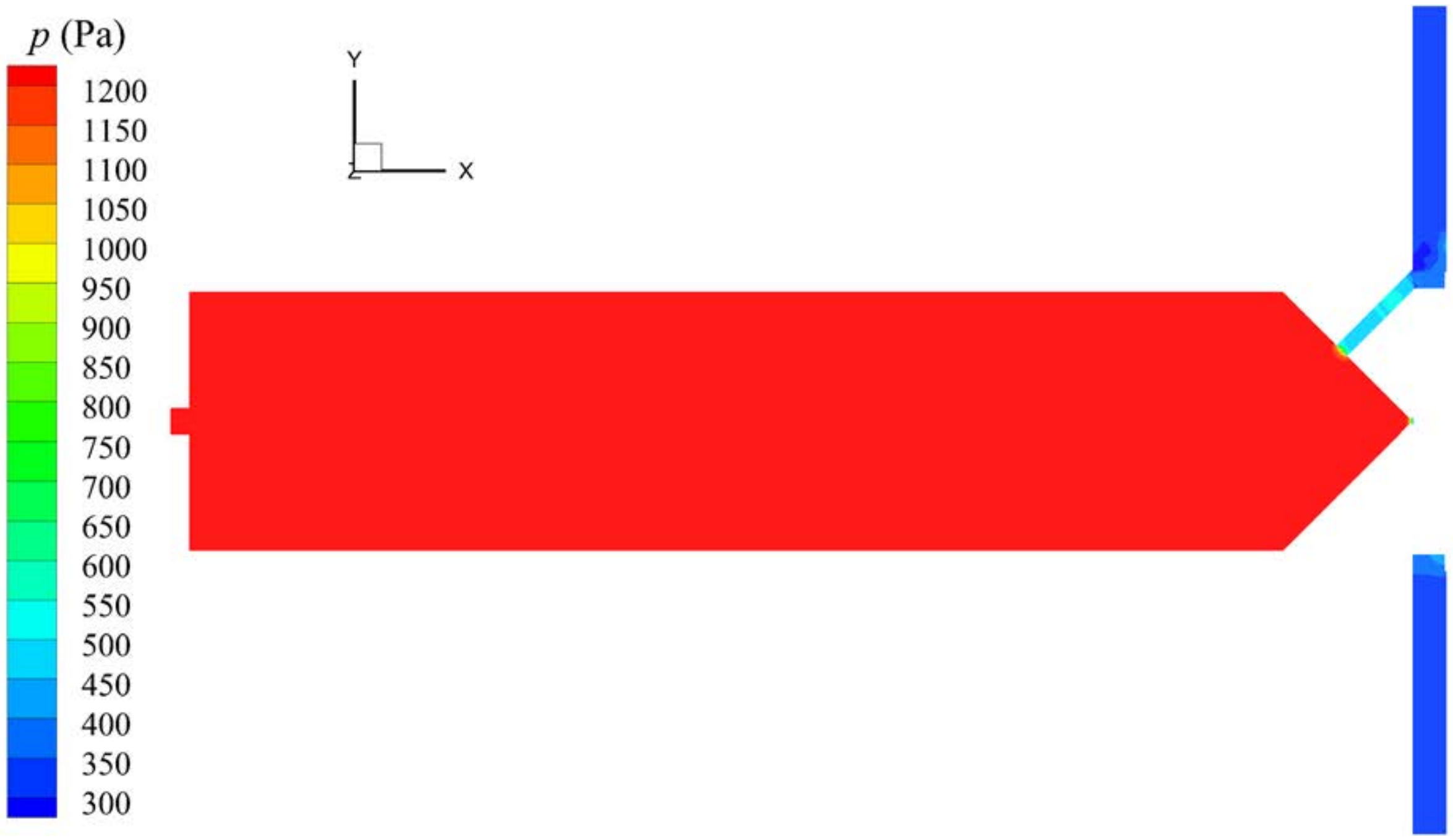

**Figure 22. Pressure distribution along the centerplane of cold pure hydrogen flow in the RFD chamber, nozzle and bypass system of the HCNHIS-2 at inlet flow rate 1000 sccm and gas temperature 300 K.**

The pressure distribution along the symmetry plane is shown in Figure 22 and it is nearly uniform in the RFD chamber. Most of the pressure variation occurs in the nozzle region and inside the bypass tubes. The Mach number at the centerplane is presented in Figure 23. The flow is subsonic in the RFD chamber and the Mach number is close to 0.4-0.5 in the inlet region and below 0.1 in the rest of the chamber. The simulations show that the flow in the nozzle and in the bypass tubes is choked. In the RFD chamber of the HCNHIS the hydrogen gas is heated resulting in an even higher pressure compared to the cold gas flow and, therefore, the plasma flow in the real device is expected to choke as well.

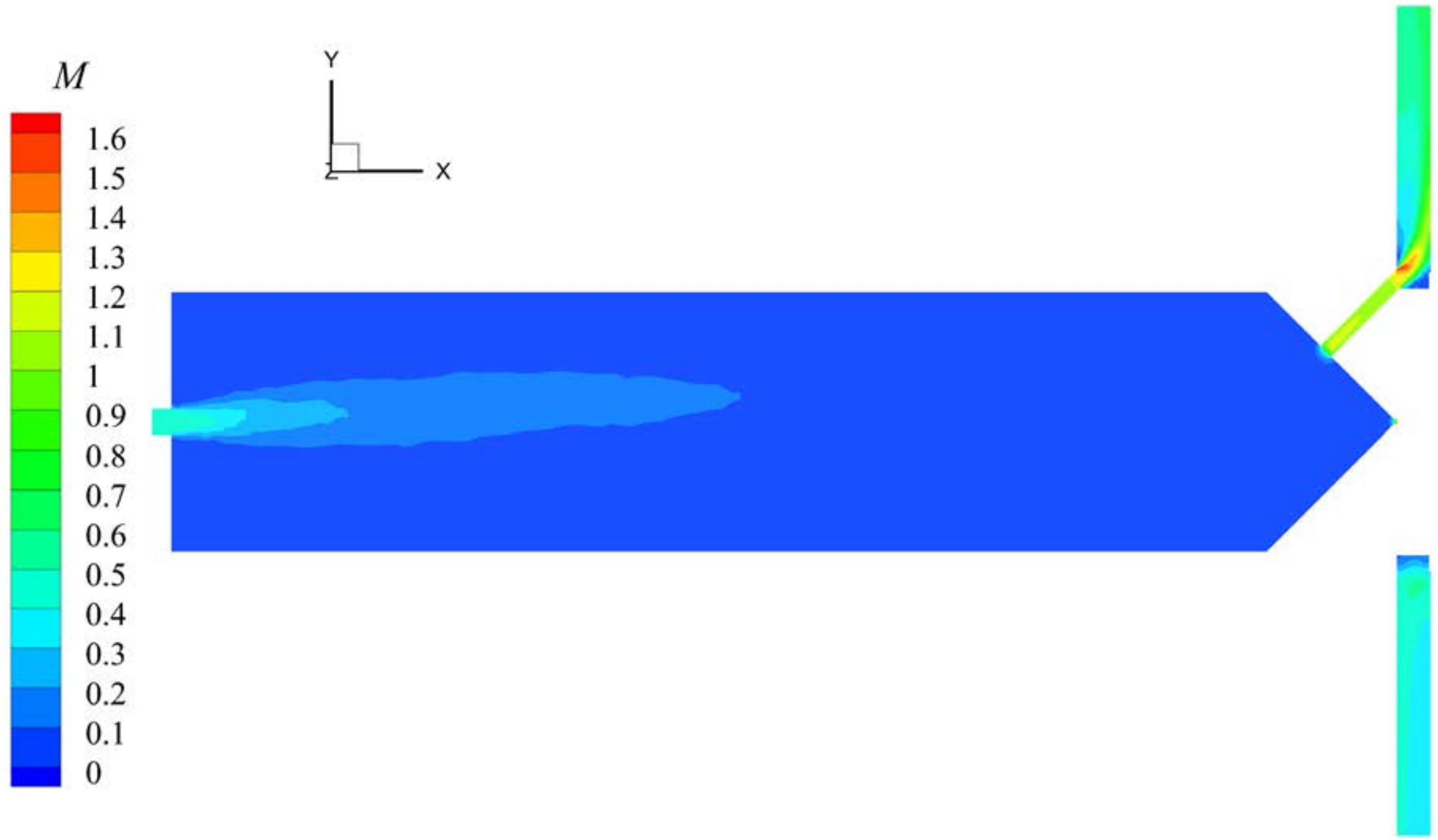

**Figure 23. Mach number distribution along the centerplane of cold pure hydrogen flow in the RFD region of the HCNHIS-2 at inlet flow rate 1000 sccm and gas temperature 300 K.**

The 3d compressible inviscid simulations show that the cold hydrogen flow is choked in the bypass tubes and nozzle as seen in Figure 24. It is therefore for computational efficiency desirable to implement an analytical isentropic frozen-flow model for the estimation of the outlet mass flow rates for the plasma species through the nozzle and bypass tubes.

The continuum nozzle flow is discussed first. Assuming that the plasma has a sonic speed at the nozzle throat and the flow is frozen and isentropic the term representing the number flux leaving the plasma in the continuity equation (3.28) can be written as (Anderson, Jr., 2000; John & Keith, 2006)

$$Q_{p,\mathrm{nozzle,out}} = \frac{A_{\mathrm{nozzle}}}{V} \frac{p_{p,0}}{\sqrt{m k_B T_{p,0}}} \sqrt{\gamma_{\mathrm{fr}}} \left( \frac{\gamma_{\mathrm{fr}} + 1}{2} \right)^{-\frac{\gamma_{\mathrm{fr}}+1}{2\left(\gamma_{\mathrm{fr}}-1\right)}} . \tag{4.1}$$

In the above expression $m = \frac{1}{n} \sum_{p=1}^{N_s} m_p n_p$ is the average particle mass of the mixture; $\gamma_{\mathrm{fr}}$ is the frozen ratio of specific heats for the plasma calculated by taking into account only translational and rotational degrees of freedoms of molecules, atoms, ions, and electrons; $p_{p,0}$ and $T_{p,0}$ are the total pressure and temperature in the RFD chamber respectively assumed to be equal to their static values due to the low Mach numbers expected in the RFD chamber.

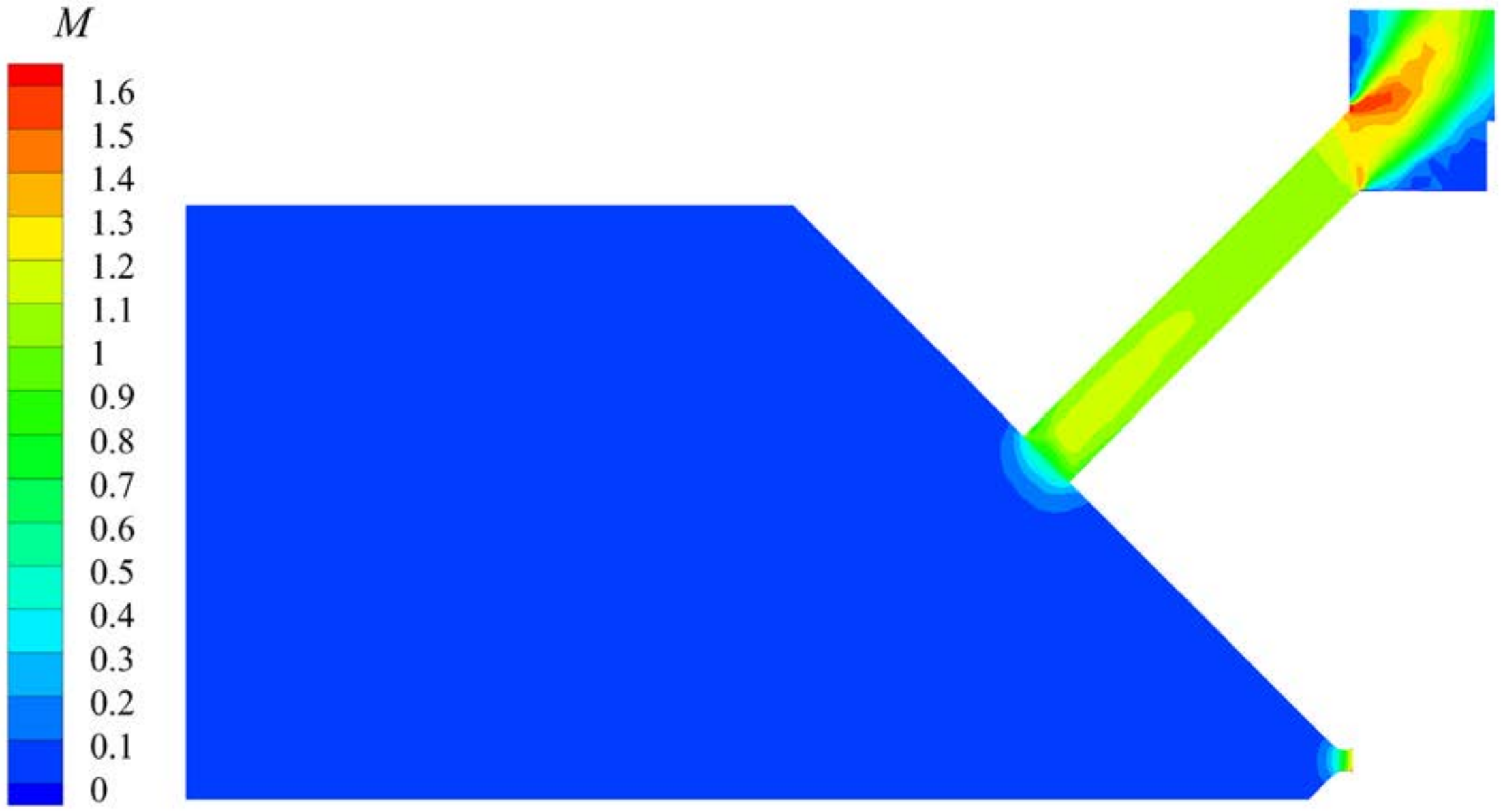

**Figure 24. Mach number distribution along the centerplane in the bypass tubes and the nozzle of cold pure hydrogen flow in the HCNHIS-2 at inlet flow rate 1000 sccm and gas temperature 300 K.**

At chamber pressures below 2 Torr corresponding to the throat Knudsen numbers less than 0.1 the flow in the nozzle throat exhibits rarefaction effects. Assuming free-molecular flow conditions (Gombosi, 1994) the outlet particle flux is equal to the random flux due to effusion and can be represented as

$$Q_{p,\text{effusion,out}} = \frac{A_{\text{nozzle}}}{V}\frac{1}{4}n_{p,0}\sqrt{\frac{8k_B T_{p,0}}{\pi m_p}}\,. \tag{4.2}$$

Assuming a low degree of ionization and dissociation in the plasma in the HCNHIS RFD chamber the dominant species is molecular hydrogen with $\gamma_{\text{fr}}=1.4$. Substituting these values into Eq. (4.1) it becomes

$$Q_{\text{H}_2,\text{nozzle,out}} = \frac{A_{\text{nozzle}}}{V}n_{\text{H}_2,0}\sqrt{\frac{k_B T_{h,0}}{m_{\text{H}_2}}}\sqrt{\gamma_{\text{fr}}}\left(\frac{\gamma_{\text{fr}}+1}{2}\right)^{-\frac{\gamma_{\text{fr}}+1}{2(\gamma_{\text{fr}}-1)}} \approx 0.685\frac{A_{\text{nozzle}}}{V}n_{\text{H}_2,0}\sqrt{\frac{k_B T_{h,0}}{m_{\text{H}_2}}}\,. \tag{4.3}$$

On the other hand, Eq. (4.2) becomes

$$Q_{\mathrm{H_2,effusion,out}} = \frac{1}{\sqrt{2\pi}} \frac{A_{\mathrm{nozzle}}}{V} n_{\mathrm{H_2},0} \sqrt{\frac{k_B T_{h,0}}{m_{\mathrm{H_2}}}} \approx 0.4 \frac{A_{\mathrm{nozzle}}}{V} n_{\mathrm{H_2},0} \sqrt{\frac{k_B T_{h,0}}{m_{\mathrm{H_2}}}}. \tag{4.4}$$

Thus, the particle flux through isentropic nozzle is approximately one and a half times larger than the flux due to effusion. To take into account the rarefaction effect the approach similar to the one used in writing Eq. (3.82) can be adopted resulting in the following expression for the outlet particle flux through the nozzle

$$Q_{p,\mathrm{nozzle,out}} = \frac{A_{\mathrm{nozzle}}}{V} n_{p,0} \sqrt{\frac{k_B T_{p,0}}{m_p}} \left[ \sqrt{\gamma_{\mathrm{fr}}} \left( \frac{\gamma_{\mathrm{fr}} + 1}{2} \right)^{-\frac{\gamma_{\mathrm{fr}}+1}{2(\gamma_{\mathrm{fr}}-1)}} \frac{1}{1 + \mathrm{Kn}_{p,\mathrm{thr}}} + \frac{1}{\sqrt{2\pi}} \frac{\mathrm{Kn}_{p,\mathrm{thr}}}{1 + \mathrm{Kn}_{p,\mathrm{thr}}} \right], \tag{4.5}$$

where $\mathrm{Kn}_{p,\mathrm{thr}} = \lambda_p / 2R_{\mathrm{thr}}$ is the throat Knudsen number and $R_{\mathrm{thr}}$ is the throat radius.

The modeling of the flow in the bypass tubes requires additional consideration due to viscous and rarefaction effects. From the inviscid simulations the flow speed was found to become sonic in the bypass region close to the entrance of the tubes as shown in Figure 24. However, in the real flow the choking is expected to occur due to friction and the viscous boundary layer developing in the tubes. Thus, the bypass tube cross-sectional area accessible for the flow and mass flow rate is reduced compared to the area used in the inviscid simulations. In order to take this effect into account the Fanno flow theory for the vibrationally and chemically frozen flow with friction is used (Anderson, Jr., 2000; John & Keith, 2006). In addition, the flow is considered to choke at the end of all bypass tubes. According to Fanno flow theory to predict the outlet particle flux it is necessary to find the inlet Mach number $M_{\mathrm{b,in}}$ from the following implicit equation, which is solved numerically by the Newton-Raphson method,

$$\frac{fL_{\mathrm{b}}}{2R_{\mathrm{b}}} = \frac{\gamma_{\mathrm{fr}} + 1}{2\gamma_{\mathrm{fr}}} \ln \left[ \frac{(\gamma_{\mathrm{fr}} + 1) M_{\mathrm{b,in}}^2}{2 + (\gamma_{\mathrm{fr}} - 1) M_{\mathrm{b,in}}^2} \right] - \frac{1}{\gamma_{\mathrm{fr}}} \left( 1 - \frac{1}{M_{\mathrm{b,in}}^2} \right), \tag{4.6}$$

where $L_{\mathrm{b}}$ and $R_{\mathrm{b}}$ are the length and radius of the bypass tubes, $f$ is the Darcy friction factor.

The diameters of the bypass tubes are 0.0762 mm and 1.016 mm for the HCNHIS-1 and HCNHIS-2 respectively resulting in the Reynolds numbers below 100 and Knudsen numbers 0.01-0.2. Therefore, the Darcy factor for the bypass tubes is calculated through the expression for laminar Poiseuille flow with rarefaction effects taken into account as (Valougeorgis, 2007)

$$f = \frac{64}{\mathrm{Re}_{D_\mathrm{b}}\left(1 + \dfrac{2-\sigma_v}{\sigma_v} 8\mathrm{Kn}_{D_\mathrm{b}}\right)}. \tag{4.7}$$

The Reynolds number $\mathrm{Re}_{D_\mathrm{b}}$ in Eq. (4.7) is calculated by

$$\mathrm{Re}_{D_\mathrm{b}} = \frac{\rho a D_\mathrm{b}}{\mu}, \tag{4.8}$$

where $D_\mathrm{b} = 2R_\mathrm{b}$ is the bypass tube diameter, $\rho$ is the density, $a = \sqrt{\gamma_\mathrm{fr} k_B T_{p,0} / m}$ is the speed of sound, and $\mu$ is the viscosity of the plasma in the discharge chamber approximated as the viscosity of pure molecular hydrogen gas due to low degree of dissociation and ionization in the RFD chamber.

The Knudsen number in Eq. (4.7) is given by

$$\mathrm{Kn}_{D_\mathrm{b}} = \frac{\lambda_{\mathrm{H}_2}}{D_\mathrm{b}}, \tag{4.9}$$

where $\lambda_{\mathrm{H}_2} = \left[\sqrt{2}\pi d^2_{\mathrm{ref,H}_2} n_{\mathrm{H}_2} \left(T_{\mathrm{ref,H}_2} / T\right)^{\omega_{\mathrm{ref,H}_2} - \frac{1}{2}}\right]^{-1}$ is the mean free path of molecular hydrogen given by VHS model for the collision cross section (Bird, 1994), $d_{\mathrm{ref,H}_2}$, $T_{\mathrm{ref,H}_2}$, and $\omega_{\mathrm{ref,H}_2}$ are reference diameter, reference temperature, and viscosity index of molecular hydrogen in VHS model.

Once the bypass inlet Mach number is calculated the particle flux at the outlet is evaluated by (John & Keith, 2006)

$$Q_{p,\mathrm{bypass,out}} = \frac{A_\mathrm{b}}{V} \frac{p_{p,0}}{\sqrt{m k_B T_{p,0}}} \sqrt{\gamma_\mathrm{fr}} M_\mathrm{b,in} \left(1 + \frac{\gamma_\mathrm{fr} + 1}{2} M^2_\mathrm{b,in}\right)^{-\frac{\gamma_\mathrm{fr}+1}{2(\gamma_\mathrm{fr}-1)}}. \tag{4.10}$$

The outlet fluxes given by Eq. (4.5) and (4.10) are used for all species in the continuity equations. However, the relative contribution of these fluxes in the continuity equations is typically below 0.1 % for trace neutral species and ions compared to volumetric and surface chemical reactions under operating conditions of the HCNHIS designs listed in Table 13. It is ground state atomic and molecular hydrogen for which the contribution of the outflow fluxes in continuity equations is considerable and plays an important role in the chamber pressure establishment.

### 4.2.3 Wall Material Properties Used in the GEVKM Simulations of the HCNHIS

The input parameters for the GEVKM simulations of the RFD chamber of the HCNHIS configurations include the geometric dimensions of the discharge chamber, the nozzle, and by-pass system, the inlet flow rate, the absorbed power, and material properties of the walls of the chamber. The wall material of the RFD chamber in the HCNHIS setups is boron nitride. The material specific values such as the atomic hydrogen recombination coefficient $\gamma_{\mathrm{H,rec}}$, the molecular $\alpha_{\mathrm{H_2}}$ and atomic hydrogen $\alpha_{\mathrm{H}}$ thermal accommodation coefficients, the emissivity of boron nitride $\epsilon_{\mathrm{BN}}$, and the tangential momentum accommodation coefficient $\sigma_v$ used in the GEVKM simulations of the HCNHIS designs are reported in Table 14. Due to the lack of the experimental data the recombination coefficient of atomic hydrogen is approximated based on the typical values for other materials. The thermal accommodation coefficients of $\mathrm{H}$ and $\mathrm{H_2}$ are approximated using the expressions given by Eq. (3.129). The emissivity of boron nitride is based on the experimental measurements of Matlock et al. (2007). The tangential momentum coefficient is assumed to be 0.065. This value is close to the tangential momentum coefficient of hydrogen molecules on graphite surfaces at wall temperatures above 1000 K as was calculated by Kovalev et al. (2011) and was chosen in order to better represent measured chamber pressures.

**Table 14. Boron nitride properties used in the GEVKM simulations of the HCNHIS-1 and HCNHIS-2.**

| Parameters | $\gamma_{\mathrm{H,rec}}$ | $\alpha_{\mathrm{H_2}}$ | $\alpha_{\mathrm{H}}$ | $\epsilon_{\mathrm{BN}}$ | $\sigma_v$ |
|---|---|---|---|---|---|
| Numerical values | 0.1 | Eq. (3.129) | Eq. (3.129) | 0.85 | 0.065 |

### 4.2.4 GEVKM Simulations of the RFD Chamber of the HCNHIS-1: Validation of the Inlet\outlet Fluxes Model

In order to validate the inlet\outlet flux model, a series of simulations of the HCNHIS-1 are performed using the GEVKM for a range of absorbed power from 440 to 600 W and inlet flow rate of 3000 sccm corresponding to the operational parameters shown in Table 13. Figure 25 shows the predicted RFD chamber pressure of the HCNHIS-1 compared to the measurements obtained at the pressure port located near the inlet of the RFD chamber as shown in Figure 1. The flow model for the nozzle and bypass outlet predicts well the pressure in the RFD chamber of the HCNHIS-1. Since the RFD chamber pressure depends on the neutral gas temperature predicted by the GEVKM, this result serves as a validation of the temperature equation models (3.106) of the GEVKM.

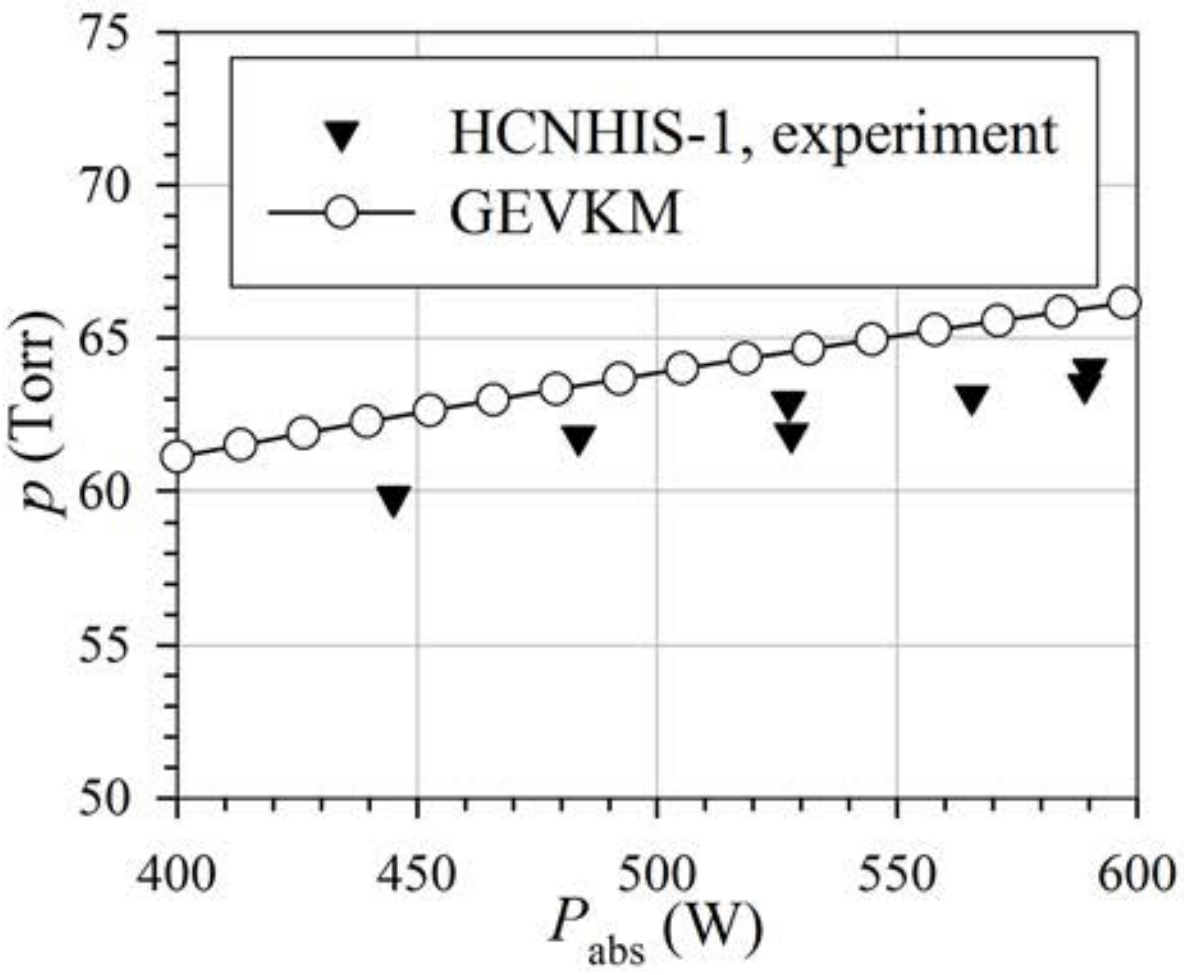

**Figure 25. Pressure in the RFD chamber of the HCNHIS-1 as a function of the absorbed power at fixed hydrogen inlet flow rate 3000 sccm.**

There were no reliable measurements of the RFD chamber pressure during the experiments with the HCNHIS-2 configuration. However, the total bypass area is rather close in the HCNHIS-1 and HCNHIS-2 configurations. Figure 26 shows the calculated chamber pressure for two HCNHIS design configurations as a function of the inlet flow rate at constant absorbed power 341 W. The difference between pressures in these two configurations is very small for a wide range of the operational parameters.

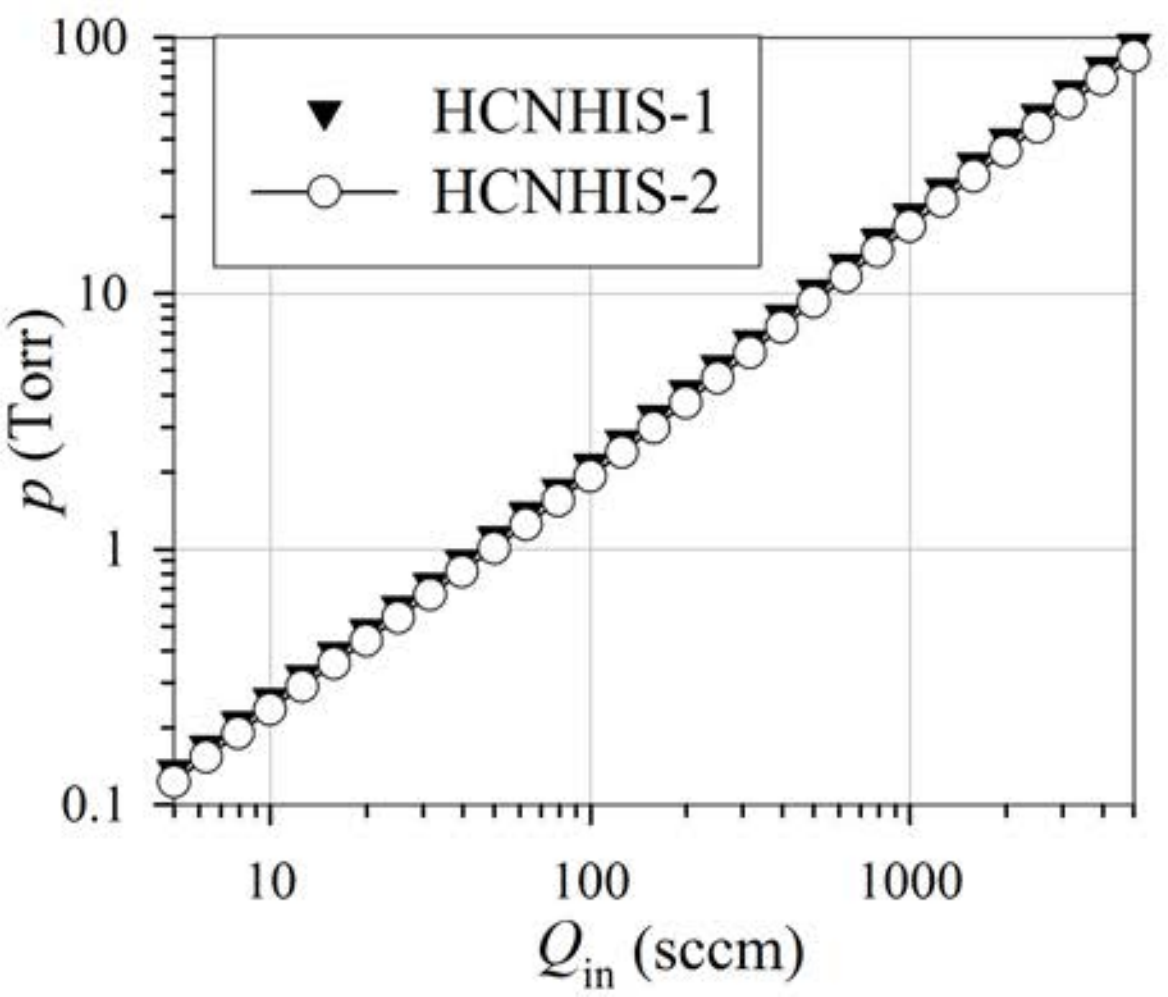

**Figure 26. Pressure in the RFD chamber of the HCNHIS-1 and HCNHIS-2 as a function of hydrogen inlet flow rate at constant absorbed power 341 W.**

#### 4.2.5 GEVKM Simulation of the RFD Chamber: Results and Discussion

A series of GEVKM simulations are performed using input conditions shown in Table 15 with the wall material parameters listed in Table 14 covering the entire regime of operation of the device given in Table 13. These simulations provide the chemical composition, electron and heavy particles temperature as well as the wall temperature in the RFD chamber, outflow fluxes to the NIP Region of the HCNHIS, and investigate the effects of the absorbed power and the inlet flow rate. The HCNHIS-2 is chosen for these simulations because its reduced NIP region allows immediate evaluation of the maximum extractable negative hydrogen ion current and as such can be compared to the experimentally measured current shown in Figure 20. The expanded NIP region in HCNHIS-1 requires modeling of hot rarefied plasma, which is beyond the scope of this work.

**Table 15. Input parameters used in the GEVKM simulations of the HCNHIS-1 and the HCNHIS-2 designs.**

| Parameters | HCNHIS-1 | HCNHIS-2 |
|---|---|---|
| $Q_{\text{in}}(\text{sccm})$ | 300-3000 | 5-5000 |
| $P_{\text{abs}}(W)$ | 430-600 | 200-1000 |

#### 4.2.5.1 Effects of the Inlet Flow Rate: $P_{abs}$=341 W, $Q_{in}$=5-5000 sccm

The effects of the inlet flow rate on the chemical composition and electron and heavy particles temperature are first examined by a simulating the HCNHIS-2 setup at absorbed power 341 W and inlet flow rate 5-5000 sccm corresponding to the operational parameters from Table 13. This regime also covers the operational parameters of conventional low-pressure NHIS. The wide range of inlet flow rates considered allows the establishment of the optimum inlet flow rate at which the production of vibrationally excited molecules and hydrogen anions is maximized.

Figure 27 shows the plasma composition as a function of the RFD chamber pressure at constant absorbed power of 341 W. For reference, the inlet flow rates are shown on the top axis in the subsequent plots with the corresponding RFD pressures on the bottom axis. The dominant positive ion species at considered pressures is $H_3^+$ and its number density is very close to the number density of electrons. The main production channels of the $H_3^+$ are the ion conversion (reaction 17 in Table 4) and collisions of electronically excited atoms with hydrogen molecules (reaction 50 in Table 8). Other positive ions have number densities, which are almost always one order of magnitude smaller. Atomic hydrogen concentration weakly varies with the pressure. The $H^-$ number density increases at low pressures and then slowly decreases with increasing the pressure. From Figure 27 it can be seen that $H^-$ number density has a maximum value around $10^{16}$ $m^{-3}$ at a point where the pressure approximately equals to 1 Torr which corresponds to the inlet flow rate 50 sccm.

In order to explain this trend in the hydrogen anion number density, the production and destruction rates of $H^-$ for different processes in the RFD chamber are plotted in Figure 28. At low pressures corresponding to low hydrogen inlet flow rates the production through DEA to high-lying vibrational states dominates. As the pressure increases the production of $H^-$ from high-lying vibrational states increases, reaches the maximum, and at higher pressures decreases. The explanation of this phenomenon is that at low pressures and consequently high electron temperatures, high-lying vibrationally excited states are effectively populated by EV collisions (reaction 34 from Table 7). At high pressures, on the other hand, the vibrationally excited molecules are effectively quenched by the VTm (reactions 39 and 40 from Table 7) and more importantly by the VTa (reaction 42 from Table 7) processes. It can be seen from Figure 29, which shows the VDF at different pressures in the discharge chamber at a constant absorbed power 341 W. At very low pressures the VDF substantially deviates from the Boltzmann equilibrium distribution and resembles Bray distribution (Fridman, 2008). At high pressures the distribution looks more like an equilibrium one. An important feature of the VDF at low pressures is the existence of the plateau at vibrational levels from 4-10. As it was mentioned in Section 1.2 this plateau was experimentally observed. At very high pressures corresponding to high hydrogen inlet flow rates the main contribution of $H^-$ is through DEA to ground state and the first vibrationally excited state of hydrogen molecules. As for $H^-$ destruction rates which are shown in Figure 28(b), the main chemical reactions are mutual neutralization with $H_3^+$ and collisions with $H$. More interestingly, the destruction rates decrease with increasing the chamber pressure. For $H_3^+$ the reduction of the destruction efficiency is due to the decreasing of its density and increasing of the heavy particles temperature, which according to Figure 3 leads to larger destruction mean free paths. For atomic hydrogen, the situation is slightly different. The destruction mean free path is also increasing with the temperature increase but this effect is partially compensated by the number density increase. Therefore, the destruction rate due to atomic hydrogen is decreasing slower than that of $H_3^+$.

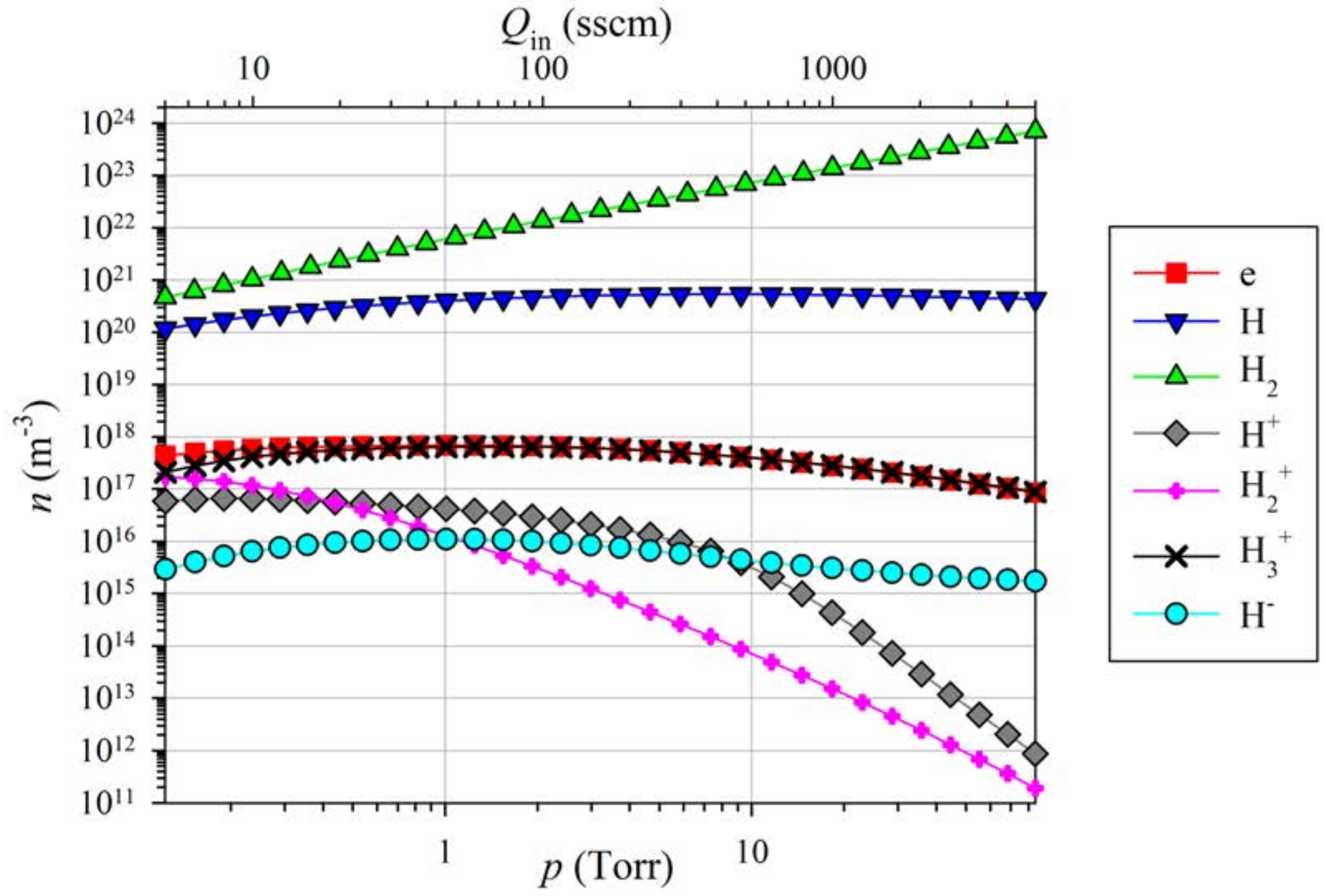

**Figure 27. Plasma composition in the RFD chamber of the HCNHIS-2 as a function of the chamber pressure at fixed absorbed power of 341 W.**

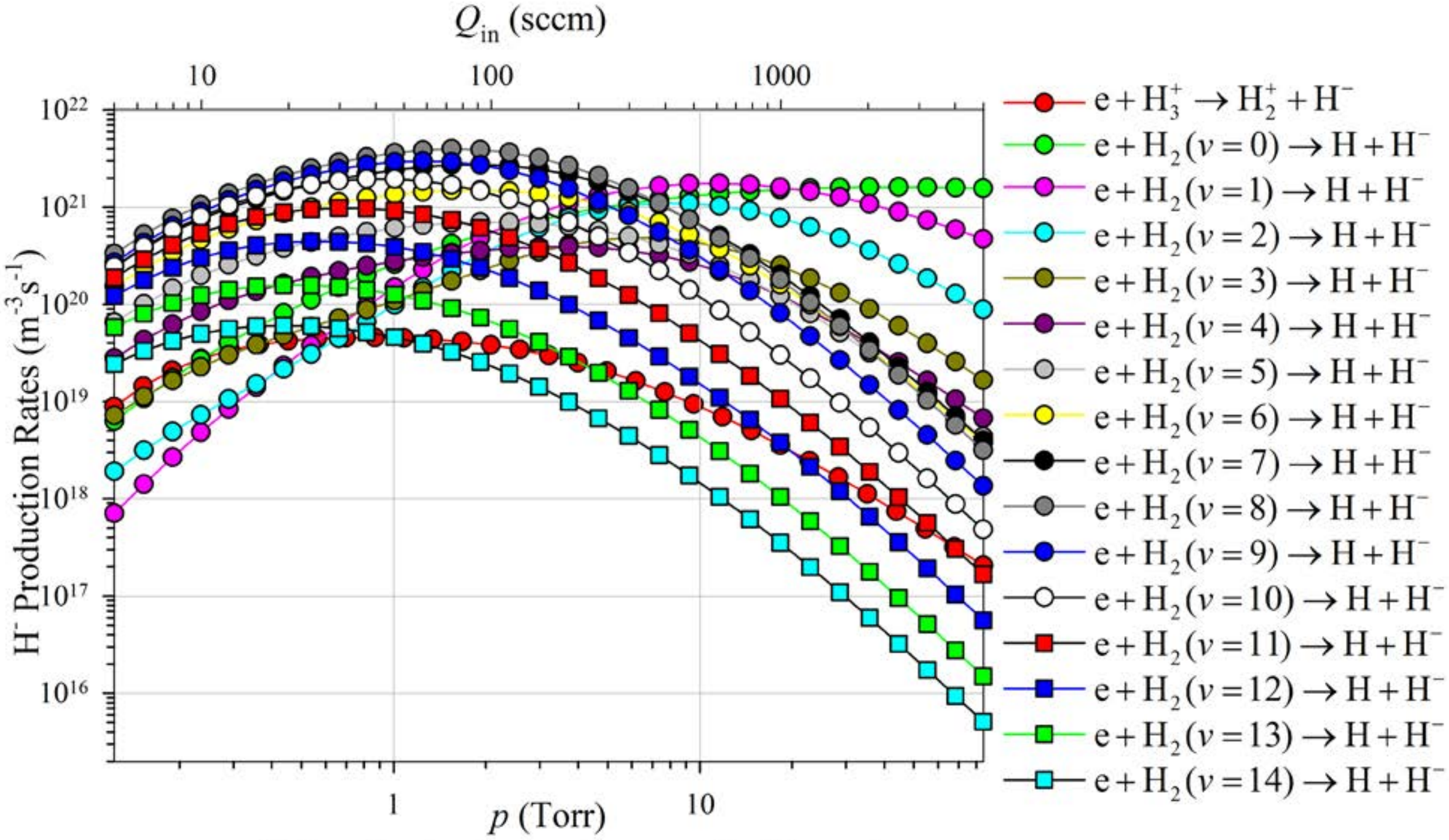

**a) Negative hydrogen ions production rates**

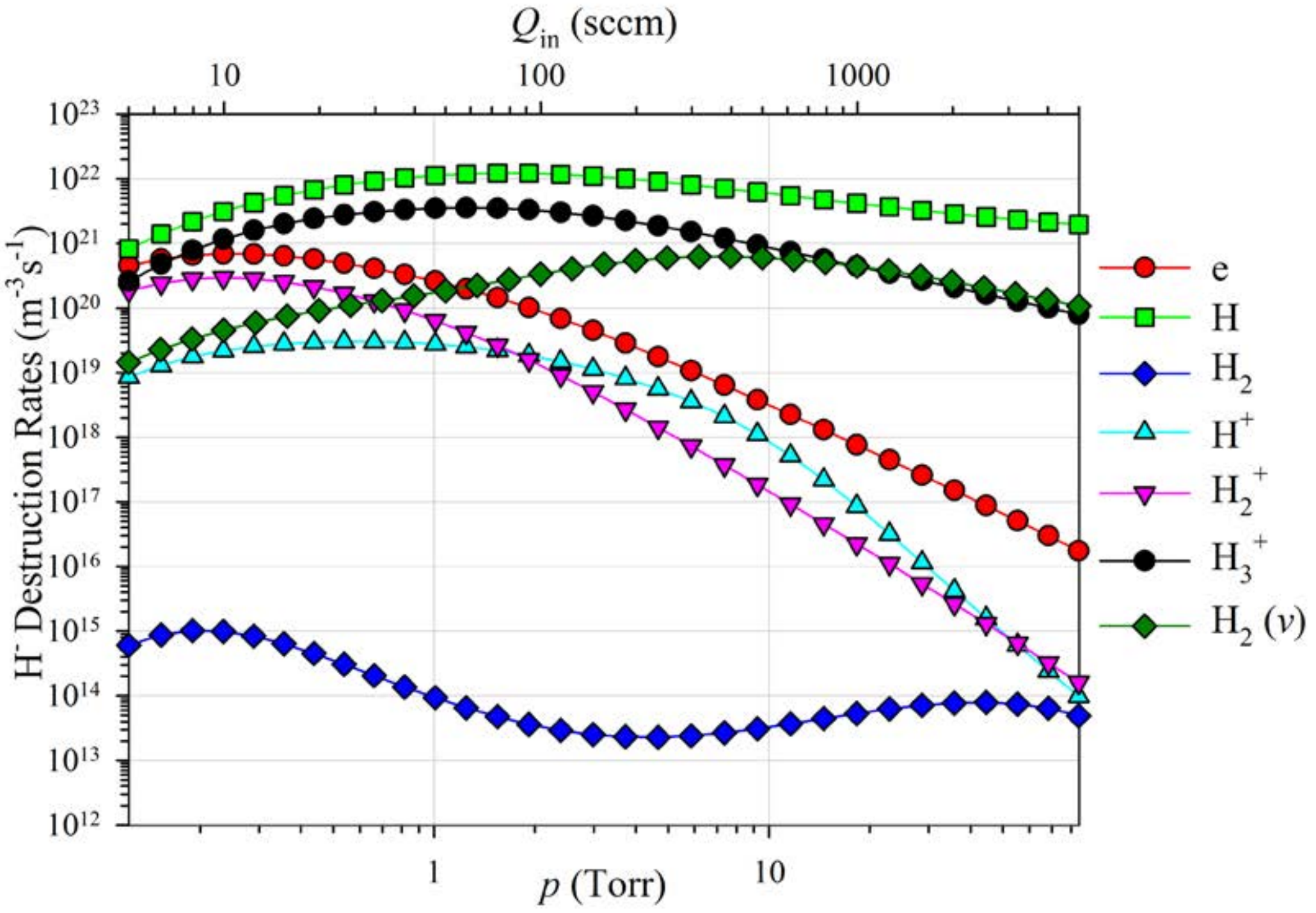

**b) Negative hydrogen ions destruction rates**

**Figure 28. Negative hydrogen ion production (a) and destruction (b) rates in the RFD chamber of the HCNHIS-2 as a function of the chamber pressure at fixed absorbed power 341W.**

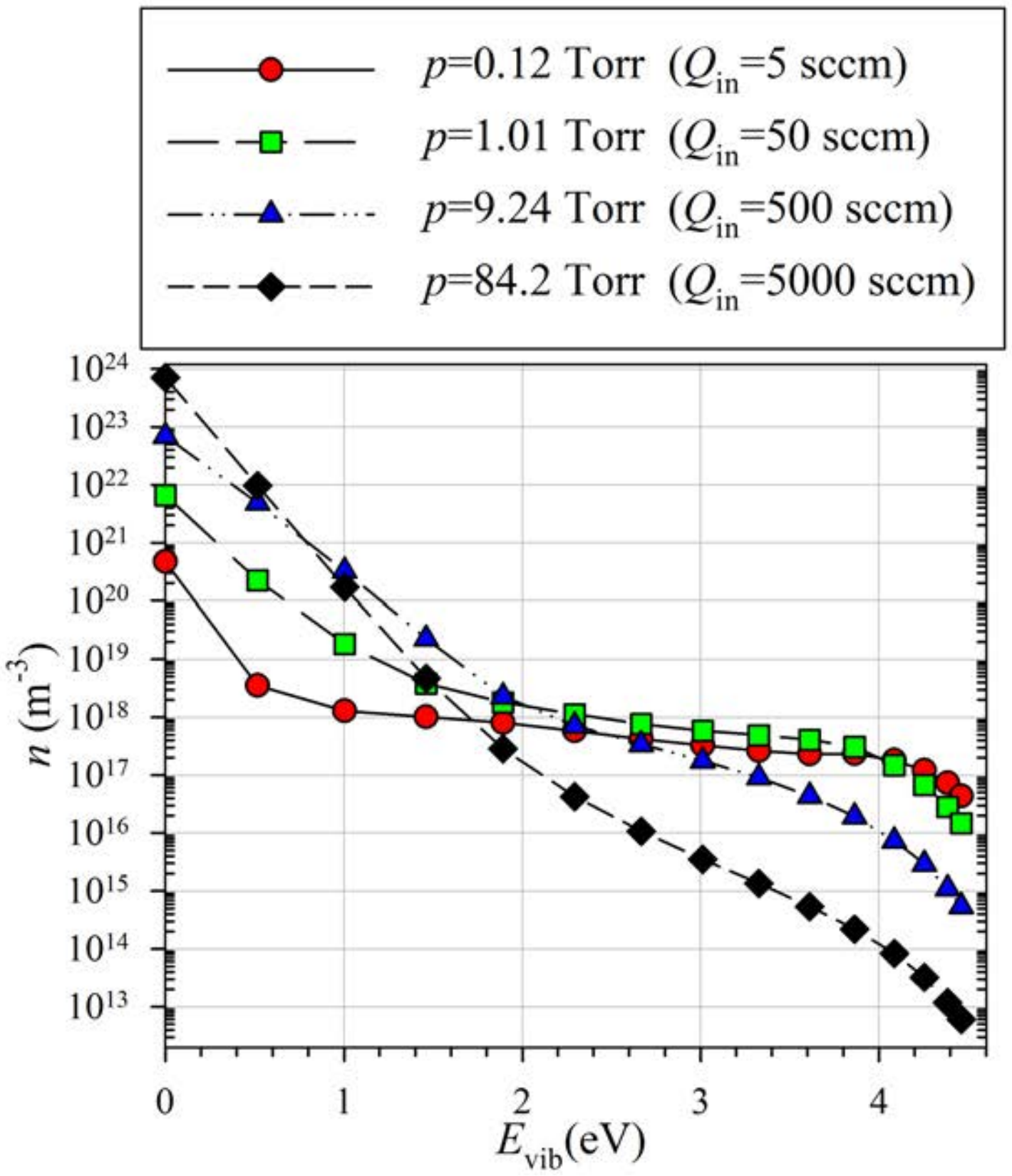

**Figure 29. VDF of molecular hydrogen in the RFD chamber of the HCNHIS-2 at different chamber pressures at constant absorbed power 341 W.**

The electron, heavy particles and wall temperatures as a function of the chamber pressure are shown in Figure 30(a) and Figure 30(b) respectively. The electron temperature decreases as the chamber pressure increases reaching the temperature around 1 eV at the pressures corresponding to the operational conditions of the HCNHIS-2. The heavy-particle temperature has a maximum of 2000 K near the pressure 0.1 Torr while the wall temperature stays in the range 900-975 K. This high values of heavy-particle temperature at low pressures can be explained in the same way as it was done in Sections 3.3.1 and 3.3.2 in the discussion of the verification and validation of the GEVKM. Each electron impact dissociation reaction has the threshold energy around 10 eV for ground state molecules and goes through the electronic excitation of hydrogen molecule. On the other hand, the dissociation energy of hydrogen molecule is only 4.52 eV. Thus two resulting hydrogen atoms gain additional kinetic energy of about 5.48 eV. At high pressures and low degree of dissociation, the atoms effectively transfer this energy to molecules. At low pressures they cannot effectively transfer it to the molecules but directly bring it to the walls.

Therefore, the molecular hydrogen gas is kept relatively cold. These are the hydrogen atoms that have high energy and consequently contribute to the average heavy particles temperature, which is calculated in the GEVKM. In other words, the assumption of the unique temperature for heavy species breaks down at low pressures and for each species distinct temperature should be considered.

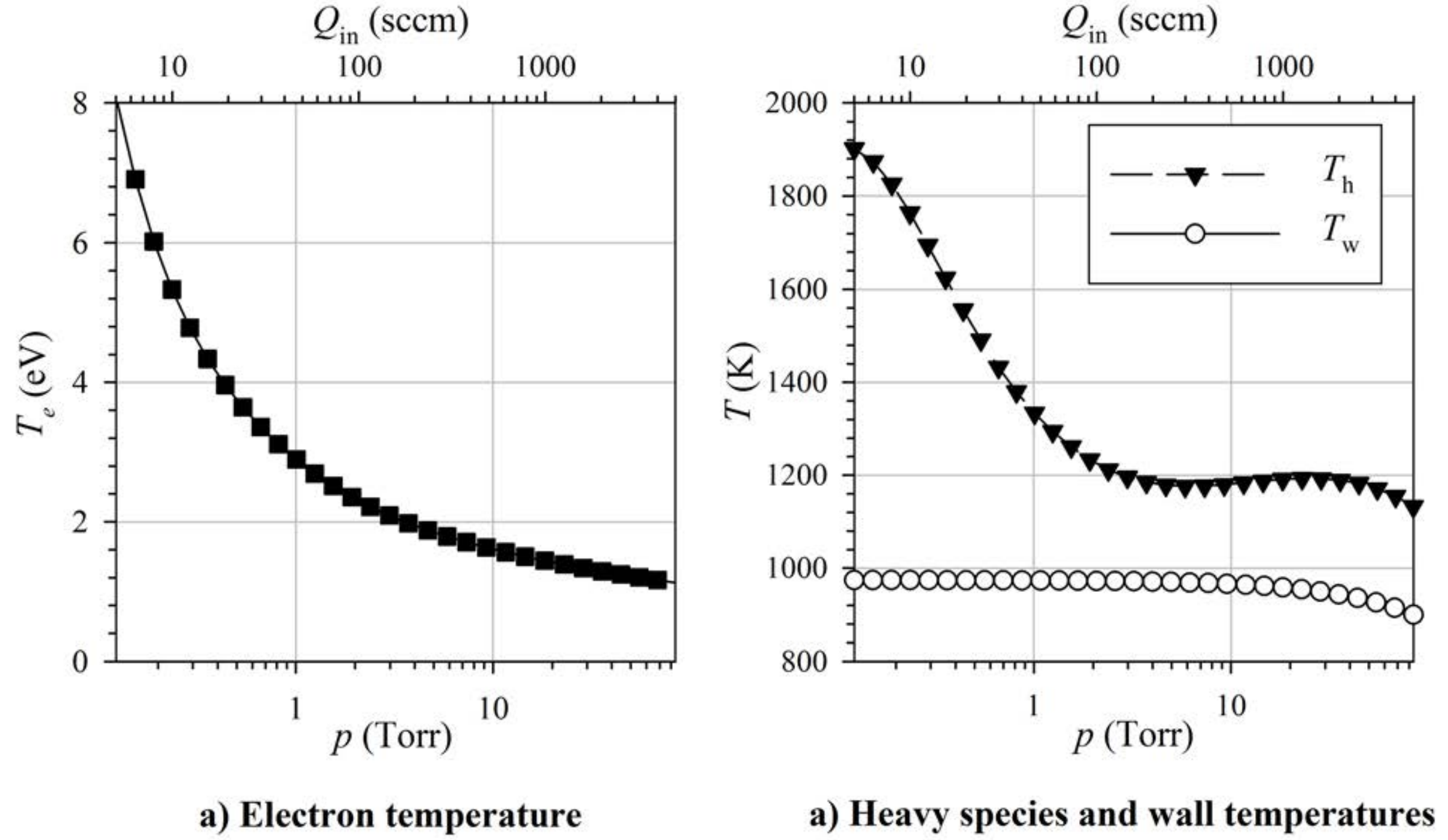

**Figure 30. Electron (a), heavy particles and wall temperatures (b) in the RFD chamber of the HCNHIS-2 as a function of the chamber pressure at fixed absorbed power 341W.**

Different electron losses contributions are shown in Figure 31. At low pressures the electrons lose their energy primarily to the walls to support sheath potential accelerating ion species. In addition, electrons lose their energy for vibrational and electronic excitation and dissociation of hydrogen molecules. At high pressures the main losses are vibrational excitation and dissociation of hydrogen molecules and elastic collisions.

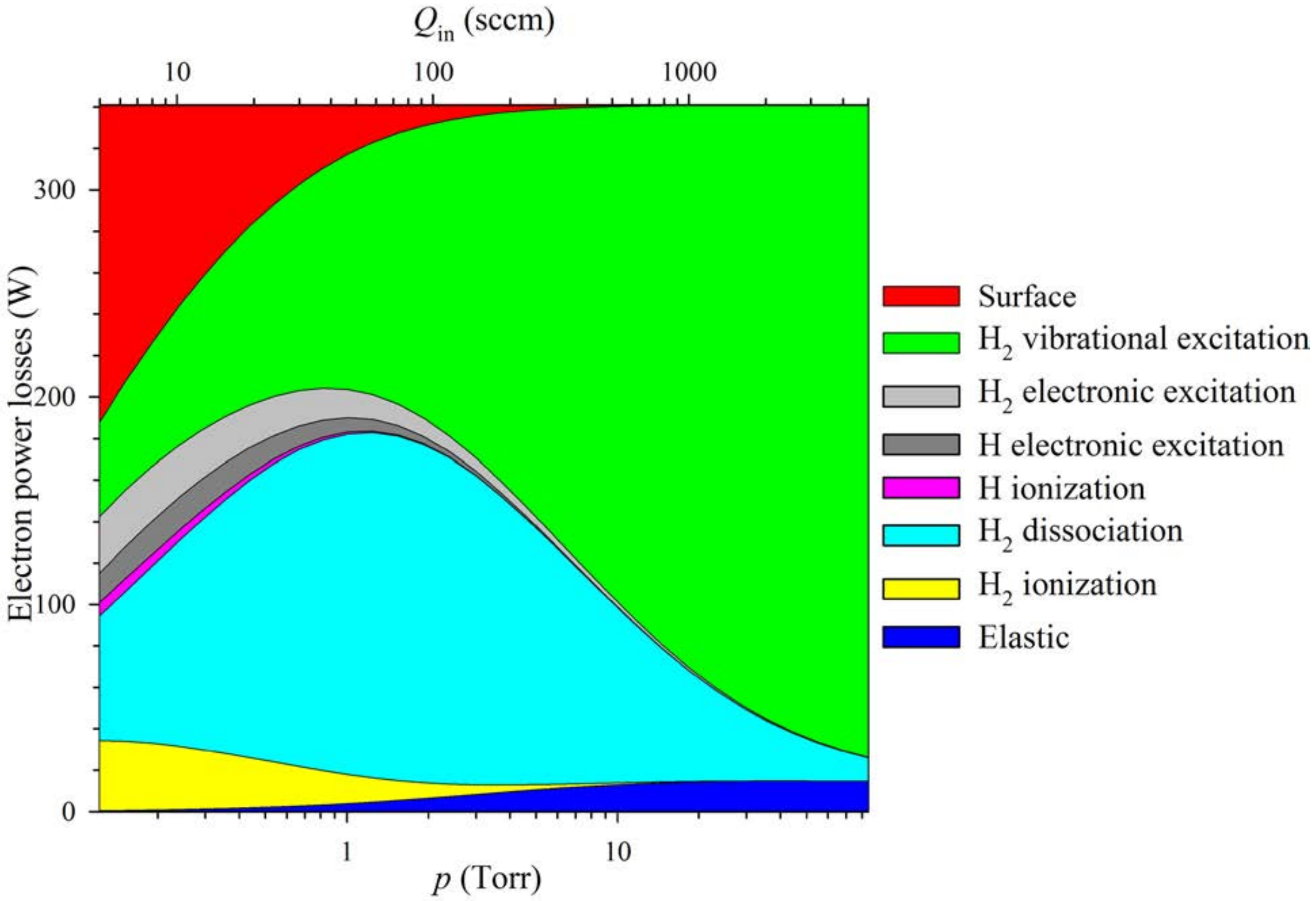

**Figure 31. Electron power losses channels in the RFD chamber of the HCNHIS-2 as a function of the chamber pressure at fixed absorbed power 341 W.**

#### *4.2.5.2 Effects of the Absorbed Power: $P_{abs}$=100-1000 W, $Q_{in}$=1000 sccm*

In order to investigate the effects of the absorbed power a series of simulations are performed at a fixed inlet flow rate of 1000 sccm and absorbed power in the range of 200-1000 W covering the regime of operation of HCNHIS-2.

In Figure 32 the plasma composition is shown as a function of absorbed power at constant inlet flow rate 1000 sccm. Even though the flow rate is kept constant the chamber pressure changes as the absorbed power changes due to the gas heating effect. The number density of the plasma components varies slightly with the change of absorbed power considered. At these pressures and absorbed powers the dominant positive ion species is $H_3^+$. Its number density increases as absorbed power increases. The $H^-$ and $H$ number densities also increase as the absorbed power increases.

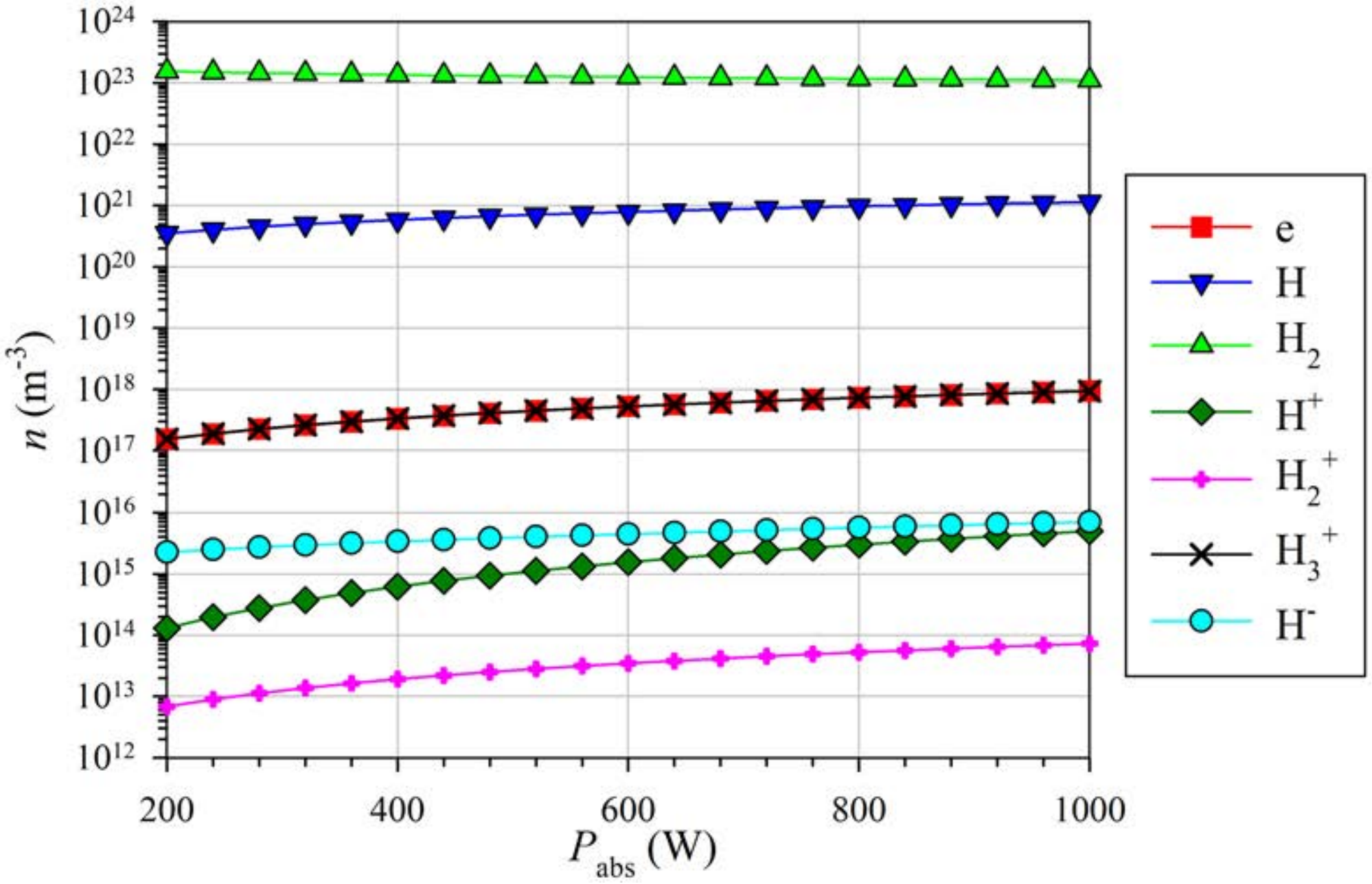

**Figure 32. Plasma composition in the RFD chamber of the HCNHIS-2 as a function of the absorbed power at fixed inlet flow rate of 1000 sccm.**

Figure 33 depicts the production and destruction rates as a function of absorbed power at fixed hydrogen inlet flow rate. Since hydrogen flow rate corresponds to pressures 18-21 Torr depending on the absorbed power the main production mechanism is DEA to ground and low-lying vibrational states of hydrogen molecules. As it was pointed out earlier, it indicates quenching of high-lying vibrational states by VTa and VTm reactions due to rather high chamber pressure caused by initially high inlet flow rate. The production rates increase monotonically with increasing absorbed power. The main $H^-$ destruction mechanism as shown in Figure 33 is the electron detachment in collisions with hydrogen atoms. The destruction rate of this reaction is almost one order of magnitude higher than the mutual neutralization of $H^-$ and $H_3^+$ and electron detachment in collisions of negative hydrogen ions with vibrationally excited hydrogen molecules. Other destruction processes contribute less than 0.1% to the $H^-$ destruction. As absorbed power increases all destruction reaction rates monotonically increase.

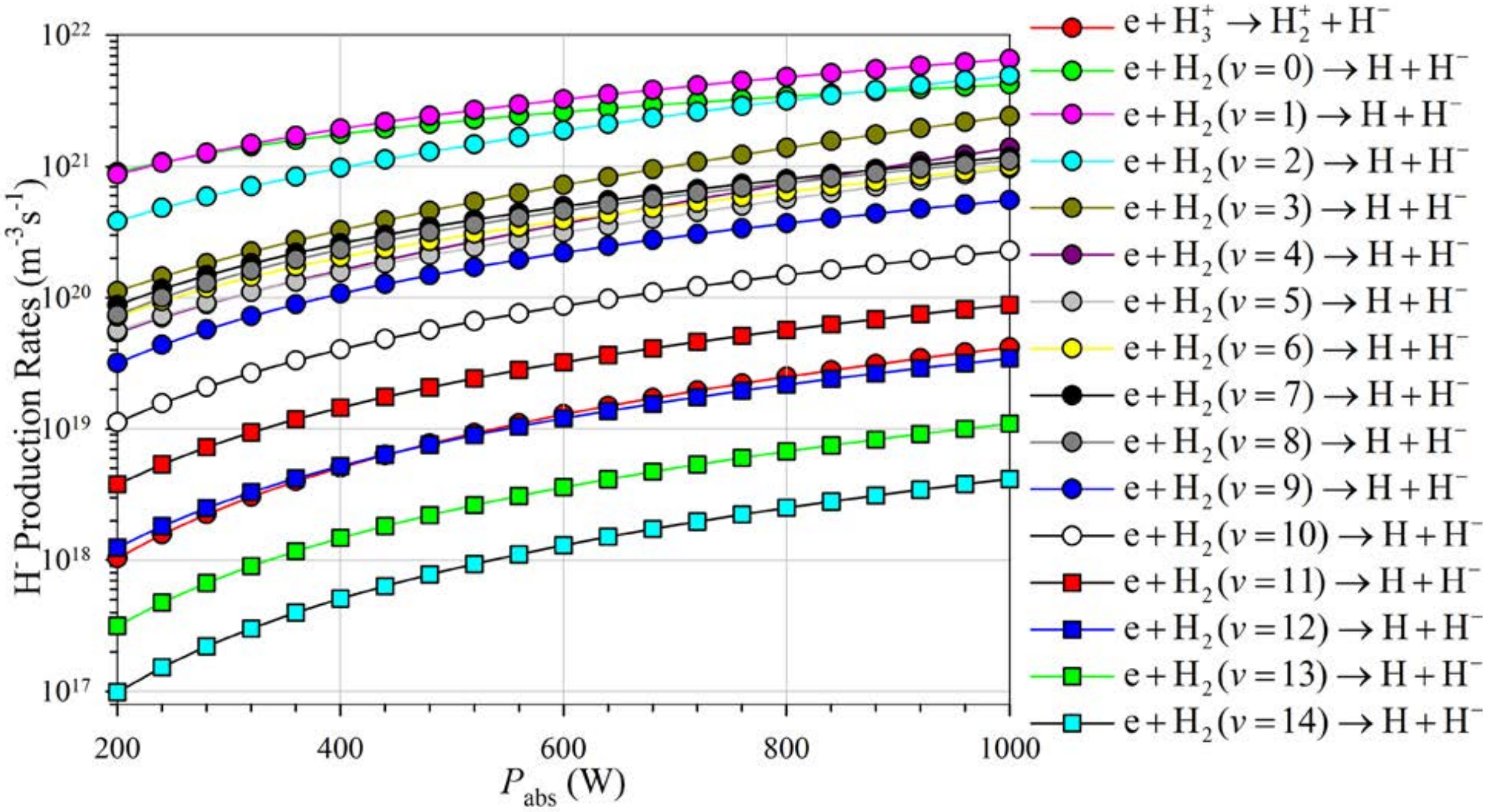

**a) Negative hydrogen ions production rates**

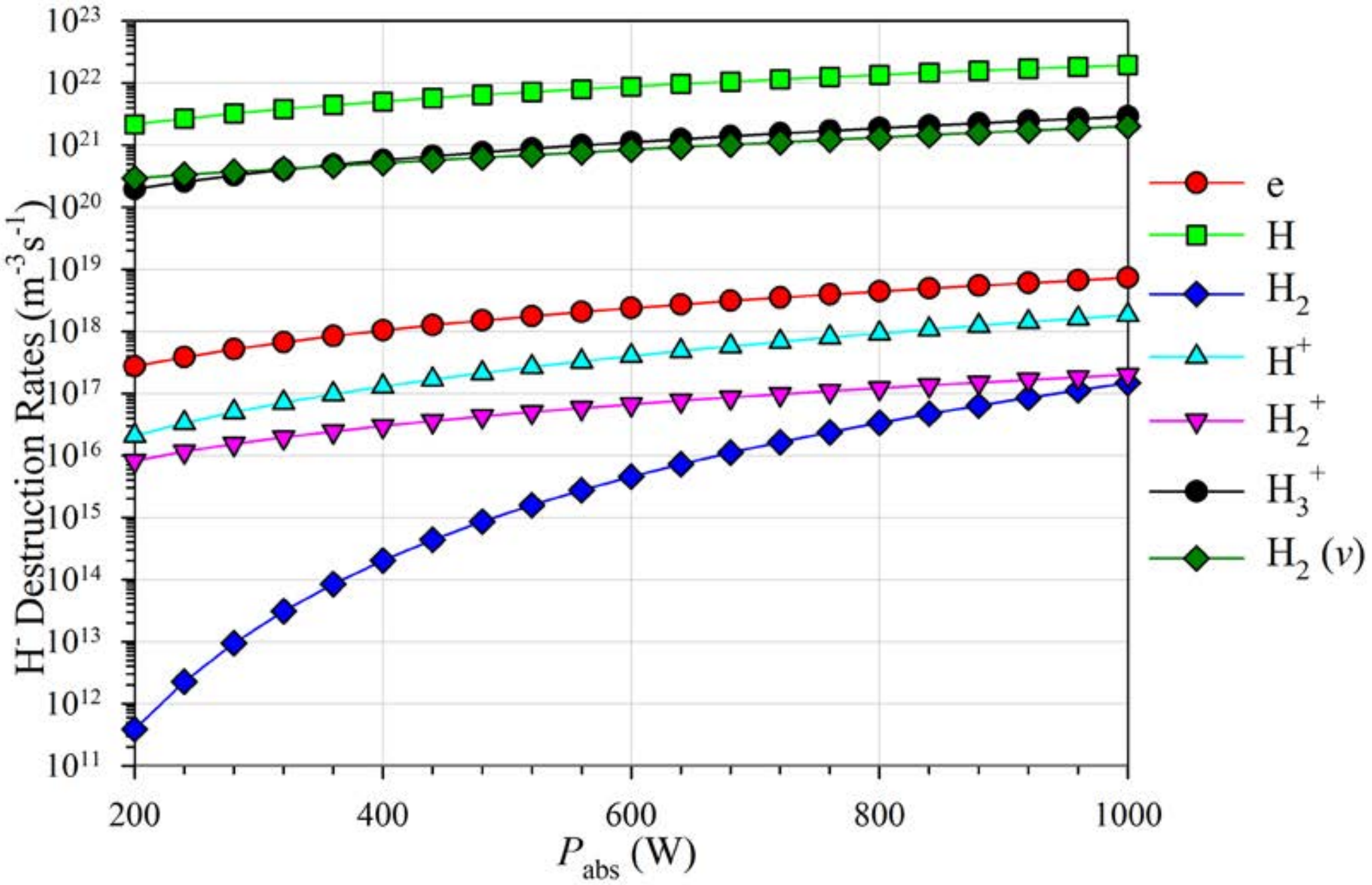

**b) Negative hydrogen ions destruction rates**

**Figure 33. Negative hydrogen ion production (a) and destruction (b) rates in the RFD chamber of the HCNHIS-2 as a function of the absorbed power at fixed hydrogen inlet flow rate of 1000 sccm.**

Electron, heavy particles and wall temperature as a function of absorbed power are shown in Figure 34. As absorbed power increases all temperatures increase monotonically. The electron power losses distribution normalized by the absorbed power is shown in Figure 35. At these moderate pressures (18-21 Torr) the main losses are vibrational excitation and dissociation of molecular hydrogen as well as elastic collisions. Increase in the absorbed power only slightly enhances the electron power losses to the dissociation by decrease of the power losses resulting in the vibrational excitation. On the other hand, the VDF shown in Figure 36 has the same shape at all considered absorbed powers. The increase in the absorbed power results in the slight increase in the population of the high-lying vibrational states. However, this increase is rather small compared to the effect of the inlet flow rate variation as it was discussed earlier. The main reason for such small impact of the absorbed power is due to rather high chamber pressure. VT and VV processes effectively redistribute vibrational energy at these conditions leading to the gas heating and vibrational equilibration rather than vibrational excitation of high-lying levels.

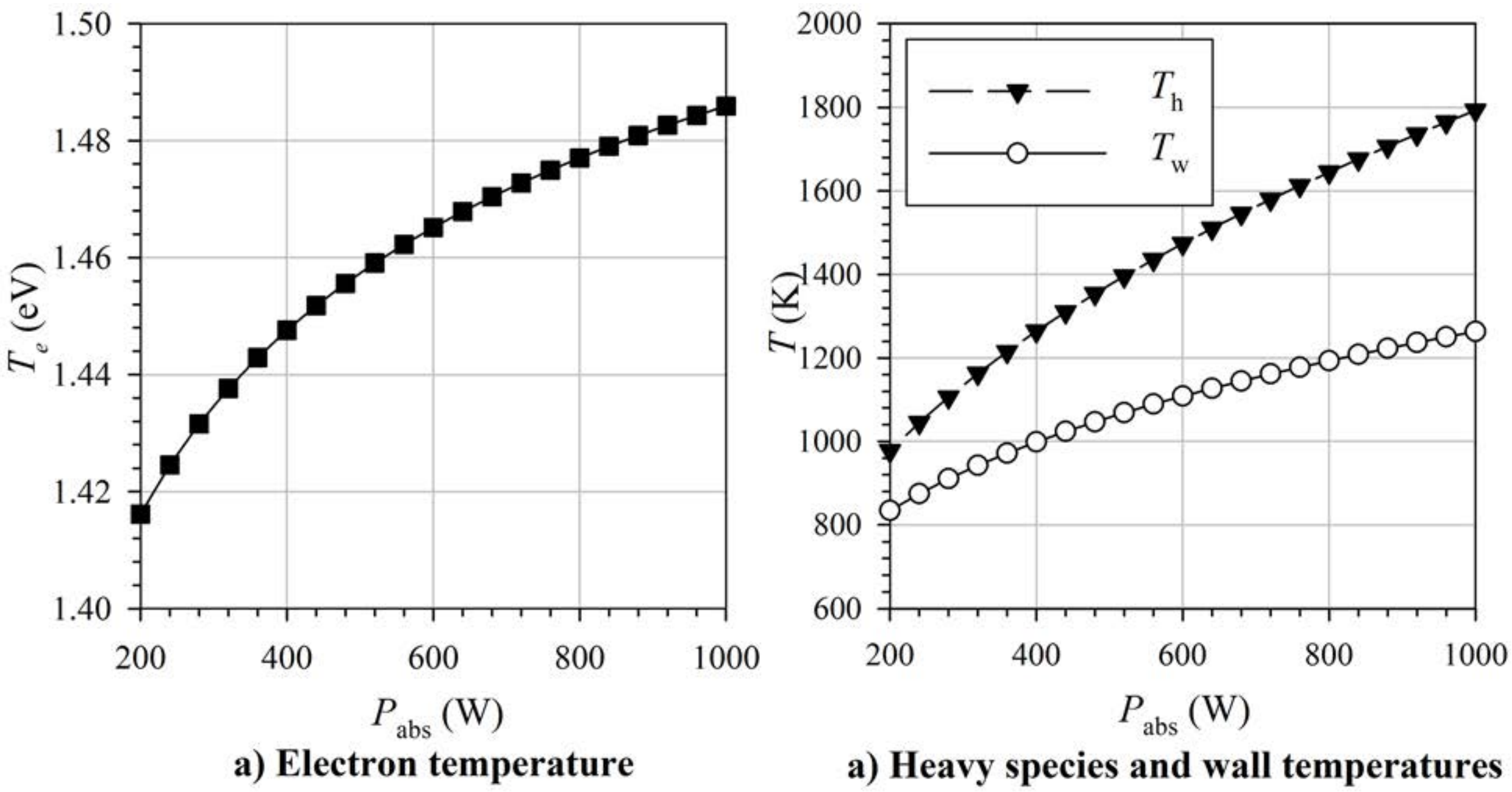

**Figure 34. Electron (a), heavy-particle and wall (b) temperatures in the RFD chamber of the HCNHIS-2 as a function of the absorbed power at fixed inlet flow rate 1000 sccm.**

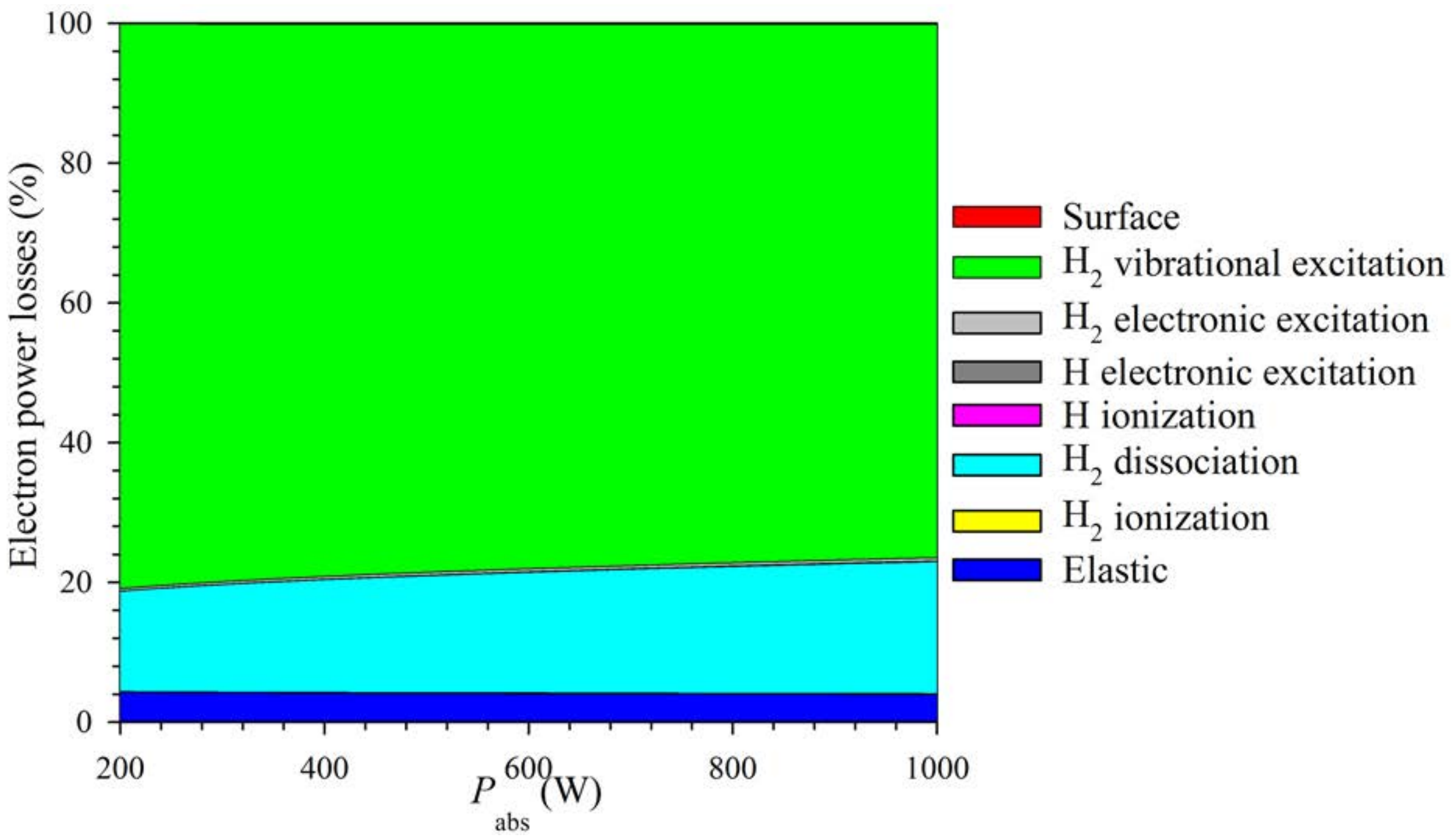

**Figure 35. Electron power losses channels in the RFD chamber of the HCNHIS-2 as a function of the absorbed power at fixed inlet flow rate 1000 sccm.**

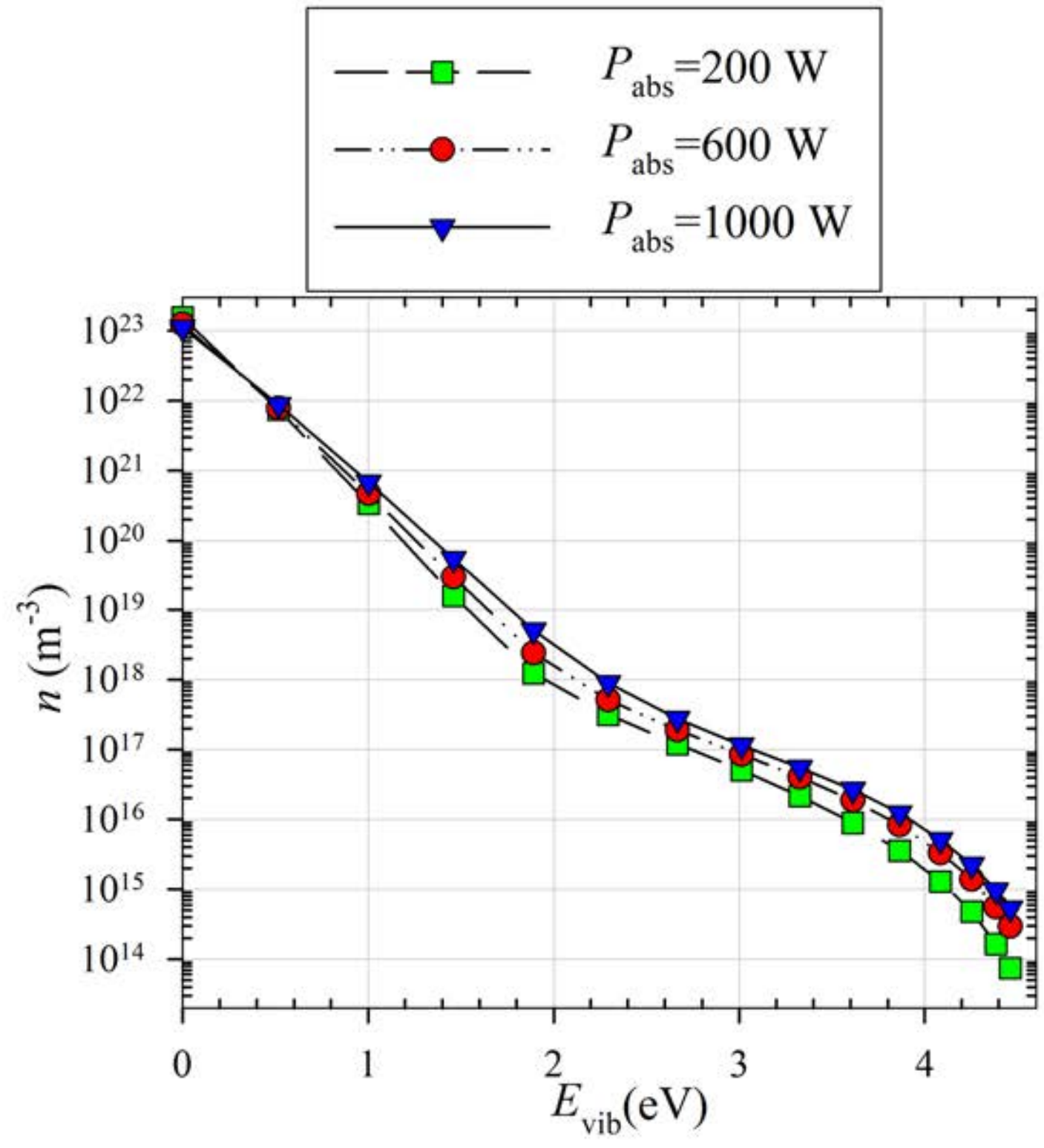

**Figure 36. VDF of molecular hydrogen in the RFD chamber of the HCNHIS-2 at different absorbed powers and constant inlet flow rate 1000 sccm.**

### 4.2.5.3 *Combined Effects of the Absorbed Power and Inlet Flow Rates: $P_{abs}$=200-1000 W, $Q_{in}$=5-1000 sccm*

Due to high computational efficiency of the GEVKM, it is possible to run multidimensional parametric studies. Figure 37 contains $H^-$ number density in the RFD chamber of the HCNHIS-2 setup at different absorbed powers and chamber pressures corresponding to hydrogen inlet flow rates 5-1000 sccm and absorbed powers 200-1000 W. This plot composed of around 1000 individual GEVKM simulations each took less than a couple seconds thanks to the computational efficiency. It can be seen that there is a narrow peak in the negative hydrogen ion number density at pressures about 1 Torr at different absorbed powers. As the absorbed power increases, the negative hydrogen ion number density also increases. This peak corresponds to the optimum parameters for negative hydrogen ion production. However, it does not guarantee the optimum negative hydrogen ion current because the hydrogen anions can be very effectively destroyed by hydrogen atoms and positive ions during the extraction.

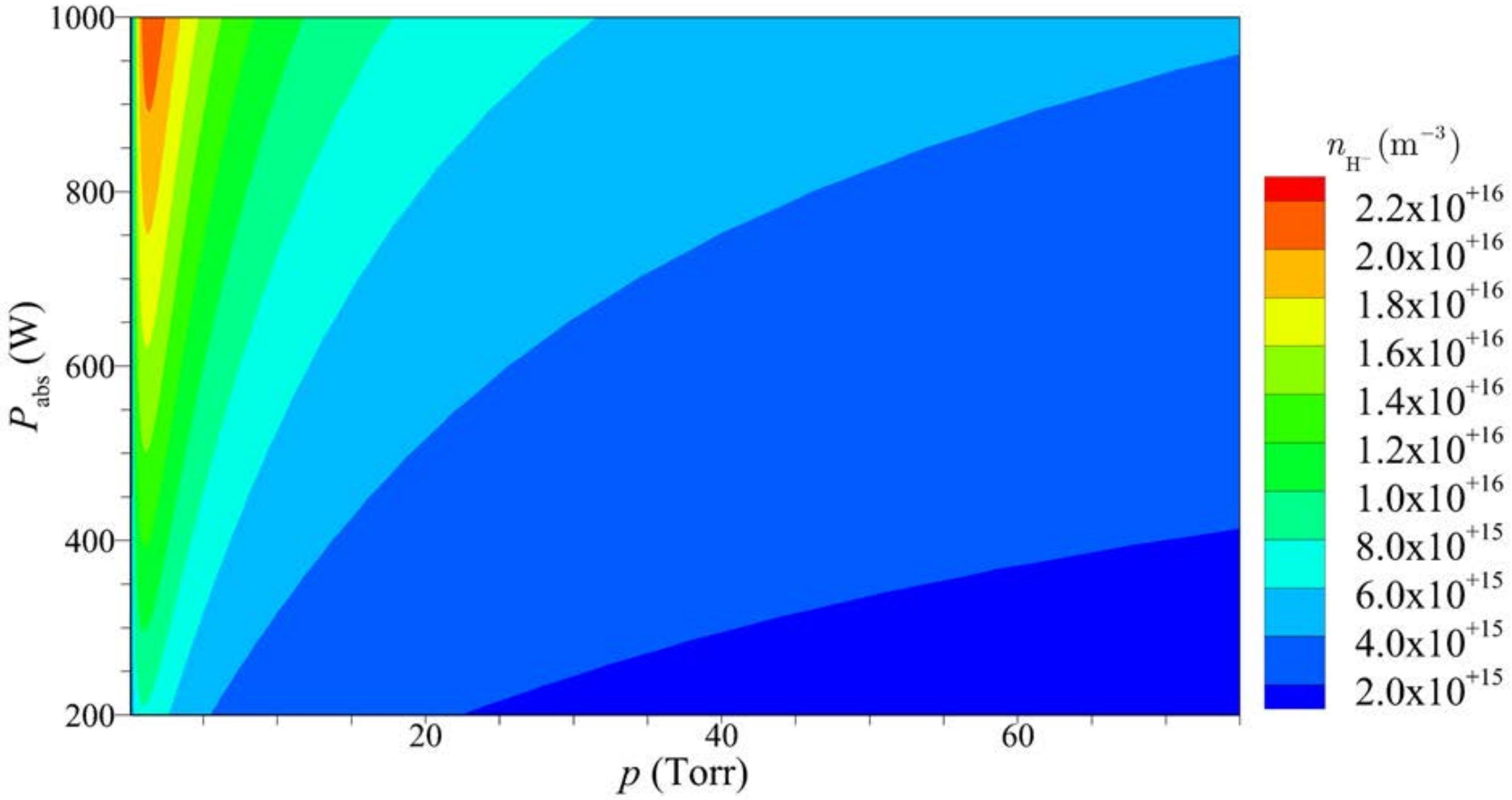

**Figure 37. Negative hydrogen ion number density in the RFD chamber of the HCNHIS-2 as a function of the chamber pressure and absorbed power.**

In order to better understand the physics of this peak Figure 38 shows average hydrogen vibrational energy per absorbed power. At fixed absorbed power the average hydrogen vibrational energy per absorbed power has a local minimum at pressure 1 Torr (50 sccm of the inlet flow rate) which corresponds to the maximum number density of $H^-$. It means that the DEA mechanism at this particular geometry works most effectively at this flow rate. On the other hand, there are two local maxima of the average vibrational energy per absorbed watt one of them is at very low pressures below 0.1 Torr and the other is located at about 5 Torr. For the design of the HCNHIS the maximum located at 5 Torr corresponds to the optimum operation of the RFD chamber since it yields the maximum vibrational excitation required for the negative hydrogen ion production. The maximum located at pressures below 0.1 Torr corresponds to the optimum parameters for vibrational excitation in conventional NHISs working based on the volume production principle such as magnetically filtered multicusp volume negative ion sources discussed in Section 1.3.

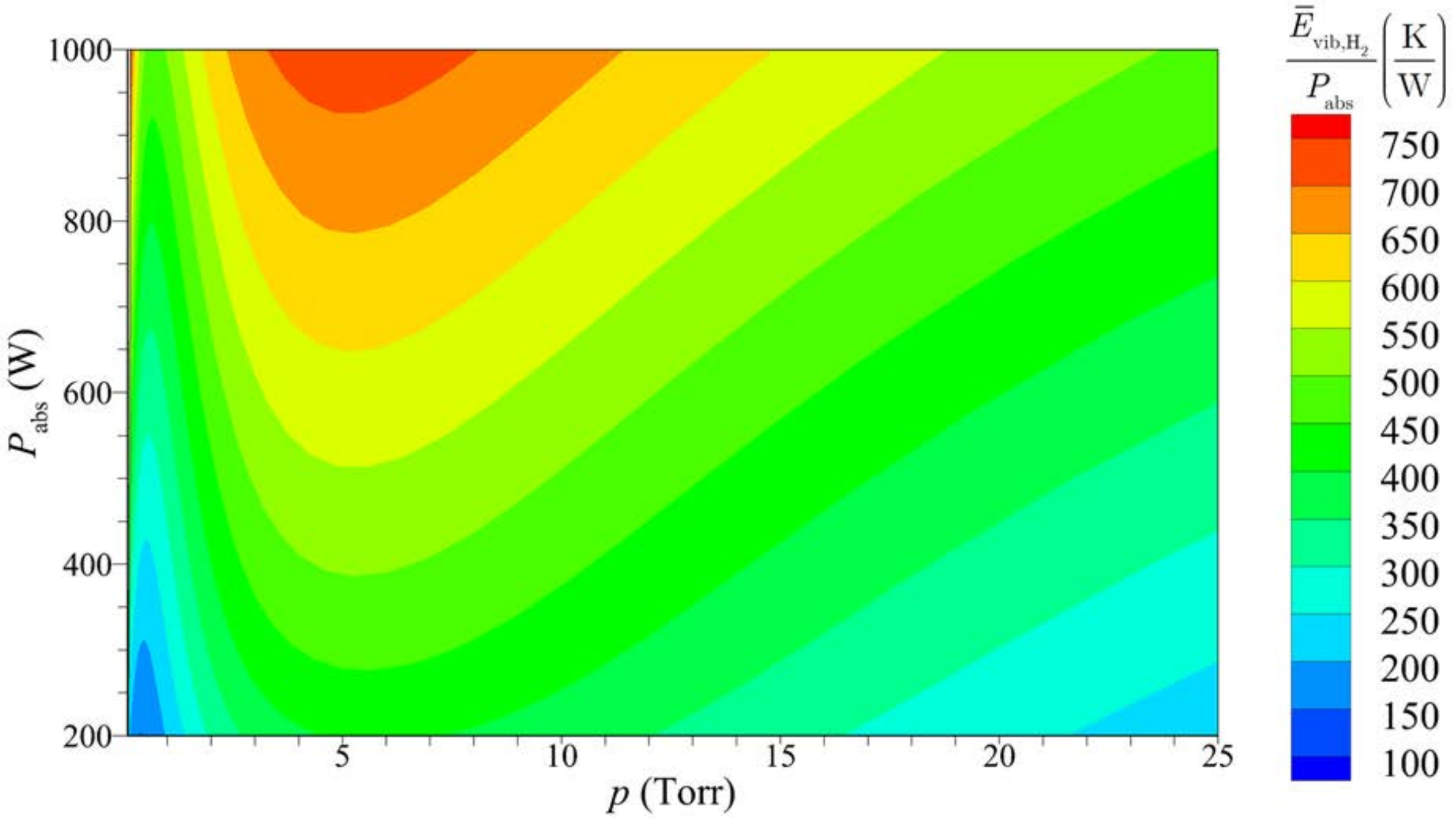

**Figure 38. Average hydrogen vibrational energy per the absorbed power in the RFD chamber of the HCNHIS-2 as a function of the chamber pressure and absorbed power.**

### 4.2.6 Comparisons of the Maximum Extractable Negative Hydrogen Ion Current Calculated by the GEVKM with the Faraday Cup Measurements

The HCNHIS-2 contains a short NIP region, which is used in conjunction with the extraction grids in the Faraday cup experiments as shown in Figure 19. Due to very small sizes of the NIP region it is unlikely that volumetric processes leading to production or destruction of negative ions take place there. In addition, as it was outlined in Section 4.2.2 the relative contribution of the outlet fluxes in the continuity equations for charged species is much lower than other volumetric and surface processes. Therefore, in order to estimate maximum extractable negative hydrogen ion current it is assumed that the negative hydrogen ion number density is the same in the NIP region as in the RFD chamber. This assumption can be justified by the fact that negative ions are repelled from negatively charged walls and have large diffusion coefficient.

With the calculated negative hydrogen ion density the negative hydrogen ion current can be estimated from the Bohm flux to the extraction grid as (Brown, 2004)

$$I_{\mathrm{H}^-} = -en_{\mathrm{H}^-} u_{\mathrm{B,H}^-} A_{\mathrm{aperture}}, \tag{4.11}$$

where $A_{\mathrm{aperture}}$ is the extraction aperture area and $u_{\mathrm{B,H}^-} = \sqrt{k_B T_e / m_{\mathrm{H}^-}}$ is the Bohm speed of $\mathrm{H}^-$.

Equation (4.11) assumes that the current is emission limited and, therefore, gives the maximum extractable negative hydrogen ion current. If the current is space-charge limited then it can be calculated from the Child-Langmuir law as (Brown, 2004)

$$I_{CL} = \frac{4}{9}\epsilon_0 \sqrt{\frac{2e}{m_{\mathrm{H}^-}}} \frac{A_{\mathrm{aperture}}}{d^2} V^{3/2}, \tag{4.12}$$

where $V$ is the potential drop between plasma and grounded electrode, $d$ is the distance between two electrodes.

Figure 39 shows the maximum extractable negative hydrogen ion current calculated by the GEVKM using Eq. (4.11) as a function of absorbed power at fixed hydrogen inlet flow rate 1000 sccm corresponding to the experimental conditions at which the negative hydrogen ion current was measured. The experimentally measured current is very close to the current predicted by the GEVKM.

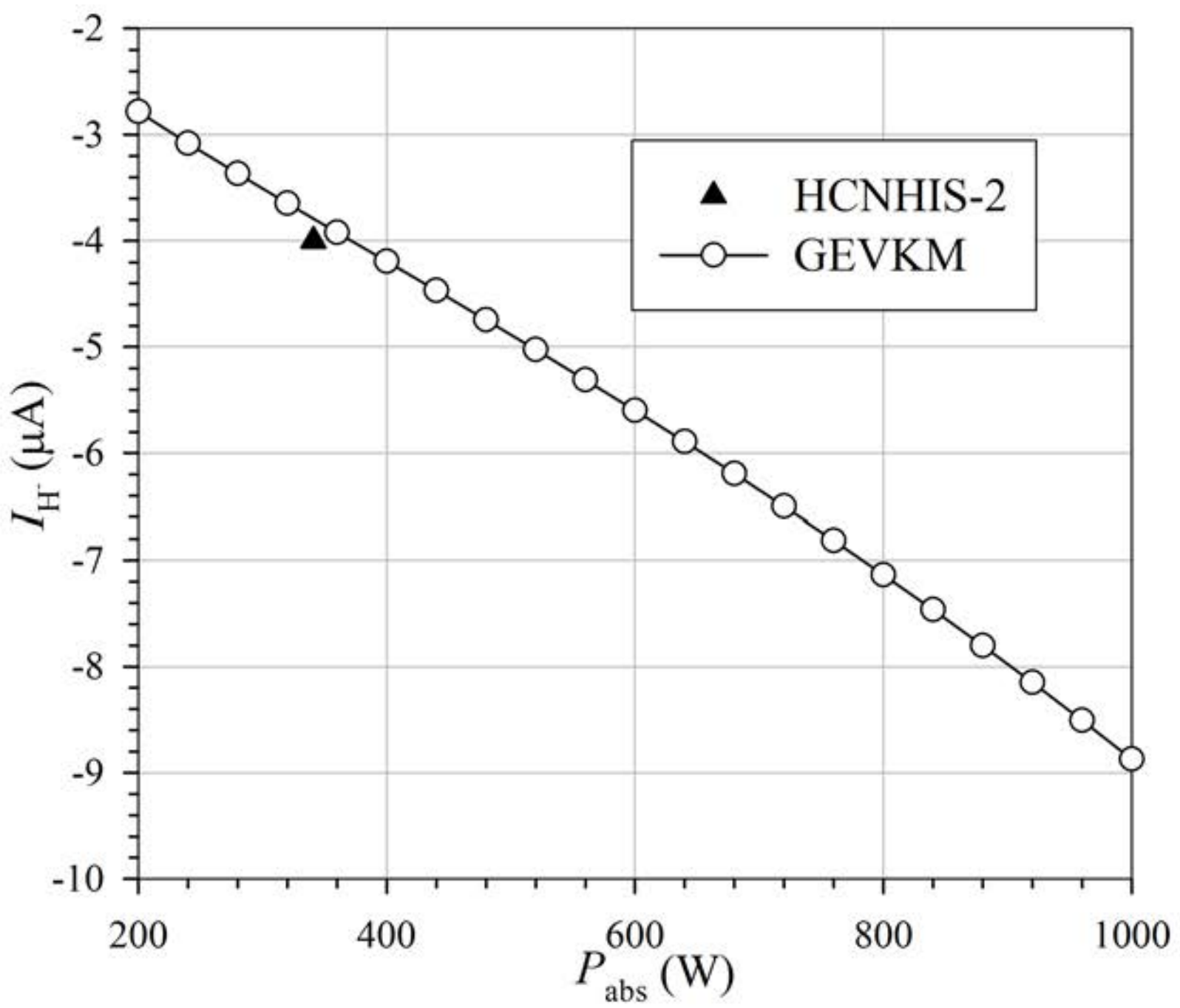

**Figure 39. Negative hydrogen ion current extracted from the HCNHIS-2 setup as a function of the absorbed power at fixed hydrogen inlet flow rate 1000 sccm.**

Figure 40 shows the maximum extractable negative hydrogen ion current calculated by the GEVKM as a function of absorbed power and chamber pressure. At constant pressures the current monotonically increases as the absorbed power increases. Similar to negative hydrogen ion number density there is a narrow peak of $H^-$ current at low pressures and fixed absorbed power. However, this peak current is not at 1 Torr as one might expected from $H^-$ number density plotted in Figure 37 but shifted to the pressure 0.9 Torr. The explanation of this phenomenon is that the emission limited extracted current given by Eq. (4.11) depends also on the electron temperature. The electron temperature in the RFD chamber of the HCNHIS-2 setup as a function of absorbed power and chamber pressure is shown in Figure 41. As the pressure in the discharge chamber is decreased, the electron temperature is monotonically increased. Therefore, the negative hydrogen ion current is shifted to lower pressures.

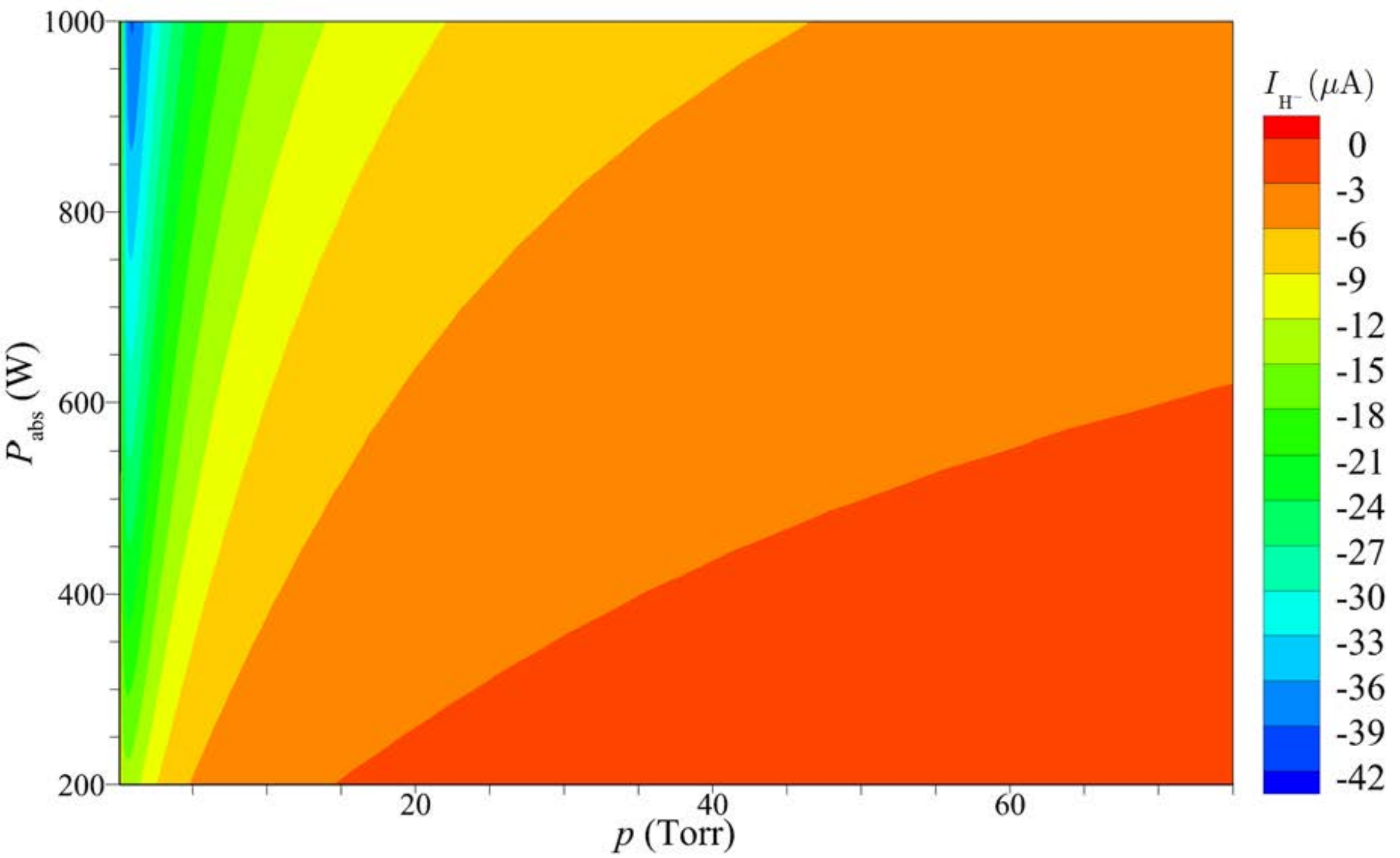

**Figure 40. Negative hydrogen ion current extracted from the HCNHIS-2 setup as a function of the absorbed power and chamber pressure.**

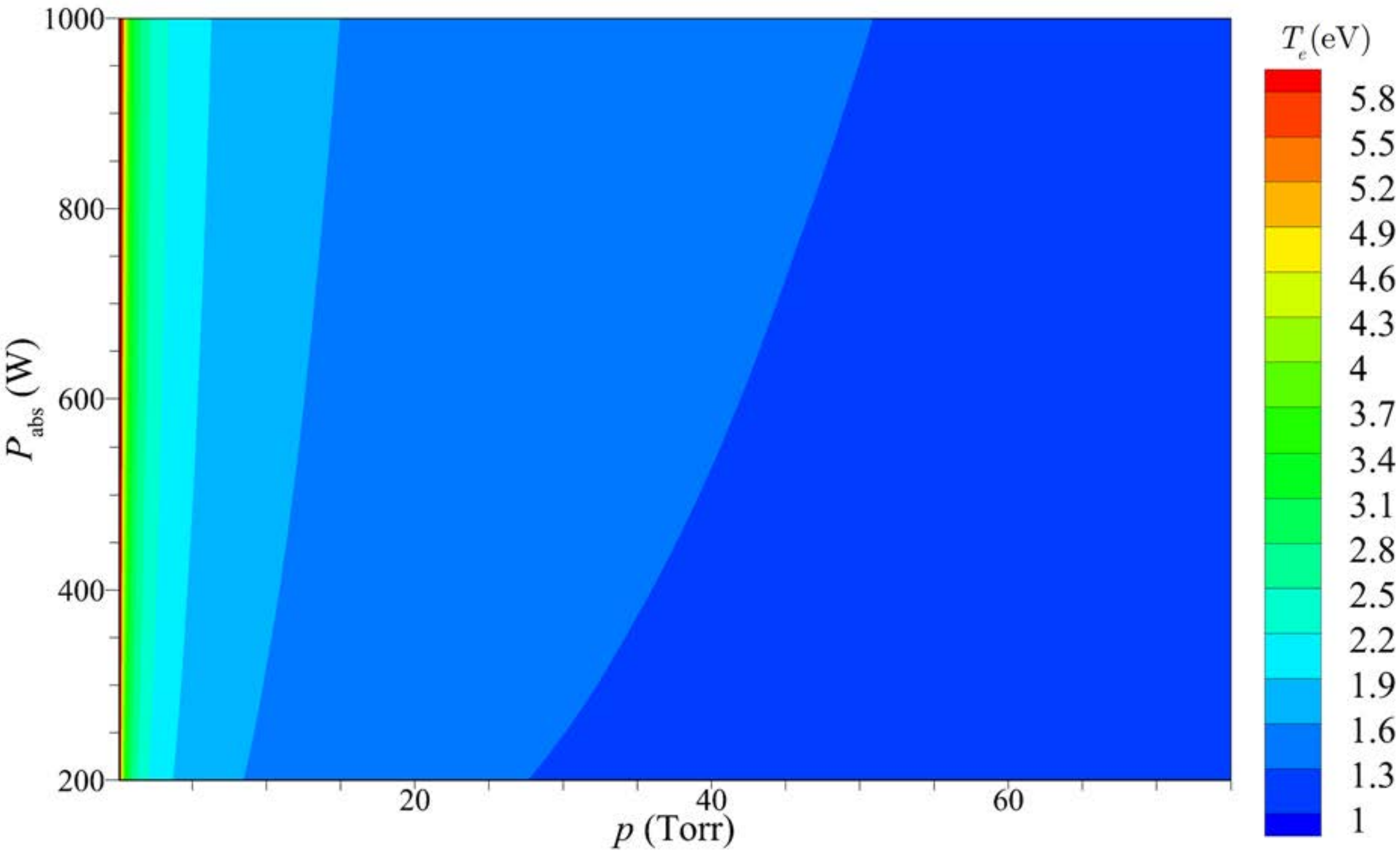

**Figure 41. Electron temperature in the RFD chamber of the HCNHIS-2 as a function of the absorbed power and chamber pressure.**

# 5 Summary, Conclusions and Recommendations

## 5.1 Summary and Conclusions

This work was devoted to modeling of a new High Current Negative Hydrogen Ion Source developed by Busek Co. Inc. and Worcester Polytechnic Institute. The HCNHIS consists of the high-pressure radio-frequency discharge chamber and low-pressure negative hydrogen ion production region. The hydrogen molecules are vibrationally excited in the high-pressure chamber. Then the part of the plasma goes through the nozzle from the RFD chamber into the NIP region. In this chamber more $H^-$ ions are produced by dissociative attachment of low energy electrons to rovibrationally excited hydrogen molecules. Finally, the negative hydrogen ions and electrons exit through the extraction grid assembly. The goal of this dissertation was to develop a comprehensive theoretical and computational model of the chemically reacting plasmadynamics processes in the RFD chamber of the HCNHIS.

The Global Enhanced Vibrational Kinetic (GEVKM) model was developed for the RFD chamber of the HCNHIS. It is based on the state-to-state moment equations for multi-temperature chemically reacting plasmas derived from the Wang Chang-Uhlenbeck equations. The transport properties, transport collision integrals and boundary conditions for the state-to-state moment equations for multi-temperature chemically reacting plasmas were also reviewed in this work. The GEVKM was developed for a cylindrical geometry of an inductively coupled discharge chamber. The species included in the model are ground state hydrogen atoms $\mathrm{H}$ and molecules $\mathrm{H}_2$, 14 vibrationally excited hydrogen molecules $\mathrm{H}_2(v)$, $v = 1 - 14$, electronically excited hydrogen atoms $\mathrm{H}(2)$, $\mathrm{H}(3)$, ground state positive ions $\mathrm{H}^+$, $\mathrm{H}_2^+$, $\mathrm{H}_3^+$, ground state anions $\mathrm{H}^-$, and electrons $\mathrm{e}$. Electrons and translational-rotational degrees of freedom of heavy particles (ions, atoms and molecules) are assumed to obey equilibrium distribution functions. The heavy particles are assumed to have the same temperature which is different from the electron temperature. The power deposition is assumed to be primarily due to Joule heating of electrons by RF electric field while the stochastic heating is disregarded. The species temperature in the GEVKM is considered to be uniform in the plasma reactor. The spatial variation of the number densities of the plasma components is assumed to follow the product of two one-dimensional dis-tributions corresponding to infinite long cylinder and two infinite plates. Heuristic expressions derived from the exact and numerical solutions of the momentum and continuity equations covering low to high pressure regimes are used in order to link together wall, center and average number densities of plasma components. The volume-averaged steady-state continuity equations and quasi-neutrality in the bulk coupled with the electron and total energy equations and the heat transfer to the chamber walls are solved simultaneously in order to obtain volume-averaged number densities of plasma components as well as electron and heavy-particle temperatures and the wall temperatures. The model contains comprehensive set of surface and volumetric processes governing vibrational and ionization kinetics of hydrogen plasmas. The input conditions to the model are the inlet flow rate of the feedstock gas, absorbed power, geometry configuration and material properties of the plasma reactor.

The GEVKM was implemented into a robust computational tool written in Fortran 90. It is configured through text configuration files for greater flexibility. The reliability of the computational code is achieved by enhanced unit testing and usage of the version control system. The GEVKM consists of the non-linear algebraic system of equations solver which utilizes the Newton-Raphson method. The GEVKM includes also a solver framework for the continuity and energy equations with self-consistency checks to guarantee conservation of charge, particles and energy in the system. The GEVKM is suitable for simulating arbitrary cylindrical hydrogen plasma reactors with various components and chemical reactions. The GEVKM includes also a simulation tool for the calculation of reaction rates from cross-section data assuming Maxwellian distribution functions for colliding partners and fitting it into analytical representations.

The GEVKM was verified and validated in a low-pressure (0.2-100 mTorr) and low absorbed power density (0.053-0.32 W/cm$^3$) regime by comparing with the simulations and experimental measurements of Zorat et al. (2000) and Zorat and Vender (2000). The electron temperature and number density calculated by the GEVKM agree very well with the Langmuir probe measurements. Moreover, compared to the global model from Zorat et al. (2000) the GEVKM was able to capture the dependence of the electron number density on the discharge chamber pressure. The GEVKM simulations and experiments show that the electron number density is increasing with increasing chamber pressure while the global model of Zorat et al. (2000) predicts a slight decrease of the electron number density. The differences in the number densities of positive ions calculated in this work and those calculated by Zorat et al. (2000) and Zorat and Vender (2000) were attributed to the differences in the wall fluxes estimations as well as the inclusion of the ionization through electronic states of hydrogen atoms into the GEVKM as was suggested in Hassouni et al. (1999). The GEVKM simulations showed that the heavy-particle temperature decreases as the discharge pressure increases at these low pressures (0.2-100 mTorr) and low absorbed power densities (0.053-0.32 W/cm$^3$).

In the intermediate to high-pressure (1-100 Torr) and high absorbed power density (8.26-22 W/cm$^3$) regime the GEVKM was verified and validated by comparisons with the numerical simulations and experimental measurements of Chen et al. (1999). The GEVKM shows a rather good agreement with the gas temperature measurements. Electron and positive ions number density predictions of the GEVKM are also in accordance with the simulation results of Chen et al. (1999). The electron temperature predicted by the GEVKM is about 0.5 eV smaller than values predicted by the calculations of Chen et al. (1999). However, Chen et al. (1999) stated that that their electron temperatures were overpredicted.

The GEVKM was applied to the simulation of the RFD chamber of the HCNHIS operating at inlet flow rates 300-3000 sccm and absorbed powers 430-600 W in its baseline configuration and inlet flow rate 1000 sccm and absorbed power 341 W in the alternative configuration. In order to take into account the impact of the nozzle and bypass system of the HCNHIS, analytic outlet boundary conditions were derived which include compressibility, viscous and rarefaction effects. ANSYS Fluent simulations were performed for the HCNHIS-2 configuration and inlet mass flow rate of 1000 sccm assuming that the plasma behaves as cold fluid with the temperature 300 K. The ANSYS Fluent results show that the flow is choked in the bypass tubes and in the throat of the nozzle. Based on the ANSYS Fluent simulation results and for computational efficiency, analytic boundary conditions were developed based on the Fanno flow theory for the flow in the bypass tubes with the modifications due to rarefactions effects. Analytic outlet conditions for the nozzle flow were also developed based on isentropic flow theory with corrections for high Knudsen numbers effects. These analytical outlet boundary conditions were validated by comparison of the pressures predicted by the GEVKM with the pressure measurements of the HCNHIS-1 undertaken at Busek Co. Inc. The GEVKM was used next for simulations of the HCNHIS-2 at inlet flow rate of 1000 sccm and absorbed power of 341 W. The HCNHIS-2 includes a short NIP region and was used in the experiments that measured the negative current downstream the extraction region (see Figure 19) by Faraday Cup. The GEVKM predictions of negative hydrogen ions number densities and electron temperatures in the RFD chamber of the HCNHIS-2 were used to estimate the negative hydrogen ion current using the Bohm flux approximation. The estimated negative current compares well with the Faraday Cup measurements and provides validation of the model. The GEVKM was used in a parametric investigation of the HCNHIS-2 at inlet mass flow rates of 5-5000 sccm and absorbed powers of 200-1000 W covering the regime of operation of the HCNHIS-2 as well as the conventional NHIS. These simulations examined the effects of the inlet flow rate and absorbed power on the production and destruction of vibrationally excited hydrogen molecules, the plasma composition in the ion source, the production and destruction of negative hydrogen ions, the electron and heavy particles temperature, the maximum extractable negative hydrogen ion current. It was found that the inlet flow rate has major impact on the number densities and temperatures in the HCNHIS-2. The main production of negative ions is the dissociative electron attachment to high-lying vibrational states $7 \leq v \leq 11$ at pressures below 10 Torr and absorbed power 341 W. At pressures above 10 Torr the main production is due to dissociative electron attachment to first two vibrational states. The main destruction channels of $\mathrm{H}^-$ are the mutual neutralization with $\mathrm{H}_3^+$ and the electron detachment in collisions with $\mathrm{H}$. Based on the parametric study the optimum operational parameters for production of negative hydrogen ions and vibrationally excited hydrogen molecules were identified. Due to the small differences in the geometry of the bypass and nozzle system between HCNHIS-1 and HCNHIS-2, similar trends are expected in the RFD chamber of HCNHIS-1.

## 5.2 Recommendations for Future Work

The GEVKM is the first step towards understanding the physical phenomena in the HCNHIS. A number of simplifying assumptions were used in deriving the governing equations of the GEVKM. In addition, the actual device has rather complicated geometry which was modeled in simplified manner. There are different approaches to further improve the modeling of the HCNHIS. The first group of approaches includes enhancements related to the GEVKM of the RFD chamber:

1. The inclusion of the space homogeneous steady-state Boltzmann equation for the electron energy distribution function (EEDF) instead of assuming Maxwellian electrons and utilizing energy equation for electrons allows predicting non-equilibrium effects in the EEDF as well as the coupling of the EEDF and VDF. The implementation can be based on the state-to-state approach for the electrons as it was done in Pagano et al. (2007). The use of the Boltzmann equation is necessary at low operating pressures when the EEDF deviates from Maxwellian. It gives electron transport coefficients as well as the rate coefficients for the electron impact chemical reactions. These coefficients are necessary for fluid models.

2. In order to properly take into account the radiation losses in the total energy equation it is necessary to consider detailed collisional-radiative model for the electronic states of molecular and atomic hydrogen. Such model can be based on the work of Hassouni et al. (1999). The inclusion of the extended set of electronic states is crucial in the EEDF calculations.
3. In the current approach, the fluxes into and out of the nozzle and bypass tubes are treated based on the analytic solutions assuming one-dimensional frozen flow. In order to properly take into account the chemical reactions in the conical part of the nozzle it is necessary to include one-dimensional nozzle model of the chemically reacting partially ionized plasma. This 1D model should be coupled with the GEVKM. It will improve the outlet boundary fluxes which are the boundary conditions for the modeling of the NIP region.

One of the main limitations of the GEVKM is the assumption of the specified plasma number density profiles which represent heuristic patching of the exact and numerical solutions of the continuity and momentum equations at different pressures and the isothermal approximation. In order to properly take into account non-uniformity of the species number densities and temperatures in the model of the RFD chamber it is necessary to add the following components:

1. A model that allows the determination of the spatial distribution of the absorbed power as it is required by the proper calculation of the electron heating and, as a result, electron and heavy particles temperatures. This model could be based on the electromagnetic model described in Section 3.1.3 or in Lymberopoulos and Economou (1995). The main outcome of such model is the prediction of the non-uniformity of the electron heating which is essential in determining non-uniformity of number densities and temperatures of plasma components.

2. A model that allows the calculation of the two- or three-dimensional variations of species number densities and temperatures in the RFD chamber of the HCNHIS. This model can be based on the solution of diffusion equations for every species in the bulk of the plasma including temperature equations. In this case the absorbed power distribution is obtained from the solution of Maxwell's equations coupled with continuity and energy equations (Ramamurthi & Economou, 2002). In this model the transport is ambipolar and momentum equations are eliminated. The main achievement will be estimation of non-uniformity effects. However this kind of models is limited to the situations when drift-diffusion approximation is valid (see Section 3.2.1.1). For example, in the choked flow in the nozzle of the HCNHIS the first condition of Eq. (3.40) is not satisfied. Therefore the drift-diffusion approximation should be used with caution.
3. A model that allows estimating the two- or three-dimensional variations of species number densities and temperatures in the conical part of the RFD chamber of the HCNHIS including the nozzle. Based on the GEVKM results it is possible to extract most important chemical reactions as well as the estimation for the VDF shape. With this information it can be possible to model the conical part of the RFD chamber of the HCNHIS by the full multi-temperature multi-fluid moment equations coupled with the Maxwell's equations as was described in Hagelaar et al. (2011). This information will provide boundary conditions for the model of the NIP region.

The results of the extensive fluid simulations can be used later in improving the GEVKM predictions of the wall fluxes and non-uniformity profiles which opens possibility for the optimization of the HCNHIS.

The production and destruction of negative hydrogen ions in the NIP region of the HCNHIS was beyond the scope of this work. As the next step in the modeling the HCNHIS it is necessary to make a model for the NIP region. Preliminary results of the DSMC simulations of the NIP region performed by Gatsonis et al. (2012) and Averkin et al. (2012) revealed that rarefaction effects play an important role in the NIP region. In addition, it was found that there are large gradients in plasma species number densities. Therefore the model for the NIP region should take into account these effects. More complicated and more rigorous approach will be to use hybrid continuum-particle simulation (Gatsonis & Yin, 2001). In this case some of the species treated as particles and others as fluids. All other procedures follow those of PIC method.

The main difficulty however is that the Debye length and mean-free-path are rather small in the HCNHIS so instead of Poisson equation it is possible to use alternative technique based on quasi-neutrality.

Another aspect of the operation of the HCNHIS which was not an objective of this work is the extraction of the negative hydrogen ion beam. The model for the extraction physics should take into account the geometric configuration of plasma grids and magnetic filter effects. The usual approaches to the simulation of the extraction physics are:

1. The gun-type simulation schemes in which trajectories of real particles are simulated (Brown, 2004). These codes include capabilities for starting beams from the charge compensated plasma.
2. The PIC solver in which the electrostatic Vlasov-Maxwell system of equations is self-consistently solved (Gatsonis & Spirkin, 2009). This model gives the space distribution of the beam including phase-space information which is necessary in calculating optical properties of the beam.

# Appendix A. Reaction Rates Used in the Global Enhanced Vibrational Kinetic Model

In this part the reaction rates, their calculation (integration of cross sections), interpolation and code implementation will be discussed.

## A.1 Reaction Rates for a Binary Collision of Particles Obeying Maxwellian Distributions with Different Temperatures

The reaction rates in the particle continuity equations for binary reactions of species 1 and 2 with the total cross sections $\sigma_T \equiv \sigma_{12}^{(n)}$ can be calculated by

$$k = \left\langle \sigma_T \left( g_r \right) g_r \right\rangle = \iint \sigma_T \left( g_r \right) g_r f_1 f_2 d\mathbf{v}_1 d\mathbf{v}_2 \,. \tag{A.1}$$

This equation can be simplified if isotropic Maxwellian – Boltzmann distributions are assumed. In general in plasma it is possible for species to have their own distributions with different temperatures. Therefore Eq. (A.1) can be rewritten as

$$k = \left( \frac{m_1}{2\pi k_B T_1} \frac{m_2}{2\pi k_B T_2} \right)^{3/2} \int\limits_{\mathbb{R}^3} \int\limits_{\mathbb{R}^3} \sigma_T \left( g_r \right) g_r e^{-\frac{m_1 \mathbf{v}_1^2}{2k_B T_1} - \frac{m_2 \mathbf{v}_2^2}{2k_B T_2}} d\mathbf{v}_1 d\mathbf{v}_2 \,. \tag{A.2}$$

Introducing new variables

$$\begin{aligned} \mathbf{v}_m &= \frac{\frac{m_1}{T_1}\mathbf{v}_1 + \frac{m_2}{T_2}\mathbf{v}_2}{\frac{m_1}{T_1} + \frac{m_2}{T_2}}, \\ \mathbf{g}_r &= \mathbf{v}_1 - \mathbf{v}_2 \end{aligned} \tag{A.3}$$

with Jacobian

$$\frac{\partial\left(\mathbf{g}_r, \mathbf{v}_m\right)}{\partial\left(\mathbf{v}_1, \mathbf{v}_2\right)} = \begin{vmatrix} \frac{\partial \mathbf{g}_r}{\partial \mathbf{v}_1} & \frac{\partial \mathbf{g}_r}{\partial \mathbf{v}_2} \\ \frac{\partial \mathbf{v}_m}{\partial \mathbf{v}_1} & \frac{\partial \mathbf{v}_m}{\partial \mathbf{v}_2} \end{vmatrix} = \begin{vmatrix} 1 & -1 \\ \frac{\frac{m_1}{T_1}}{\frac{m_1}{T_1} + \frac{m_2}{T_2}} & \frac{\frac{m_2}{T_2}}{\frac{m_1}{T_1} + \frac{m_2}{T_2}} \end{vmatrix} = 1 \tag{A.4}$$

Eq. (A.2) can be rewritten as

$$k = \frac{1}{\left(2\pi k_B\right)^3}\left(\frac{m_1 m_2}{T_1 T_2}\right)^{3/2}\int_{\mathbb{R}^3} e^{-\left(\frac{m_1}{T_1}+\frac{m_2}{T_2}\right)\frac{\mathbf{v}_m^2}{2k_B}} d\mathbf{v}_m \int_{\mathbb{R}^3} \sigma_T\left(g_r\right) g_r e^{-\frac{\frac{m_1 m_2}{T_1 T_2}}{\frac{m_1}{T_1}+\frac{m_2}{T_2}}\frac{\mathbf{g}_r^2}{2k_B}} d\mathbf{g}_r\,. \tag{A.5}$$

The first integral is merely Gaussian integral and can be easily calculated

$$\int_{\mathbb{R}^3} e^{-\left(\frac{m_1}{T_1}+\frac{m_2}{T_2}\right)\frac{\mathbf{v}_m^2}{2k_B}} d\mathbf{v}_m = \left(\int_{-\infty}^{+\infty} e^{-\left(\frac{m_1}{T_1}+\frac{m_2}{T_2}\right)\frac{v_{m,x}^2}{2k_B}} dv_{m,x}\right)^3$$

$$= \frac{\left(2k_B\right)^{3/2}}{\left(\frac{m_1}{T_1}+\frac{m_2}{T_2}\right)^{3/2}}\left(\int_{-\infty}^{+\infty} e^{-\left[\sqrt{\frac{m_1}{T_1}+\frac{m_2}{T_2}}\frac{v_{m,x}}{\sqrt{k_B}}\right]^2} d\sqrt{\frac{m_1}{T_1}+\frac{m_2}{T_2}}\frac{v_{m,x}}{\sqrt{k_B}}\right)^3 = \left(\frac{2\pi k_B}{\frac{m_1}{T_1}+\frac{m_2}{T_2}}\right)^{3/2}. \tag{A.6}$$

Due to independence of the total cross section on the angles the second integral can be integrated over all angles to give

$$\int_{\mathbb{R}^3} \sigma_T\left(g_r\right) g_r e^{-\frac{\frac{m_1 m_2}{T_1 T_2}}{\frac{m_1}{T_1}+\frac{m_2}{T_2}}\frac{\mathbf{g}_r^2}{2k_B}} d\mathbf{g}_r = 4\pi\int_0^{+\infty} \sigma_T\left(g_r\right) g_r^3 e^{-\frac{g_r^2}{2k_B\left(\frac{T_1}{m_1}+\frac{T_2}{m_2}\right)}} dg_r\,. \tag{A.7}$$

Finally the reaction rate for a binary collision of two particles obeying Maxwellian distributions with different temperatures can be written as

$$k = \sqrt{\frac{2}{\pi}}\frac{1}{k_B^{3/2}\left(\frac{T_1}{m_1}+\frac{T_2}{m_2}\right)^{3/2}}\int_0^{+\infty} \sigma_T\left(g_r\right) g_r^3 e^{-\frac{g_r^2}{2k_B\left(\frac{T_1}{m_1}+\frac{T_2}{m_2}\right)}} dg_r\,. \tag{A.8}$$

Introducing an effective temperature $T_{eff} = \frac{m_1 T_2 + m_2 T_1}{m_1 + m_2}$ and reduced mass $\mu = \frac{m_1 m_2}{m_1 + m_2}$ the above equation could be rewritten as

$$k = \sqrt{\frac{2}{\pi}}\left(\frac{\mu}{k_B T_{eff}}\right)^{3/2}\int_0^{+\infty} \sigma_T\left(g_r\right) g_r^3 e^{-\frac{\mu g_r^2}{2k_B T_{eff}}} dg_r\,. \tag{A.9}$$

Equation (A.9) is exactly the same as for the reaction rate of two particles of the same species and temperature and was also derived by Light et al. (1969). It could be further simplified to

$$k = \sqrt{\frac{8k_B T_{eff}}{\pi \mu}} \int_0^{+\infty} \sigma_T \left( \sqrt{\frac{2k_B T_{eff}}{\mu}} \xi \right) \xi^3 e^{-\xi^2} d\xi \tag{A.10}$$

where $\xi = \sqrt{\frac{\mu}{2k_B T_{eff}}} g_r$ is the nondimensional relative velocity. In the above equation the first term represents the average thermal velocity of a particle which mass and temperature equal to reduced mass of the colliding particles and their effective temperature respectively. The second term has dimensions of cross section and represents the effective averaged cross section of the collision. In other words

$$k = \left\langle \sigma v \right\rangle = \left\langle \sigma_{eff} \right\rangle \left\langle v_{eff} \right\rangle . \tag{A.11}$$

Sometimes total cross sections $\sigma_T = \sigma_T(\mathcal{E}_r)$ are given as a function of relative energy $\mathcal{E}_r$ rather than relative speed $g_r$. Introducing nondimensional relative energy $\eta = \frac{\mathcal{E}_r}{k_B T_{eff}} = \xi^2$ the Eq. (A.10) can be modified to

$$k = \sqrt{\frac{2k_B T_{eff}}{\pi \mu}} \int_0^{+\infty} \sigma_T \left( \eta \right) \eta e^{-\eta} d\eta \tag{A.12}$$

There are certain extreme cases which are widely used and allow further simplifying the expression for the reaction rate. If one of the species is electron $m_1 = m_e$ then $m_e \ll m_2$ and in most cases $T_e \geq T_2$. The reduced mass and effective temperature are $\mu = \frac{m_e m_2}{m_e + m_2} \simeq m_e$ and $T_{eff} = \frac{m_e T_2 + m_2 T_e}{m_e + m_2} \simeq \frac{m_e}{m_2} T_2 + T_e \simeq T_e$ respectively. For electrons the cross section is usually given in terms of electron energy

$$\sigma_{T,e} = \sigma_{T,e} \left( \mathcal{E}_e \right), \tag{A.13}$$

where $\mathcal{E}_e = \frac{m_e v_e^2}{2}$. Also the relative speed of the electrons colliding with heavy particles is mostly determined by them. Therefore

$$\eta \simeq \frac{\mathcal{E}_e}{k_B T_e}. \tag{A.14}$$

Substituting the definition of $\eta$ in terms of $\mathcal{E}_e$ from Eq. (A.13) and using it in Eq. (A.12) gives the following formula for the reaction rate of electron collisions with heavy particles

$$k = \sqrt{\frac{2}{\pi m_e}} \frac{1}{\left(k_B T_e\right)^{3/2}} \int_0^{+\infty} \sigma_{T,e}\left(\mathcal{E}_e\right) \mathcal{E}_e e^{-\frac{\mathcal{E}_e}{k_B T_e}} d\mathcal{E}_e. \tag{A.15}$$

Equation (A.10) is also useful in evaluating the mean free path

$$\lambda_{12} = \frac{\left\langle v_1 \right\rangle}{\left\langle \sigma_{T,12} g_{r,12} \right\rangle n_2} = \frac{\sqrt{\frac{T_1 \mu}{T_{eff} m_1}}}{n_2 \int_0^{+\infty} \sigma_T \left( \sqrt{\frac{2 k_B T_{eff}}{\mu}} \xi \right) \xi^3 e^{-\xi^2} d\xi}. \tag{A.16}$$

When the temperature is the same for both species $T_1 = T_2 = T_{eff} = T$ the above relation simplifies to

$$\lambda_{12} = \frac{1}{\sqrt{1 + \frac{m_1}{m_2}} n_2 \int_0^{+\infty} \sigma_T \left( \sqrt{\frac{2 k_B T}{\mu}} \xi \right) \xi^3 e^{-\xi^2} d\xi}. \tag{A.17}$$

As an example for VHS molecules with the same masses $m_1 = m_2 = 2\mu$ and $\sigma_T = \sigma_{T,ref} \left( \frac{g_r}{g_{r,ref}} \right)^{1-2\omega}$ (Bird, 1994) the mean free path is

$$\lambda_{11} = \frac{1}{\sqrt{2} n_1 \int_0^{+\infty} \sigma_{T,ref} \left( \sqrt{\frac{4 k_B T_1}{m_1}} \xi \right)^{1-2\omega} \xi^3 e^{-\xi^2} d\xi} = \frac{g_{r,ref}^{1-2\omega} \left( \frac{4 k_B T_1}{m_1} \right)^{\omega - \frac{1}{2}}}{\sqrt{2} n_1 \sigma_{T,ref} \Gamma\left( \frac{5}{2} - \omega \right)}. \tag{A.18}$$

Substituting the definition of $g_{r,ref}=\sqrt{\frac{4k_BT_{ref}}{m_1}}\left[\Gamma\left(\frac{5}{2}-\omega\right)\right]^{-\frac{1}{2\omega-1}}$ and $\sigma_{T,ref}=\pi d_{ref}^2$ the above equation can be further simplified to

$$\lambda_{11}=\frac{1}{\sqrt{2}n_1\pi d_{ref}^2\left(\frac{T_1}{T_{ref}}\right)^{\frac{1}{2}-\omega}}, \tag{A.19}$$

which is exactly the same formula obtained by Bird (1994) for a single Maxwellian distribution.

### A.2 Numerical Calculation of Reaction Rates

Reaction rates are numerically calculated by using Eq. (A.10) or Eq.(A.12). The integrals are calculated by using Simpson's rule (Teukolsky, et al., 1996). In principle the numerical values can be used during simulation as is in the lookup tables. However, lookup tables are less efficient than the analytic expression. Therefore calculated reaction rates should be fitted to a relatively simple analytic expression.

One of the common choices is the Arrhenius type fit given as

$$k=\sigma T^{\alpha}e^{-\frac{E_a}{T}}, \tag{A.20}$$

where $\sigma$, $\alpha$, and $E_a$ are constants.

Calculated reaction rate is fitted to Eq. (A.20) by using least square method. Suppose that the reaction rate was calculated at $n$ temperature points $\left\{T_i\right\}_{i=1}^{n}$ and corresponding reaction rates $\left\{k_i\right\}_{i=1}^{n}$. The coefficients $\sigma$, $\alpha$, and $E_a$ are found from the solution of the following system of linear equations

$$\begin{cases}\sigma n+\alpha\sum_{i=1}^{n}\ln T_i-E_a\sum_{i=1}^{n}\frac{1}{T_i}=\sum_{i=1}^{n}\ln k_i,\\ \sigma\sum_{i=1}^{n}\ln T_i+\alpha\sum_{i=1}^{n}\ln^2 T_i-E_a\sum_{i=1}^{n}\frac{1}{T_i}\ln T_i=\sum_{i=1}^{n}\ln k_i\ln T_i,\\ \sigma\sum_{i=1}^{n}\frac{1}{T_i}+\alpha\sum_{i=1}^{n}\frac{1}{T_i}\ln T_i-E_a\sum_{i=1}^{n}\frac{1}{T_i^2}=\sum_{i=1}^{n}\frac{1}{T_i}\ln k_i.\end{cases} \tag{A.21}$$

The above system can be solved analytically using Cramer's rule.

The relative error of the fitting formula is given by

$$\epsilon = \frac{1}{n}\left[\sum_{i=1}^{n}\left(k_i - \sigma T_i^{\alpha} e^{-\frac{E_a}{T_i}}\right)^2\right]^{1/2}. \qquad \text{(A.22)}$$

Another fitting formula used is the one used in Janev et al. (1987) and is given as

$$k = \exp\left(\sum_{j=0}^{N} a_j T^j\right). \qquad \text{(A.23)}$$

Similar to the Arrhenius fitting the reaction rate is fitted to the above expression by using the least square methods. With given number of polynomials $N$ the coefficients can be found from the solution of the following system of linear equations

$$\sum_{j=0}^{N}\sum_{i=1}^{n} a_j T_i^j T_i^l = \sum_{i=1}^{n} T_i^k \ln k_i,\ l = 0,\dots,N. \qquad \text{(A.24)}$$

Since the number of polynomials is typically higher than 8 the system is faster to solve using iterative method such as Jacobi method (Teukolsky, et al., 1996). From practical point of view this fitting gives error below 0.1% for most of the chemical reactions used in the GEVKM.

# Appendix B. Numerical Values for Reaction Rates in Hydrogen Plasmas

In this appendix all chemical reactions along with their reaction rates are listed. For simplicity the numbering of reactions follow the numbering from Table 3-Table 9. All reaction rates are given in $s^{-1}$, $m^3s^{-1}$, and $m^6s^{-1}$ for unimolecular, binary, and ternary reactions respectively. Electron temperature is given in eV while heavy particles temperature denoted by $T_h$ is given in Kelvins. The references to the sources of cross-sections or reaction rates are given in Table 3-Table 9. Wherever possible the reaction rates from the original sources are used.

$$k_1 = 1.1 \times 10^{-14} T_e^{0.42} e^{-16.05/T_e}$$

$$k_2 = 4.73 \times 10^{-14} T_e^{-0.23} e^{-10.09/T_e}$$

$$k_3 = \frac{6.3498 \times 10^{-19} \times \sqrt{T_e}}{1.0 + T_e / 0.45} \times \exp\left(-\frac{3.72}{T_e}\right) \times \left(\frac{3.72}{T_e} + \frac{1.0}{1.0 + T_e / 0.45}\right)$$

$$k_4 = \exp\left(\sum_{i=0}^{8} a_i T_e^i\right)$$

$$\begin{aligned} a_0 &= -3.834597006782\text{e}1; \; a_1 = 1.426322356722\text{e}1; \; a_2 = -5.826468569506 \\ a_3 &= 1.727940947913; \; a_4 = -3.598120866343\text{e-}1; \; a_5 = 4.822199350494\text{e-}2 \\ a_6 &= 3.909402993006\text{e-}3; \; a_6 = 1.738776657690\text{e-}4; \; a_6 = -3.252844486351\text{e-}6 \end{aligned}$$

$$k_5 = 7.89 \times 10^{-15} T_e^{0.41} e^{-14.23/T_e}$$

$$k_6 = 1.24 \times 10^{-12} T_e^{0.03} e^{-10.44/T_e}$$

$$k_7 = 2.35 \times 10^{-14} T_e^{0.4}$$

$$k_8 = 1.88 \times 10^{-13} T_e^{-0.39} e^{-28.82/T_e}$$

$$k_9 = 2.19 \times 10^{-15} T_e^{0.8}$$

$$k_{10} = 7.3 \times 10^{-16} T_e^{0.8}$$

$$k_{11} = 1.93 \times 10^{-16} T_e^{-1.07} e^{-2.7/T_e} + 2.59 \times 10^{-15} T_e^{-1.27} e^{-6.45/(T_e+0.1)}$$

$$k_{12} = 1.0 \times 10^{-13} T_e^{0.37} e^{-14.46/T_e}$$

$$k_{13} = 4.1073 \times 10^{-34} T_h^{3.6917} e^{-55576.9255/T_h}$$

$$k_{14} = \exp(a_6 + a_5 T_h + a_4 T_h^2 + a_3 T_h^3 + a_2 \ln(T_h) + a_1 e^{T_h/11600.0}) \times 10^{-6}$$

$a_1 = 113.761$; $a_2 = 47.387$; $a_3 = \text{-}4.327\text{e - }11$;
$a_4 = 4.729\text{e - }7$; $a_5 = \text{-}2.37\text{e - }2$; $a_6 = \text{-}496.794$

$$k_{15} = 2.6775 \times 10^{-25} (T_h / 11600.0 + 4.9 \times 11600.0 / T_h)^{1.9}$$

$$k_{16} = 1.4816 \times 10^{-14} T_h^{-0.3775} e^{-28914.5046/T_h}$$

$$k_{17} = \exp\left(\sum_{i=0}^{9} a_i T_h^i\right)$$

$a_0 = \text{-}0.216029102\text{e}3$; $a_1 = 0.121978583\text{e}3$; $a_2 = \text{-}0.299905577\text{e}2$; $a_3 = 0.261291326\text{e}1$
$a_4 = 0.143511763$; $a_5 = \text{-}0.3455493090\text{e-}1$; $a_6 = \text{-}0.855726245\text{e-}3$; $a_7 = 0.516639237\text{e-}3$;
$a_8 = \text{-}0.391600571\text{e-}4$; $a_9 = 0.974410646\text{e-}6$

$$k_{18} = 6.4 \times 10^{-16}$$

$$k_{19} = \exp\left(\sum_{i=0}^{7} a_i T_h^i\right)$$

$a_0 = \text{-}0.241516512\text{e}3$; $a_1$=0.169825321e3; $a_2$=-0.600824859e2; $a_3$=0.120348001e2;
$a_4$=-0.147853482e1; $a_5$=0.111213169; $a_6$=-0.472096194e-2; $a_7$0.868175182e-4

$$k_{20} = \exp\left(\sum_{i=0}^{10} a_i T_h^i\right)$$

$a_0 = \text{-}0.506644368\text{e}3$; $a_1 = 0.422452027\text{e}3$; $a_2 = \text{-}0.151802463\text{e}3$; $a_3 = 0.231901722\text{e}2$
$a_4 = \text{-}0.510835519\text{e-}1$; $a_5 = \text{-}0.420844927$; $a_6 = 0.3062554780\text{e-}1$; $a_7 = 0.494889891\text{e-}2$
$a_8 = \text{-}0.967756739\text{e-}3$; $a_9 = 0.618900281\text{e-}4$; $a_{10} = \text{-}0.143957321\text{e-}5$

$$k_{21} = \exp\left(\sum_{i=0}^{10} a_i T_h^i\right)$$

$a_0$ = -0.663670587e4; $a_1$ = 0.322616347e4; $a_2$ = -0.507939826e3; $a_3$ = 0.274314923e1
$a_4$ = 0.641868815e1; $a_5$ = -0.230281260; $a_6$ = -0.681246157e-1; $a_7$ = 0.477710488e-2
$a_8$ = 0.330007636e-3; $a_9$ = -0.440107270e-4; $a_{10}$ = 0.125949732e-5

$$k_{22} = 1.28 \times 10^{-13} \sqrt{300 / T_h}$$

$$k_{23} = \exp\left(\sum_{i=0}^{9} a_i T_h^i\right)$$

$a_0$ = -0.338977878e4; $a_1$ = 0.197001860e4; $a_2$ = -0.438560571e3; $a_3$ = 0.371807870e2
$a_4$ = 0.807180037; $a_5$ = -0.172733665; $a_6$ = -0.428904443e-1; $a_7$ = 0.840963351e-2
$a_8$ = -0.543091805e-3; $a_9$ = 0.126331589e-4

$$k_{24} = 1.28 \times 10^{-13} \sqrt{300 / T_h}$$

$$k_{25} = 8.29 \times 10^{-13} \sqrt{300 / T_h}$$

$$k_{26} = \exp\left(\sum_{i=0}^{9} a_i T_h^i\right)$$

$a_0$ = 0.4178545e3; $a_1$ = 0.237326922e3; $a_2$ = -0.555810680e2; $a_3$ = 0.468004282e1
$a_4$ = 0.256016343e0; $a_5$ = -0.588543185e-1; $a_6$ = -0.223996560e-2; $a_7$ = 0.102508118e-2
$a_8$ = -0.777506642e-4; $a_9$ = 0.197441344e-5

$$k_{27} = 10^{-13}$$

$$k_{28} = 2.0 \times 10^{-13} \sqrt{300 / T_h}$$

$$k_{29} = 8.29 \times 10^{-13} \sqrt{300 / T_h}$$

$$k_{30} = 2.68 \times 10^{-37} T_h^{-0.6}$$

$$k_{31} = 8.04 \times 10^{-37} T_h^{-0.6}$$

$$k_{32} = 3.1 \times 10^{-35}\sqrt{300 / T_h}$$

$$k_{33}(v) = \exp\left(\sum_{i=0}^{10} a_i T_h^i\right)$$

$v = 0$

| $v'$ | $a_0$ | $a_1$ | $a_2$ | $a_3$ | $a_4$ | $a_5$ | $a_6$ | $a_7$ | $a_8$ | $a_9$ | $a_{10}$ |
|---|---|---|---|---|---|---|---|---|---|---|---|
| 1 | -34.004 | 1.171 | -0.613 | 0.049 | 9.393e-3 | 3.971e-3 | -2.698e-3 | 2.185e-4 | 4.894e-5 | -6.611e-7 | -1.030e-6 |
| 2 | -37.629 | 2.129 | -1.134 | 0.149 | 8.156e-3 | 1.437e-3 | -3.742e-3 | 7.658e-4 | 3.790e-6 | -8.963e-6 | 1.234e-7 |
| 3 | -39.999 | 2.384 | -1.344 | 0.241 | -0.010 | -8.142e-4 | -3.573e-3 | 1.406e-3 | -1.006e-4 | -2.482e-5 | 3.219e-6 |
| 4 | -41.895 | 2.466 | -1.448 | 0.342 | -0.065 | 8.397e-3 | -1.124e-3 | 5.969e-4 | -1.521e-4 | 9.723e-6 | 4.375e-7 |
| 5 | -43.382 | 2.503 | -1.502 | 0.365 | -0.099 | 0.027 | -2.251e-3 | -6.777e-4 | 9.698e-7 | 4.868e-5 | -5.595e-6 |
| 6 | -44.596 | 2.482 | -1.574 | 0.415 | -0.106 | 0.028 | -4.847e-3 | 2.962e-4 | 6.026e-6 | 5.069e-6 | -8.649e-7 |
| 7 | -45.856 | 2.682 | -1.710 | 0.452 | -0.108 | 0.028 | -5.675e-3 | 5.821e-4 | -2.911e-5 | 6.057e-6 | -8.033e-7 |
| 8 | -47.202 | 3.028 | -1.936 | 0.518 | -0.105 | 0.025 | -7.772e-3 | 1.448e-3 | -2.232e-5 | -3.131e-5 | 3.118e-6 |
| 9 | -48.287 | 3.223 | -2.079 | 0.591 | -0.129 | 0.026 | -5.946e-3 | 1.108e-3 | -1.039e-4 | -1.355e-6 | 6.847e-7 |
| 10 | -49.359 | 3.463 | -2.225 | 0.652 | -0.149 | 0.031 | -6.296e-3 | 1.006e-3 | -9.380e-5 | 2.311e-6 | 1.897e-7 |
| 11 | -48.191 | 4.456 | -2.372 | 0.652 | -0.148 | 0.034 | -6.934e-3 | 9.427e-4 | -9.710e-5 | 9.870e-6 | -5.944e-7 |
| 12 | -48.991 | 4.619 | -2.455 | 0.681 | -0.155 | 0.036 | -7.158e-3 | 9.780e-4 | -1.020e-4 | 1.029e-5 | -6.095e-7 |
| 13 | -49.781 | 4.745 | -2.520 | 0.703 | -0.161 | 0.037 | -7.264e-3 | 9.894e-4 | -1.098e-4 | 1.232e-5 | -7.718e-7 |
| 14 | -50.552 | 4.820 | -2.558 | 0.715 | -0.164 | 0.038 | -7.367e-3 | 1.006e-3 | -1.120e-4 | 1.250e-5 | -7.779e-7 |

$v = 1$

| $v'$ | $a_0$ | $a_1$ | $a_2$ | $a_3$ | $a_4$ | $a_5$ | $a_6$ | $a_7$ | $a_8$ | $a_9$ | $a_{10}$ |
|---|---|---|---|---|---|---|---|---|---|---|---|
| 0 | -33.432 | 0.661 | -0.514 | -0.021 | 0.029 | -1.335e-3 | -1.279e-3 | 1.186e-4 | 2.889e-5 | -2.539e-6 | -2.322e-7 |
| 2 | -33.045 | 0.865 | -0.724 | 0.115 | -7.064e-3 | -2.429e-3 | -9.117e-5 | 4.736e-4 | -8.317e-5 | -4.582e-6 | 1.328e-6 |
| 3 | -36.981 | 1.534 | -0.839 | 0.118 | -0.010 | 0.010 | -3.385e-3 | 2.060e-4 | 3.538e-5 | 2.120e-6 | -9.044e-7 |
| 4 | -39.316 | 1.906 | -1.039 | 0.192 | -0.030 | 0.013 | -3.725e-3 | 3.278e-4 | 8.465e-6 | 2.302e-6 | -6.153e-7 |
| 5 | -41.037 | 2.266 | -1.232 | 0.262 | -0.048 | 0.016 | -4.126e-3 | 4.752e-4 | -1.542e-5 | 1.172e-6 | -2.378e-7 |
| 6 | -42.393 | 2.611 | -1.415 | 0.326 | -0.065 | 0.018 | -4.584e-3 | 5.875e-4 | -3.098e-5 | 7.983e-7 | -5.899e-8 |
| 7 | -43.528 | 2.933 | -1.585 | 0.385 | -0.079 | 0.021 | -5.033e-3 | 6.779e-4 | -4.279e-5 | 8.687e-7 | 2.453e-8 |
| 8 | -44.528 | 3.231 | -1.740 | 0.438 | -0.093 | 0.024 | -5.435e-3 | 7.416e-4 | -5.243e-5 | 1.784e-6 | -1.056e-8 |
| 9 | -45.440 | 3.502 | -1.880 | 0.486 | -0.105 | 0.026 | -5.776e-3 | 7.856e-4 | -6.151e-5 | 3.413e-6 | -1.296e-7 |
| 10 | -46.297 | 3.743 | -2.005 | 0.528 | -0.116 | 0.028 | -6.102e-3 | 8.364e-4 | -6.911e-5 | 4.174e-6 | -1.631e-7 |
| 11 | -47.121 | 3.957 | -2.116 | 0.566 | -0.126 | 0.030 | -6.360e-3 | 8.686e-4 | -7.686e-5 | 5.724e-6 | -2.812e-7 |
| 12 | -47.905 | 4.119 | -2.199 | 0.594 | -0.133 | 0.032 | -6.533e-3 | 8.871e-4 | -8.385e-5 | 7.438e-6 | -4.199e-7 |
| 13 | -48.672 | 4.245 | -2.264 | 0.616 | -0.139 | 0.033 | -6.705e-3 | 9.141e-4 | -8.768e-5 | 7.783e-6 | -4.331e-7 |
| 14 | -49.417 | 4.319 | -2.301 | 0.629 | -0.142 | 0.033 | -6.747e-3 | 9.131e-4 | -9.300e-5 | 9.518e-6 | -5.819e-7 |

$v = 2$

| $v'$ | $a_0$ | $a_1$ | $a_2$ | $a_3$ | $a_4$ | $a_5$ | $a_6$ | $a_7$ | $a_8$ | $a_9$ | $a_{10}$ |
|---|---|---|---|---|---|---|---|---|---|---|---|
| 0 | -36.623 | 1.139 | -0.645 | -0.028 | 0.056 | -4.592e-3 | -3.391e-3 | 4.039e-4 | 9.780e-5 | -8.277e-6 | -1.153e-6 |
| 1 | -32.541 | 0.364 | -0.464 | 0.035 | 7.776e-3 | -5.641e-3 | 1.033e-3 | 2.671e-4 | -7.395e-5 | -4.451e-6 | 1.433e-6 |
| 3 | -32.455 | 0.540 | -0.613 | 0.143 | -0.030 | -4.632e-3 | 2.487e-3 | 5.002e-4 | -1.948e-4 | -4.517e-6 | 3.165e-6 |
| 4 | -36.946 | 1.485 | -0.813 | 0.109 | -7.440e-3 | 9.126e-3 | -3.316e-3 | 2.004e-4 | 3.676e-5 | 1.793e-6 | -8.789e-7 |
| 5 | -39.260 | 1.830 | -0.999 | 0.178 | -0.027 | 0.012 | -3.622e-3 | 3.166e-4 | 1.077e-5 | 1.865e-6 | -5.830e-7 |
| 6 | -40.957 | 2.163 | -1.178 | 0.243 | -0.043 | 0.015 | -3.988e-3 | 4.576e-4 | -1.214e-5 | 6.520e-7 | -2.030e-7 |
| 7 | -42.288 | 2.479 | -1.346 | 0.302 | -0.058 | 0.017 | -4.409e-3 | 5.630e-4 | -2.671e-5 | 2.119e-7 | -2.350e-8 |
| 8 | -43.398 | 2.772 | -1.500 | 0.356 | -0.072 | 0.020 | -4.808e-3 | 6.356e-4 | -3.719e-5 | 7.264e-7 | -1.237e-9 |
| 9 | -44.371 | 3.039 | -1.640 | 0.404 | -0.084 | 0.022 | -5.178e-3 | 7.027e-4 | -4.624e-5 | 1.085e-6 | 2.479e-8 |
| 10 | -45.255 | 3.278 | -1.764 | 0.446 | -0.095 | 0.024 | -5.493e-3 | 7.482e-4 | -5.397e-5 | 2.114e-6 | -3.687e-8 |
| 11 | -46.087 | 3.490 | -1.874 | 0.484 | -0.105 | 0.026 | -5.760e-3 | 7.832e-4 | -6.115e-5 | 3.377e-6 | -1.280e-7 |
| 12 | -46.864 | 3.651 | -1.958 | 0.512 | -0.112 | 0.027 | -5.978e-3 | 8.170e-4 | -6.622e-5 | 3.886e-6 | -1.504e-7 |
| 13 | -47.616 | 3.775 | -2.022 | 0.534 | -0.118 | 0.029 | -6.113e-3 | 8.299e-4 | -7.122e-5 | 5.189e-6 | -2.592e-7 |
| 14 | -48.339 | 3.849 | -2.060 | 0.547 | -0.121 | 0.029 | -6.214e-3 | 8.456e-4 | -7.352e-5 | 5.407e-6 | -2.682e-7 |

$v = 3$

| $v'$ | $a_0$ | $a_1$ | $a_2$ | $a_3$ | $a_4$ | $a_5$ | $a_6$ | $a_7$ | $a_8$ | $a_9$ | $a_{10}$ |
|---|---|---|---|---|---|---|---|---|---|---|---|
| 0 | -38.533 | 0.960 | -0.637 | -0.035 | 0.064 | -5.060e-3 | -4.033e-3 | 4.361e-4 | 1.262e-4 | -9.449e-6 | -1.549e-6 |
| 1 | -36.036 | 0.589 | -0.367 | -0.039 | 0.029 | 1.784e-3 | -2.110e-3 | 3.666e-5 | 6.016e-5 | -1.637e-6 | -6.259e-7 |
| 2 | -31.997 | 0.084 | -0.385 | 0.064 | -0.011 | -7.670e-3 | 2.925e-3 | 3.684e-4 | -1.689e-4 | -5.525e-6 | 3.027e-6 |
| 4 | -32.546 | 0.671 | -0.460 | 0.078 | -0.010 | -1.998e-3 | -3.867e-4 | 4.292e-4 | -4.094e-5 | -4.962e-6 | 5.376e-7 |
| 5 | -33.912 | 0.731 | -0.666 | 0.187 | -0.059 | 2.576e-3 | 2.603e-3 | 2.463e-4 | -1.964e-4 | 3.693e-6 | 2.505e-6 |
| 6 | -35.115 | 0.816 | -0.764 | 0.192 | -0.062 | 0.015 | -9.971e-4 | -2.492e-4 | -1.154e-5 | 1.663e-5 | -1.628e-6 |
| 7 | -36.537 | 1.075 | -0.891 | 0.208 | -0.059 | 0.018 | -2.884e-3 | 1.096e-4 | -4.564e-6 | 8.559e-6 | -1.041e-6 |
| 8 | -37.983 | 1.483 | -1.107 | 0.260 | -0.067 | 0.022 | -4.023e-3 | 8.258e-5 | 3.092e-5 | 9.707e-6 | -1.701e-6 |
| 9 | -39.147 | 1.753 | -1.298 | 0.334 | -0.074 | 0.017 | -3.757e-3 | 5.860e-4 | -5.311e-5 | 1.850e-6 | 3.165e-8 |
| 10 | -44.214 | 2.845 | -1.538 | 0.369 | -0.075 | 0.020 | -4.915e-3 | 6.605e-4 | -3.992e-5 | 5.224e-7 | 4.326e-8 |
| 11 | -45.073 | 3.055 | -1.648 | 0.407 | -0.085 | 0.022 | -5.200e-3 | 7.059e-4 | -4.676e-5 | 1.144e-6 | 2.177e-8 |
| 12 | -45.859 | 3.214 | -1.731 | 0.435 | -0.092 | 0.024 | -5.407e-3 | 7.350e-4 | -5.193e-5 | 1.895e-6 | -2.647e-8 |
| 13 | -46.605 | 3.338 | -1.795 | 0.457 | -0.098 | 0.025 | -5.573e-3 | 7.606e-4 | -5.589e-5 | 2.318e-6 | -4.656e-8 |
| 14 | -47.313 | 3.411 | -1.833 | 0.470 | -0.101 | 0.025 | -5.672e-3 | 7.758e-4 | -5.822e-5 | 2.566e-6 | -5.826e-8 |

$v = 4$

| $v'$ | $a_0$ | $a_1$ | $a_2$ | $a_3$ | $a_4$ | $a_5$ | $a_6$ | $a_7$ | $a_8$ | $a_9$ | $a_{10}$ |
|---|---|---|---|---|---|---|---|---|---|---|---|
| 0 | -39.999 | 0.575 | -0.520 | 0.033 | 0.020 | -0.010 | 9.041e-4 | 5.749e-4 | -1.034e-4 | -1.042e-5 | 2.283e-6 |
| 1 | -37.941 | 0.533 | -0.353 | -0.039 | 0.027 | 1.997e-3 | -1.998e-3 | 2.352e-5 | 5.675e-5 | -1.390e-6 | -5.845e-7 |
| 2 | -36.058 | 0.597 | -0.370 | -0.039 | 0.029 | 1.732e-3 | -2.125e-3 | 3.957e-5 | 6.056e-5 | -1.690e-6 | -6.300e-7 |
| 3 | -32.111 | 0.242 | -0.248 | -2.343e-3 | 0.014 | -6.430e-3 | -1.438e-4 | 4.313e-4 | -2.488e-5 | -8.390e-6 | 7.008e-7 |
| 5 | -33.790 | 1.091 | -0.602 | 0.035 | 0.015 | 3.717e-3 | -2.568e-3 | 1.895e-4 | 3.820e-5 | -1.702e-6 | -4.464e-7 |
| 6 | -36.878 | 1.391 | -0.763 | 0.091 | -2.405e-3 | 7.955e-3 | -3.151e-3 | 1.951e-4 | 3.751e-5 | 1.015e-6 | -7.837e-7 |
| 7 | -39.149 | 1.684 | -0.920 | 0.148 | -0.019 | 0.011 | -3.518e-3 | 2.489e-4 | 2.469e-5 | 2.417e-6 | -8.117e-7 |

| $v'$ | $a_0$ | $a_1$ | $a_2$ | $a_3$ | $a_4$ | $a_5$ | $a_6$ | $a_7$ | $a_8$ | $a_9$ | $a_{10}$ |
|---|---|---|---|---|---|---|---|---|---|---|---|
| 8 | -40.800 | 1.959 | -1.068 | 0.203 | -0.033 | 0.013 | -3.758e-3 | 3.649e-4 | 2.063e-6 | 1.662e-6 | -4.665e-7 |
| 9 | -42.081 | 2.214 | -1.205 | 0.252 | -0.046 | 0.015 | -4.056e-3 | 4.662e-4 | -1.376e-5 | 9.116e-7 | -2.206e-7 |
| 10 | -43.138 | 2.445 | -1.327 | 0.296 | -0.057 | 0.017 | -4.363e-3 | 5.568e-4 | -2.560e-5 | 5.615e-8 | -1.390e-8 |
| 11 | -44.059 | 2.651 | -1.437 | 0.334 | -0.066 | 0.019 | -4.648e-3 | 6.125e-4 | -3.328e-5 | 2.223e-7 | 2.768e-8 |
| 12 | -44.874 | 2.808 | -1.519 | 0.362 | -0.074 | 0.020 | -4.866e-3 | 6.533e-4 | -3.874e-5 | 3.771e-7 | 5.119e-8 |
| 13 | -45.628 | 2.930 | -1.583 | 0.384 | -0.079 | 0.021 | -5.029e-3 | 6.774e-4 | -4.270e-5 | 8.577e-7 | 2.512e-8 |
| 14 | -46.332 | 3.003 | -1.621 | 0.397 | -0.083 | 0.022 | -5.130e-3 | 6.954e-4 | -4.508e-5 | 9.526e-7 | 3.156e-8 |

$v = 5$

| $v'$ | $a_0$ | $a_1$ | $a_2$ | $a_3$ | $a_4$ | $a_5$ | $a_6$ | $a_7$ | $a_8$ | $a_9$ | $a_{10}$ |
|---|---|---|---|---|---|---|---|---|---|---|---|
| 0 | -41.091 | 0.166 | -0.347 | 0.031 | -0.012 | -4.821e-3 | 3.304e-3 | 1.672e-4 | -1.804e-4 | -1.498e-6 | 3.188e-6 |
| 1 | -39.260 | 0.494 | -0.345 | -0.038 | 0.026 | 2.053e-3 | -1.913e-3 | 1.894e-5 | 5.391e-5 | -1.297e-6 | -5.463e-7 |
| 2 | -37.969 | 0.543 | -0.355 | -0.039 | 0.028 | 1.972e-3 | -2.018e-3 | 2.521e-5 | 5.742e-5 | -1.422e-6 | -5.932e-7 |
| 3 | -33.082 | -0.100 | -0.245 | 0.049 | -0.028 | -4.571e-3 | 4.258e-3 | 1.034e-4 | -2.087e-4 | 2.486e-7 | 3.501e-6 |
| 4 | -33.388 | 0.691 | -0.403 | -0.033 | 0.032 | 7.705e-4 | -2.195e-3 | 8.876e-5 | 5.974e-5 | -2.542e-6 | -5.837e-7 |
| 6 | -33.776 | 1.069 | -0.591 | 0.031 | 0.016 | 3.371e-3 | -2.520e-3 | 1.898e-4 | 3.815e-5 | -1.906e-6 | -4.215e-7 |
| 7 | -36.844 | 1.344 | -0.737 | 0.082 | 2.640e-4 | 7.377e-3 | -3.082e-3 | 1.906e-4 | 3.883e-5 | 6.734e-7 | -7.572e-7 |
| 8 | -39.093 | 1.609 | -0.879 | 0.134 | -0.015 | 0.010 | -3.413e-3 | 2.392e-4 | 2.688e-5 | 1.936e-6 | -7.740e-7 |
| 9 | -40.720 | 1.855 | -1.012 | 0.182 | -0.028 | 0.012 | -3.656e-3 | 3.202e-4 | 1.002e-5 | 2.011e-6 | -5.939e-7 |
| 10 | -41.976 | 2.079 | -1.132 | 0.226 | -0.039 | 0.014 | -3.891e-3 | 4.141e-4 | -5.938e-6 | 1.279e-6 | -3.393e-7 |
| 11 | -43.009 | 2.281 | -1.240 | 0.265 | -0.049 | 0.016 | -4.140e-3 | 5.047e-4 | -1.854e-5 | 2.051e-7 | -9.134e-8 |
| 12 | -43.887 | 2.435 | -1.322 | 0.294 | -0.056 | 0.017 | -4.350e-3 | 5.549e-4 | -2.527e-5 | 9.642e-9 | -1.103e-8 |
| 13 | -44.671 | 2.555 | -1.386 | 0.316 | -0.062 | 0.018 | -4.510e-3 | 5.771e-4 | -2.917e-5 | 5.526e-7 | -4.422e-8 |
| 14 | -45.384 | 2.626 | -1.423 | 0.329 | -0.065 | 0.018 | -4.614e-3 | 6.077e-4 | -3.246e-5 | 1.153e-7 | 3.389e-8 |

$v = 6$

| $v'$ | $a_0$ | $a_1$ | $a_2$ | $a_3$ | $a_4$ | $a_5$ | $a_6$ | $a_7$ | $a_8$ | $a_9$ | $a_{10}$ |
|---|---|---|---|---|---|---|---|---|---|---|---|
| 0 | -41.930 | -0.195 | -0.214 | -0.025 | -8.022e-3 | 5.868e-3 | 8.920e-4 | -4.242e-4 | -2.267e-5 | 9.411e-6 | 9.452e-8 |
| 1 | -40.242 | 0.465 | -0.340 | -0.037 | 0.025 | 2.055e-3 | -1.852e-3 | 1.760e-5 | 5.177e-5 | -1.266e-6 | -5.167e-7 |
| 2 | -39.294 | 0.504 | -0.347 | -0.038 | 0.026 | 2.045e-3 | -1.935e-3 | 1.980e-5 | 5.465e-5 | -1.315e-6 | -5.564e-7 |
| 3 | -33.913 | -0.399 | -0.144 | -8.851e-4 | -0.021 | 3.310e-3 | 2.300e-3 | -3.063e-4 | -9.341e-5 | 7.327e-6 | 1.414e-6 |
| 4 | -36.102 | 0.616 | -0.376 | -0.038 | 0.030 | 1.606e-3 | -2.155e-3 | 4.655e-5 | 6.124e-5 | -1.816e-6 | -6.354e-7 |
| 5 | -33.401 | 0.697 | -0.406 | -0.033 | 0.032 | 6.669e-4 | -2.188e-3 | 9.363e-5 | 5.909e-5 | -2.621e-6 | -5.706e-7 |
| 7 | -33.761 | 1.047 | -0.580 | 0.027 | 0.017 | 3.006e-3 | -2.469e-3 | 1.904e-4 | 3.795e-5 | -2.121e-6 | -3.935e-7 |
| 8 | -36.810 | 1.295 | -0.710 | 0.073 | 2.512e-3 | 6.613e-3 | -2.925e-3 | 1.981e-4 | 3.471e-5 | 1.969e-8 | -6.088e-7 |
| 9 | -39.037 | 1.532 | -0.839 | 0.118 | -0.010 | 0.010 | -3.351e-3 | 2.144e-4 | 3.294e-5 | 1.874e-6 | -8.402e-7 |
| 10 | -40.640 | 1.749 | -0.955 | 0.161 | -0.022 | 0.012 | -3.557e-3 | 2.790e-4 | 1.814e-5 | 2.181e-6 | -7.051e-7 |
| 11 | -41.872 | 1.944 | -1.060 | 0.200 | -0.033 | 0.013 | -3.737e-3 | 3.625e-4 | 2.533e-6 | 1.576e-6 | -4.603e-7 |
| 12 | -42.864 | 2.094 | -1.140 | 0.229 | -0.040 | 0.014 | -3.911e-3 | 4.166e-4 | -6.421e-6 | 1.361e-6 | -3.450e-7 |
| 13 | -43.712 | 2.212 | -1.203 | 0.252 | -0.046 | 0.015 | -4.053e-3 | 4.658e-4 | -1.368e-5 | 8.977e-7 | -2.196e-7 |
| 14 | -44.456 | 2.281 | -1.241 | 0.266 | -0.049 | 0.016 | -4.142e-3 | 5.048e-4 | -1.857e-5 | 2.096e-7 | -9.164e-8 |

$v = 7$

| $v'$ | $a_0$ | $a_1$ | $a_2$ | $a_3$ | $a_4$ | $a_5$ | $a_6$ | $a_7$ | $a_8$ | $a_9$ | $a_{10}$ |
|---|---|---|---|---|---|---|---|---|---|---|---|
| 0 | -42.829 | -0.347 | -0.183 | -0.057 | 0.014 | 4.842e-3 | -1.235e-3 | -1.864e-4 | 4.994e-5 | 2.878e-6 | -7.839e-7 |
| 1 | -41.033 | 0.443 | -0.337 | -0.036 | 0.025 | 2.038e-3 | -1.807e-3 | 1.758e-5 | 5.017e-5 | -1.261e-6 | -4.943e-7 |
| 2 | -40.280 | 0.475 | -0.341 | -0.037 | 0.026 | 2.058e-3 | -1.874e-3 | 1.790e-5 | 5.253e-5 | -1.274e-6 | -5.272e-7 |
| 3 | -34.961 | -0.496 | -0.099 | -0.055 | 3.610e-3 | 4.994e-3 | -3.704e-4 | -2.036e-4 | 1.162e-5 | 3.323e-6 | -1.123e-7 |
| 4 | -38.028 | 0.564 | -0.360 | -0.039 | 0.028 | 1.901e-3 | -2.062e-3 | 2.971e-5 | 5.878e-5 | -1.508e-6 | -6.102e-7 |
| 5 | -36.125 | 0.626 | -0.379 | -0.038 | 0.030 | 1.527e-3 | -2.169e-3 | 5.085e-5 | 6.148e-5 | -1.893e-6 | -6.361e-7 |
| 6 | -33.415 | 0.704 | -0.409 | -0.032 | 0.032 | 5.510e-4 | -2.178e-3 | 9.899e-5 | 5.827e-5 | -2.708e-6 | -5.545e-7 |
| 8 | -33.745 | 1.025 | -0.568 | 0.023 | 0.018 | 2.618e-3 | -2.411e-3 | 1.913e-4 | 3.757e-5 | -2.345e-6 | -3.617e-7 |
| 9 | -36.775 | 1.246 | -0.684 | 0.064 | 5.386e-3 | 5.966e-3 | -2.848e-3 | 1.944e-4 | 3.598e-5 | -3.627e-7 | -5.782e-7 |
| 10 | -38.979 | 1.454 | -0.797 | 0.103 | -5.727e-3 | 8.761e-3 | -3.271e-3 | 1.970e-4 | 3.762e-5 | 1.576e-6 | -8.613e-7 |
| 11 | -40.561 | 1.643 | -0.898 | 0.140 | -0.016 | 0.011 | -3.462e-3 | 2.436e-4 | 2.589e-5 | 2.159e-6 | -7.917e-7 |
| 12 | -41.754 | 1.788 | -0.976 | 0.170 | -0.024 | 0.012 | -3.564e-3 | 3.105e-4 | 1.204e-5 | 1.612e-6 | -5.638e-7 |
| 13 | -42.717 | 1.902 | -1.037 | 0.192 | -0.030 | 0.013 | -3.680e-3 | 3.562e-4 | 3.820e-6 | 1.338e-6 | -4.428e-7 |
| 14 | -43.526 | 1.970 | -1.074 | 0.206 | -0.034 | 0.013 | -3.744e-3 | 3.967e-4 | -2.545e-6 | 6.851e-7 | -2.971e-7 |

$v = 8$

| $v'$ | $a_0$ | $a_1$ | $a_2$ | $a_3$ | $a_4$ | $a_5$ | $a_6$ | $a_7$ | $a_8$ | $a_9$ | $a_{10}$ |
|---|---|---|---|---|---|---|---|---|---|---|---|
| 0 | -43.889 | -0.261 | -0.267 | -0.062 | 0.028 | 6.895e-3 | -2.991e-3 | -3.513e-4 | 1.530e-4 | 6.405e-6 | -2.834e-6 |
| 1 | -41.717 | 0.426 | -0.335 | -0.036 | 0.024 | 2.016e-3 | -1.774e-3 | 1.810e-5 | 4.896e-5 | -1.266e-6 | -4.774e-7 |
| 2 | -41.075 | 0.454 | -0.338 | -0.037 | 0.025 | 2.048e-3 | -1.829e-3 | 1.749e-5 | 5.092e-5 | -1.261e-6 | -5.049e-7 |
| 3 | -36.112 | -0.390 | -0.163 | -0.052 | 7.890e-3 | 6.378e-3 | -1.023e-3 | -3.592e-4 | 6.633e-5 | 7.027e-6 | -1.412e-6 |
| 4 | -39.364 | 0.527 | -0.351 | -0.039 | 0.027 | 2.011e-3 | -1.983e-3 | 2.250e-5 | 5.629e-5 | -1.370e-6 | -5.784e-7 |
| 5 | -38.059 | 0.576 | -0.364 | -0.039 | 0.029 | 1.850e-3 | -2.085e-3 | 3.280e-5 | 5.947e-5 | -1.565e-6 | -6.184e-7 |
| 6 | -36.150 | 0.637 | -0.383 | -0.037 | 0.031 | 1.430e-3 | -2.182e-3 | 5.595e-5 | 6.163e-5 | -1.983e-6 | -6.349e-7 |
| 7 | -33.430 | 0.712 | -0.412 | -0.031 | 0.032 | 4.211e-4 | -2.163e-3 | 1.049e-4 | 5.724e-5 | -2.801e-6 | -5.349e-7 |
| 9 | -33.730 | 1.002 | -0.556 | 0.019 | 0.020 | 2.208e-3 | -2.347e-3 | 1.926e-4 | 3.696e-5 | -2.579e-6 | -3.256e-7 |
| 10 | -36.740 | 1.195 | -0.657 | 0.054 | 8.347e-3 | 5.276e-3 | -2.766e-3 | 1.916e-4 | 3.707e-5 | -7.731e-7 | -5.425e-7 |
| 11 | -38.924 | 1.377 | -0.755 | 0.088 | -1.609e-3 | 7.784e-3 | -3.131e-3 | 1.937e-4 | 3.791e-5 | 9.141e-7 | -7.760e-7 |
| 12 | -40.468 | 1.518 | -0.831 | 0.116 | -9.416e-3 | 9.404e-3 | -3.330e-3 | 2.128e-4 | 3.333e-5 | 1.775e-6 | -8.322e-7 |
| 13 | -41.634 | 1.629 | -0.890 | 0.137 | -0.016 | 0.010 | -3.441e-3 | 2.417e-4 | 2.632e-5 | 2.065e-6 | -7.843e-7 |
| 14 | -42.560 | 1.694 | -0.926 | 0.151 | -0.019 | 0.011 | -3.482e-3 | 2.717e-4 | 1.974e-5 | 1.845e-6 | -6.790e-7 |

$v = 9$

| $v'$ | $a_0$ | $a_1$ | $a_2$ | $a_3$ | $a_4$ | $a_5$ | $a_6$ | $a_7$ | $a_8$ | $a_9$ | $a_{10}$ |
|---|---|---|---|---|---|---|---|---|---|---|---|
| 0 | -44.173 | -0.790 | -0.042 | -0.055 | -4.342e-3 | 7.520e-3 | 1.273e-4 | -3.787e-4 | -6.674e-6 | 6.725e-6 | 2.096e-7 |
| 1 | -42.345 | 0.412 | -0.333 | -0.035 | 0.024 | 1.992e-3 | -1.748e-3 | 1.882e-5 | 4.804e-5 | -1.276e-6 | -4.645e-7 |
| 2 | -41.763 | 0.437 | -0.336 | -0.036 | 0.025 | 2.031e-3 | -1.795e-3 | 1.772e-5 | 4.971e-5 | -1.262e-6 | -4.879e-7 |
| 3 | -36.954 | -0.428 | -0.192 | -0.052 | 0.022 | 2.083e-3 | -1.603e-3 | -2.230e-5 | 5.642e-5 | -1.881e-7 | -7.960e-7 |
| 4 | -40.361 | 0.499 | -0.346 | -0.038 | 0.026 | 2.049e-3 | -1.924e-3 | 1.934e-5 | 5.428e-5 | -1.306e-6 | -5.514e-7 |
| 5 | -39.402 | 0.540 | -0.354 | -0.039 | 0.028 | 1.981e-3 | -2.011e-3 | 2.462e-5 | 5.720e-5 | -1.411e-6 | -5.903e-7 |
| 6 | -38.093 | 0.589 | -0.367 | -0.039 | 0.029 | 1.783e-3 | -2.110e-3 | 3.671e-5 | 6.016e-5 | -1.638e-6 | -6.260e-7 |

| $v'$ | $a_0$ | $a_1$ | $a_2$ | $a_3$ | $a_4$ | $a_5$ | $a_6$ | $a_7$ | $a_8$ | $a_9$ | $a_{10}$ |
|---|---|---|---|---|---|---|---|---|---|---|---|
| 7 | -36.176 | 0.648 | -0.387 | -0.037 | 0.031 | 1.313e-3 | -2.193e-3 | 6.203e-5 | 6.164e-5 | -2.089e-6 | -6.307e-7 |
| 8 | -33.445 | 0.720 | -0.416 | -0.030 | 0.032 | 2.758e-4 | -2.143e-3 | 1.113e-4 | 5.594e-5 | -2.901e-6 | -5.111e-7 |
| 10 | -33.714 | 0.978 | -0.544 | 0.015 | 0.021 | 1.773e-3 | -2.275e-3 | 1.941e-4 | 3.607e-5 | -2.821e-6 | -2.843e-7 |
| 11 | -36.707 | 1.147 | -0.632 | 0.045 | 0.011 | 4.584e-3 | -2.680e-3 | 1.900e-4 | 3.783e-5 | -1.186e-6 | -5.028e-7 |
| 12 | -38.857 | 1.281 | -0.703 | 0.071 | 3.313e-3 | 6.435e-3 | -2.904e-3 | 1.970e-4 | 3.508e-5 | -8.557e-8 | -6.006e-7 |
| 13 | -40.375 | 1.389 | -0.762 | 0.090 | -2.062e-3 | 7.999e-3 | -3.182e-3 | 1.890e-4 | 3.987e-5 | 1.150e-6 | -8.353e-7 |
| 14 | -41.505 | 1.453 | -0.796 | 0.103 | -5.680e-3 | 8.751e-3 | -3.269e-3 | 1.969e-4 | 3.765e-5 | 1.570e-6 | -8.609e-7 |

$v = 10$

| $v'$ | $a_0$ | $a_1$ | $a_2$ | $a_3$ | $a_4$ | $a_5$ | $a_6$ | $a_7$ | $a_8$ | $a_9$ | $a_{10}$ |
|---|---|---|---|---|---|---|---|---|---|---|---|
| 0 | -45.488 | -0.396 | -0.293 | 1.699e-3 | 0.012 | 1.633e-3 | -1.221e-3 | -2.293e-5 | 4.575e-5 | 5.160e-7 | -7.664e-7 |
| 1 | -42.950 | 0.402 | -0.332 | -0.035 | 0.024 | 1.971e-3 | -1.729e-3 | 1.957e-5 | 4.734e-5 | -1.288e-6 | -4.547e-7 |
| 2 | -42.396 | 0.423 | -0.334 | -0.036 | 0.024 | 2.012e-3 | -1.769e-3 | 1.822e-5 | 4.879e-5 | -1.268e-6 | -4.750e-7 |
| 3 | -41.812 | 0.449 | -0.337 | -0.037 | 0.025 | 2.044e-3 | -1.819e-3 | 1.751e-5 | 5.057e-5 | -1.261e-6 | -5.000e-7 |
| 4 | -41.167 | 0.478 | -0.342 | -0.038 | 0.026 | 2.058e-3 | -1.880e-3 | 1.801e-5 | 5.273e-5 | -1.277e-6 | -5.301e-7 |
| 5 | -40.406 | 0.513 | -0.348 | -0.038 | 0.027 | 2.034e-3 | -1.954e-3 | 2.073e-5 | 5.531e-5 | -1.334e-6 | -5.653e-7 |
| 6 | -39.444 | 0.554 | -0.358 | -0.039 | 0.028 | 1.937e-3 | -2.042e-3 | 2.750e-5 | 5.817e-5 | -1.466e-6 | -6.027e-7 |
| 7 | -38.128 | 0.603 | -0.372 | -0.039 | 0.030 | 1.695e-3 | -2.135e-3 | 4.168e-5 | 6.081e-5 | -1.728e-6 | -6.322e-7 |
| 8 | -36.204 | 0.661 | -0.391 | -0.036 | 0.031 | 1.170e-3 | -2.201e-3 | 6.929e-5 | 6.141e-5 | -2.215e-6 | -6.224e-7 |
| 9 | -33.462 | 0.728 | -0.420 | -0.028 | 0.032 | 1.134e-4 | -2.116e-3 | 1.183e-4 | 5.431e-5 | -3.006e-6 | -4.820e-7 |
| 11 | -33.700 | 0.957 | -0.533 | 0.012 | 0.022 | 1.370e-3 | -2.204e-3 | 1.956e-4 | 3.501e-5 | -3.038e-6 | -2.434e-7 |
| 12 | -36.662 | 1.082 | -0.598 | 0.033 | 0.015 | 3.582e-3 | -2.549e-3 | 1.896e-4 | 3.820e-5 | -1.782e-6 | -4.369e-7 |
| 13 | -38.788 | 1.183 | -0.651 | 0.052 | 9.050e-3 | 5.109e-3 | -2.745e-3 | 1.911e-4 | 3.729e-5 | -8.729e-7 | -5.333e-7 |
| 14 | -40.273 | 1.245 | -0.684 | 0.064 | 5.433e-3 | 5.955e-3 | -2.847e-3 | 1.944e-4 | 3.600e-5 | -3.693e-7 | -5.776e-7 |

$v = 11$

| $v'$ | $a_0$ | $a_1$ | $a_2$ | $a_3$ | $a_4$ | $a_5$ | $a_6$ | $a_7$ | $a_8$ | $a_9$ | $a_{10}$ |
|---|---|---|---|---|---|---|---|---|---|---|---|
| 0 | -44.105 | 0.376 | -0.330 | -0.034 | 0.023 | 1.909e-3 | -1.684e-3 | 2.200e-5 | 4.573e-5 | -1.328e-6 | -4.324e-7 |
| 1 | -43.551 | 0.393 | -0.331 | -0.034 | 0.023 | 1.952e-3 | -1.713e-3 | 2.028e-5 | 4.679e-5 | -1.299e-6 | -4.471e-7 |
| 2 | -43.004 | 0.413 | -0.333 | -0.035 | 0.024 | 1.993e-3 | -1.749e-3 | 1.879e-5 | 4.808e-5 | -1.276e-6 | -4.651e-7 |
| 3 | -42.449 | 0.436 | -0.336 | -0.036 | 0.024 | 2.030e-3 | -1.793e-3 | 1.775e-5 | 4.964e-5 | -1.262e-6 | -4.870e-7 |
| 4 | -41.864 | 0.462 | -0.339 | -0.037 | 0.025 | 2.054e-3 | -1.846e-3 | 1.755e-5 | 5.154e-5 | -1.264e-6 | -5.136e-7 |
| 5 | -41.217 | 0.493 | -0.344 | -0.038 | 0.026 | 2.054e-3 | -1.910e-3 | 1.884e-5 | 5.381e-5 | -1.295e-6 | -5.449e-7 |
| 6 | -40.453 | 0.529 | -0.352 | -0.039 | 0.027 | 2.007e-3 | -1.987e-3 | 2.277e-5 | 5.642e-5 | -1.375e-6 | -5.801e-7 |
| 7 | -39.487 | 0.570 | -0.362 | -0.039 | 0.029 | 1.875e-3 | -2.074e-3 | 3.130e-5 | 5.916e-5 | -1.538e-6 | -6.147e-7 |
| 8 | -38.165 | 0.619 | -0.376 | -0.038 | 0.030 | 1.584e-3 | -2.160e-3 | 4.777e-5 | 6.132e-5 | -1.838e-6 | -6.357e-7 |
| 9 | -36.231 | 0.674 | -0.397 | -0.035 | 0.032 | 1.006e-3 | -2.203e-3 | 7.741e-5 | 6.090e-5 | -2.353e-6 | -6.090e-7 |
| 10 | -33.476 | 0.736 | -0.423 | -0.027 | 0.032 | -4.478e-5 | -2.085e-3 | 1.248e-4 | 5.255e-5 | -3.101e-6 | -4.513e-7 |
| 12 | -33.674 | 0.917 | -0.513 | 5.870e-3 | 0.024 | 5.853e-4 | -2.049e-3 | 1.982e-4 | 3.224e-5 | -3.424e-6 | -1.564e-7 |
| 13 | -36.614 | 1.012 | -0.561 | 0.021 | 0.019 | 2.394e-3 | -2.377e-3 | 1.920e-4 | 3.727e-5 | -2.474e-6 | -3.423e-7 |
| 14 | -38.710 | 1.070 | -0.592 | 0.031 | 0.016 | 3.384e-3 | -2.522e-3 | 1.898e-4 | 3.816e-5 | -1.899e-6 | -4.225e-7 |

$v = 12$

| $v'$ | $a_0$ | $a_1$ | $a_2$ | $a_3$ | $a_4$ | $a_5$ | $a_6$ | $a_7$ | $a_8$ | $a_9$ | $a_{10}$ |
|---|---|---|---|---|---|---|---|---|---|---|---|
| 0 | -44.738 | 0.370 | -0.330 | -0.033 | 0.023 | 1.896e-3 | -1.676e-3 | 2.256e-5 | 4.543e-5 | -1.337e-6 | -4.284e-7 |
| 1 | -44.168 | 0.387 | -0.331 | -0.034 | 0.023 | 1.938e-3 | -1.703e-3 | 2.083e-5 | 4.641e-5 | -1.308e-6 | -4.418e-7 |
| 2 | -43.614 | 0.405 | -0.332 | -0.035 | 0.024 | 1.979e-3 | -1.736e-3 | 1.928e-5 | 4.759e-5 | -1.283e-6 | -4.583e-7 |
| 3 | -43.067 | 0.427 | -0.335 | -0.036 | 0.024 | 2.017e-3 | -1.776e-3 | 1.807e-5 | 4.902e-5 | -1.266e-6 | -4.783e-7 |
| 4 | -42.511 | 0.451 | -0.338 | -0.037 | 0.025 | 2.046e-3 | -1.824e-3 | 1.750e-5 | 5.074e-5 | -1.261e-6 | -5.024e-7 |
| 5 | -41.926 | 0.479 | -0.342 | -0.038 | 0.026 | 2.058e-3 | -1.882e-3 | 1.805e-5 | 5.280e-5 | -1.278e-6 | -5.310e-7 |
| 6 | -41.277 | 0.511 | -0.348 | -0.038 | 0.027 | 2.036e-3 | -1.951e-3 | 2.054e-5 | 5.519e-5 | -1.331e-6 | -5.636e-7 |
| 7 | -40.512 | 0.549 | -0.356 | -0.039 | 0.028 | 1.955e-3 | -2.030e-3 | 2.632e-5 | 5.780e-5 | -1.444e-6 | -5.980e-7 |
| 8 | -39.541 | 0.591 | -0.368 | -0.039 | 0.029 | 1.768e-3 | -2.115e-3 | 3.754e-5 | 6.029e-5 | -1.653e-6 | -6.273e-7 |
| 9 | -38.213 | 0.640 | -0.384 | -0.037 | 0.031 | 1.399e-3 | -2.186e-3 | 5.756e-5 | 6.165e-5 | -2.011e-6 | -6.340e-7 |
| 10 | -36.271 | 0.693 | -0.404 | -0.033 | 0.032 | 7.308e-4 | -2.193e-3 | 9.064e-5 | 5.950e-5 | -2.573e-6 | -5.788e-7 |
| 11 | -33.505 | 0.751 | -0.431 | -0.024 | 0.032 | -3.717e-4 | -2.005e-3 | 1.374e-4 | 4.833e-5 | -3.268e-6 | -3.806e-7 |
| 13 | -33.656 | 0.890 | -0.499 | 2.121e-3 | 0.025 | 3.889e-5 | -1.926e-3 | 1.989e-4 | 2.977e-5 | -3.649e-6 | -9.122e-8 |
| 14 | -36.570 | 0.945 | -0.527 | 0.010 | 0.023 | 1.144e-3 | -2.162e-3 | 1.965e-4 | 3.430e-5 | -3.154e-6 | -2.193e-7 |

$v = 13$

| $v'$ | $a_0$ | $a_1$ | $a_2$ | $a_3$ | $a_4$ | $a_5$ | $a_6$ | $a_7$ | $a_8$ | $a_9$ | $a_{10}$ |
|---|---|---|---|---|---|---|---|---|---|---|---|
| 0 | -45.397 | 0.367 | -0.330 | -0.033 | 0.023 | 1.885e-3 | -1.670e-3 | 2.299e-5 | 4.522e-5 | -1.345e-6 | -4.255e-7 |
| 1 | -44.805 | 0.382 | -0.331 | -0.034 | 0.023 | 1.927e-3 | -1.696e-3 | 2.127e-5 | 4.614e-5 | -1.316e-6 | -4.381e-7 |
| 2 | -44.236 | 0.400 | -0.332 | -0.035 | 0.024 | 1.968e-3 | -1.726e-3 | 1.967e-5 | 4.725e-5 | -1.290e-6 | -4.535e-7 |
| 3 | -43.683 | 0.420 | -0.334 | -0.036 | 0.024 | 2.007e-3 | -1.763e-3 | 1.837e-5 | 4.858e-5 | -1.270e-6 | -4.721e-7 |
| 4 | -43.136 | 0.443 | -0.337 | -0.036 | 0.025 | 2.039e-3 | -1.808e-3 | 1.758e-5 | 5.018e-5 | -1.260e-6 | -4.945e-7 |
| 5 | -42.581 | 0.469 | -0.340 | -0.037 | 0.025 | 2.057e-3 | -1.861e-3 | 1.770e-5 | 5.208e-5 | -1.269e-6 | -5.210e-7 |
| 6 | -41.995 | 0.499 | -0.346 | -0.038 | 0.026 | 2.049e-3 | -1.925e-3 | 1.936e-5 | 5.430e-5 | -1.306e-6 | -5.516e-7 |
| 7 | -41.345 | 0.534 | -0.353 | -0.039 | 0.027 | 1.996e-3 | -1.998e-3 | 2.358e-5 | 5.678e-5 | -1.391e-6 | -5.848e-7 |
| 8 | -40.577 | 0.573 | -0.363 | -0.039 | 0.029 | 1.865e-3 | -2.079e-3 | 3.192e-5 | 5.929e-5 | -1.549e-6 | -6.163e-7 |
| 9 | -39.602 | 0.616 | -0.376 | -0.038 | 0.030 | 1.603e-3 | -2.156e-3 | 4.676e-5 | 6.125e-5 | -1.820e-6 | -6.354e-7 |
| 10 | -38.267 | 0.664 | -0.393 | -0.036 | 0.031 | 1.132e-3 | -2.202e-3 | 7.118e-5 | 6.132e-5 | -2.247e-6 | -6.196e-7 |
| 11 | -36.316 | 0.716 | -0.414 | -0.030 | 0.032 | 3.429e-4 | -2.153e-3 | 1.083e-4 | 5.656e-5 | -2.855e-6 | -5.223e-7 |
| 12 | -33.526 | 0.762 | -0.436 | -0.021 | 0.032 | -6.121e-4 | -1.930e-3 | 1.456e-4 | 4.465e-5 | -3.357e-6 | -3.219e-7 |
| 14 | -33.633 | 0.852 | -0.480 | -2.582e-3 | 0.027 | -6.851e-4 | -1.732e-3 | 1.946e-4 | 2.603e-5 | -3.817e-6 | -8.126e-9 |

$v = 14$

| $v'$ | $a_0$ | $a_1$ | $a_2$ | $a_3$ | $a_4$ | $a_5$ | $a_6$ | $a_7$ | $a_8$ | $a_9$ | $a_{10}$ |
|---|---|---|---|---|---|---|---|---|---|---|---|
| 0 | -46.091 | 0.365 | -0.329 | -0.033 | 0.023 | 1.880e-3 | -1.666e-3 | 2.325e-5 | 4.510e-5 | -1.349e-6 | -4.239e-7 |
| 1 | -45.472 | 0.380 | -0.330 | -0.034 | 0.023 | 1.921e-3 | -1.691e-3 | 2.153e-5 | 4.599e-5 | -1.320e-6 | -4.360e-7 |
| 2 | -44.881 | 0.397 | -0.332 | -0.035 | 0.024 | 1.962e-3 | -1.721e-3 | 1.992e-5 | 4.705e-5 | -1.294e-6 | -4.508e-7 |
| 3 | -44.314 | 0.417 | -0.334 | -0.035 | 0.024 | 2.000e-3 | -1.756e-3 | 1.856e-5 | 4.833e-5 | -1.273e-6 | -4.686e-7 |
| 4 | -43.762 | 0.439 | -0.336 | -0.036 | 0.025 | 2.033e-3 | -1.799e-3 | 1.767e-5 | 4.986e-5 | -1.261e-6 | -4.901e-7 |
| 5 | -43.217 | 0.464 | -0.340 | -0.037 | 0.025 | 2.055e-3 | -1.850e-3 | 1.758e-5 | 5.168e-5 | -1.265e-6 | -5.155e-7 |
| 6 | -42.663 | 0.493 | -0.344 | -0.038 | 0.026 | 2.054e-3 | -1.910e-3 | 1.884e-5 | 5.380e-5 | -1.295e-6 | -5.448e-7 |

| $v'$ | $a_0$ | $a_1$ | $a_2$ | $a_3$ | $a_4$ | $a_5$ | $a_6$ | $a_7$ | $a_8$ | $a_9$ | $a_{10}$ |
|---|---|---|---|---|---|---|---|---|---|---|---|
| 7 | -42.077 | 0.525 | -0.351 | -0.039 | 0.027 | 2.014e-3 | -1.980e-3 | 2.231e-5 | 5.619e-5 | -1.366e-6 | -5.771e-7 |
| 8 | -41.426 | 0.562 | -0.360 | -0.039 | 0.028 | 1.908e-3 | -2.058e-3 | 2.930e-5 | 5.867e-5 | -1.500e-6 | -6.089e-7 |
| 9 | -40.655 | 0.604 | -0.372 | -0.039 | 0.030 | 1.694e-3 | -2.136e-3 | 4.174e-5 | 6.081e-5 | -1.729e-6 | -6.323e-7 |
| 10 | -39.675 | 0.649 | -0.387 | -0.037 | 0.031 | 1.311e-3 | -2.194e-3 | 6.214e-5 | 6.163e-5 | -2.091e-6 | -6.306e-7 |
| 11 | -38.335 | 0.697 | -0.406 | -0.033 | 0.032 | 6.707e-4 | -2.189e-3 | 9.345e-5 | 5.912e-5 | -2.619e-6 | -5.711e-7 |
| 12 | -36.362 | 0.740 | -0.426 | -0.026 | 0.032 | -1.365e-4 | -2.065e-3 | 1.285e-4 | 5.145e-5 | -3.152e-6 | -4.325e-7 |
| 13 | -33.555 | 0.777 | -0.444 | -0.017 | 0.031 | -9.397e-4 | -1.792e-3 | 1.537e-4 | 3.852e-5 | -3.386e-6 | -2.313e-7 |

$$k_{34}(v,v') = \begin{cases} \exp\left(b_1(1.6T_h)^{-b_2} + b_3(1.6T_h)^{-b_4} + b_5\exp(-b_6\ln(1.6T_h)^2)\right)\times 10^{-6}, \\ \exp\left(\sum_{i=0}^{10} a_i T_h^i\right). \end{cases}$$

$v=0$

| $v'$ | $b_1$ | $b_2$ | $b_3$ | $b_4$ | $b_5$ | $b_6$ |
|---|---|---|---|---|---|---|
| 0 | -22.1030 | 0.0251 | -121.1300 | 1.0583 | -6.5889 | 0.2873 |
| 1 | -124.8000 | 1.0277 | -21.1960 | 0.0123 | -3.4388 | 0.2746 |
| 2 | -121.3500 | 1.0563 | -22.6420 | 0.0241 | -6.3281 | 0.2890 |
| 3 | -22.1280 | 0.0193 | -122.4600 | 1.0473 | -5.4407 | 0.2831 |
| 4 | -22.1730 | 0.0195 | -122.4800 | 1.0471 | -5.4929 | 0.2779 |
| 5 | -22.4920 | 0.0218 | -121.6000 | 1.0545 | -6.1702 | 0.2887 |
| 6 | -21.5540 | 0.0125 | -124.3400 | 1.0319 | -4.1247 | 0.2553 |
| 7 | -125.5000 | 1.0231 | -21.0430 | 6.3836e-3 | -3.6812 | 0.2098 |
| 8 | -20.3250 | -1.1112e-3 | -127.6100 | 1.0059 | -1.9403 | 0.1603 |
| 9 | -20.8490 | 2.8543e-3 | -127.2400 | 1.0081 | -1.8860 | 0.1951 |
| 10 | -127.7800 | 1.0030 | -21.0110 | 4.0391e-3 | -0.8570 | 0.2796 |
| 11 | -21.1640 | 4.7984e-3 | -127.3900 | 1.0061 | -1.0805 | 0.3118 |
| 12 | -126.6900 | 1.0117 | -21.4710 | 7.0803e-3 | -1.3133 | 0.3942 |
| 13 | -118.8200 | 1.0733 | -25.0500 | 0.0370 | -8.4809 | 0.3313 |
| 14 | -25.4110 | 0.0354 | -118.7700 | 1.0739 | -8.6323 | 0.3331 |

$v=1$

| $v'$ | $a_0$ | $a_1$ | $a_2$ | $a_3$ | $a_4$ | $a_5$ | $a_6$ | $a_7$ | $a_8$ | $a_9$ | $a_{10}$ |
|---|---|---|---|---|---|---|---|---|---|---|---|
| 0 | -44.4659 | 10.9935 | -5.5318 | 1.7971 | -0.4059 | 0.0854 | -0.0205 | 3.1518e-3 | 2.7523e-5 | -7.1266e-5 | 6.0935e-6 |
| 1 | -43.5157 | 10.5148 | -5.3211 | 1.7454 | -0.3937 | 0.0802 | -0.0189 | 3.0279e-3 | -3.0168e-5 | -5.6637e-5 | 5.0708e-6 |
| 2 | -43.4968 | 10.3170 | -5.3433 | 1.8111 | -0.4230 | 0.0844 | -0.0177 | 2.4700e-3 | 2.5349e-5 | -5.1949e-5 | 4.2547e-6 |
| 3 | -44.5171 | 11.0973 | -5.6295 | 1.8489 | -0.4193 | 0.0845 | -0.0193 | 3.1644e-3 | -9.4044e-5 | -4.4245e-5 | 4.2386e-6 |
| 4 | -44.8043 | 11.1110 | -5.6264 | 1.8499 | -0.4209 | 0.0844 | -0.0192 | 3.1644e-3 | -1.0576e-4 | -4.2303e-5 | 4.1530e-6 |
| 9 | -44.8507 | 10.9023 | -5.5280 | 1.8512 | -0.4430 | 0.0909 | -0.0180 | 2.3252e-3 | 1.7439e-5 | -4.2826e-5 | 3.4038e-6 |
| 10 | -44.8507 | 10.9023 | -5.5280 | 1.8512 | -0.4430 | 0.0909 | -0.0180 | 2.3252e-3 | 1.7439e-5 | -4.2826e-5 | 3.4038e-6 |

$v = 2$

| $v'$ | $a_0$ | $a_1$ | $a_2$ | $a_3$ | $a_4$ | $a_5$ | $a_6$ | $a_7$ | $a_8$ | $a_9$ | $a_{10}$ |
|---|---|---|---|---|---|---|---|---|---|---|---|
| 0 | -43.5984 | 10.1204 | -5.1013 | 1.6522 | -0.3678 | 0.0785 | -0.0196 | 2.9494e-3 | 6.6130e-5 | -7.3665e-5 | 6.0766e-6 |
| 2 | -43.2092 | 10.3285 | -5.2363 | 1.7256 | -0.3881 | 0.0765 | -0.0179 | 3.1323e-3 | -1.2376e-4 | -4.0459e-5 | 4.1440e-6 |
| 3 | -43.5022 | 10.0946 | -5.0910 | 1.6647 | -0.3727 | 0.0752 | -0.0183 | 3.1182e-3 | -6.4879e-5 | -5.4265e-5 | 5.1710e-6 |
| 4 | -43.8875 | 10.1946 | -5.1248 | 1.6732 | -0.3780 | 0.0769 | -0.0182 | 3.0218e-3 | -7.0696e-5 | -4.8503e-5 | 4.6101e-6 |
| 5 | -43.8875 | 10.1946 | -5.1248 | 1.6732 | -0.3780 | 0.0769 | -0.0182 | 3.0218e-3 | -7.0696e-5 | -4.8503e-5 | 4.6101e-6 |
| 7 | -43.8875 | 10.1946 | -5.1248 | 1.6732 | -0.3780 | 0.0769 | -0.0182 | 3.0218e-3 | -7.0696e-5 | -4.8503e-5 | 4.6101e-6 |
| 6 | -43.7636 | 10.1205 | -5.1086 | 1.6890 | -0.3869 | 0.0759 | -0.0166 | 2.8061e-3 | -1.5711e-4 | -2.2042e-5 | 2.6162e-6 |

$v = 3$

| $v'$ | $b_1$ | $b_2$ | $b_3$ | $b_4$ | $b_5$ | $b_6$ |
|---|---|---|---|---|---|---|
| 0 | -23.5360 | 0.0358 | -95.8550 | 1.0939 | -8.5219 | 0.2800 |
| 1 | -21.3750 | 0.0185 | -101.1000 | 1.0380 | -3.7274 | 0.2687 |
| 2 | -20.6770 | 0.0119 | -102.8300 | 1.0205 | -1.8410 | 0.3417 |
| 3 | -20.1960 | 0.0108 | -103.9100 | 1.0091 | -0.9341 | 0.3768 |
| 4 | -20.3780 | 0.0108 | -103.9800 | 1.0081 | -0.7401 | 0.4650 |
| 5 | -20.5020 | 0.0107 | -104.0700 | 1.0072 | -0.6866 | 0.4684 |
| 6 | -20.8870 | 0.0116 | -103.2300 | 1.0159 | -1.2099 | 0.4534 |
| 7 | -21.2260 | 0.0120 | -102.8500 | 1.0198 | -1.4422 | 0.4495 |
| 8 | -21.1990 | 9.7491e-3 | -104.1700 | 1.0064 | -0.7561 | 0.3813 |
| 9 | -21.5940 | 0.0101 | -103.6700 | 1.0119 | -2.1076 | 0.1865 |
| 10 | -24.8750 | 0.0357 | -97.6300 | 1.0687 | -7.1231 | 0.2896 |
| 11 | -19.3380 | -0.0146 | -113.0400 | 0.9533 | -3.0441 | 1.2462 |
| 12 | -19.1750 | -0.0161 | -113.1100 | 0.9602 | -2.9959 | 1.1425 |
| 13 | -20.8380 | -4.2019e-3 | -114.9800 | 0.9932 | -1.4961 | 0.7835 |
| 14 | -22.4250 | 3.5418e-3 | -116.7400 | 0.9913 | -0.3656 | 0.5245 |

$v = 4$

| $v'$ | $a_0$ | $a_1$ | $a_2$ | $a_3$ | $a_4$ | $a_5$ | $a_6$ | $a_7$ | $a_8$ | $a_9$ | $a_{10}$ |
|---|---|---|---|---|---|---|---|---|---|---|---|
| 0 | -42.4612 | 9.1602 | -4.5977 | 1.5113 | -0.3474 | 0.0734 | -0.0168 | 2.2934e-3 | 7.2209e-5 | -5.6647e-5 | 4.3695e-6 |
| 1 | -42.8081 | 9.4760 | -4.7802 | 1.5795 | -0.3573 | 0.0702 | -0.0164 | 2.8501e-3 | -8.6599e-5 | -4.5012e-5 | 4.4673e-6 |
| 2 | -42.8721 | 9.4547 | -4.7558 | 1.5850 | -0.3676 | 0.0705 | -0.0148 | 2.5726e-3 | -1.7132e-4 | -1.6111e-5 | 2.2119e-6 |
| 3 | -42.3009 | 9.3317 | -4.7296 | 1.5781 | -0.3604 | 0.0686 | -0.0152 | 2.7415e-3 | -1.4953e-4 | -2.8212e-5 | 3.2583e-6 |
| 4 | -41.9583 | 9.2451 | -4.6324 | 1.5172 | -0.3409 | 0.0678 | -0.0164 | 2.8796e-3 | -7.9738e-5 | -4.7691e-5 | 4.7120e-6 |
| 5 | -42.2472 | 9.1685 | -4.6459 | 1.5565 | -0.3572 | 0.0668 | -0.0143 | 2.6528e-3 | -1.9512e-4 | -1.5541e-5 | 2.3781e-6 |
| 6 | -42.8580 | 9.4446 | -4.7216 | 1.5594 | -0.3598 | 0.0707 | -0.0156 | 2.6545e-3 | -1.1028e-4 | -3.3746e-5 | 3.5465e-6 |

$v = 5$

| $v'$ | $a_0$ | $a_1$ | $a_2$ | $a_3$ | $a_4$ | $a_5$ | $a_6$ | $a_7$ | $a_8$ | $a_9$ | $a_{10}$ |
|---|---|---|---|---|---|---|---|---|---|---|---|
| 0 | -42.2248 | 9.0246 | -4.4875 | 1.4440 | -0.3207 | 0.0692 | -0.0180 | 2.7657e-3 | 9.5896e-5 | -8.1149e-5 | 6.7472e-6 |
| 1 | -42.2538 | 8.7414 | -4.4122 | 1.4928 | -0.3504 | 0.0658 | -0.0132 | 2.2712e-3 | -1.5719e-4 | -1.3392e-5 | 1.8884e-6 |
| 2 | -41.4962 | 8.1866 | -4.1545 | 1.4066 | -0.3234 | 0.0595 | -0.0126 | 2.3225e-3 | -1.6629e-4 | -1.3830e-5 | 1.9718e-6 |

| $v'$ | $a_0$ | $a_1$ | $a_2$ | $a_3$ | $a_4$ | $a_5$ | $a_6$ | $a_7$ | $a_8$ | $a_9$ | $a_{10}$ |
|---|---|---|---|---|---|---|---|---|---|---|---|
| 3 | -42.1556 | 9.0816 | -4.4512 | 1.4322 | -0.3293 | 0.0705 | -0.0168 | 2.5476e-3 | 4.3371e-5 | -6.3907e-5 | 5.4878e-6 |
| 4 | -42.0390 | 9.0332 | -4.4619 | 1.4481 | -0.3306 | 0.0677 | -0.0161 | 2.6939e-3 | -3.7947e-5 | -5.2588e-5 | 4.9645e-6 |
| 5 | -41.4596 | 8.6264 | -4.2231 | 1.3537 | -0.3070 | 0.0666 | -0.0167 | 2.5235e-3 | 8.0437e-5 | -7.2003e-5 | 5.9989e-6 |
| 6 | -41.6710 | 8.4137 | -4.2056 | 1.3667 | -0.3005 | 0.0613 | -0.0162 | 2.8026e-3 | 8.3692e-6 | -6.7910e-5 | 6.1282e-6 |

$v = 6$

| $v'$ | $b_1$ | $b_2$ | $b_3$ | $b_4$ | $b_5$ | $b_6$ |
|---|---|---|---|---|---|---|
| 0 | -21.6060 | 0.0215 | -87.8890 | 1.0641 | -5.8349 | 0.2330 |
| 1 | -20.9790 | 0.0141 | -91.9180 | 1.0137 | -1.0005 | 0.4291 |
| 2 | -20.4590 | 0.0134 | -92.0090 | 1.0128 | -0.8702 | 0.4936 |
| 3 | -20.5640 | 0.0141 | -91.5580 | 1.0180 | -1.1663 | 0.4695 |
| 4 | -20.4680 | 0.0131 | -92.0710 | 1.0122 | -0.9036 | 0.4496 |
| 5 | -20.3770 | 0.0125 | -92.7020 | 1.0047 | -0.4358 | 0.5732 |
| 6 | -19.9810 | 0.0109 | -93.3420 | 0.9979 | -0.1372 | 0.5521 |
| 7 | -20.3690 | 0.0120 | -92.6370 | 1.0056 | -0.4888 | 0.5620 |
| 8 | -20.7160 | 0.0126 | -92.4200 | 1.0081 | -0.6746 | 0.4607 |
| 9 | -21.3640 | 0.0129 | -92.0890 | 1.0119 | -0.9077 | 0.4328 |
| 10 | -21.7420 | 0.0144 | -93.0070 | 1.0004 | -0.2817 | 0.4172 |
| 11 | -23.7780 | 0.0342 | -86.7590 | 1.0714 | -5.8675 | 0.2910 |
| 12 | -19.8660 | -4.5927e-3 | -96.9010 | 0.9474 | -3.7898 | 0.7412 |
| 13 | -93.3860 | 0.9812 | -22.9410 | 0.0107 | -4.5976 | 0.5078 |
| 14 | -19.8210 | -0.0152 | -99.0870 | 0.9382 | -3.8688 | 0.9995 |

$v = 7$

| $v'$ | $a_0$ | $a_1$ | $a_2$ | $a_3$ | $a_4$ | $a_5$ | $a_6$ | $a_7$ | $a_8$ | $a_9$ | $a_{10}$ |
|---|---|---|---|---|---|---|---|---|---|---|---|
| 0 | -41.9387 | 8.8081 | -4.3557 | 1.3950 | -0.3089 | 0.0674 | -0.0180 | 2.7841e-3 | 1.3205e-4 | -9.2300e-5 | 7.6684e-6 |
| 1 | -41.3596 | 8.0536 | -4.0853 | 1.3980 | -0.3277 | 0.0582 | -0.0112 | 2.1835e-3 | -2.4184e-4 | 5.1316e-6 | 7.3117e-7 |
| 2 | -41.7447 | 8.7152 | -4.2362 | 1.3591 | -0.3167 | 0.0681 | -0.0158 | 2.3238e-3 | 3.4140e-5 | -5.5790e-5 | 4.7869e-6 |
| 3 | -41.4123 | 8.2819 | -4.2114 | 1.4378 | -0.3379 | 0.0599 | -0.0113 | 2.2107e-3 | -2.7381e-4 | 1.2351e-5 | 2.8059e-7 |
| 6 | -41.7323 | 8.6983 | -4.2661 | 1.3713 | -0.3126 | 0.0668 | -0.0164 | 2.5360e-3 | 5.1283e-5 | -6.7268e-5 | 5.8064e-6 |
| 7 | -41.0493 | 8.3623 | -4.1443 | 1.3990 | -0.3445 | 0.0687 | -0.0123 | 1.6407e-3 | -9.0474e-5 | -4.6901e-6 | 5.3485e-7 |
| 8 | -41.2934 | 8.1545 | -4.1541 | 1.4064 | -0.3230 | 0.0591 | -0.0124 | 2.3168e-3 | -1.8379e-4 | -9.8098e-6 | 1.7039e-6 |

$v = 8$

| $v'$ | $a_0$ | $a_1$ | $a_2$ | $a_3$ | $a_4$ | $a_5$ | $a_6$ | $a_7$ | $a_8$ | $a_9$ | $a_{10}$ |
|---|---|---|---|---|---|---|---|---|---|---|---|
| 0 | -41.5943 | 8.3779 | -4.0786 | 1.3022 | -0.2952 | 0.0651 | -0.0167 | 2.4632e-3 | 1.2292e-4 | -8.0693e-5 | 6.5761e-6 |
| 10 | -41.9988 | 8.3693 | -4.0396 | 1.2844 | -0.2987 | 0.0669 | -0.0158 | 2.1508e-3 | 9.8417e-5 | -6.2498e-5 | 4.9044e-6 |
| 2 | -41.6092 | 8.5604 | -4.1626 | 1.3396 | -0.3129 | 0.0658 | -0.0150 | 2.3409e-3 | -1.7734e-5 | -4.6776e-5 | 4.2786e-6 |
| 3 | -41.8626 | 8.4520 | -4.0831 | 1.3054 | -0.3068 | 0.0667 | -0.0150 | 2.1370e-3 | 2.0835e-5 | -4.5723e-5 | 3.8935e-6 |
| 5 | -41.9425 | 8.4831 | -4.1254 | 1.3228 | -0.3065 | 0.0658 | -0.0154 | 2.3170e-3 | 2.3504e-5 | -5.3454e-5 | 4.6043e-6 |
| 8 | -40.8718 | 8.1223 | -3.9466 | 1.2996 | -0.2912 | 0.0381 | -3.1985e-3 | 1.8780e-3 | -8.9203e-4 | 1.6903e-4 | -1.1424e-5 |

$v = 9$

| $v'$ | $b_1$ | $b_2$ | $b_3$ | $b_4$ | $b_5$ | $b_6$ |
|---|---|---|---|---|---|---|
| 0 | -20.9900 | 0.0170 | -78.6510 | 1.0395 | -3.5278 | 0.2210 |
| 1 | -20.3810 | 0.0148 | -80.8800 | 1.0079 | -0.6562 | 0.4009 |
| 2 | -20.9490 | 0.0167 | -79.5100 | 1.0260 | -1.3543 | 0.4776 |
| 3 | -52.6320 | 0.8598 | -31.0850 | 1.1793 | -19.3080 | -2.5748e-4 |
| 4 | -21.6110 | 0.0172 | -80.1260 | 1.0174 | -1.4768 | 0.2796 |
| 5 | -20.6080 | 0.0138 | -80.8750 | 1.0082 | -0.5853 | 0.4879 |
| 6 | -6.8740 | -0.0596 | -83.3160 | 0.9858 | -12.2980 | 7.0748e-3 |
| 7 | -20.4920 | 0.0107 | -80.5290 | 1.0147 | -1.9074 | 0.1869 |
| 8 | -20.4460 | 0.0145 | -80.6200 | 1.0113 | -0.6890 | 0.5223 |
| 9 | -18.5450 | 2.2565e-3 | -82.1160 | 0.9955 | -0.9527 | 0.0668 |
| 10 | -20.4240 | 0.0124 | -80.7810 | 1.0105 | -1.3618 | 0.2072 |
| 11 | -24.3830 | 0.0446 | -73.3660 | 1.1024 | -7.4579 | 0.3296 |
| 12 | -20.0460 | 1.0244e-3 | -85.5970 | 0.9341 | -2.9283 | 0.6471 |
| 13 | -19.3520 | -9.1361e-3 | -89.8070 | 0.9566 | -2.8593 | 1.1490 |
| 14 | -19.2910 | -0.0180 | -89.6780 | 0.9036 | -3.0580 | 1.1906 |

$v = 10$

| $v'$ | $a_0$ | $a_1$ | $a_2$ | $a_3$ | $a_4$ | $a_5$ | $a_6$ | $a_7$ | $a_8$ | $a_9$ | $a_{10}$ |
|---|---|---|---|---|---|---|---|---|---|---|---|
| 0 | -40.6451 | 7.4488 | -3.7092 | 1.2894 | -0.3263 | 0.0637 | -0.0109 | 1.5573e-3 | -1.4462e-4 | 6.7024e-6 | -1.8083e-7 |
| 10 | -40.0725 | 7.7618 | -3.8297 | 1.2617 | -0.3006 | 0.0647 | -0.0140 | 1.7925e-3 | 6.8595e-5 | -4.4993e-5 | 3.3491e-6 |
| 11 | -40.7325 | 7.5653 | -3.7799 | 1.2894 | -0.3164 | 0.0628 | -0.0114 | 1.5206e-3 | -7.7311e-5 | -4.6604e-6 | 3.7855e-7 |
| 1 | -40.4539 | 7.6338 | -3.8342 | 1.3196 | -0.3256 | 0.0640 | -0.0121 | 1.7902e-3 | -8.3103e-5 | -1.6291e-5 | 1.7349e-6 |
| 9 | -40.4317 | 7.2137 | -3.5656 | 1.2444 | -0.3197 | 0.0613 | -9.2083e-3 | 1.1968e-3 | -1.7578e-4 | 2.5930e-5 | -1.8972e-6 |

$v = 11$

| $v'$ | $a_0$ | $a_1$ | $a_2$ | $a_3$ | $a_4$ | $a_5$ | $a_6$ | $a_7$ | $a_8$ | $a_9$ | $a_{10}$ |
|---|---|---|---|---|---|---|---|---|---|---|---|
| | -40.5083 | 7.5105 | -3.8246 | 1.3245 | -0.3197 | 0.0605 | -0.0117 | 1.8920e-3 | -9.8191e-5 | -1.8289e-5 | 2.0664e-6 |
| | -40.5179 | 7.4805 | -3.8096 | 1.3174 | -0.3148 | 0.0580 | -0.0108 | 1.7949e-3 | -1.5839e-4 | 9.6940e-7 | 5.1220e-7 |
| | -39.9721 | 7.8453 | -3.8620 | 1.2703 | | | -0.0150 | 1.7045e-3 | 1.7739e-4 | -6.6948e-5 | 4.8095e-6 |
| | -41.0086 | 7.8600 | -3.9325 | 1.2771 | -0.2780 | 0.0562 | -0.0152 | 2.7280e-3 | 6.7979e-6 | -6.8093e-5 | 6.2222e-6 |
| | -40.3053 | 7.3619 | -3.6811 | 1.2855 | -0.3233 | 0.0604 | -9.5768e-3 | 1.4385e-3 | -1.9060e-4 | 1.9297e-5 | -1.0642e-6 |

$v = 12$

| $v'$ | $a_0$ | $a_1$ | $a_2$ | $a_3$ | $a_4$ | $a_5$ | $a_6$ | $a_7$ | $a_8$ | $a_9$ | $a_{10}$ |
|---|---|---|---|---|---|---|---|---|---|---|---|
| 0 | -40.2328 | 7.1672 | -3.6337 | 1.2664 | -0.3020 | 0.0525 | -9.6814e-3 | 1.9144e-3 | -2.0915e-4 | 1.1671e-6 | 9.9672e-7 |
| 10 | -40.5234 | 7.0375 | -3.6367 | 1.2837 | -0.3030 | 0.0530 | -0.0108 | 2.1343e-3 | -1.5498e-4 | -2.0009e-5 | 2.7696e-6 |
| 11 | -40.4554 | 7.1807 | -3.5879 | 1.2327 | -0.3009 | 0.0568 | -9.5918e-3 | 1.3864e-3 | -1.6222e-4 | 1.7679e-5 | -1.1973e-6 |
| 12 | -39.4197 | 7.4170 | -3.6118 | 1.1834 | -0.2890 | 0.0653 | -0.0134 | 1.3263e-3 | 1.2650e-4 | -3.8672e-5 | 2.1807e-6 |
| 13 | -40.7272 | 6.8954 | -3.4607 | 1.2305 | -0.3077 | 0.0559 | -9.4329e-3 | 1.6517e-3 | -2.0604e-4 | 1.2043e-5 | -1.6982e-7 |
| 1 | -40.3929 | 7.1466 | -3.5544 | 1.2312 | -0.3069 | 0.0576 | -9.1997e-3 | 1.3177e-3 | -1.6901e-4 | 1.9201e-5 | -1.2741e-6 |
| 3 | -40.4687 | 6.9629 | -3.5932 | 1.2642 | -0.2965 | 0.0512 | -0.0103 | 2.0639e-3 | -1.6242e-4 | -1.4486e-5 | 2.1942e-6 |

$v = 13$

| $v'$ | $a_0$ | $a_1$ | | $a_3$ | $a_4$ | $a_5$ | $a_6$ | $a_7$ | $a_8$ | $a_9$ | $a_{10}$ |
|---|---|---|---|---|---|---|---|---|---|---|---|
| 0 | -40.4617 | 7.3059 | -3.7325 | 1.3010 | -0.3052 | 0.0525 | -0.0102 | 2.0770e-3 | -2.0765e-4 | -3.4418e-6 | 1.4039e-6 |
| 13 | -39.7141 | 7.6737 | -3.7066 | 1.1991 | -0.2952 | 0.0702 | -0.0145 | 1.1973e-3 | 1.9819e-4 | -4.5564e-5 | 2.2388e-6 |
| 1 | -41.5462 | 7.8626 | -3.9986 | 1.3699 | -0.3337 | 0.0686 | -0.0136 | 1.8220e-3 | -3.5662e-5 | -2.0258e-5 | 1.5692e-6 |
| 5 | -41.5462 | 7.8626 | -3.9986 | 1.3699 | -0.3337 | 0.0686 | -0.0136 | 1.8220e-3 | -3.5662e-5 | -2.0258e-5 | 1.5692e-6 |
| 12 | -41.5462 | 7.8626 | -3.9986 | 1.3699 | -0.3337 | 0.0686 | -0.0136 | 1.8220e-3 | -3.5662e-5 | -2.0258e-5 | 1.5692e-6 |
| 7 | -41.1046 | 7.3134 | -3.6726 | 1.2886 | -0.3171 | 0.0567 | -9.1980e-3 | 1.5522e-3 | -2.1646e-4 | 2.0032e-5 | -9.5247e-7 |

$v = 14$

| $v'$ | $a_0$ | $a_1$ | $a_2$ | $a_3$ | $a_4$ | $a_5$ | $a_6$ | $a_7$ | $a_8$ | $a_9$ | $a_{10}$ |
|---|---|---|---|---|---|---|---|---|---|---|---|
| 0 | -40.7748 | 7.1397 | -3.6241 | 1.2506 | -0.2900 | 0.0529 | -0.0117 | 2.1852e-3 | -9.6292e-5 | -3.2111e-5 | 3.4706e-6 |
| 12 | -41.7370 | 7.3711 | -3.6439 | 1.2348 | -0.2994 | 0.0603 | -0.0116 | 1.5632e-3 | -5.3367e-5 | -1.1242e-5 | 8.3365e-7 |
| 13 | -41.5623 | 7.5324 | -3.5959 | 1.1812 | -0.3091 | 0.0748 | -0.0135 | 8.0536e-4 | 1.3421e-4 | -1.3083e-5 | -6.3658e-7 |
| 14 | -40.9163 | 7.9734 | -3.7024 | 1.1628 | -0.3054 | 0.0812 | -0.0152 | 4.4377e-4 | 3.3093e-4 | -3.9182e-5 | 3.8949e-7 |
| 2 | -41.4969 | 7.3714 | -3.6452 | 1.2419 | -0.3004 | 0.0580 | -0.0108 | 1.6311e-3 | -1.1763e-4 | -1.4129e-6 | 3.7721e-7 |

$$k_{35}(v) = \exp\Big(b_1(1.6T_h)^{-b_2} + b_3(1.6T_h)^{-b_4} + b_5\exp(-b_6\ln(1.6T_h)^2)\Big)\times 10^{-6}$$

| $v$ | $b_1$ | $b_2$ | $b_3$ | $b_4$ | $b_5$ | $b_6$ |
|---|---|---|---|---|---|---|
| 0 | -58.1640 | 0.9249 | -28.1020 | -0.0452 | 0.4644 | 0.8795 |
| 1 | -52.0880 | 0.9163 | -24.8730 | -0.0499 | 0.4529 | 0.8760 |
| 2 | -43.5500 | 0.9674 | -23.1680 | -0.0485 | -1.7222 | 0.1986 |
| 3 | -38.1390 | 0.9639 | -21.2640 | -0.0517 | -1.8121 | 0.1928 |
| 4 | -18.8520 | -0.0622 | -35.3970 | 0.8529 | -0.5660 | 8.8997 |
| 5 | 5.5609 | 0.9110 | -33.8610 | 0.9110 | -20.8520 | -6.7487e-3 |
| 6 | -0.3982 | 3.1731 | -23.7510 | 0.8733 | -19.6260 | -7.2046e-3 |
| 7 | -19.7450 | 0.8337 | -0.3807 | 3.4841 | -18.7000 | -7.4195e-3 |
| 8 | -0.1005 | 5.0051 | -18.2780 | 0.7777 | -17.7540 | -7.9751e-3 |
| 9 | -13.5890 | 0.6515 | -0.7750 | 2.6711 | -16.8500 | -8.5243e-3 |
| 10 | -11.5040 | 0.8451 | -14.6030 | -0.0677 | -3.2615 | 0.1367 |
| 11 | -19.7470 | 0.3101 | -7.0442 | -0.1689 | -3.6709e-5 | -6.5292e-3 |
| 12 | -0.2613 | 4.2504 | -15.6420 | 0.1525 | -7.4971 | -0.0223 |

$$k_{36}(v) = \exp\left(b_1(1.6T_h)^{-b_2} + b_3(1.6T_h)^{-b_4} + b_5\exp(-b_6\ln(1.6T_h)^2)\right)\times 10^{-6}$$

| $v$ | $b_1$ | $b_2$ | $b_3$ | $b_4$ | $b_5$ | $b_6$ |
|---|---|---|---|---|---|---|
| 0 | -58.1640 | 0.9249 | -28.1020 | -0.0452 | 0.4644 | 0.8795 |
| 3 | -52.0880 | 0.9163 | -24.8730 | -0.0499 | 0.4529 | 0.8760 |
| 6 | -43.5500 | 0.9674 | -23.1680 | -0.0485 | -1.7222 | 0.1986 |
| 9 | -38.1390 | 0.9639 | -21.2640 | -0.0517 | -1.8121 | 0.1928 |
| 12 | -18.8520 | -0.0622 | -35.3970 | 0.8529 | -0.5660 | 8.8997 |

In between these states the following fitting is used

$$k_{36}(3l+1) = \exp\left[\frac{2}{3}\ln\left(k_{36}(3l)\right) + \frac{1}{3}\ln\left(k_{36}(3l+3)\right)\right],$$
$$k_{36}(3l+2) = \exp\left[\frac{1}{3}\ln\left(k_{36}(3l)\right) + \frac{2}{3}\ln\left(k_{36}(3l+3)\right)\right],$$

where $0 \le l \le 3$ and $k_{36}(3l)$ and $k_{36}(3l+3)$ are calculated by the formula given above with the coefficients from the table.

For $v = 13$ and $v = 14$ the following expressions are used respectively

$$k_{36}(13) = \exp\left[-\frac{1}{3}\ln\left(k_{36}(9)\right) + \frac{4}{3}\ln\left(k_{36}(12)\right)\right],$$
$$k_{36}(14) = \exp\left[-\frac{2}{3}\ln\left(k_{36}(9)\right) + \frac{5}{3}\ln\left(k_{36}(12)\right)\right].$$

$$k_{37}(v) = \exp\left(b_1(1.6T_h)^{-b_2} + b_3(1.6T_h)^{-b_4} + b_5\exp(-b_6\ln(1.6T_h)^2)\right)\times 10^{-6}$$

| $v$ | $b_1$ | $b_2$ | $b_3$ | $b_4$ | $b_5$ | $b_6$ |
|---|---|---|---|---|---|---|
| 0 | -11.5650 | -0.0760 | -78.4330 | 0.7496 | -2.2126 | 0.2201 |
| 1 | -12.0350 | -0.0661 | -67.8060 | 0.7240 | -1.5419 | 1.5195 |
| 2 | -13.5660 | -0.0437 | -55.9330 | 0.7229 | -2.3103 | 1.5844 |
| 3 | -46.6640 | 0.7412 | -15.2970 | -0.0224 | -1.3674 | 1.3621 |
| 4 | -37.4630 | 0.8176 | -0.4037 | -0.4585 | -18.0930 | 0.0115 |
| 5 | -28.2830 | 0.9905 | -10.3770 | -0.0856 | -11.0530 | 0.0673 |
| 6 | -23.7240 | 1.0112 | -2.9905 | -0.2479 | -17.9310 | 0.0344 |

| $v$ | $b_1$ | $b_2$ | $b_3$ | $b_4$ | $b_5$ | $b_6$ |
|---|---|---|---|---|---|---|
| 7 | -19.5470 | 1.0224 | -1.7489 | -0.3141 | -19.4080 | 0.0286 |
| 8 | -15.9360 | 1.0213 | -1.0175 | -0.3804 | -20.2400 | 0.0242 |
| 9 | -12.7120 | 1.0212 | -0.6040 | -0.4457 | -20.7660 | 0.0212 |
| 10 | -0.4056 | -0.4972 | -9.9025 | 1.0212 | -21.0310 | 0.0194 |

$$k_{38}(v) = \left[1.972\sigma_{v_0} \frac{\sqrt{T_e}}{1.0 + T_e / 0.45} \exp\left(-\frac{\left|E_{th}\right|}{T_e}\right)\left(\frac{\left|E_{th}\right|}{T_e} + \frac{1.0}{1.0 + T_e / 0.45} \times 10^{-8}\right)\right] \times 10^{-6}$$

| $v$ | $\sigma_{v_0}$ | $E_{th}$ |
|---|---|---|
| 0 | 3.2200e-5 | 3.7200 |
| 1 | 5.1800e-4 | 3.2100 |
| 2 | 4.1600e-3 | 2.7200 |
| 3 | 0.0220 | 2.2600 |
| 4 | 0.1220 | 1.8300 |
| 5 | 0.4530 | 1.4300 |
| 6 | 1.5100 | 1.3600 |
| 7 | 4.4800 | 0.7130 |
| 8 | 10.1000 | 0.3970 |
| 9 | 13.9000 | 0.1130 |
| 10 | 11.8000 | -0.1390 |
| 11 | 8.8700 | -0.3540 |
| 12 | 7.1100 | -0.5290 |
| 13 | 5.0000 | -0.6590 |
| 14 | 3.3500 | -0.7360 |

$$k_{39}(v,w) = \begin{cases} k(0,1)(v+1)w\left(\frac{3}{2} - \frac{1}{2}e^{-\delta(v-w+1)}\right)e^{\Delta_1(v-w+1)-\Delta_2(v-w+1)^2}, v > w-1, \\ k_{39}(w-1, v+1)e^{\frac{E_v + E_w - E_{v+1} - E_{w-1}}{k_B T_h}}, v \le w-1, \end{cases}$$

where

$$k(0,1) = 4.23 \times 10^{-21} \left( \frac{300}{T_h} \right)^{1/3},\ \delta = 0.21 \left( \frac{300}{T_h} \right)^{1/2},\ \Delta_1 = 0.236 \left( \frac{T}{300} \right)^{1/4},\ \Delta_2 = 0.0572 \left( \frac{300}{T} \right)^{1/3},$$

and $E_v$ is the excitation energy of $v$th vibrational level of $\mathrm{H}_2$.

$$k_{40}(v,w) = \begin{cases} k_{40}^{v \to v+1}(v,w), \\ k_{40}^{v \to v-1}(v,w), \end{cases}$$

where $k_{40}^{v \to v+1}$ and $k_{40}^{v \to v-1}$ correspond to excitation and deexcitation respectively and are given by

$$k_{40}^{v \to v-1}(v,w) = 7.47 \times 10^{-18} \sqrt{T_h} e^{-93.87 T_h^{-1/3}} v e^{\delta_{\mathrm{VT}}(v-1) + \delta'_{\mathrm{VT}} w},$$

$$\delta_{\mathrm{VT}} = 0.97 \left( \frac{300}{T_h} \right)^{1/3},\ \delta_{\mathrm{VT}} = 0.287 \sqrt{\frac{300}{T_h}},$$

and

$$k_{40}^{v \to v+1}(v,w) = k_{40}^{v \to v-1}(v+1,w) e$$

$$k_{41} = k_0 e^{-E_a/T_h}$$

| $v$ | $k_0$ | $E_a$ |
|---|---|---|
| 10 | 1.3010e-16 | 7472.8000 |
| 11 | 3.9570e-16 | 5838.2000 |
| 12 | 5.8440e-16 | 3649.2000 |
| 13 | 1.3620e-15 | 3350.1000 |
| 14 | 3.1070e-15 | 2962.1000 |

$$k_{42}(v,v') = \begin{cases} \sigma_{\mathrm{react}} e^{-4184 \frac{E_{\mathrm{react}}}{T}} + \sigma_{\mathrm{nreact}} e^{-4184 \frac{E_{\mathrm{nreact}}}{T}}, v' < v \le 9, \\ \tilde{k}_{42}(v,v') + \delta_{vv'+1} \hat{k}(v), v' < v, v > 9, \\ k_{42}(v',v) e^{\frac{E_v - E_{v'}}{T}}, v' > v, \end{cases}$$

where $\sigma_{\text{react}}$, $E_{\text{react}}$, $\sigma_{\text{nreact}}$, and $E_{\text{nreact}}$ are given in tables below, $\delta_{vv'+1} = \begin{cases} 1, v = v' + 1, \\ 0, v \neq v' + 1. \end{cases}$. Other rate coefficients are defined below and are based on the rate coefficients from Capitelli et al. (2000).

| $v \setminus v'$ | | 0 | 1 | 2 | 3 | 4 | 5 | 6 | 7 | 8 |
|---|---|---|---|---|---|---|---|---|---|---|
| 1 | $E_{\text{react}}$ | 4.253 | | | | | | | | |
| | $\sigma_{\text{react}}$ | 3.707e-17 | | | | | | | | |
| 2 | $E_{\text{react}}$ | 2.397 | 3.293 | | | | | | | |
| | $\sigma_{\text{react}}$ | 2.227e-17 | 6.071e-17 | | | | | | | |
| 3 | $E_{\text{react}}$ | 1.627 | 1.813 | 2.121 | | | | | | |
| | $\sigma_{\text{react}}$ | 2.169e-17 | 2.732e-17 | 5.389e-17 | | | | | | |
| 4 | $E_{\text{react}}$ | 1.019 | 1.179 | 1.264 | 1.459 | | | | | |
| | $\sigma_{\text{react}}$ | 2.110e-17 | 2.637e-17 | 3.130e-17 | 4.780e-17 | | | | | |
| 5 | $E_{\text{react}}$ | 1.438 | 1.185 | 0.828 | 1.039 | 1.287 | | | | |
| | $\sigma_{\text{react}}$ | 1.993e-17 | 2.603e-17 | 2.418e-17 | 3.430e-17 | 2.816e-17 | | | | |
| 6 | $E_{\text{react}}$ | 1.235 | 0.342 | 0.670 | 0.559 | 0.852 | 0.662 | | | |
| | $\sigma_{\text{react}}$ | 2.283e-17 | 6.844e-18 | 1.466e-17 | 1.809e-17 | 3.928e-17 | 3.165e-17 | | | |
| 7 | $E_{\text{react}}$ | 1.014 | 0.756 | 0.525 | 0.670 | 0.763 | 0.524 | 0.840 | | |
| | $\sigma_{\text{react}}$ | 1.293e-17 | 1.069e-17 | 1.291e-17 | 1.598e-17 | 2.462e-17 | 2.544e-17 | 4.019e-17 | | |
| 8 | $E_{\text{react}}$ | 0.395 | 0.823 | 1.078 | 0.997 | 0.595 | 0.494 | 0.556 | 0.353 | |
| | $\sigma_{\text{react}}$ | 6.037e-18 | 1.105e-17 | 1.529e-17 | 1.597e-17 | 1.437e-17 | 1.768e-17 | 2.496e-17 | 2.348e-17 | |
| 9 | $E_{\text{react}}$ | 1.691 | 1.090 | 0.822 | 0.924 | 0.695 | 0.657 | 0.350 | 0.232 | 0.345 |
| | $\sigma_{\text{react}}$ | 6.902e-18 | 1.023e-17 | 8.771e-18 | 1.386e-17 | 1.568e-17 | 1.711e-17 | 1.771e-17 | 1.622e-17 | 2.287e-17 |

| $v \setminus v'$ | | 0 | 1 | 2 | 3 | 4 | 5 | 6 | 7 | 8 |
|---|---|---|---|---|---|---|---|---|---|---|
| 1 | $E_{\text{nreact}}$ | 4.356 | | | | | | | | |
| | $\sigma_{\text{nreact}}$ | 7.682e-17 | | | | | | | | |
| 2 | $E_{\text{nreact}}$ | 2.628 | 2.908 | | | | | | | |
| | $\sigma_{\text{nreact}}$ | 5.755e-17 | 1.166e-16 | | | | | | | |
| 3 | $E_{\text{nreact}}$ | 2.017 | 2.138 | 1.893 | | | | | | |
| | $\sigma_{\text{nreact}}$ | 5.635e-17 | 9.184e-17 | 1.379e-16 | | | | | | |
| 4 | $E_{\text{nreact}}$ | 1.240 | 1.689 | 1.641 | 1.497 | | | | | |
| | $\sigma_{\text{nreact}}$ | 3.852e-17 | 7.542e-17 | 1.192e-16 | 1.638e-16 | | | | | |

| $v \setminus v'$ | | 0 | 1 | 2 | 3 | 4 | 5 | 6 | 7 | 8 |
|---|---|---|---|---|---|---|---|---|---|---|
| 5 | $E_{\mathrm{nreact}}$ | 1.862 | 1.464 | 1.479 | 1.458 | 1.090 | | | | |
| | $\sigma_{\mathrm{nreact}}$ | 4.378e-17 | 7.054e-17 | 9.250e-17 | 1.456e-16 | 1.719e-16 | | | | |
| 6 | $E_{\mathrm{nreact}}$ | 1.049 | 1.538 | 1.587 | 1.254 | 1.268 | 1.023 | | | |
| | $\sigma_{\mathrm{nreact}}$ | 2.564e-17 | 6.390e-17 | 8.766e-17 | 1.026e-16 | 1.523e-16 | 1.970e-16 | | | |
| 7 | $E_{\mathrm{nreact}}$ | 0.8265 | 1.515 | 1.475 | 1.252 | 1.233 | 0.9715 | 0.6984 | | |
| | $\sigma_{\mathrm{nreact}}$ | 2.169e-17 | 5.194e-17 | 6.410e-17 | 8.713e-17 | 1.247e-16 | 1.300e-16 | 1.531e-16 | | |
| 8 | $E_{\mathrm{nreact}}$ | 1.106 | 1.496 | 1.382 | 1.475 | 1.164 | 1.307 | 0.8242 | 0.6326 | |
| | $\sigma_{\mathrm{nreact}}$ | 1.384e-17 | 4.723e-17 | 4.799e-17 | 7.861e-17 | 8.159e-17 | 1.385e-16 | 1.180e-16 | 1.539e-16 | |
| 9 | $E_{\mathrm{nreact}}$ | 1.559 | 1.328 | 1.678 | 1.541 | 1.576 | 1.182 | 1.186 | 0.7444 | 0.4683 |
| | $\sigma_{\mathrm{nreact}}$ | 1.379e-17 | 2.839e-17 | 5.027e-17 | 6.499e-17 | 9.414e-17 | 9.037e-17 | 1.342e-16 | 1.264e-16 | 1.456e-16 |

$$\tilde{k}_{42}(v,v') = \begin{cases} \frac{1}{v} K_v, 1 \le v \le 5, \\ \frac{1}{5} K_v, v > 5, v - v' \le 5, \\ 0, v \ge 5, v - v' > 5, \end{cases}$$

where

$$K_v = \begin{cases} 2.4 \times 10^{-17} v \exp\left( -\frac{2280}{T} + \frac{0.115 E_v}{T} \right), E_v \le 1982.6 \text{ K}, \\ 2.4 \times 10^{-17}, E_v > 1982.6 \text{ K}. \end{cases}$$

The rate constant for adiabatic one-quantum transitions is taken from Gordiets et al. (1998).

$$\hat{k}(v) = 2.4 \times 10^{-14} (v+1)(1 + 2.92 \times 10^{-2} v)$$
$$\times (1 - 5.76 \times 10^{-2} v)^{2.66} \exp\left( -\frac{162.6}{T^{1/3}} (1 - 5.76 \times 10^{-2} v)^{0.681} \right)$$

$$k_{43} = \exp\left( c_6 + c_5 T_h + c_4 T_h^2 + c_3 T_h^3 + c_2 \ln(T_h) + c_1 \exp(T_h / 11600.0) \right) \times 10^{-6}$$

| $v$ | $c_1$ | $c_2$ | $c_3$ | $c_4$ | $c_5$ | $c_6$ |
|---|---|---|---|---|---|---|
| 0 | 113.7610 | 47.3870 | -4.3270e-11 | 4.7290e-7 | -0.0237 | -496.7940 |
| 1 | 186.0410 | 53.8920 | -6.7610e-11 | 5.9750e-7 | -0.0344 | -609.2430 |
| 2 | 130.7710 | 42.9360 | -4.8240e-11 | 4.6130e-7 | -0.0252 | -474.4880 |
| 3 | 173.6110 | 44.3470 | -6.2640e-11 | 5.2290e-7 | -0.0310 | -522.1510 |
| 4 | 212.4250 | 42.2650 | -7.4490e-11 | 5.2330e-7 | -0.0351 | -540.7590 |
| 5 | 207.3050 | 38.1420 | -7.2140e-11 | 4.8170e-7 | -0.0336 | -503.6070 |
| 6 | 208.5320 | 33.9250 | -7.1570e-11 | 4.3290e-7 | -0.0327 | -472.2830 |
| 7 | 329.3480 | 37.6430 | -1.1050e-10 | 5.3740e-7 | -0.0479 | -610.9340 |
| 8 | 228.9980 | 28.1180 | -7.7400e-11 | 4.0120e-7 | -0.0339 | -445.3490 |
| 9 | 224.7890 | 22.6140 | -7.4440e-11 | 3.1470e-7 | -0.0317 | -400.9630 |
| 10 | 232.7150 | 18.4820 | -7.5580e-11 | 2.5250e-7 | -0.0313 | -377.8420 |

$$k_{44}(v) = \exp\left(\sum_{i=0}^{10} a_i T_h^i\right)$$

| $v$ | $a_0$ | $a_1$ | $a_2$ | $a_3$ | $a_4$ | $a_5$ | $a_6$ | $a_7$ | $a_8$ | $a_9$ | $a_{10}$ |
|---|---|---|---|---|---|---|---|---|---|---|---|
| 1 | -7.914e3 | 3.652e3 | -4.828e2 | -2.276e1 | 8.551 | 1.321e-1 | -1.155e-1 | -6.155e-4 | 1.664e-3 | -1.358e-4 | 3.476e-6 |
| 2 | -1.192e4 | 7.620e3 | -1.848e3 | 1.666e2 | 6.628 | -1.817 | -7.525e-2 | 3.300e-2 | -2.477e-3 | 6.217e-5 | 0.000 |
| 3 | -5.825e3 | 3.505e3 | -8.351e2 | 8.765e1 | -2.368 | 7.009e-2 | -1.003e-1 | 1.618e-2 | -1.021e-3 | 2.384e-5 | 0.000 |
| 4 | -1.585e4 | 1.475e4 | -5.405e3 | 8.553e2 | -1.018e1 | -1.435e1 | 1.179 | 1.556e-1 | -3.300e-2 | 2.169e-3 | -5.142e-5 |
| 5 | -8.050e2 | 6.029e2 | -2.106e2 | 4.243e1 | -5.297 | 4.076e-1 | -1.779e-2 | 3.372e-4 | 0.000 | 0.000 | 0.000 |
| 6 | -7.761e2 | 5.319e2 | -1.518e2 | 1.800e1 | 3.920e-1 | -3.147e-1 | 1.432e-2 | 4.172e-3 | -6.644e-4 | 3.924e-5 | -8.689e-7 |
| 7 | -7.760e2 | 5.317e2 | -1.517e2 | 1.798e1 | 3.959e-1 | -3.149e-1 | 1.430e-2 | 4.178e-3 | -6.649e-4 | 3.926e-5 | -8.693e-7 |
| 8 | -7.759e2 | 5.317e2 | -1.516e2 | 1.793e1 | 4.007e-1 | -3.139e-1 | 1.409e-2 | 4.175e-3 | -6.613e-4 | 3.894e-5 | -8.600e-7 |
| 9 | -7.761e2 | 5.318e2 | -1.516e2 | 1.793e1 | 4.004e-1 | -3.140e-1 | 1.409e-2 | 4.176e-3 | -6.616e-4 | 3.895e-5 | -8.604e-7 |
| 10 | -7.765e2 | 5.321e2 | -1.517e2 | 1.793e1 | 4.030e-1 | -3.143e-1 | 1.406e-2 | 4.184e-3 | -6.621e-4 | 3.897e-5 | -8.606e-7 |
| 11 | -7.768e2 | 5.323e2 | -1.518e2 | 1.796e1 | 3.994e-1 | -3.141e-1 | 1.409e-2 | 4.179e-3 | -6.617e-4 | 3.895e-5 | -8.603e-7 |
| 12 | -7.772e2 | 5.322e2 | -1.517e2 | 1.793e1 | 4.045e-1 | -3.146e-1 | 1.410e-2 | 4.184e-3 | -6.623e-4 | 3.899e-5 | -8.610e-7 |
| 13 | -7.777e2 | 5.323e2 | -1.518e2 | 1.795e1 | 4.021e-1 | -3.144e-1 | 1.410e-2 | 4.182e-3 | -6.622e-4 | 3.898e-5 | -8.609e-7 |
| 14 | -7.778e2 | 5.322e2 | -1.517e2 | 1.794e1 | 4.024e-1 | -3.144e-1 | 1.410e-2 | 4.179e-3 | -6.617e-4 | 3.896e-5 | -8.603e-7 |

$k_{45}(v) = k_0$

| $v$ | $k_0$ |
|---|---|
| 2 | 3.4000e-16 |
| 3 | 8.6000e-16 |
| 4 | 1.6000e-15 |
| 5 | 2.2000e-15 |
| 6 | 2.7000e-15 |

$$k_{46}(n) = \begin{cases} \exp\left(\sum_{i=0}^{8} a_i T_e^i\right) \times 10^{-6} + \exp\left(\sum_{i=0}^{8} a_i' T_e^i\right) \times 10^{-6}, \\ \exp\left(\sum_{i=0}^{8} a_i T_e^i\right) \times 10^{-6}. \end{cases}$$

$n = 2$

$a_0$ = -2.814949375869e1; $a_1$ = 1.009828023274e1; $a_2$ = -4.771961915818

$a_3$ = 1.467805963618; $a_4$ = -2.979799374553e-1; $a_5$ = 3.861631407174e-2

$a_6$ = -3.051685780771e-3; $a_7$ = 1.335472720988e-4; $a_8$ = -2.476088392502e-6

$a_0'$ = -2.833259375256e1; $a_1'$ = 9.587356325603; $a_2'$ = -4.833579851041

$a_3'$ = 1.415863373520; $a_4'$ = -2.537887918825e-1; $a_5'$ = 2.800713977946e-2

$a_6'$ = -1.871408172571e-3; $a_7'$ = 6.986668318407e-5; $a_8'$ = -1.123758504195e-6

$n = 3$

$a_0$ = -3.113714569232e1; $a_1$ = 1.170494035550e1; $a_2$ = -5.598117886823

$a_3$ = 1.668467661343; $a_4$ = -3.186788446245e-1; $a_5$ = 3.851704802605e-2

$a_6$ = -2.845199866183e-3; $a_7$ = 1.171512424827e-4; $a_8$ = -2.059295818495e-6

$$k_{47}(n) = \exp\left(\sum_{i=0}^{8} a_i T_e^i\right) \times 10^{-6}$$

$n = 2$

$a_0$ = -3.4541755913671; $a_1$ = 1.4126559112801; $a_2$ = -6.004466156761

$a_3$ = 1.589476697488; $a_4$ = -2.775796909649e-1; $a_5$ = 3.152736888124e-2

$a_6$ = -2.229578042005e-3; $a_7$ = 8.890114963166e-5; $a_8$ = -1.523912962346e-6

$n = 3$

$a_0$ = -3.884976142596e1; $a_1$ = 1.520368281111e1; $a_2$ = -6.078494762845

$a_3$ = 1.535455119900; $a_4$ = -2.628667482712e-1; $a_5$ = 2.994456451213e-2

$a_6$ = -2.156175515382e-3; $a_7$ = 8.826547202670e-5; $a_8$ = -1.558890013181e-6

$$k_{48}(n) = \exp\left(\sum_{i=0}^{8} a_i T_e^i\right) \times 10^{-6}$$

$n = 2$

$a_0$ = -1.973476726029e1; $a_1$ = 3.992702671457; $a_2$ = -1.773436308973

$a_3$ = 5.331949621358e-1; $a_4$ = -1.181042453190e-1; $a_5$ = 1.763136575032e-2

$a_6$ = -1.616005335321e-3; $a_7$ = 8.093908992682e-5; $a_8$ = -1.686664454913e-6

$n = 3$

$a_0$ = -1.566968719411e1; $a_1$ = 1.719661170920; $a_2$ = -8.365041963678e-1

$a_3$ = 2.642794957304e-1; $a_4$ = -6.527754894629e-2; $a_5$ = 1.066883130107e-2

$a_6$ = -1.041488149422e-3; $a_7$ = 5.457216484634e-5; $a_8$ = -1.177539827071e-6

$$k_{49} = 6.58 \times 10^{7}$$

$$k_{50}(n = 2,3) = 2.6724 \times 10^{-17} T^{0.5}$$

$$k_{51}(n = 2,3) = 2.6724 \times 10^{-17} T^{0.5}$$